\documentclass{aa}  
\usepackage{hyperref}
\usepackage{graphicx}
\usepackage{txfonts}
\usepackage{adjustbox}
\usepackage{lineno}
\usepackage{rotating}

\begin{document} 

\title{\textit{Swift}-XRT follow-up analysis of \\  unidentified hard X-ray sources}
\subtitle{Searching for soft X-ray counterparts of unidentified hard X-ray sources}

   \author{M. Kosiba\inst{1, 2}
          \and
          F. Massaro\inst{2, 4, 5}
          \and
          A. Paggi\inst{2, 3, 4}
          \and 
          H. A. Pe\~{n}a-Herazo\inst{3}
          \and \\
          N. Masetti\inst{6, 7}
          \and
          V. Chavushyan\inst{8}
          \and
          E. Bottacini\inst{9, 10}
          \and
          N. Werner\inst{1}
          }

    \institute{Department of Theoretical Physics and Astrophysics, Faculty of Science, Masaryk                    University, Kotl\'a\v rsk\'a 2, Brno, 611 37, Czech \\
               Republic
        \and
            Dipartimento di Fisica, Universit\`{a} degli Studi di Torino, via Pietro Giuria 1, I-10125 Torino, Italy
        \and
            East Asian Observatory, 660 North A'oh{\=o}k{\=u} Place, Hilo, Hawaii 96720, USA
        \and
            Istituto Nazionale di Fisica Nucleare, Sezione di Torino, I-10125 Torino, Italy
        \and
            INAF–Osservatorio Astrofisico di Torino, via Osservatorio 20, I-10025 Pino Torinese, Italy
        \and
            INAF - Osservatorio di Astrofisica e Scienza dello Spazio, via Piero Gobetti 101, 40129 Bologna, Italy
        \and
            Departamento de Ciencias F\'isicas, Universidad Andr\'es Bello, Fern\'andez Concha 700, Las Condes, Santiago, Chile
        \and
            Instituto Nacional de Astrof\'{i}sica, \'{O}ptica y Electr\'{o}nica, Luis Enrique Erro 1, Tonantzintla, Puebla 72840, M\'{e}xico 
        \and 
            Dipartimento di Fisica e Astronomia G. Galilei, Univerist\`a di Padova, Padova, Italy
        \and
            Eureka Scientific, 2452 Delmer Street Suite 100, Oakland, CA 94602-3017, USA
        }

\date{Received; accepted}

\abstract
{It is currently established that the sources contributing to the cosmic X-ray background (CXB) emission are mainly nearby active galactic nuclei (AGN), in particular those that are obscured. Thus, it is important to fully identify the hard X-ray sky source population to accurately characterize the individual contribution of different AGNs to the overall CXB emission.}
{We present a follow-up analysis of all the 218 sources marked as unidentified in our previous revision of the third release of the Palermo \textit{Swift}-BAT hard X-ray catalog (3PBC) based on our multifrequency classification scheme. These 218 sources were classified as unidentified in our previous analyses because they lack an assigned low-energy counterpart.}
{We searched for soft X-ray counterparts of these 218 3PBC sources in archival \textit{Swift}-XRT observations obtained between 2005 January 1st and 2018 August 1st. In particular, we found 1213 archival \textit{Swift}-XRT observations for 192 of the 218 unidentified sources.}
{We found 93 possible \textit{Swift}-XRT counterparts inside of the \textit{Swift}-BAT positional uncertainty regions. These correspond to 73 3PBC sources, where 60 have only a single \textit{Swift}-XRT detection, and 13 sources have multiple detections.  We present a catalog of all the detected possible counterparts of the yet unidentified hard X-ray sources to the community as a catalog for future spectroscopic follow-up targets, together with a short catalog of our classification of the ten sources for which we found available spectra.}
{}


\maketitle


\section{Introduction}
\label{sec:intro}
The hard X-ray sky, at energies greater than $\sim$10 keV, was systematically observed by several telescopes in the past decades. \textit{Uhuru} was the first X-ray satellite \citep{Giacconi1971}, launched in 1970, which delivered an all-sky hard X-ray survey \citep{Forman1978} in 2 -- 20 keV band listing 339 sources. Subsequently, \citet{Levine1984}, thanks to the X-ray and Gamma-ray detector HEAO-A4 on board the \textit{HEAO~1} satellite \citep{Rothschild1979} conducted an all-sky survey in 13--180~keV range providing 77 newly detected sources. The INTErnational Gamma-Ray Astrophysics Laboratory \textit{INTEGRAL} \citep{Winkler2003} with its Imager on Board the \textit{INTEGRAL} Satellite \textit{IBIS} \citep{Ubertini2003} was launched in 2002, observing in energy range from 15 keV up to 10 MeV. Finally, the Neil Gehrels \textit{Swift} Observatory \citep{Gehrels2004}, launched in 2004, carried an all-sky hard X-ray survey at 14 -- 195 keV using the Burst Alert Telescope (BAT) \citep{Barthelmy2004}. Significant improvements in the soft X-ray background were possible mainly thanks to NASA's HEAO-2 \textit{Einstein} observatory \citep{Giacconi1979}, the German-US-UK X-ray observatory \textit{ROSAT} (Röntgen Satellite) \citep{Hasinger1999}, the \textit{Chandra} X-ray observatory \citep{Weisskopf2000} and the \textit{XMM-Newton} observatory \citep{Jansen2001}.

The \textit{INTEGRAL} and \textit{Swift} have allowed for the creation of many catalogs focusing on the hard X-ray sky \citep[see, e.g.,][]{Markwardt2005, Beckmann2006, Churazov2007, Krivonos2007b, Sazonov2007, Tueller2008, Cusumano2010, Bottacini2012, Bird2016, Mereminskiy2016, Krivonos2017, Oh2018, Krivonos2021, Krivonos2022} and catalogs providing the association of hard X-ray sources with their low-energy counterparts \citep[e.g.,][]{Malizia2010, Koss2019, Bar2019, Smith2020}. They were also necessary for the optical spectroscopy follow-up observations \citep[e.g.,][]{Masetti2006a, Masetti2006b, Masetti2008, Masetti2012, Masetti2013, Parisi2014, Rojas2017, Marchesini2019}. These missions are still operational nowadays and deliver new scientific results, e.g., the \textit{INTEGRAL}-IBIS 17-yr hard X-ray all-sky survey \citep{Krivonos2022}, the AGN catalog and optical spectroscopy for the second data release of the \textit{Swift}-BAT AGN Spectroscopic Survey (BASS DR2) \citep{Koss2022} and the upcoming catalog based on the \textit{Swift}-BAT 157-month survey \citep{Lien2023}.


Multiple catalogs based on the Burst Alert Telescope (BAT) on board the \textit{Swift} observatory exist. \citet{Koss2022} constructed the AGN catalog and optical spectroscopy for the second data release of the \textit{Swift}-BAT AGN Spectroscopic Survey (BASS DR2). They provide 1449 optical spectra corresponding to 858 hard-X-ray-selected AGNs in the \textit{Swift}-BAT 70-month observations. Their AGN sample is spectroscopically complete, with 857/858 AGNs reported with redshifts.
\citet{Oh2018} created the 105-month \textit{Swift}-BAT catalog of hard X-ray sources. 
This catalog covers over 90\,\% of the sky with a sensitivity of 8.40\,$\times$\,10$^{-12}$ erg s$^{-1}$ cm$^{-2}$ and 7.24\,$\times$\,10$^{-12}$ erg\,s$^{-1}$ cm$^{-2}$ over 50\,\% of the sky in the 14 -- 195 keV band providing 1632 hard X-ray detections above 4.8\,$\sigma$ significance threshold. 
\citet{Cusumano2010} created the Palermo \textit{Swift}-BAT Hard X-ray catalog. This work links to the 2PBC catalog release after 54 months of sky survey. The 3$^{rd}$ release of the Palermo \textit{Swift}-BAT hard X-ray catalog (3PBC), which is based on 66 months of sky survey, is currently ongoing and is available only online\footnote{\href{http://bat.ifc.inaf.it/bat\_catalog\_web/66m\_bat\_catalog.html}{http://bat.ifc.inaf.it/bat\_catalog\_web/66m\_bat\_catalog.html}}. This is the catalog version we analyze in this study. The 3PBC lists 1256 sources detected above 4.8 $\sigma$ level of significance in the 15 -- 150 keV energy range. Their number increases to 1593 total sources when considering a threshold on the signal-to-noise ratio above 3.8. The catalog covers nearly 90\,\% of the sky to a flux limit of 1.1\,$\times$ 10$^{-11}$\,erg\,cm$^{-2}$\,s$^{-1}$, decreasing to $\sim$50\,\% when decreasing this flux threshold to 0.9\,$\times$ 10$^{-11}$\,erg\,cm$^{-2}$\,s$^{-1}$.


We recently conducted a refined analysis (\cite{Kosiba2023}, hereinafter paper\,I) of all sources listed in the 3PBC catalog. Our refined analysis is based on the multifrequency classification scheme we developed to analyze hard X-ray sources, mainly focusing on extragalactic source populations \citep{Herazo2022}. Findings of our refined analysis of the 3PBC were also based on results reported in the 105-month \textit{Swift}-BAT catalog \citep{Oh2018} and the INTEGRAL hard X-ray survey above 100 keV with its 11-year release \citep{Krivonos2015}, which we used to cross-match the 3PBC sources for counterparts to obtain luminosities and spectra if available.

We found that approximately 57\,\% of the sources listed in the 3PBC have an extragalactic origin, while 19\,\% belong to our Milky Way. The remaining 24\,\% are yet unknown. In particular, our final revised version of the 3PBC catalog lists 1176 classified sources, 218 unidentified, and 199 unclassified. Of the 1176 classified sources, 820 have an extragalactic origin, and 356 have a Galactic origin. Compared to the original 3PBC, which has 233 unidentified and 300 unclassified sources, we decreased these fractions from $\sim$33\% (533 sources) to $\sim$26\% (417 sources). 

Our study also allowed us to discover new Seyfert galaxies included in the Turin-SyCAT $2^{nd}$ release (paper\,I). In the $2^{nd}$ release of the Turin-SyCAT, there are 633 Seyfert galaxies, 282 new ones added thanks to our refined analysis and corresponding to an increase of $\sim$80\% with respect to the Turin-SyCAT $1^{st}$ release \citep{Herazo2022}. 

Moreover, trends between the hard X-ray and the gamma-ray emissions of those blazars listed in the 3PBC with a counterpart in the second release of the fourth \textit{Fermi}-LAT catalog (4FGL-DR2) were also found (paper\,I), as expected from emission models widely adopted to explain their broadband SED \citep[e.g.,][]{Marscher1985, Marscher1996, Massaro2006}.

In this work, we examine the population of the 218 unidentified hard X-ray sources listed in our revised version of the 3PBC, i.e., those lacking an assigned counterpart at lower energies than the BAT energy range. We analyzed all soft X-ray observations (between 0.5 and 10 keV) available in the archive of the X-ray telescope (XRT) \citep{Burrows2005} on board the \textit{Swift} Observatory and found available data for 192 of the 218 3PBC sources, which is the sample we further analyze in this study.
This analysis aims to search for potential counterparts in the soft X-ray data of the Swift-XRT for the 192 yet unidentified 3PBC sources. The final goal of the present analysis is to provide a catalog of all unidentified hard X-ray sources having at least one candidate counterpart if detected in the soft X-ray band, that could be targeted with follow-up spectroscopic observations to obtain its final classification, as successfully carried out in the last decades \citep[e.g.,][]{Masetti2006a, Masetti2006b, Masetti2008, Masetti2012, Masetti2013, Parisi2014, Rojas2017, Koss2017, Landi2017, Marchesini2019, Tomsick2020}. 

The paper is organized as follows. Section \ref{sec:xrays} describes the \textit{Swift}-XRT data reduction and data analysis procedure. Then Section \ref{sec:results} is devoted to our results while details on the multifrequency comparison are illustrated in Section \ref{sec:compare}. Finally, Section \ref{sec:summary} is dedicated to our summary, conclusions, and future perspectives. X-ray images for all analyzed BAT sources are reported in the Appendix.

As previously adopted in the paper I, we used cgs units unless stated otherwise. We also adopted $\Lambda$CDM cosmology with $\Omega_M = 0.286$, and Hubble constant $H_0 = 69.6$\,km\,$s^{-1}\,Mpc^{-1}$ \citep{Bennett2014} to compute cosmological corrections through the whole manuscript.

\section{\textit{Swift}-XRT Observations}
\label{sec:xrays}

\subsection{Sample selection}
\label{sec:sample_selection}
The 3PBC catalog lists 1\,593 hard X-ray sources, all detected with a signal-to-noise ratio (S/N) above 3.8 in the 15--150 keV energy range. Our analysis, presented in paper I, identified 218 3PBC sources lacking an assigned counterpart at lower energies. 

\begin{figure*}[!th]
\begin{center}
    \includegraphics[height=6cm,width=8cm,angle=0]{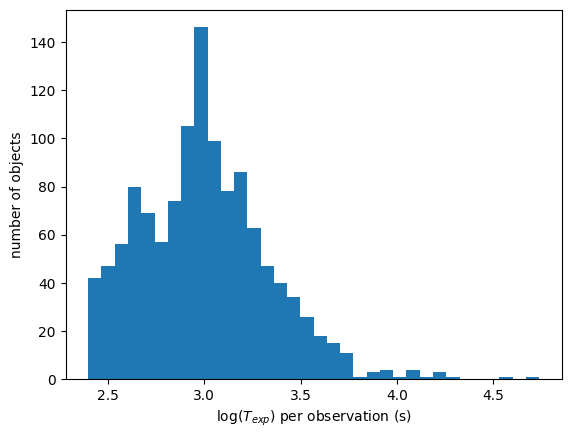}
    \includegraphics[height=6cm,width=8cm,angle=0]{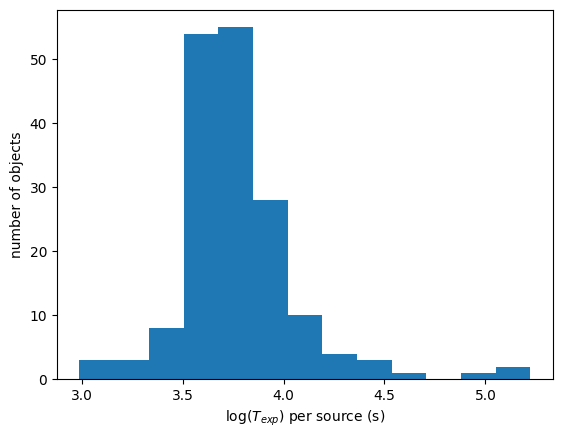}
      \caption{Distribution of $T_{exp}$ of the 1213 \textit{Swift}-XRT observations analyzed in this work, with mean 1462 s (left panel) and the total $T_{exp}$ per source of the 192 sources for which we found the \textit{Swift}-XRT observations with mean 8864 s (right panel).}
      \label{fig:hist_T_exp}
\end{center}
\end{figure*}

In this work, we searched the \textit{Swift}-XRT archive, and we found that 192 out of these 218 hard X-ray sources have at least one X-ray observation with exposure time larger than 250\,sec in the 0.5--10 keV energy range. We found a total of 1213 such observations that have been reduced and analyzed here according to the standard procedures described below. In Fig. \ref{fig:hist_T_exp}, we report the distribution of the exposure time for all selected observations. The 1213 \textit{Swift}-XRT observations have a mean of 1462 s and variance $\sim\,6\times10^6$ s with a total exposure time of 1.77\,$\times$\,10$^6$ s.

All observations we reduced and analyzed in this study were performed between April 2005 and December 2022.

\subsection{Data Reduction}
Data reduction procedures applied here for all \textit{Swift}-XRT observations are the same previously adopted for similar analyses \citep[see e.g.][]{massaro08a,massaro08b,paggi13,marchesini19,marchesini20,massaro23} and procedures of the \textit{Swift}-XRT X-Ray point source catalogs \citep{delia13,evans14,evans20}. Thus, we only describe the basic information and refer to the above references for a more detailed description. 

We downloaded raw \textit{Swift}-XRT data from the archive\footnote{https://heasarc.gsfc.nasa.gov/docs/archive.html}. Then we run the \textsc{xrtpipeline} task, developed as part of the {\it Swift} X-Ray Telescope Data Analysis Software (XRTDAS) and distributed within the HEAsoft package (version 6.30.1) \citep{HEAsoft2014}. This allowed us to obtain clean event files for all \textit{Swift}-XRT observations. The entire analysis and all X-ray images shown in the present manuscript are restricted to the 0.5-10\,keV energy range unless stated otherwise.


We subsequently calibrated these cleaned event files with the usual filtering criteria and using calibration files provided in the High Energy Astrophysics Science Archive Research Center (HEASARC) calibration database (CALDB) version (v.20220907)\footnote{https://heasarc.gsfc.nasa.gov/docs/heasarc/caldb/caldb\_supported\_\\missions.html}. Using the \textsc{xselect} task we excluded all time intervals with count rates higher than 40 photons/sec as well as those with CCD temperature exceeding -50$^\circ$C, in regions located at the edges of the XRT detector \citep{delia13}. Then, the \textsc{xselect} task was also used to merge all cleaned and filtered event files for those sources with multiple observations. Finally, it is worth mentioning that the entire analysis was carried out using the XIMAGE
software\footnote{https://heasarc.gsfc.nasa.gov/xanadu/ximage/ximage.html} to merge the corresponding exposure maps \citep{Giommi1992}.

\subsection{Data Analysis}
To detect sources, we used the sliding cell DETECT (\textsc{det}) algorithm available in the XIMAGE software package \citep{giommi92} on all merged event files as well as on the single event files for sources with only one observation. We set a threshold on the S/N equal to 3 for claiming detection of an X-ray source in the 0.5-10 keV energy range, as recently performed in the analysis of \cite{Massaro2023}.

\begin{figure*}[!th]
\begin{center}
    \includegraphics[height=4.2cm,width=6cm,angle=0]{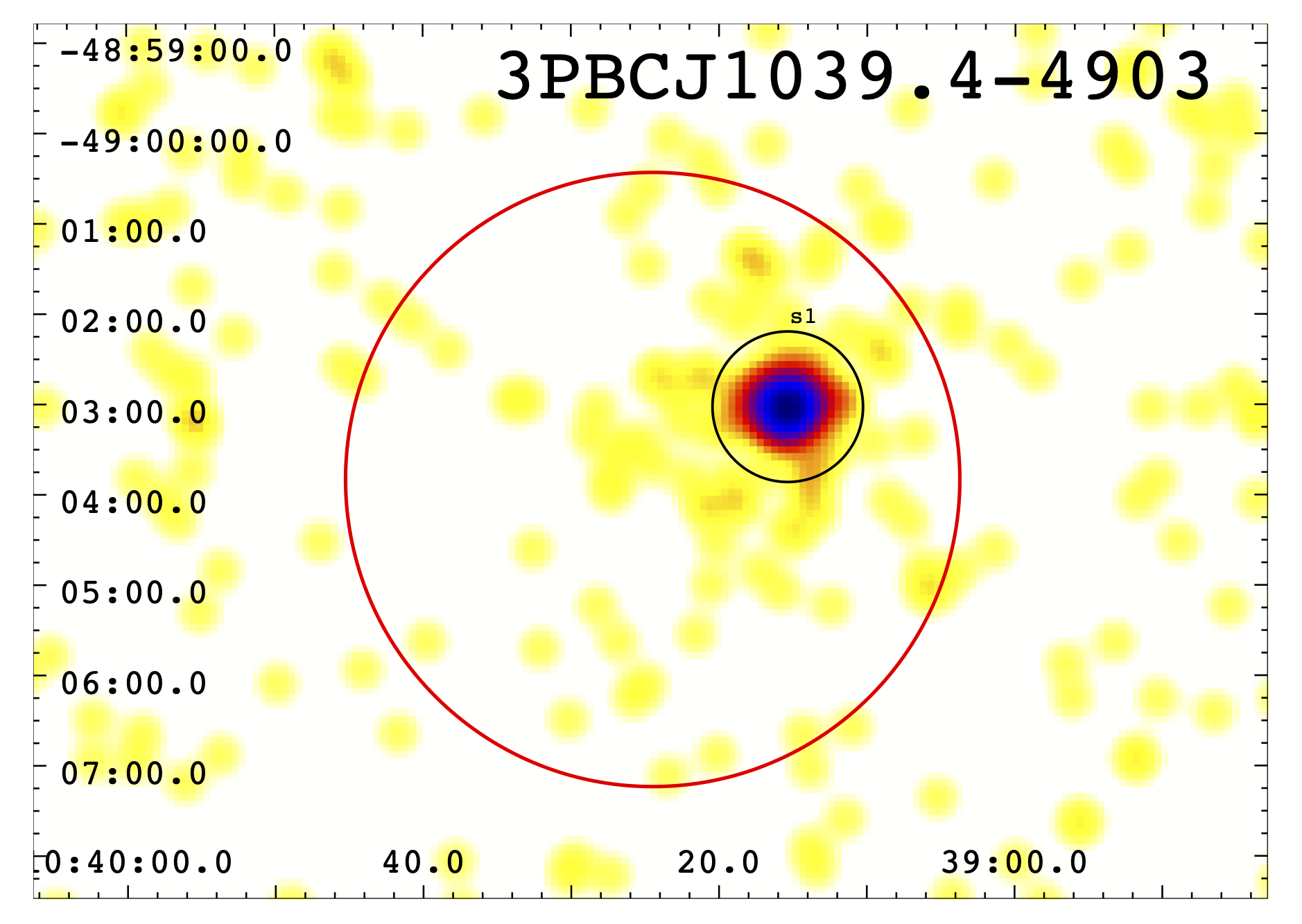}
    \includegraphics[height=4.2cm,width=6cm,angle=0]{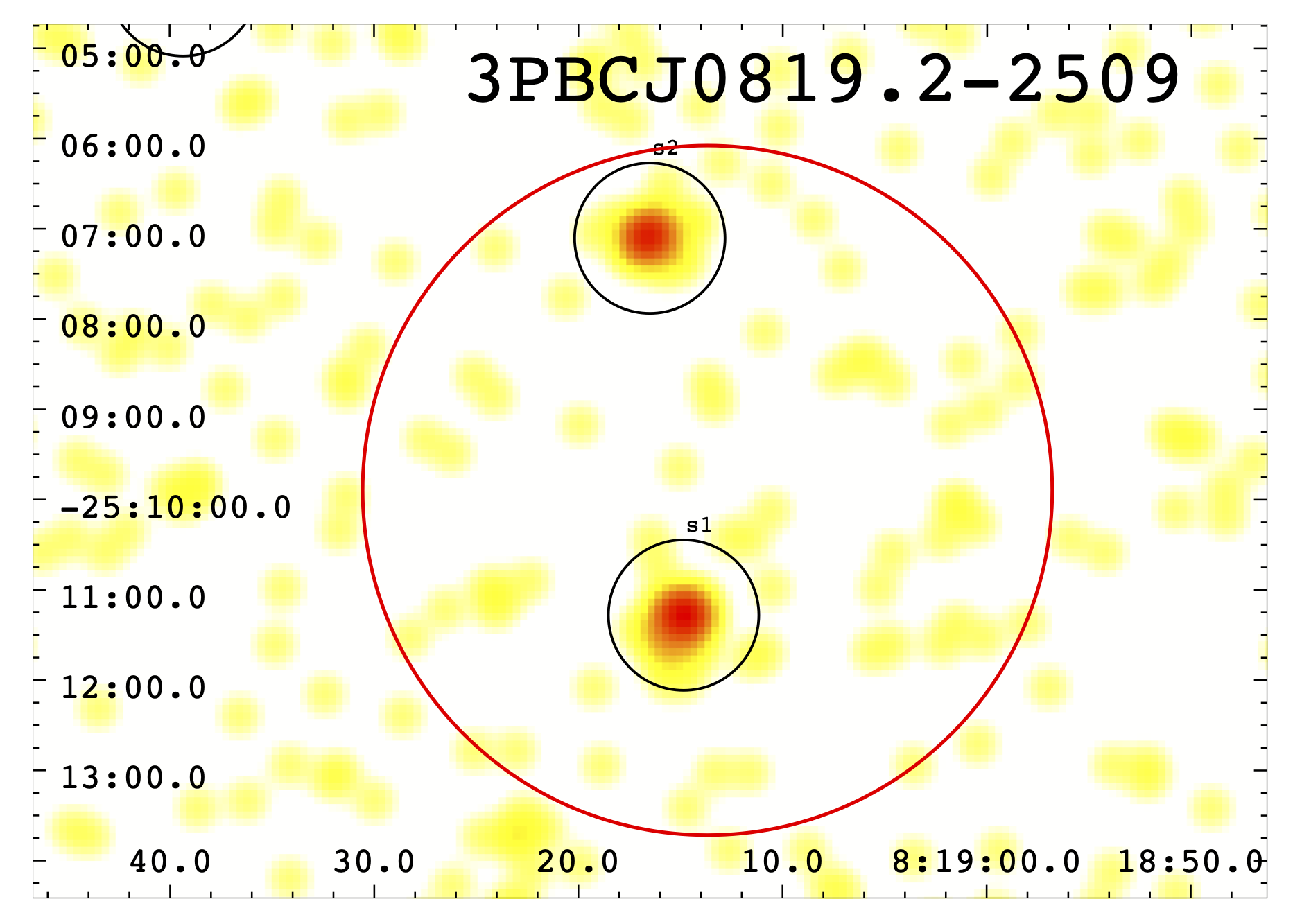}
    \includegraphics[height=4.2cm,width=6cm,angle=0]{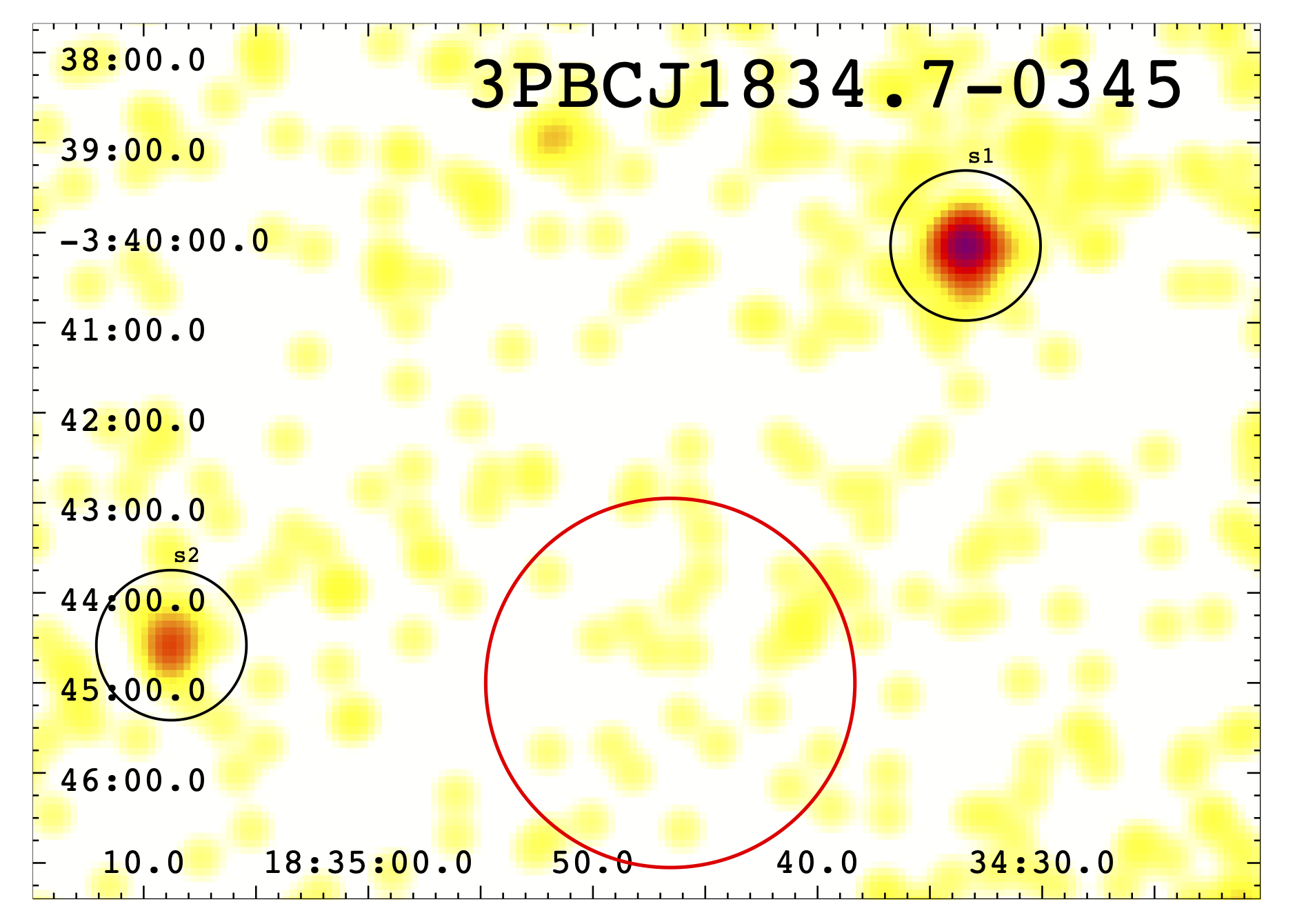}
      \caption{Each figure in this image depicts an example of our XDF flags that label the 3PBC sources. These figures are the XRT merged event files with the red circle indicating BAT positional uncertainty region and the black circles highlighting the position of a soft X-ray source detected in the \textit{Swift}-XRT archive. The left panel is an example of a 3PBC source with a single \textit{Swift}-XRT source found within the BAT positional uncertainty region (red circle), flag \textit{x}. The center panel is the case of multiple detected \textit{Swift}-XRT counterparts inside of the BAT positional uncertainty, flag \textit{m}. The right panel is an example of no \textit{Swift}-XRT counterparts detected within the BAT positional uncertainty region, flag \textit{u}. }
      \label{fig:xdf_examples}
\end{center}
\end{figure*}

Then, to identify and characterize the 3PBC sources, we labeled them using three different X-ray detection flags (XDF) on the basis of the following criteria.
\begin{itemize}
\item {\it x} flag: is used for 3PBC sources that have a single soft X-ray source within their BAT positional uncertainty region (see, e.g., 3PBC\,J1039.4-4903 shown in the left panel of Figure~\ref{fig:xdf_examples})
\item {\it m} flag: indicates 3PBC sources with more than one soft X-ray source (multiple detections) within their BAT positional uncertainty region (see, e.g., 3PBC\,J0819.2-2509 shown in the central panel of Figure~\ref{fig:xdf_examples})
\item {\it u} flag: is adopted for 3PBC sources with no soft X-ray counterparts detected in their merged event files within their BAT positional uncertainty region (see, e.g., 3PBC\,J1834.7-0345 shown in the right panel of Figure~\ref{fig:xdf_examples}).
\end{itemize}

We measured several parameters for all detected possible X-ray counterparts of the 3PBC hard X-ray sources in merged event files. In particular, we obtained coordinates of the distributions of X-ray photons from each source using the \textsc{xrtcentroid} task. We also measured $n_{90}$, the number of photons within a circular region centered on the X-ray coordinates with radius 120.207\arcsec\ (51 pixels), which is enclosing 90\% of the \textit{Swift}-XRT point spread function (PSF). 

\section{A soft X-ray perspective of the hard X-ray sky}
\label{sec:results}

\subsection{Outline of the main goal}

In our previous work \citep{Kosiba2023}, where we revised the 3PBC catalog, we found 218 sources without a low-energy counterpart, thus falling into the unidentified category. The main goal of the current paper is to search for candidate counterparts for these 218 unidentified sources in soft X-ray wavelengths of the \textit{Swift}-XRT archival data.
The final product of this analysis is a catalog of soft X-ray candidate counterparts we found in the \textit{Swift}-XRT data for the yet unidentified hard X-ray sources. We release this catalog along with this publication. This paper also describes the sources we analyze in this study.

\subsection{Overview of results}
\label{sec:overview}
We analyzed all available \textit{Swift}-XRT data, selected according to the criteria previously described, for a total of 1213 observations corresponding to 192 3PBC sources with a total exposure time 1.77\,$\times$\,10$^6$ s. Considering only the \textit{Swift}-XRT detections inside the BAT positional uncertainty region above the S/N\,=\,3 we adopted, we found 93 soft X-ray sources. These 93 soft X-ray sources correspond to 73 unique 3PBC sources. From those, 13 3PBC sources are associated with multiple soft X-ray detected sources (\textit{m} flag), and the remaining 60 3PBC sources are associated with a single soft X-ray detected source (\textit{x} flag). Those are the final results presented in this analysis.

\begin{figure*}[!th]
\begin{center}
    \includegraphics[height=6cm,width=8cm,angle=0]{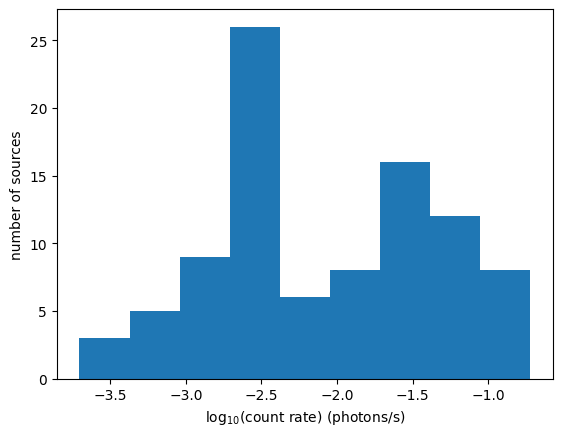}
    \includegraphics[height=6cm,width=8cm,angle=0]{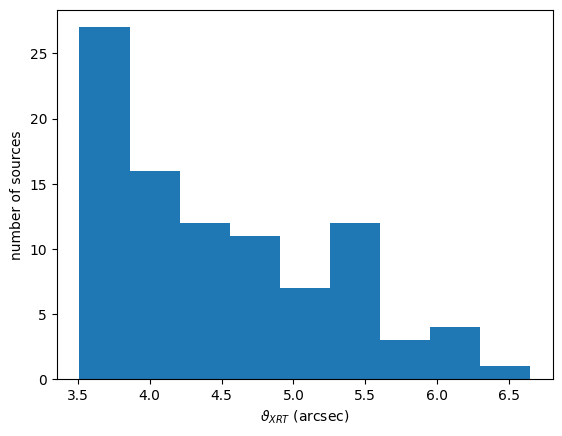}
    
    \caption{Distribution of the count rate in logarithmic scale, with mean 0.028 photons/s (left panel) and the XRT positional uncertainty ($\vartheta_{XRT}$) with mean value 4.5 arcsec (right panel). Both are in the 0.5-10\,keV energy range for all 93 \textit{Swift}-XRT detected counterparts.}
    \label{fig:hist_cntr_pos_unc}
\end{center}
\end{figure*}

We note that all 3PBC sources with at least one soft X-ray counterpart detected within their positional uncertainty region have an integrated exposure time above 975\,seconds, about four times longer than the minimum selected value. The distribution of X-ray count rates and that of their positional uncertainty in the 0.5--10 keV energy range computed using the \textsc{xrtcentroid} task for the 93 \textit{Swift}-XRT detected sources are shown in Fig. \ref{fig:hist_cntr_pos_unc} (left) and Fig. \ref{fig:hist_cntr_pos_unc} (right), respectively.

For calculating the BAT positional uncertainty region (red circles in Fig.\,\ref{fig:xdf_examples}), we took the values reported in the 3PBC catalog \citep{Cusumano2010}.

In this section, we are focusing on the \textit{m} XDF flagged sources (Fig.\,\ref{fig:m_flagged_sources}). We discuss these 13 3PBC sources separately to detail their potential soft X-ray counterparts.

In Appendix \ref{app:Appendix_1}, we report the first five columns of our final table for the first ten sources. Those are the 3PBC name, the XRT counterpart name, the XDF flag, the right ascension and declination of the centroid, and the uncertainty on the centroid's position. The remaining seven columns of our final table are also reported in the Appendix \ref{app:Appendix_1}, listing the angular separation between XRT and BAT, count rate, uncertainty on the count rate, S/N, number of observations, integrated exposure time ($T_{exp}$) and counterpart's name in WISE. 

Finally, Appendix \ref{app:Appendix_x_flag_3PBC_images} includes all X-ray images obtained from cleaned and merged event files with a red dashed circle showing the BAT positional uncertainty and, when present, black circles indicating all soft X-ray counterparts detected in our \textit{Swift}-XRT analysis. The green cross marks soft X-ray detections with a WISE counterpart. The WISE counterparts have been searched within a circular region corresponding to the positional uncertainty of the \textit{Swift}-XRT.

\subsection{3PBC sources with multiple candidate counterparts}
\label{sec:multiple}
This section shows images of 3PBC sources with a brief discussion for which we find multiple XRT PC counterparts consistent within the positional uncertainties (\textit{m} flag).

Three more sources deserve a more detailed description: 3PBC J1430.3+2303 (\textit{m} flag), 3PBC J1620.1-5001 (\textit{x} flag), and 3PBC J1730.0-3436 (\textit{x} flag). All these sources appear to have extended X-ray sources close to the BAT positional uncertainty region. Thus, we adopted the following approach to consider potential X-ray counterparts and avoid spurious detected objects.

For 3PBC J1620.1-5001 and 3PBC J1730.0-3436 (Appendix \ref{app:Appendix_x_flag_3PBC_images}), we only indicate in the main table the X-ray detected source having the highest signal-to-noise ratio among those automatically detected. In particular, both sources lie close to the Galactic plane and close to star-forming regions and could be unknown supernova remnants deserving further investigation. However, in the case of 3PBC J1730.0-3436, artifacts in the merged event file prevented us from conducting a detailed analysis. On the other hand, we noticed that for 3PBC J1620.1-5001, the BAT positional uncertainty region is not centered on the extended X-ray source, thus suggesting that it is not the soft X-ray counterpart of the hard X-ray object. A forthcoming paper will present a multifrequency analysis of 3PBC J1620.1-5001 (Kosiba et al., 2023).

\begin{figure*}[!th]
\begin{center}
    \includegraphics[height=4.2cm,width=6cm,angle=0]{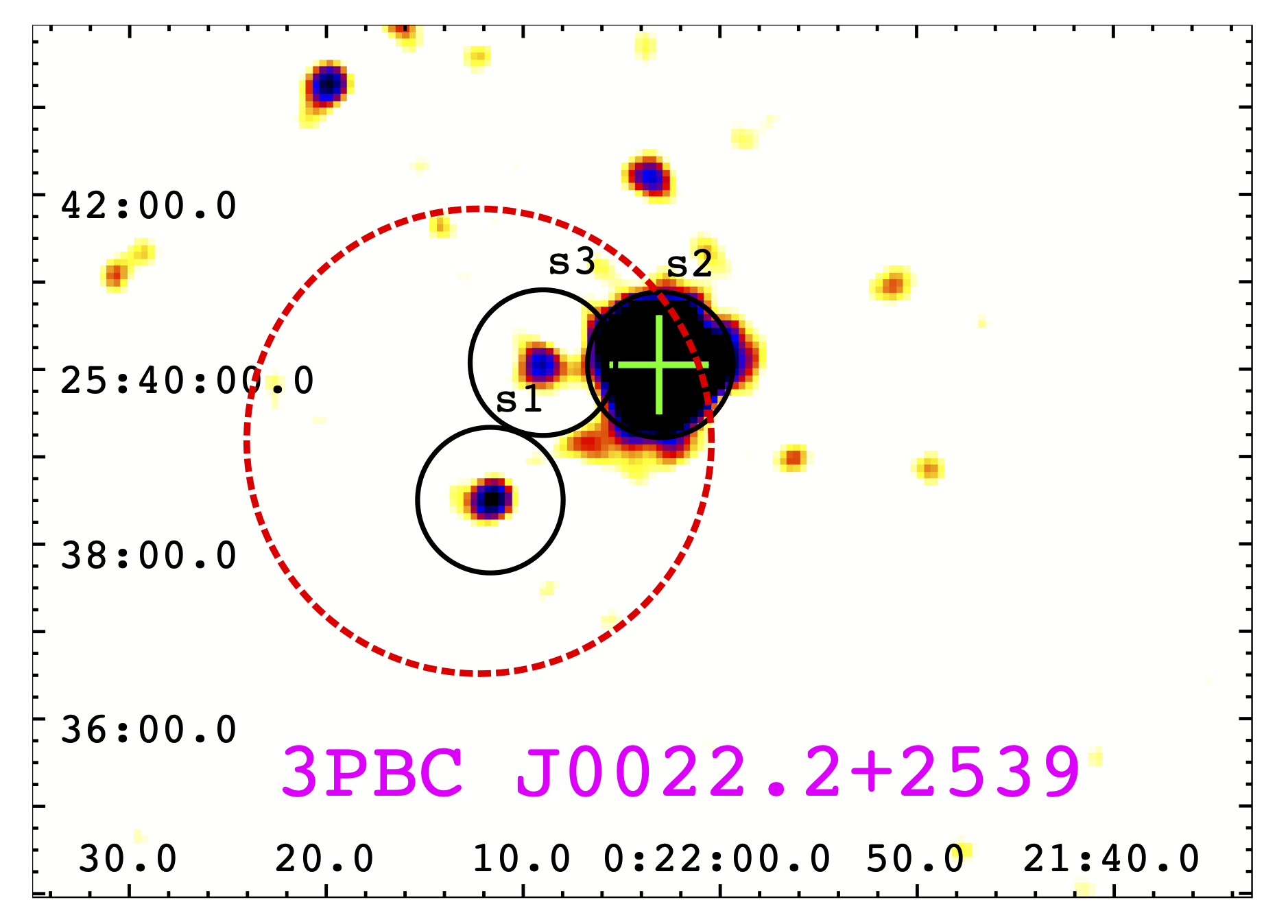}
    \includegraphics[height=4.2cm,width=6cm,angle=0]{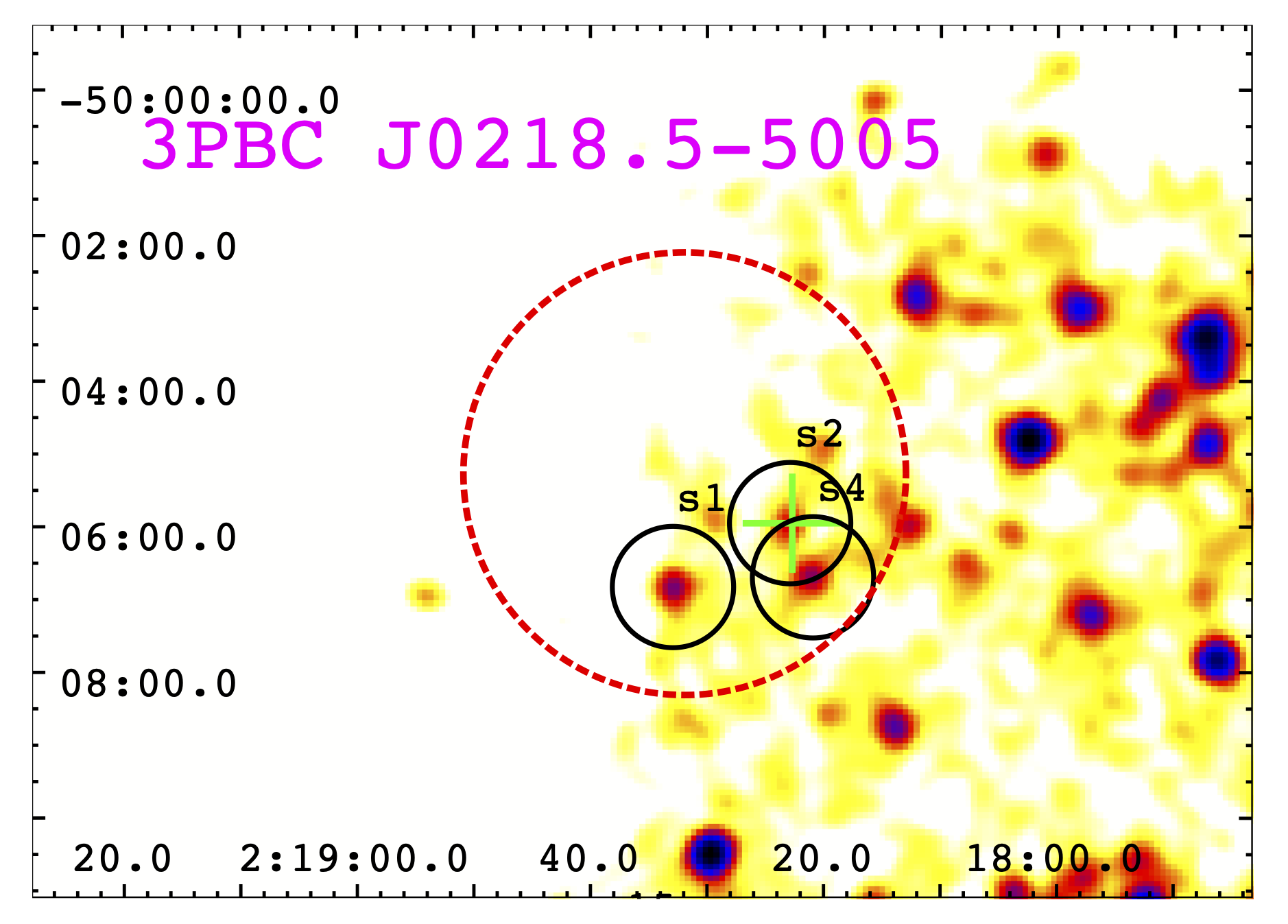}
    \includegraphics[height=4.2cm,width=6cm,angle=0]{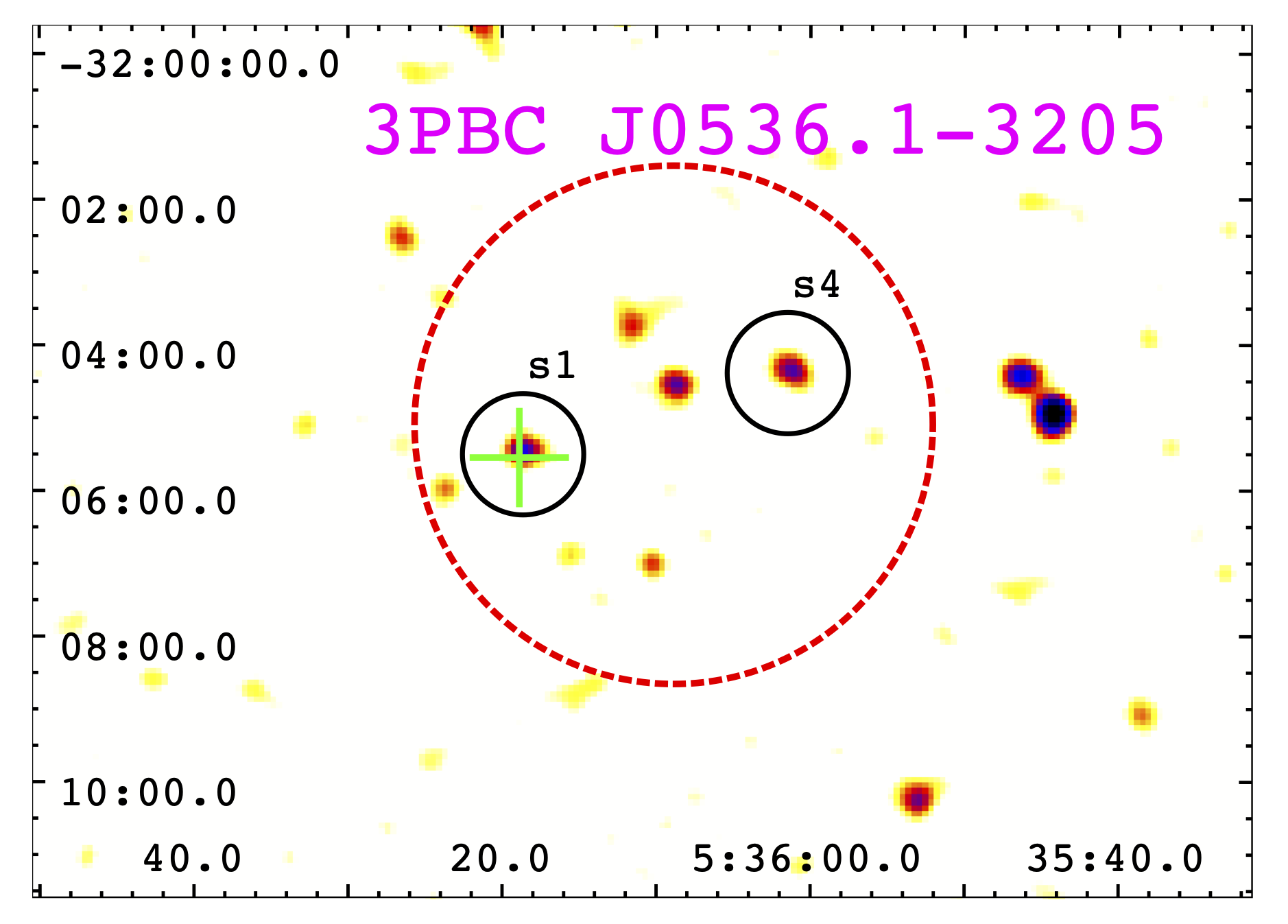}
    
    \includegraphics[height=4.2cm,width=6cm,angle=0]{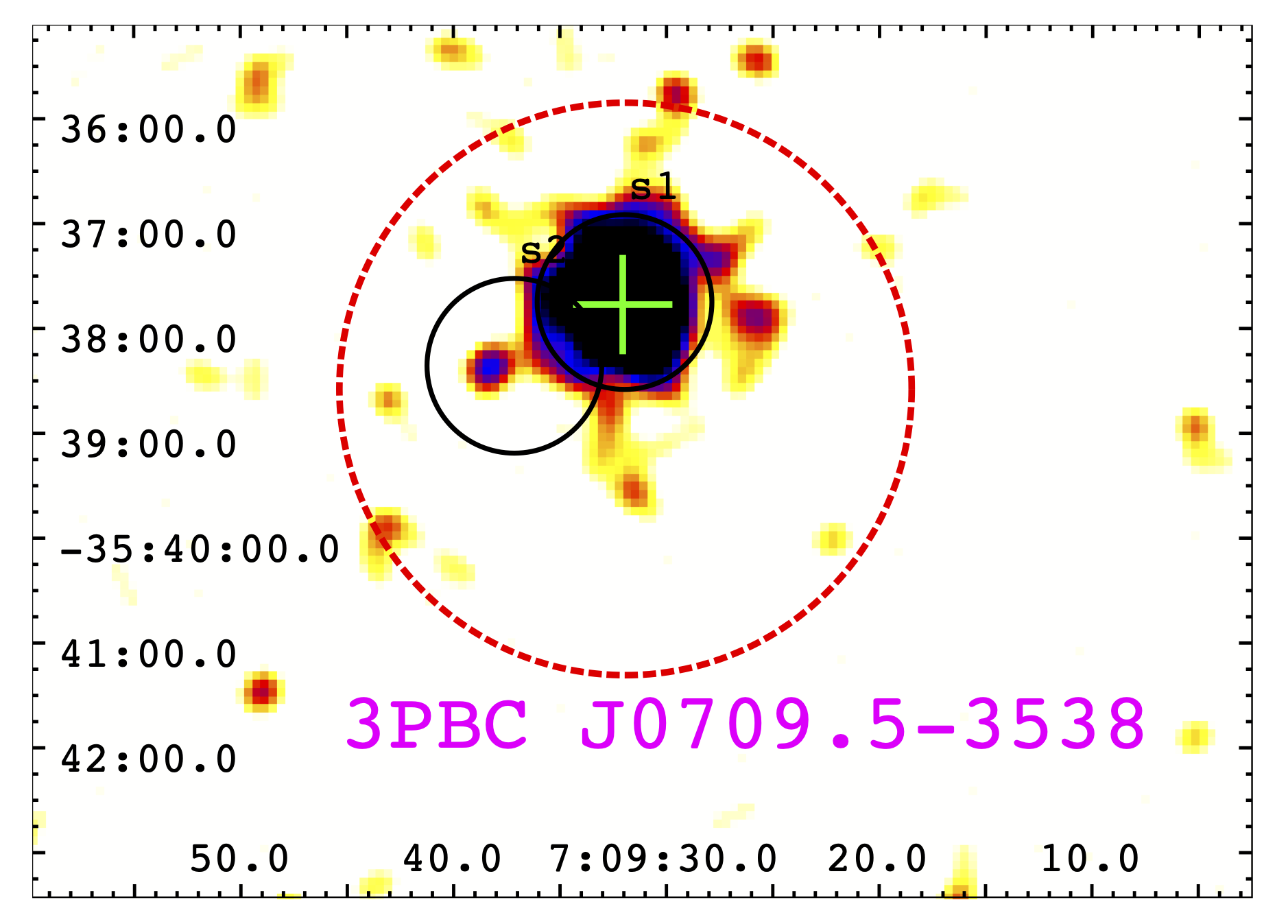}
    \includegraphics[height=4.2cm,width=6cm,angle=0]{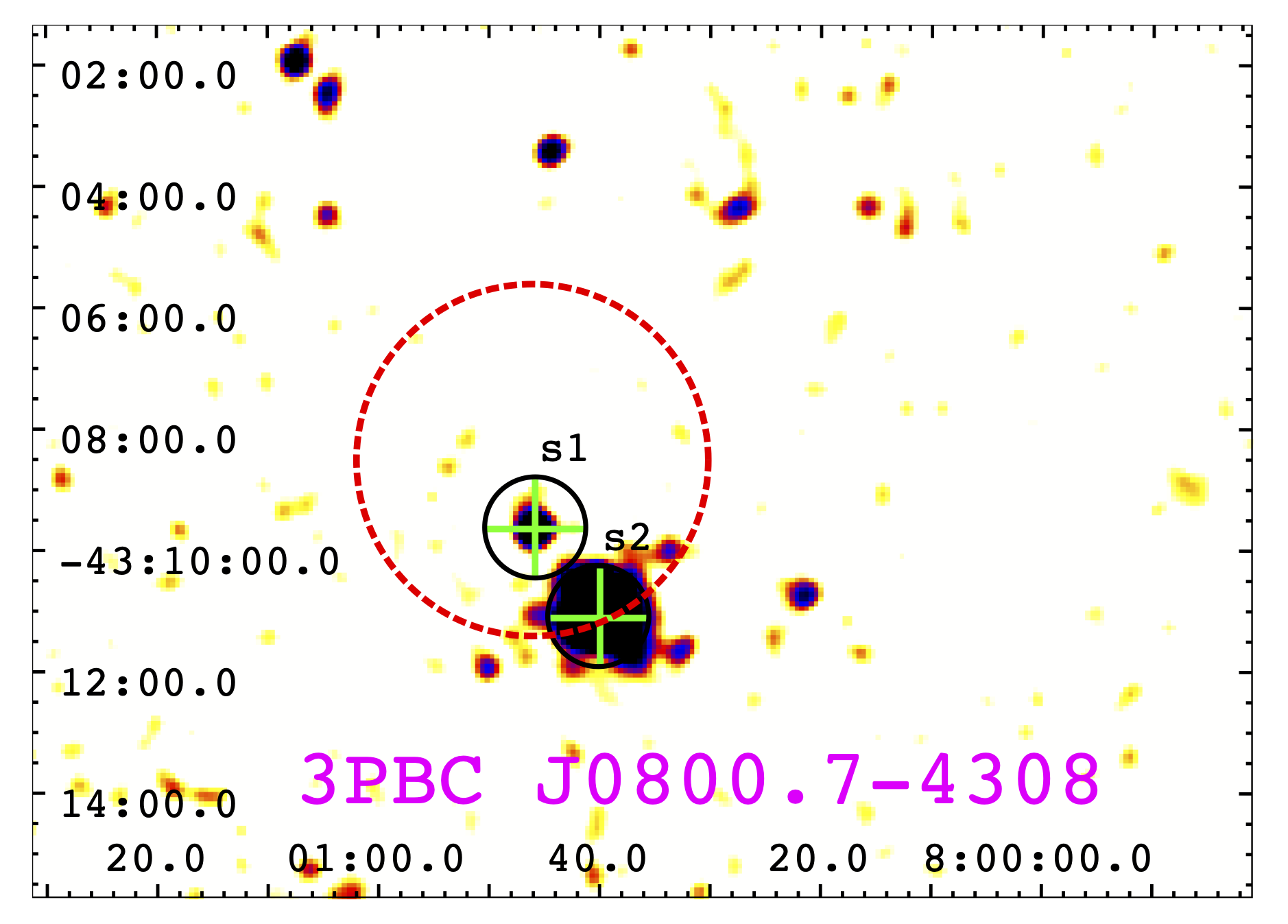}
    \includegraphics[height=4.2cm,width=6cm,angle=0]{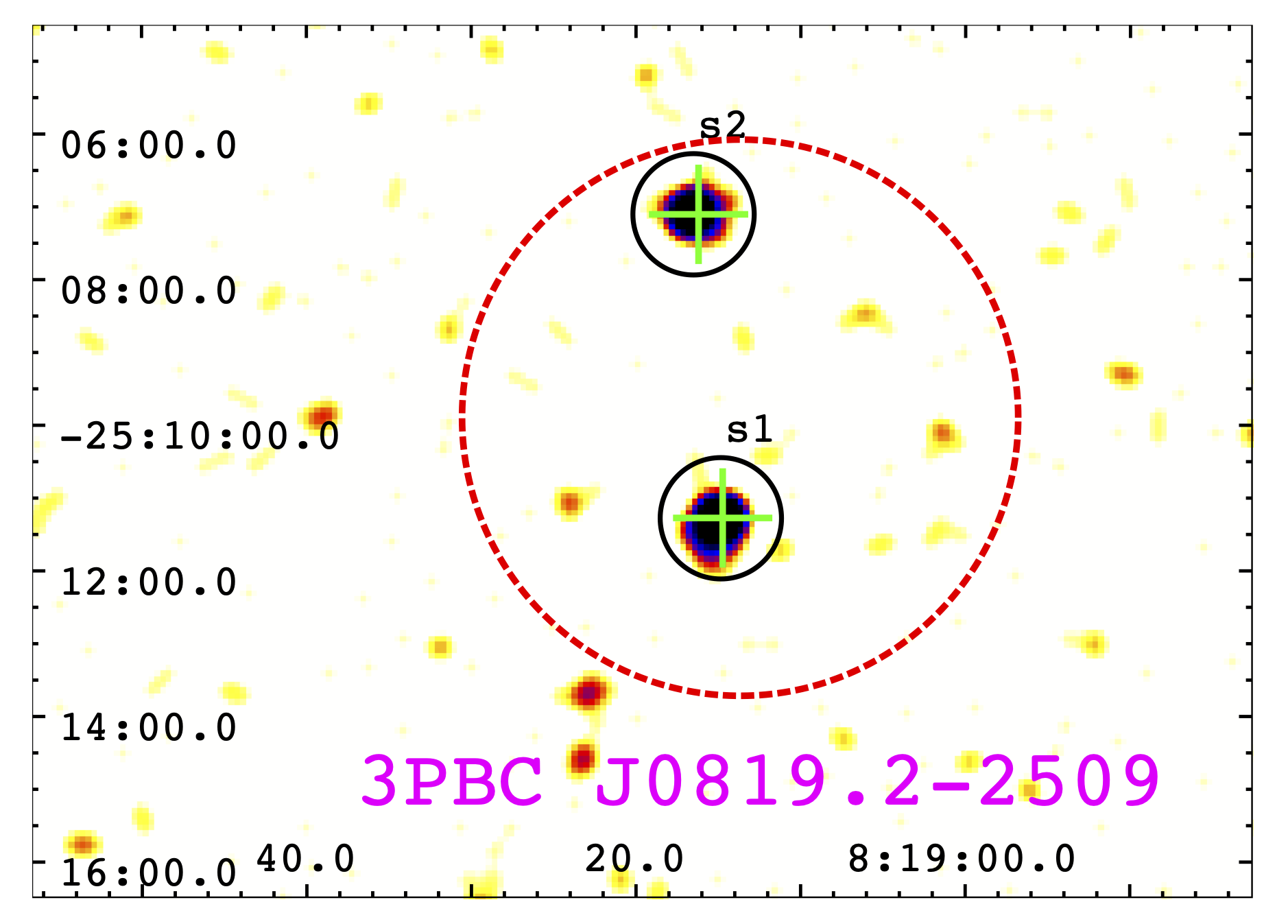}

    \includegraphics[height=4.2cm,width=6cm,angle=0]{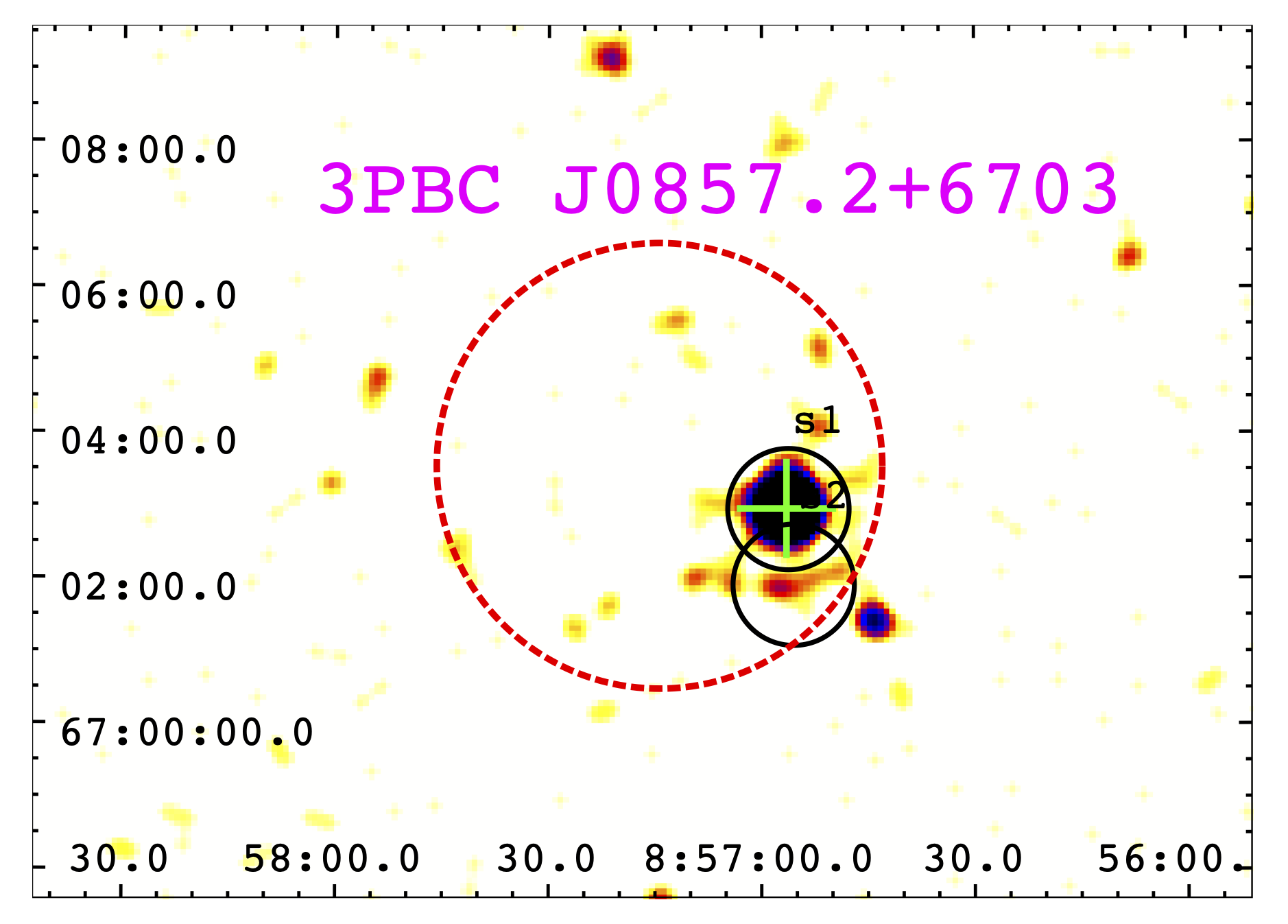}
    \includegraphics[height=4.2cm,width=6cm,angle=0]{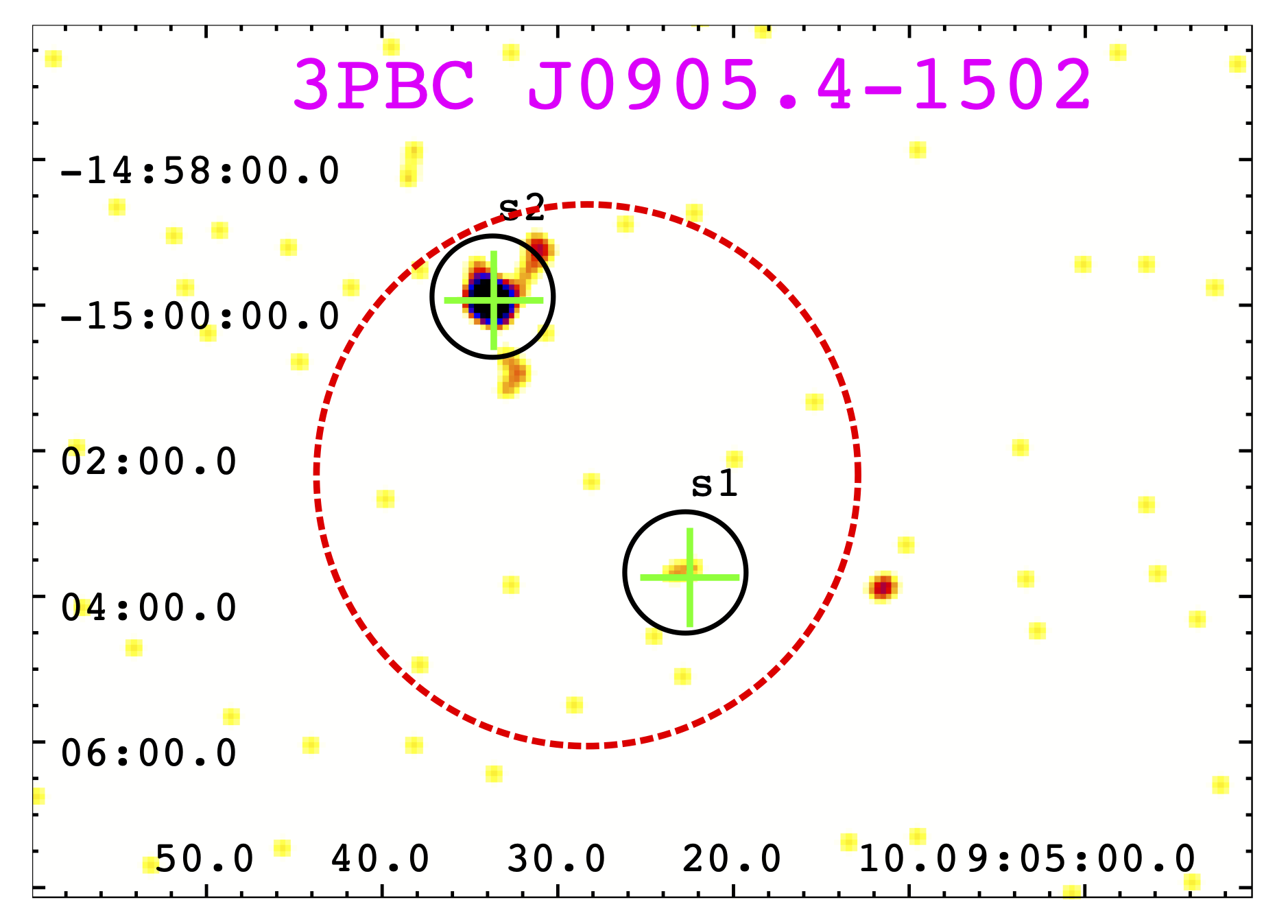}
    \includegraphics[height=4.2cm,width=6cm,angle=0]{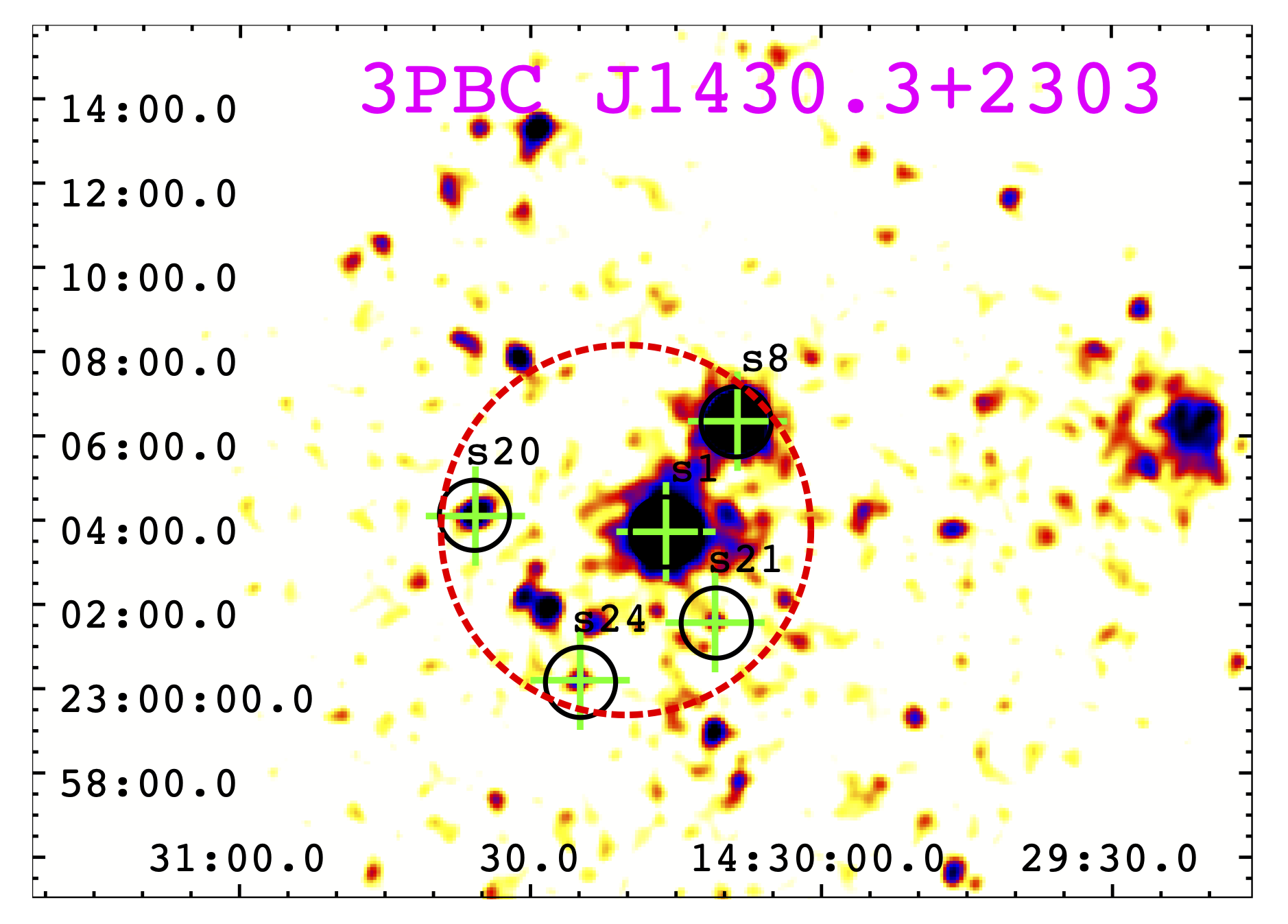}

    \includegraphics[height=4.2cm,width=6cm,angle=0]{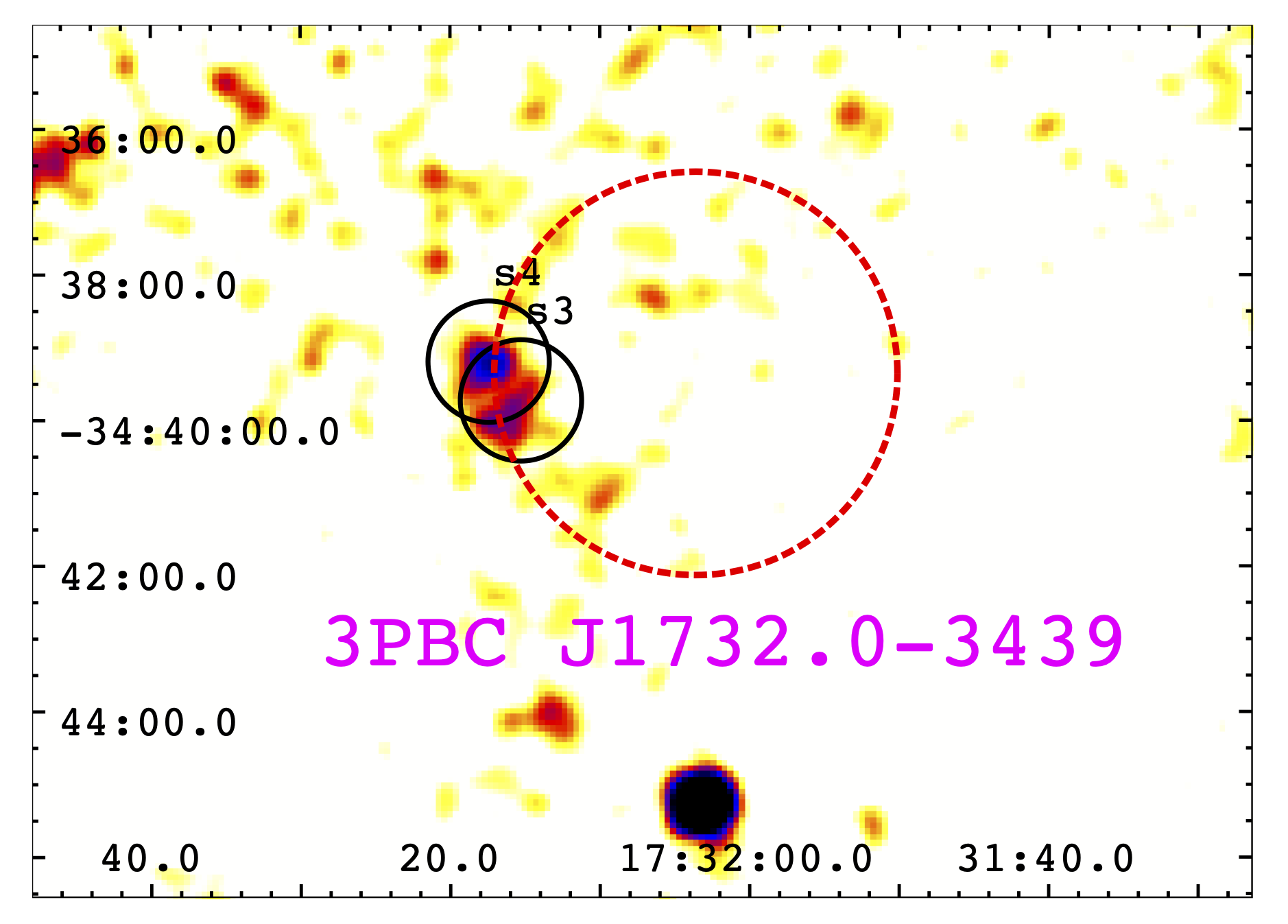}
    \includegraphics[height=4.2cm,width=6cm,angle=0]{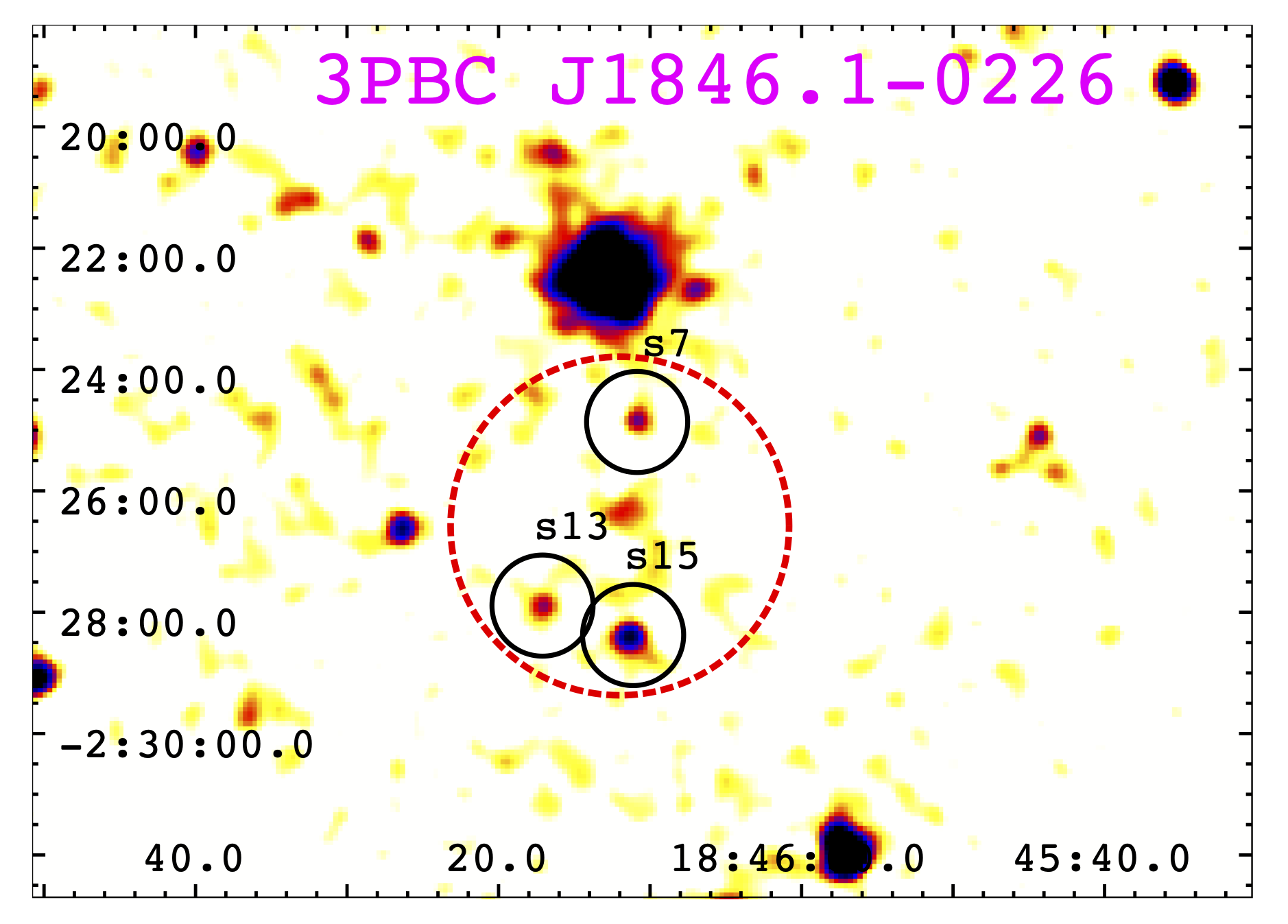}
    \includegraphics[height=4.2cm,width=6cm,angle=0]{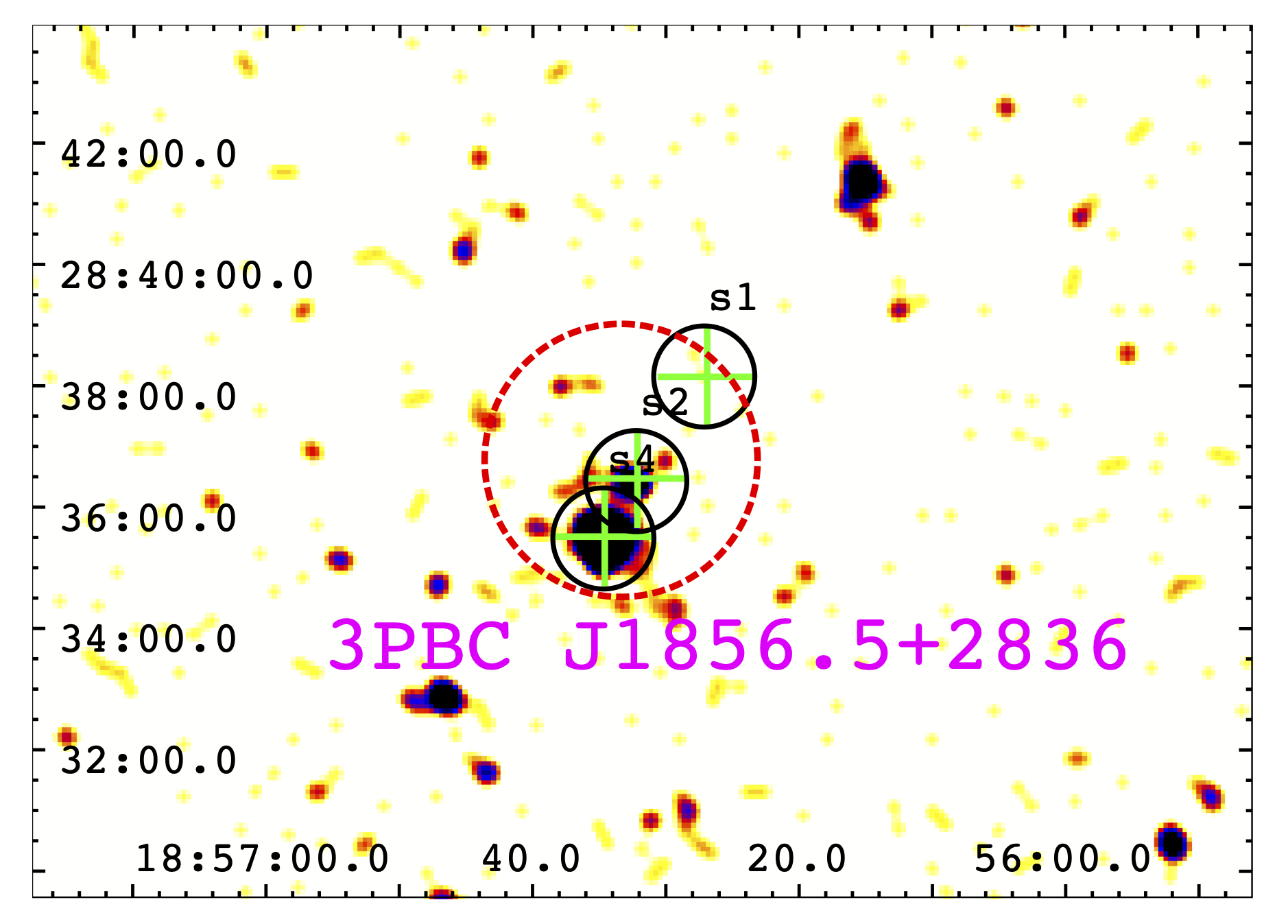}

    \includegraphics[height=4.2cm,width=6cm,angle=0]{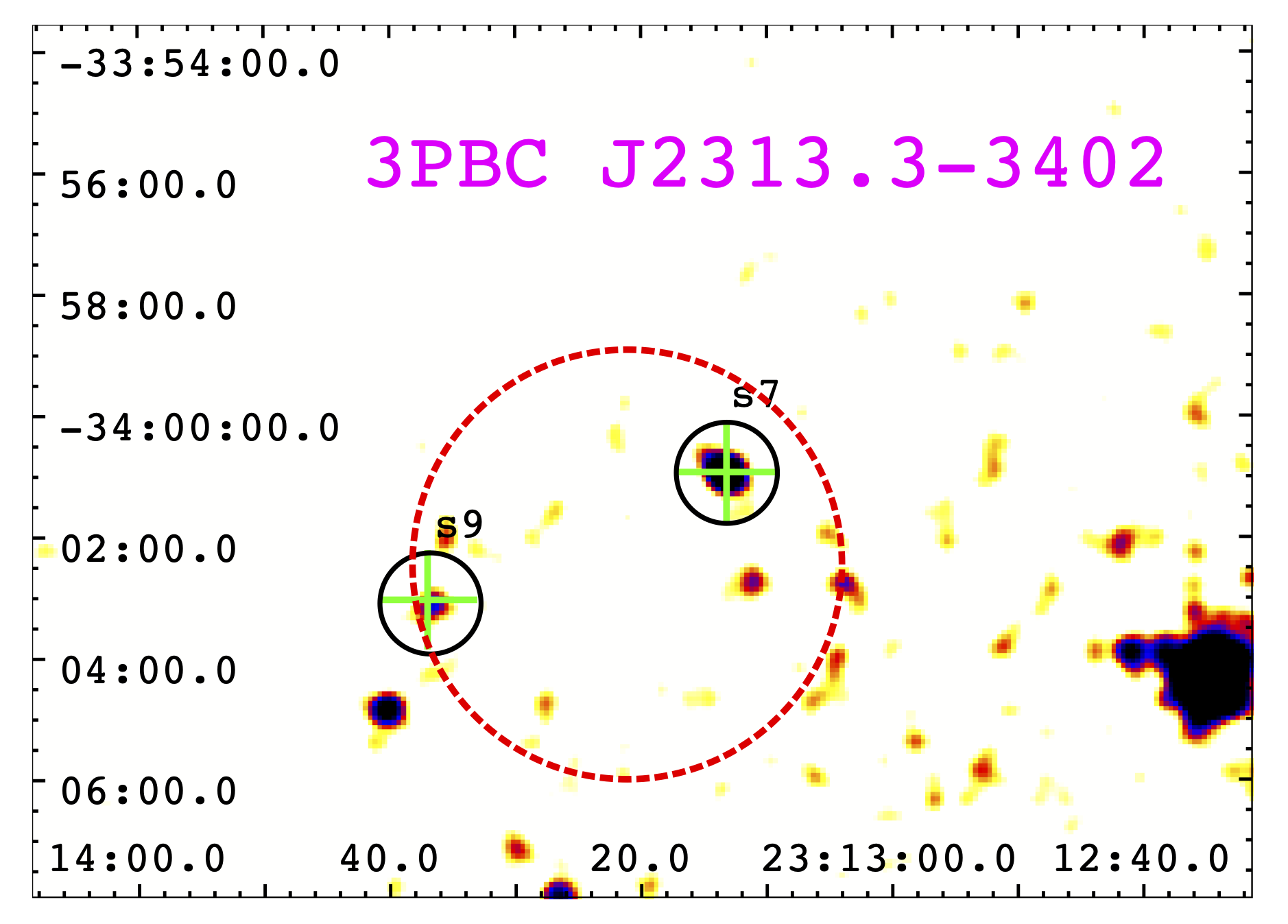}
    
    \caption{Images of the 13 3PBC sources with more than one soft \textit{Swift}-XRT source (XDF flag \textit{m}) detected inside of the BAT positional uncertainty region (red dashed circle). The soft X-ray detections are indicated with a black circle. The black circle indicates the position of the soft X-ray source, not its positional uncertainty. If the soft X-ray detection is also marked with a green cross, it indicates that it has a WISE counterpart.}
    \label{fig:m_flagged_sources}
\end{center}
\end{figure*}

Finally, we considered the case of 3PBC1430.3+2303, for which, given the diffuse X-ray emission clearly detected, we only selected as potential counterparts those targets detected as described in the previous section but also having a mid-infrared counterpart, marked with the green cross in Fig.\,\ref{fig:m_flagged_sources}. It is worth noting that among them, SWXRTJ143016.094+230343.862 seems to be associated with the galaxy cluster MSPM 05080, thus indicating that the possible origin of this extended X-ray emission is that arising from its intracluster medium.

\subsubsection{3PBC J0022.2+2539}
  This source has three XRT PC counterparts Fig.\,\ref{fig:m_flagged_sources}. While sources s1 and s3 are faint with S/Ns of \(4.7\) and \(3.1\), respectively, source s2 is much brighter with a S/N of \(37\) and a count rate of \(0.158 \pm 0.004 \text{ s}^{-1}\). In addition, source s2 is also detected by WISE (J002203.09+254003.2) and in SDSS (J002203.09+254003.1) with magnitude \(r=17.0\).

\subsubsection{3PBC J0218.5-5005}
  This source has three XRT PC counterparts Fig.\,\ref{fig:m_flagged_sources}. The brightest of these is s1 with an S/N of \(5.9\), while the faintest is s2 with an S/N of \(3.8\), also detected by WISE (J021822.70-500557.5).

\subsubsection{3PBC J0536.1-3205}
  This source has two XRT PC counterparts Fig.\,\ref{fig:m_flagged_sources}. The brightest being s1 with an S/N of \(4.1\), also detected by WISE (J053618.88-320533.0).

\subsubsection{3PBC J0709.5-3538}
  This source has two XRT PC counterparts Fig.\,\ref{fig:m_flagged_sources}, with s1 being by far the brightest, with a S/N of \(28.6\) and a countrate of \(0.176 \pm 0.006 \text{ s}^{-1}\), also detected by WISE (J070932.05-353746.5) with a spectrum reported in \cite{Rojas2017}.

\subsubsection{3PBC J0800.7-4308}
  This source also has two XRT PC counterparts Fig.\,\ref{fig:m_flagged_sources}. The source s1 is the faintest, with a S/N of \(5.8\) and a countrate 0f \(0.0036 \pm 0.0006 \text{ s}^{-1}\), while s2 has a S/N of \(27.6\) and a countrate 0f \(0.065 \pm 0.002 \text{ s}^{-1}\). Both sources have a WISE counterpart, J080045.83-430939.3, and J080039.96-431107.2, respectively.

\subsubsection{3PBC J0819.2-2509}
  This source has two XRT PC counterparts Fig.\,\ref{fig:m_flagged_sources}, with the brightest being s1 with an S/N of \(14.0\) and the faintest being s2 with an S/N of \(7.4\). Both sources have a WISE counterpart, J081914.73-251116.6, and J081916.20-250706.4, respectively, with s1 having a redshift of 0.00557 with a spectrum reported in \citep{Strauss1992}.
  

\subsubsection{3PBC J0857.2+6703}
  For this source Fig.\,\ref{fig:m_flagged_sources}, we find two XRT PC counterparts: the brightest being s1 with an S/N of \(17.6\) and the faintest being s2 with an S/N of \(3.1\). The source s1 has a WISE counterpart, namely J085656.49+670257.3.

\subsubsection{3PBC J0905.4-1502}
  This source Fig.\,\ref{fig:m_flagged_sources} has two XRT PC counterparts s1 and s2, with similar S/Ns, \(3.8\) and \(3.9\), respectively. Both are detected by WISE (J090522.48-150344.6 and J090533.64-145956.2, respectively), with s2 reported in NED with a redshift of \(0.088\).

\subsubsection{3PBC J1430.3+2303}
  For the 3PBCJ1430.3+2303, given the presence of diffuse X-ray emission, the automatic algorithm described in the previous section detected many spurious sources. We thus selected as potential soft X-ray counterparts only the sources with a mid-infrared counterpart, marked with the green cross. It is worth noting that among them, SWXRTJ143016.094+230343.862 seems to be associated with the galaxy cluster MSPM 05080, thus indicating that the possible origin of this extended X-ray emission is that arising from its intracluster medium.

\subsubsection{3PBCJ1732.0-3439}
  This source Fig.\,\ref{fig:m_flagged_sources} has two XRT PC counterparts, s3 and s4, with S/Ns of \(3.3\) and \(4.7\), respectively, and count rates of \(0.0044 \pm 0.0013\) and \(0.0087 \pm 0.0018\) ct\,s$^{-1}$, respectively.

\subsubsection{3PBCJ1846.1-0226}
  This source Fig.\,\ref{fig:m_flagged_sources} has three XRT PC counterparts, s7, s13 and s15, with similar S/Ns of \(4.2\), \(4.5\) and \(4.9\), respectively. The source s13 is also detected in SDSS (J184617.13-022753.4) with a magnitude \(r=17.6\).

\subsubsection{3PBCJ1856.5+2836}
  This source Fig.\,\ref{fig:m_flagged_sources} has three XRT PC counterparts, s1, s2 and s4, with S/Ns of \(3.6\), \(3.5\) and \(9.9\), respectively. The sources are all detected by WISE (J185626.89+283809.3, J185632.13+283628.8, and J185634.58+283531.3), although the faintest s2 only has an upper limit of \(8.3 \text{ mag}\) in \(W_4\) band.

\subsubsection{3PBCJ2313.3-3402}
  This source Fig.\,\ref{fig:m_flagged_sources} has two XRT PC counterparts, s7, and s9, with the brightest being the former, with an S/N of \(8.4\). Both sources are detected by WISE (J231313.21-340056.2 and J231337.01-340302.1), although the fainter s9 only has an upper limit in \(W_3\) and \(W_4\) bands.

\subsection{Comparison with 2SXPS}
\label{sec:2sxps}
Finally, we compared our results with those that can be obtained by simply crossmatching all 218 hard X-ray sources listed in the 3PBC with the latest release of the Second \textit{Swift}-XRT Point Source Catalog\footnote{https://heasarc.gsfc.nasa.gov/W3Browse/swift/swift2sxps.html} \citep[2SXPS][]{evans20}. The 2SXPS catalog has a sky coverage of \(3,790 \text{ deg}^2\), listing 206,335 point sources detected by XRT in the $0.3-10$\,keV energy range. Here, we briefly summarize the procedure used to build the 2SXPS.

The 2SXPS was built based on all XRT observations taken between 2005, January 1st, and 2018, August 1st, with an exposure of at least 100\,s in PC (photon-counting) mode.
Source detection was performed with the sliding-cell technique with a S/N threshold set to \(1.5\), in comparison to our choice of S/N\,=\,3, yielding the final catalog of \(206,335\) XRT PC sources listed in the 2SXPS. The catalog contains a ``clean" subsample, listing \(146,768\) sources without analysis flags (see \citep{evans20} for more details). In the following, we will consider this ``clean" subsample.

\begin{figure*}[!th]

\begin{center}
    \includegraphics[scale=0.4]{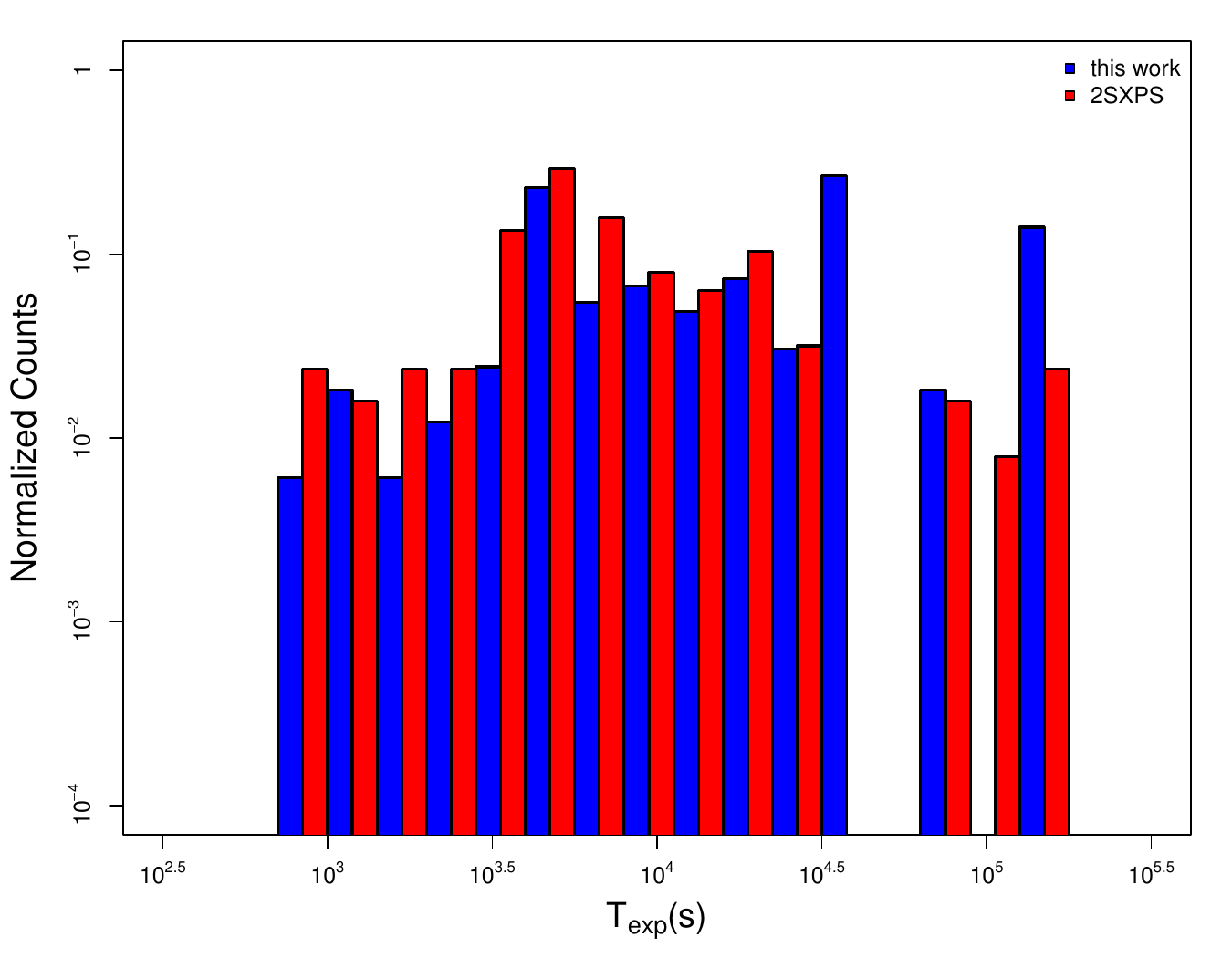}
    \includegraphics[scale=0.4]{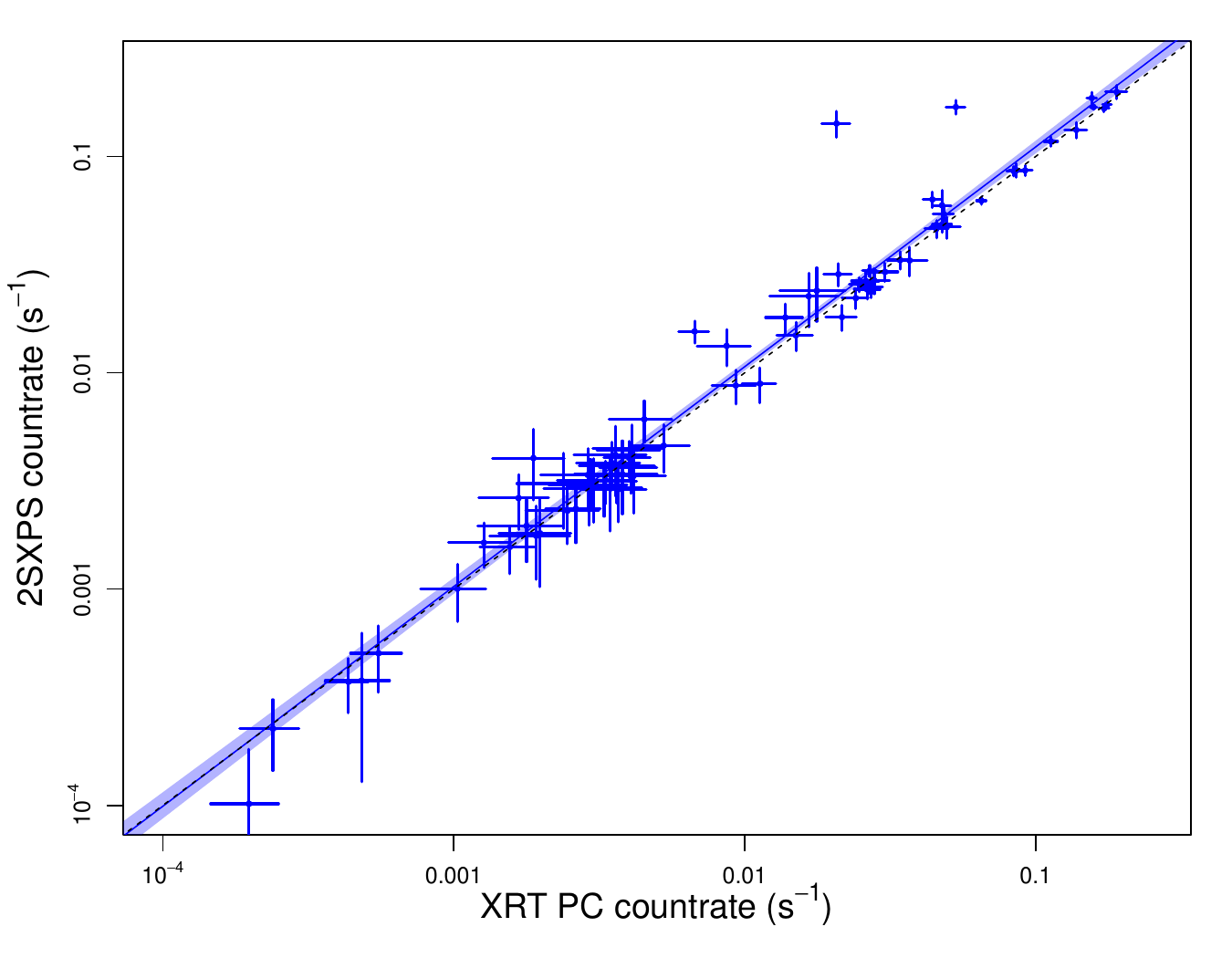}
      \caption{Left panel: the comparison between the exposure of the dataset used in the present work (blue distribution) and in the 2SXPS catalog (red distribution). Right panel: XRT PC count rates as reported in the 2SXPS catalog versus those evaluated in the present work for the 78 sources in common between the two analyses (see Sect. \ref{sec:2sxps}). The 2SXPS count rate has been rescaled to match the energy band used in the present analysis. The blue line indicates the linear regression to the logarithmic data, while the light blue shaded area represents the \(1-\sigma\) uncertainty around the best-fit relation. The black dashed line indicates the \(y=x\) relation. See Sect.\ref{sec:2sxps} for more details.}
      \label{fig:2sxps}
      
\end{center}
\end{figure*}

Due to the different dataset and analysis procedures used in the present work and in \citet{evans20}, we expect differences in the XRT PC source detections. In fact, cross-matching the 2SXPS ``clean" sample with the 3PBC sources considered here (see Sect. \ref{sec:sample_selection}), taking into account both BAT and 2SXPS positional uncertainties, we find 126 2SXPS counterparts to 90 3PBC sources, 68 3PBC sources with a single 2SXPS match and 22 3PBC sources with multiple 2SXPS matches. In the left panel of Fig. \ref{fig:2sxps}, we compare the exposures of the XRT PC observations used in the present analysis (blue distribution) and in the 2SXPS observations for which we find the 126 2SXPS counterparts to the 3PBC sources considered in this work. We see that 2SXPS datasets and the dataset used in our analysis span a similar exposure range. 

In addition, we find XRT PC counterparts to 7 3PBC sources without 2SXPS counterparts, while in the 2SXPS catalog, there are counterparts to 22 3PBC sources for which we did not find XRT PC counterparts. However, we note that the S/Ns for these 22 sources are less than \(2.5\), below the S/N threshold of \(3\) that we adopted for the present analysis. In addition, we find 78 sources positionally compatible between our sample of 93 XRT PC sources and the 126 2SXPS counterparts to the 3PBC sources. 

In the right panel of Fig. \ref{fig:2sxps}, we compare the count rate evaluated in the present analysis with the count rate reported in the 2SXPS catalog for these 78 common sources. We stress that to account for the different energy bands adopted in the present analysis (\(0.5-10 \text{ keV}\)) and in the 2SXPS catalog (\(0.3-10 \text{ keV}\)), we rescaled the count rate of the 2SXPS catalog by a factor \(0.86\) evaluated via PIMMS\footnote{https://cxc.harvard.edu/toolkit/pimms.jsp} tool assuming a power-law spectrum with a \(1.8\) slope. We see that the two estimates are in good agreement at low count rate values (\(<{10}^{-3}\,\text{cps}\)). In contrast, above \({10}^{-2}\,\text{cps}\), the 2SXPS count rates appear systematically larger than those evaluated in the present analysis. The two-count rate estimates, however, are compatible at \(2\,\sigma\) level.
The maximum deviation below 10$^{-3}$\,cps is 0.34\,$\sigma$, while above 10$^{-2}$\,cps this reaches 1.95\,$\sigma$. In addition, below 1.8$^{-4}$\,cps, the best-fit line is systematically below the $y = x$ relation, while above 1.8$^{-4}$\,cps, the best-fit line is systematically above the $y = x$ relation.

\section{Multifrequency Comparison}
\label{sec:compare}
To search for additional information regarding all detected soft X-ray candidate counterparts we crossmatched their position, derived with the \textsc{xrtcentroid} task - taking into account their positional uncertainties - with three main catalogs/surveys: (i) the NASA Extragalactic Database (NED)\footnote{http://ned.ipac.caltech.edu}; (ii) the SIMBAD Astronomical Database\footnote{http://simbad.u-strasbg.fr/simbad/} as well as (iii) the ALLWISE catalog \citep{Cutri2021} based on the all-sky survey performed with WISE telescope \citep{Wright2010} and (iv) the spectroscopic catalog of the Sloan Digital Sky Survey (SDSS) data release 16 (DR16) \citep{Blanton2017}. We used the positional uncertainty region of the \textit{Swift}-XRT for all soft X-ray detected sources to search and claim an association with its wise counterpart.

Our crossmatching analysis revealed that 84 of 93 soft X-ray potential counterparts have an identification reported in NED. In addition, 74 of the 93 have photometry available in WISE. We found that 10 out of the 93 have spectra in the SDSS DR16. 
We classified them according to their spectral characteristics and we report the results in Tab. \ref{table:spectral_classification}.

\subsection{A mid-infrared perspective}
\label{sec:wise}

\begin{figure}[!th]
\begin{center}
    \includegraphics[height=6cm,width=8cm,angle=0]{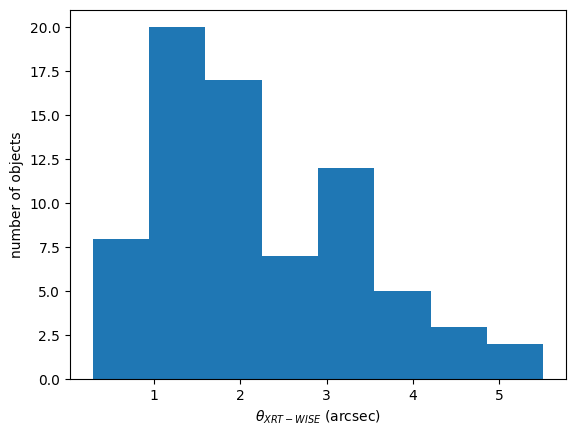}
    
    \caption{Distribution of the angular separation between the \textit{Swift}-XRT centroid and the WISE centroid ($\theta_{XRT-WISE})$ of the 74 soft X-ray detections with the WISE counterpart.}
    \label{fig:hist_angsep_XRT_WISE}
\end{center}
\end{figure}
\begin{figure}[!th]
\begin{center}
    \includegraphics[height=7cm,width=9cm,angle=0]{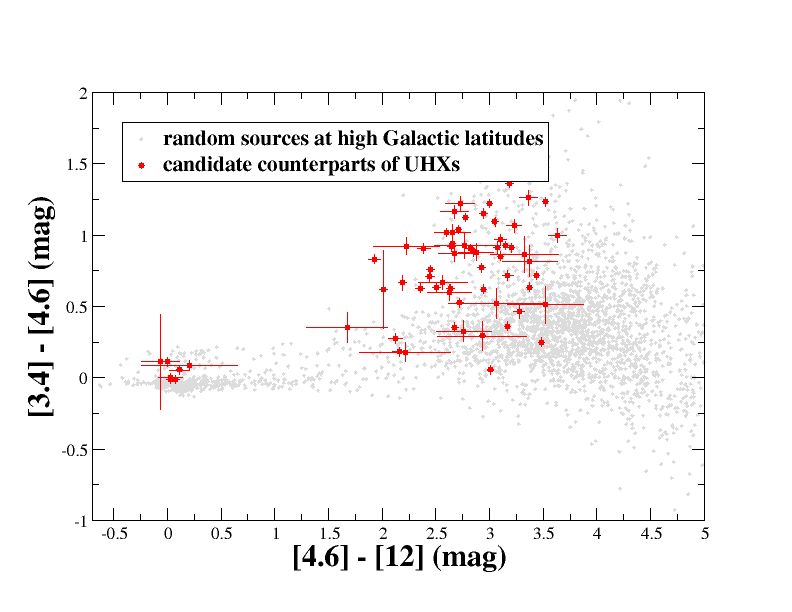}
    
    \caption{This figure shows the [3.4]-[4.6]-[12] µm color-color diagram of WISE thermal sources and blazars. The gray dots represent a sample of 3\,000 randomly selected mid-IR sources in a region of 0.5 deg radius around Galactic coordinates (50.411113,-45.668864) and (50.411113,45.668864). The 66 \textit{Swift}-XRT detected sources with available luminosities in the first three WISE bands are marked as black dots. The mid-IR colors of this sample of 66 sources do not agree with the mid-IR colors of stars. Instead, they are more consistent with AGNs, mainly Seyfert galaxies and QSOs. The sources with [4.6]-[12] mag > 2 are AGNs and QSOs, while the concentration of sources around 0 [4.6]-[12] mag are mostly normal elliptical galaxies and stars. }
    \label{fig:midIRcolorplot}
\end{center}
\end{figure}

From the 93 detected soft X-ray possible counterparts corresponding to the 73 unique 3PBC sources, 74 sources have been detected in at least one WISE band. There are 74 sources detected in both the W1 and W2 bands, 66 in W3, and 52 in W4. Fig.\,\ref{fig:hist_angsep_XRT_WISE} shows the distribution of the angular separation between the \textit{Swift}-XRT centroid and the WISE centroid of the 74 counterparts that have the WISE detection. We took all sources that are detected in the W1, W2, and W3 bands, so 66 sources, and plotted them on a color-color diagram (Fig. \ref{fig:midIRcolorplot}). The gray background sources in Fig. \ref{fig:midIRcolorplot} are 3\,000 WISE sources in the mid-IR sky selected randomly in a region of 0.5 deg radius around Galactic coordinates (50.411113,-45.668864) and (50.411113,45.668864). The black dots correspond to the 66 \textit{Swift}-XRT detected sources with reported luminosities in the first three WISE bands. The mid-IR colors of this sample of 66 sources are not in good agreement with the mid-IR colors of stars but are more consistent with AGNs, mainly Seyfert galaxies and QSOs.

\subsection{Archival optical spectra}
\label{sec:sdss}
According to previous analyses carried out during past follow-up spectroscopic campaigns \citep[see, e.g.,][]{Massaro2016, Herazo2020, Herazo2022, Kosiba2023}, we adopted a conservative criterion to provide spectroscopic identification of selected X-ray counterparts. We adopted the same classification scheme described in paper I, and we only considered reliable redshift measurements, those for which we could verify the presence of a published image of the optical spectrum or a description of the published spectrum with emission and/or absorption lines clearly reported in a table format or the publication manuscript. We found spectra in the SDSS DR16 archive for 10 soft X-ray sources. We show the spectra in Fig. \ref{fig:spectra} and the classification of those sources in Appendix \ref{app:Appendix_1}.  The most prominent spectral lines among their spectra were the H$\alpha$ + [N II], [O III], and [O II]. We classified four sources as AGN Type 2, one as AGN Type 1, four as QSOs and one as a star-forming galaxy, which agrees with the WISE colors of all sources and previous results.

\begin{figure}[!th]
\begin{center}
    \includegraphics[height=8cm,width=8cm,angle=0]{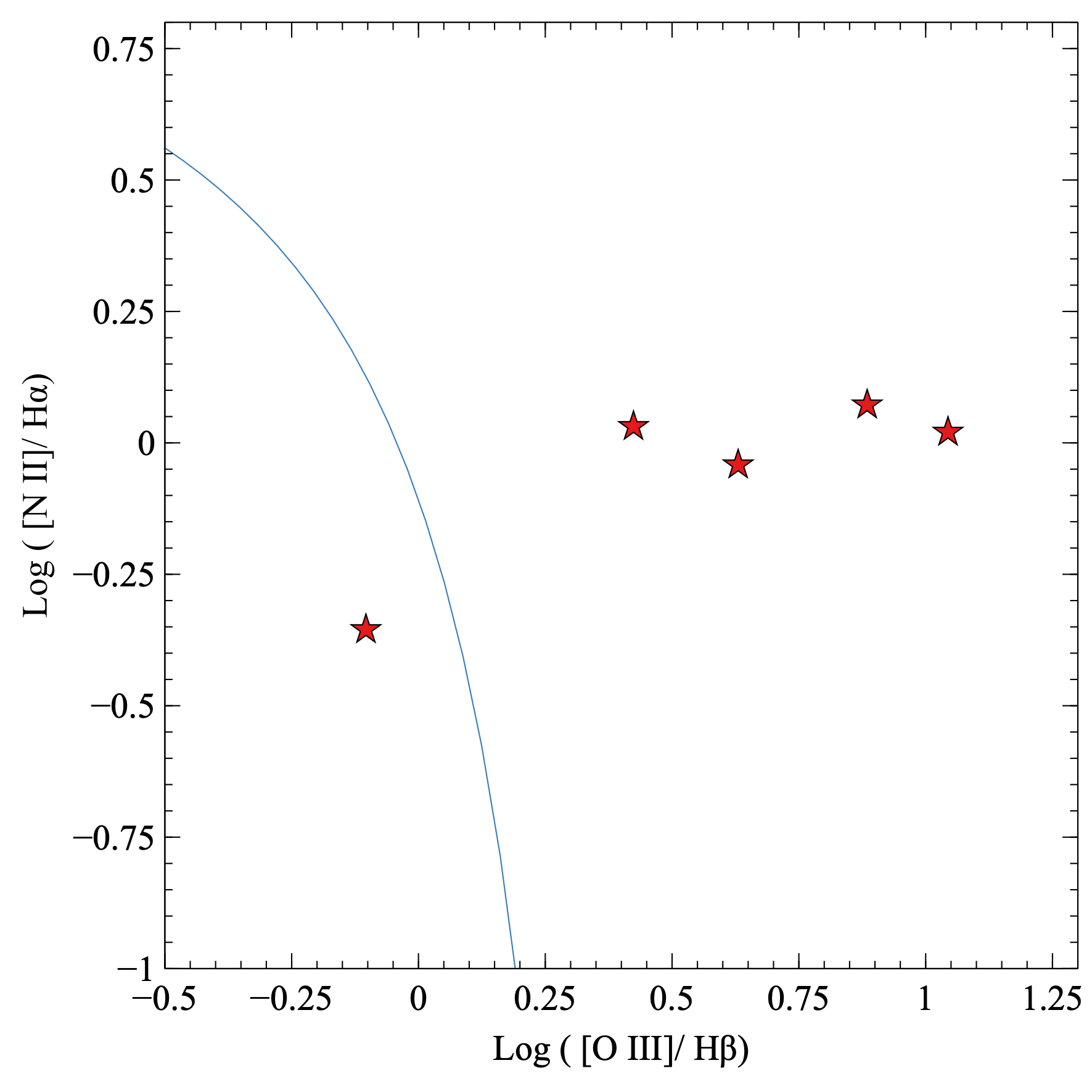}
    
    \caption{The BPT Diagram for distinguishing Type 2 and Star-Forming Galaxies. Error bars are all less than 0.009 and not visible in the plot. Note that all sources, except for SDSS J143010.96+230134.7, are above the theoretical line of \cite{Kewley2013} (right part of the graph separated by the blue line), which serves to discriminate between starburst regions and harder ionization sources.}
    \label{fig:BPT_diagram}
\end{center}
\end{figure}

Additionally, we compute the Baldwin, Phillips \& Terlevich (BPT) diagrams \citep{Baldwin1981} for sources with narrow lines with the objective of classifying them as either Seyfert 2 or star-forming galaxies. We present the BPT diagram in Figure \ref{fig:BPT_diagram}. Our findings indicate that all narrow-line sources, except for SDSS J143010.96+230134.7, fall outside the region corresponding to star-forming galaxies on the plot. This region is delineated by the theoretical line of \citep{Kewley2001, Kewley2013}. Consequently, we classify these sources as Seyfert 2 galaxies. For 3PBCJ1504.1-6019 and 3PBCJ0800.7-4308, the analysis we carried out is in agreement with that of \cite{Landi2017}, since we found the same soft X-ray sources lying within the positional uncertainty of these BAT unidentified objects. Inspecting NED and SIMBAD databases, we found the spectra for the soft X-ray counterparts of the UHXs 3PBCJ1329.7-1052 and 3PBCJ1854.4-3436. These
are MCG -02-34-058 and ESO 396- G 007, lying at z=0.021648 and z=0.019483
\citep{Jones2009}, respectively. The \textit{Swift}-XRT counterparts we assigned for the sources 3PBCJ2136.1+2002, 3PBCJ2155.3+6204, and 3PBCJ2238.8+4050 are the same as previously assigned in the literature at which optical spectroscopic follow-up observations revealed these to be three active galaxies lying at z=0.081 (Sy1), z=0.058 (Sy1) and
z=0.055 (LINER), respectively, as reported in \citep{Rojas2017}. The yet unidentified source 3PBCJ0024.1-6823 having the \textit{Swift}-XRT counterpart SWXRTJ002406.457-682052.549 could also be associated with the radio source PKS 0021-686, a gamma-ray blazar candidate selected based on its mid-IR colors \citep{D'Abrusco2012,D'Abrusco2014,D'Abrusco2019}.

\begin{figure*}[!th]
\begin{center}
    \includegraphics[height=4.2cm,width=6cm,angle=0]{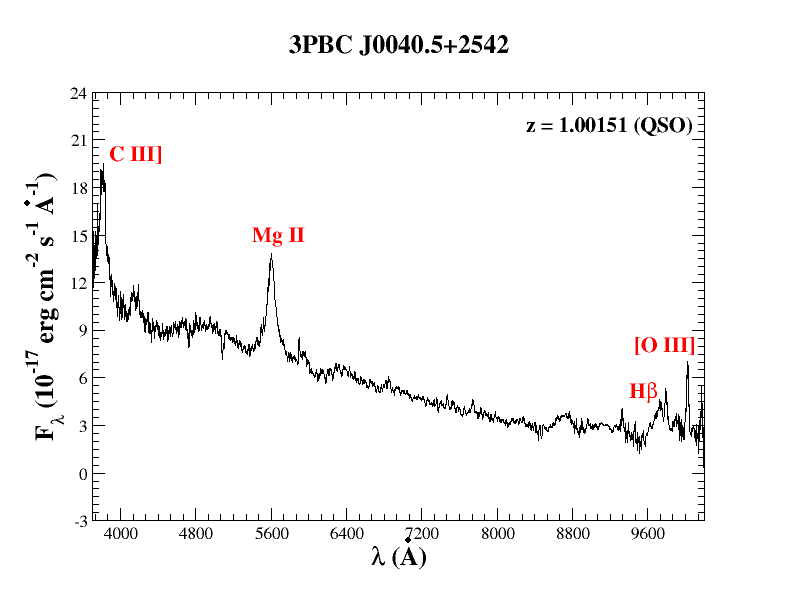}
    \includegraphics[height=4.2cm,width=6cm,angle=0]{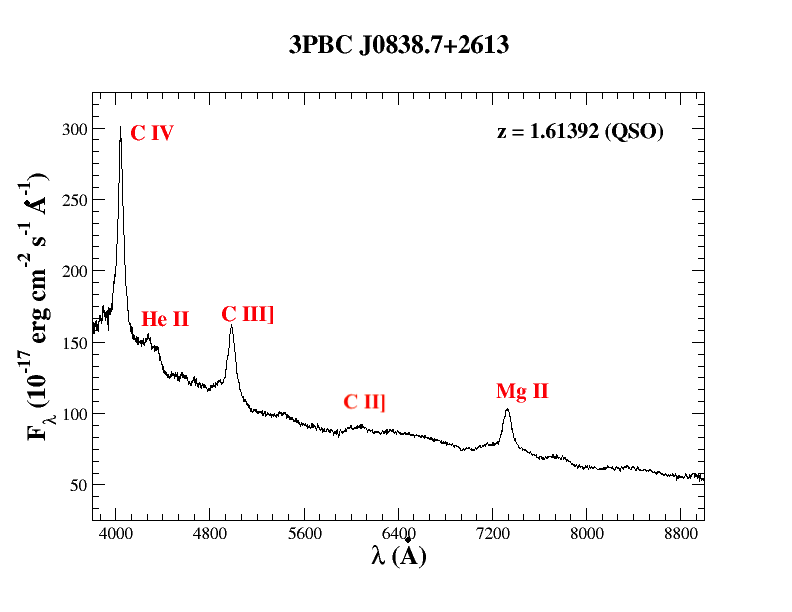}
    \includegraphics[height=4.2cm,width=6cm,angle=0]{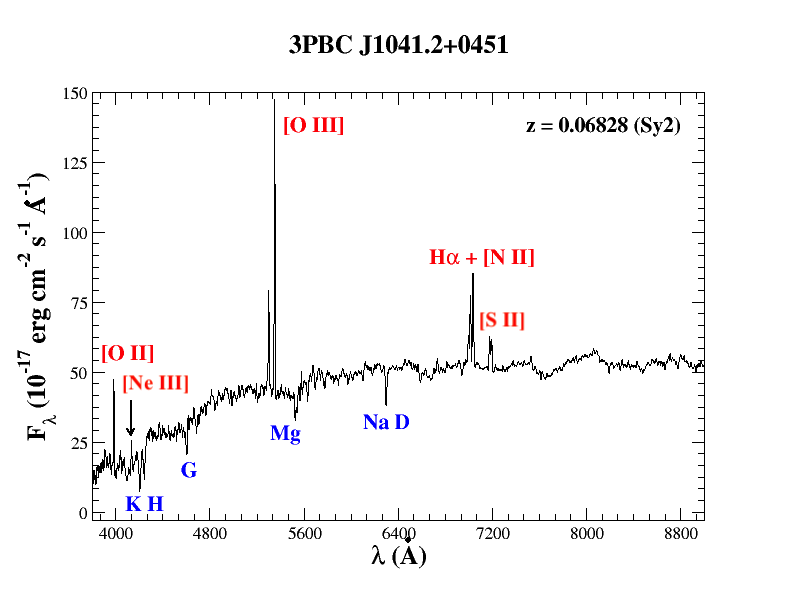}
    \includegraphics[height=4.2cm,width=6cm,angle=0]{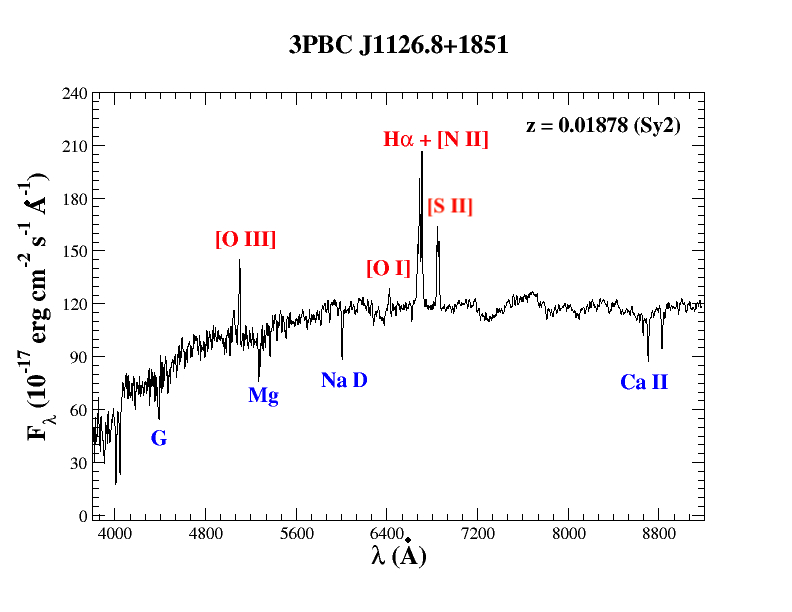}
    \includegraphics[height=4.2cm,width=6cm,angle=0]{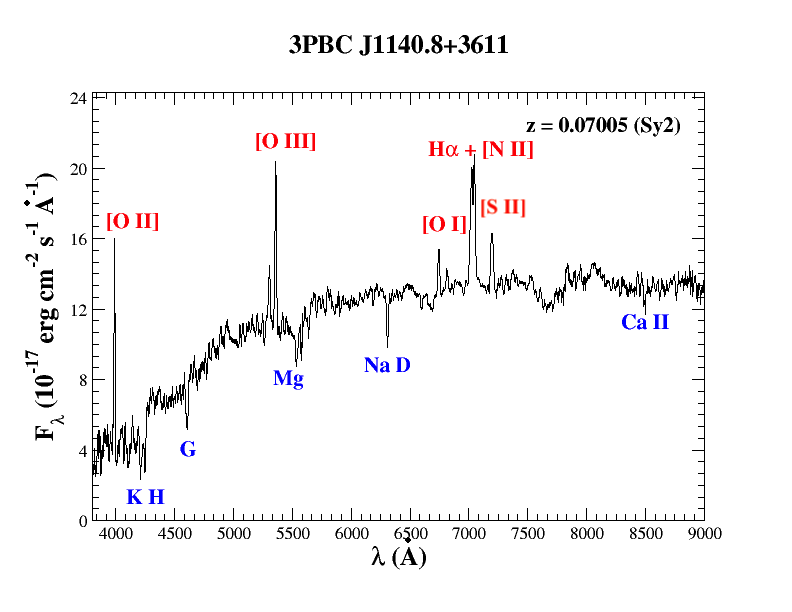}
    \includegraphics[height=4.2cm,width=6cm,angle=0]{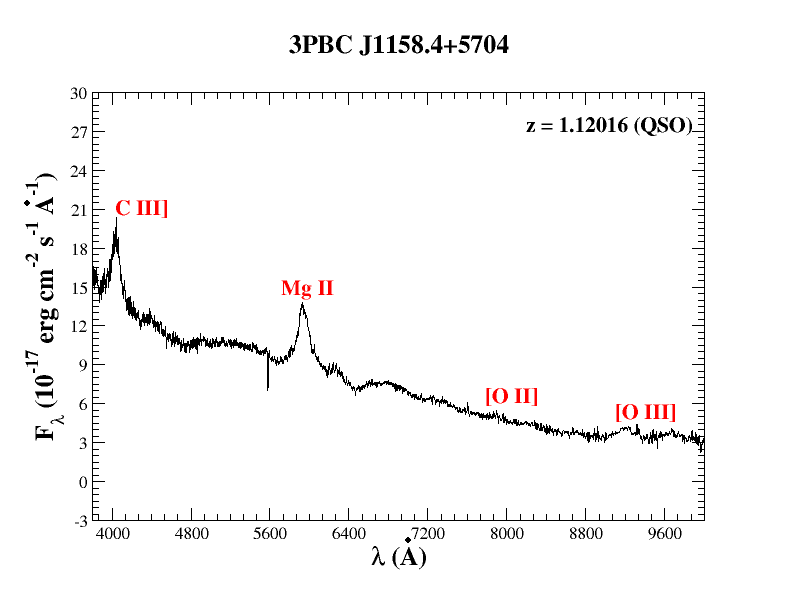}
    \includegraphics[height=4.2cm,width=6cm,angle=0]{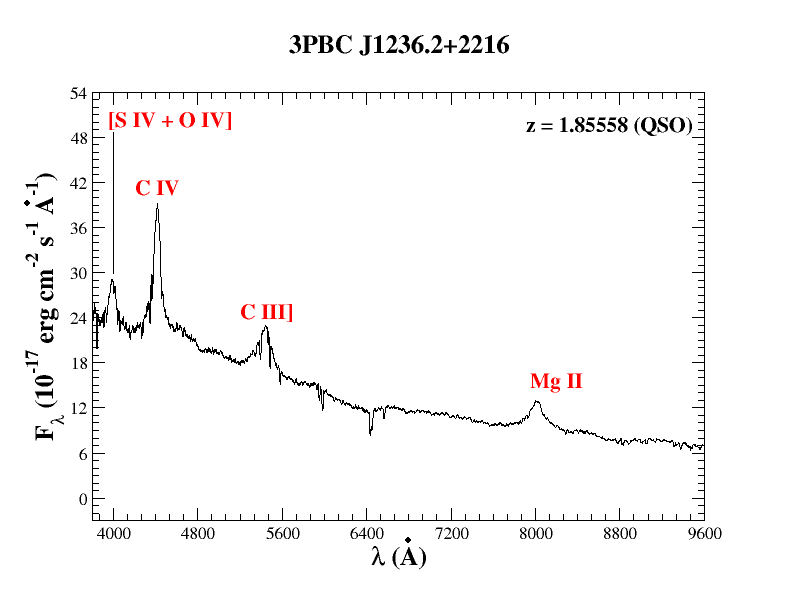}
    \includegraphics[height=4.2cm,width=6cm,angle=0]{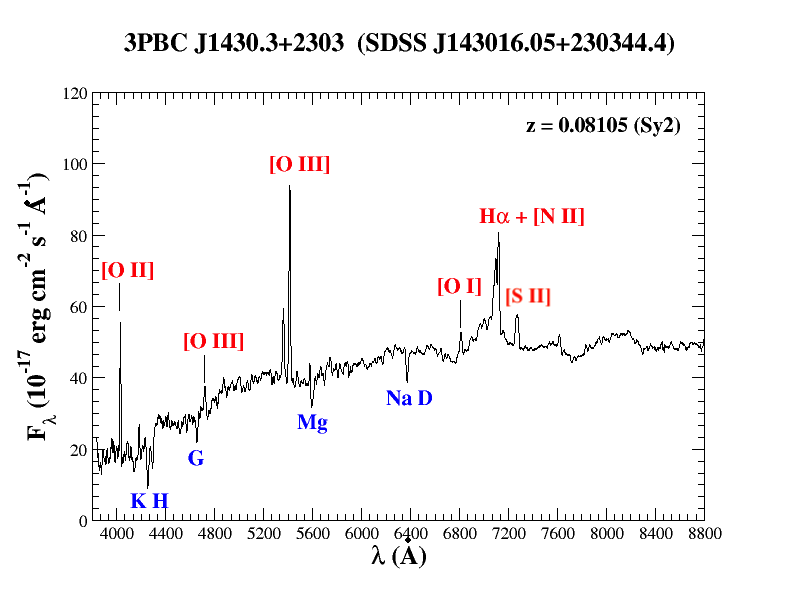}
    \includegraphics[height=4.2cm,width=6cm,angle=0]{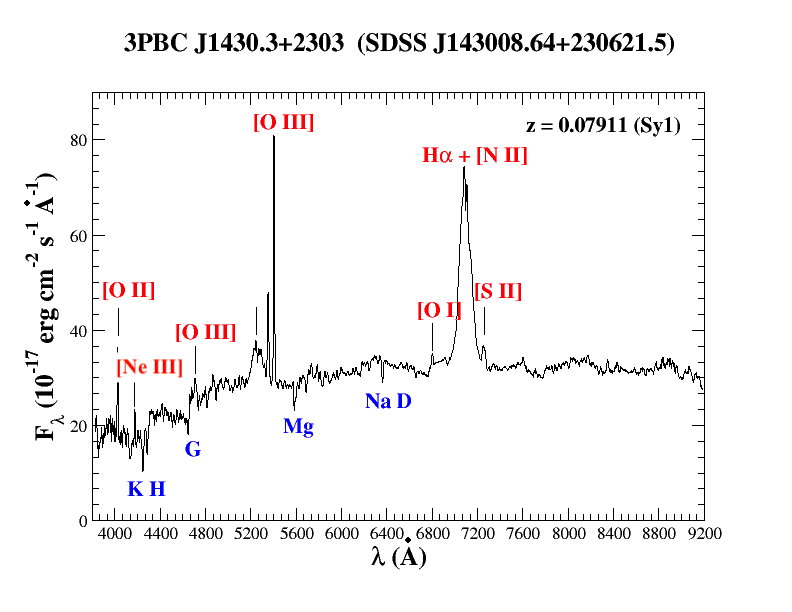}
    \includegraphics[height=4.2cm,width=6cm,angle=0]{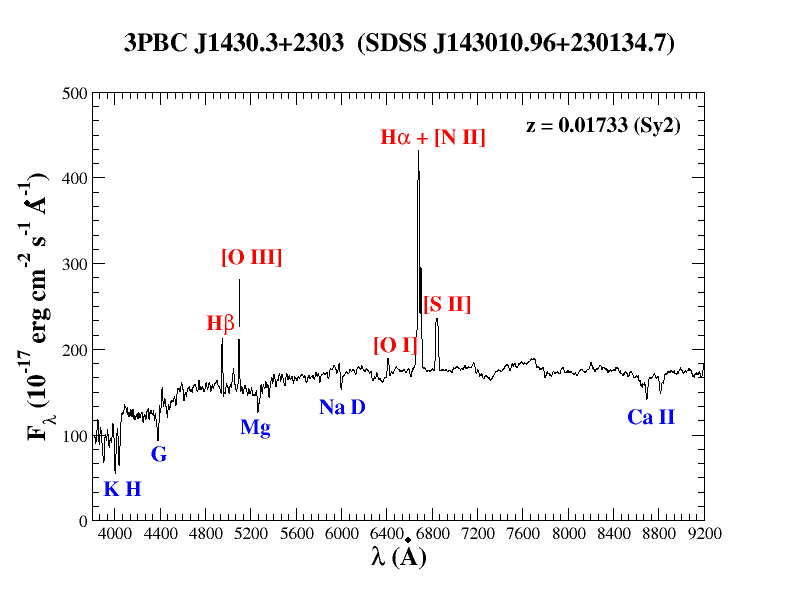}
      \caption{This figure shows the ten spectra we collected from the SDSS archive for the soft X-ray \textit{Swift}-XRT sources detected in the BAT positional uncertainty region of the 3PBC sources. The upper left spectrum (J0040.5+2542) corresponds to a quasar. The upper middle spectrum (J0838.7+2613) is a quasar spectrum as well. We classified the upper right spectrum (J1041.2+0451) as a Type 2 AGN. The second row shows spectra of Type 2 AGN (left and middle) and quasar (right). The third row shows the spectra of a quasar (left), Type 2 AGN (middle), and Type 1 AGN (right). The last spectrum image on the bottom corresponds to a star-forming galaxy. The main spectral emission and/or absorption features are marked in each figure. We are reporting the Sloan spectra with the same redshift precision since none of the fitting they performed had warnings.}
      \label{fig:spectra}
\end{center}
\end{figure*}

\section{Summary and Conclusions}
\label{sec:summary}

The main goal of the present analysis is to prepare a catalog of candidate soft X-ray \textit{Swift}-XRT counterparts detected in the 0.5--10 keV energy range to list potential targets for the optical spectroscopic campaign, aiming at the classification of the yet unknown sources in the hard X-ray sky so we could obtain a more complete overview of it. We found archival \textit{Swift}-XRT observations for 192 of the 218 3PBC sources marked as unidentified in our previous analyses. Those were the hard X-ray sources lacking an assigned low-energy counterpart. In this work, we searched for possible counterparts at soft X-ray energies for those 192 3PBC sources. If found, we carried out the literature search and multiwavelength analyses as done in paper I.


We found that only in 172 out of 192 sources, there is at least one soft X-ray detected source above our S/N threshold of 3 present in the cleaned and merged event file, and only in 73 of the 3PBC sources we find at least one soft X-ray candidate counterpart detected within the BAT positional uncertainty region. In particular, for 13 3PBC sources, there are multiple detected soft X-ray objects, while all remaining 60 3PBC sources have only a single \textit{Swift}-XRT detected object. Thus, including multiple matches, the total number of \textit{Swift}-XRT detected possible counterparts inside the BAT positional uncertainty, listed in our final catalog, is 93, sampling 73 3PBC hard X-ray sources.

Our X-ray results are in agreement with those achieved simply crossmatching the catalog of 3PBC unidentified sources with the 2SXPS, with only marginal differences, as reported in Section \ref{sec:results} mainly due to (i) longer exposure times and new observations collected after its release that were considered in our analysis and (ii) a small difference in the detection threshold chosen between the two analyses.

We found available spectra in the literature for 10 detected counterparts. For those, we carried the same multifrequency analyses as in paper I. We found four sources to be quasars, four sources to be Type 2 AGN, and one source to be Type 1 AGN, and one star-forming galaxy. The present analyses thus decreased the 218 3PBC unidentified hard X-ray sources sample to 143, which remain unidentified, lacking any low-frequency counterpart. This corresponds to a decrease by a factor of $\sim$\,34\,\%. From the 73 3PBC sources for which we found at least one assigned candidate counterpart, 10 were classified according to our multifrequency criteria, becoming identified. The remaining 65 sources are left as unclassified, indicating that they lack spectroscopic information for their classification and are thus excellent candidates for future spectroscopic follow-up observations.

Along with this publication, we provide a catalog table of all 93 soft X-ray detections and a short table with our classification of the sources with spectra. The soft X-ray sources we found in this analysis can be targets of future spectroscopic campaigns aimed at classifying them to obtain redshift values and confirm that most of those having mid-IR detection are AGNs, as expected by the mid-IR plot.

\begin{acknowledgements}

M. K. and N. W. are supported by the GACR grant 21-13491X. E. B. acknowledges NASA grant 80NSSC21K0653. M. K. was supported by the Italian Government Scholarship issued by the Italian MAECI. This work was partially supported by CONACyT (National Council of Science and Technology) research grants 280789 (Mexico)
Funding for the Sloan Digital Sky Survey V has been provided by the Alfred P. Sloan Foundation, the Heising-Simons Foundation, the National Science Foundation, and the Participating Institutions. SDSS acknowledges support and resources from the Center for High-Performance Computing at the University of Utah. The SDSS web site is \url{www.sdss.org}. SDSS is managed by the Astrophysical Research Consortium for the Participating Institutions of the SDSS Collaboration including the Brazilian Participation Group, the Carnegie Institution for Science, Carnegie Mellon University, Center for Astrophysics | Harvard \& Smithsonian (CfA), the Chilean Participation Group, the French Participation Group, Instituto de Astrofísica de Canarias, The Johns Hopkins University, Kavli Institute for the Physics and Mathematics of the Universe (IPMU) / University of Tokyo, the Korean Participation Group, Lawrence Berkeley National Laboratory, Leibniz Institut für Astrophysik Potsdam (AIP), Max-Planck-Institut für Astronomie (MPIA Heidelberg), Max-Planck-Institut für Astrophysik (MPA Garching), Max-Planck-Institut für Extraterrestrische Physik (MPE), National Astronomical Observatories of China, New Mexico State University, New York University, University of Notre Dame, Observatório Nacional / MCTI, The Ohio State University, Pennsylvania State University, Shanghai Astronomical Observatory, United Kingdom Participation Group, Universidad Nacional Autónoma de México, University of Arizona, University of Colorado Boulder, University of Oxford, University of Portsmouth, University of Utah, University of Virginia, University of Washington, University of Wisconsin, Vanderbilt University, and Yale University.

\end{acknowledgements}

\bibliographystyle{aa}
\bibliography{bibliography}

\begin{thebibliography}{77}
\expandafter\ifx\csname natexlab\endcsname\relax\def\natexlab#1{#1}\fi

\bibitem[{{Baldwin} {et~al.}(1981){Baldwin}, {Phillips}, \&
  {Terlevich}}]{Baldwin1981}
{Baldwin}, J.~A., {Phillips}, M.~M., \& {Terlevich}, R. 1981, \pasp, 93, 5

\bibitem[{{B{\"a}r} {et~al.}(2019){B{\"a}r}, {Trakhtenbrot}, {Oh}, {Koss},
  {Wong}, {Ricci}, {Schawinski}, {Weigel}, {Sartori}, {Ichikawa}, {Secrest},
  {Stern}, {Pacucci}, {Mushotzky}, {Powell}, {Ricci}, {Sani}, {Smith},
  {Harrison}, {Lamperti}, \& {Urry}}]{Bar2019}
{B{\"a}r}, R.~E., {Trakhtenbrot}, B., {Oh}, K., {et~al.} 2019, \mnras, 489,
  3073

\bibitem[{{Barthelmy}(2004)}]{Barthelmy2004}
{Barthelmy}, S.~D. 2004, in Society of Photo-Optical Instrumentation Engineers
  (SPIE) Conference Series, Vol. 5165, X-Ray and Gamma-Ray Instrumentation for
  Astronomy XIII, ed. K.~A. {Flanagan} \& O.~H.~W. {Siegmund}, 175--189

\bibitem[{{Beckmann} {et~al.}(2006){Beckmann}, {Soldi}, {Shrader}, {Gehrels},
  \& {Produit}}]{Beckmann2006}
{Beckmann}, V., {Soldi}, S., {Shrader}, C.~R., {Gehrels}, N., \& {Produit}, N.
  2006, \apj, 652, 126

\bibitem[{{Bennett} {et~al.}(2014){Bennett}, {Larson}, {Weiland}, \&
  {Hinshaw}}]{Bennett2014}
{Bennett}, C.~L., {Larson}, D., {Weiland}, J.~L., \& {Hinshaw}, G. 2014, \apj,
  794, 135

\bibitem[{{Bird} {et~al.}(2016){Bird}, {Bazzano}, {Malizia}, {Fiocchi},
  {Sguera}, {Bassani}, {Hill}, {Ubertini}, \& {Winkler}}]{Bird2016}
{Bird}, A.~J., {Bazzano}, A., {Malizia}, A., {et~al.} 2016, \apjs, 223, 15

\bibitem[{{Blanton} {et~al.}(2017){Blanton}, {Bershady}, {Abolfathi},
  {Albareti}, {Allende Prieto}, {Almeida}, {Alonso-Garc{\'\i}a}, {Anders},
  {Anderson}, {Andrews}, {Aquino-Ort{\'\i}z}, {Arag{\'o}n-Salamanca},
  {Argudo-Fern{\'a}ndez}, {Armengaud}, {Aubourg}, {Avila-Reese}, {Badenes},
  {Bailey}, {Barger}, {Barrera-Ballesteros}, {Bartosz}, {Bates}, {Baumgarten},
  {Bautista}, {Beaton}, {Beers}, {Belfiore}, {Bender}, {Berlind}, {Bernardi},
  {Beutler}, {Bird}, {Bizyaev}, {Blanc}, {Blomqvist}, {Bolton}, {Boquien},
  {Borissova}, {van den Bosch}, {Bovy}, {Brandt}, {Brinkmann}, {Brownstein},
  {Bundy}, {Burgasser}, {Burtin}, {Busca}, {Cappellari}, {Delgado Carigi},
  {Carlberg}, {Carnero Rosell}, {Carrera}, {Chanover}, {Cherinka}, {Cheung},
  {G{\'o}mez Maqueo Chew}, {Chiappini}, {Choi}, {Chojnowski}, {Chuang},
  {Chung}, {Cirolini}, {Clerc}, {Cohen}, {Comparat}, {da Costa}, {Cousinou},
  {Covey}, {Crane}, {Croft}, {Cruz-Gonzalez}, {Garrido Cuadra}, {Cunha},
  {Damke}, {Darling}, {Davies}, {Dawson}, {de la Macorra}, {Dell'Agli}, {De
  Lee}, {Delubac}, {Di Mille}, {Diamond-Stanic}, {Cano-D{\'\i}az}, {Donor},
  {Downes}, {Drory}, {du Mas des Bourboux}, {Duckworth}, {Dwelly}, {Dyer},
  {Ebelke}, {Eigenbrot}, {Eisenstein}, {Emsellem}, {Eracleous}, {Escoffier},
  {Evans}, {Fan}, {Fern{\'a}ndez-Alvar}, {Fernandez-Trincado}, {Feuillet},
  {Finoguenov}, {Fleming}, {Font-Ribera}, {Fredrickson}, {Freischlad},
  {Frinchaboy}, {Fuentes}, {Galbany}, {Garcia-Dias},
  {Garc{\'\i}a-Hern{\'a}ndez}, {Gaulme}, {Geisler}, {Gelfand},
  {Gil-Mar{\'\i}n}, {Gillespie}, {Goddard}, {Gonzalez-Perez}, {Grabowski},
  {Green}, {Grier}, {Gunn}, {Guo}, {Guy}, {Hagen}, {Hahn}, {Hall}, {Harding},
  {Hasselquist}, {Hawley}, {Hearty}, {Gonzalez Hern{\'a}ndez}, {Ho}, {Hogg},
  {Holley-Bockelmann}, {Holtzman}, {Holzer}, {Huehnerhoff}, {Hutchinson},
  {Hwang}, {Ibarra-Medel}, {da Silva Ilha}, {Ivans}, {Ivory}, {Jackson},
  {Jensen}, {Johnson}, {Jones}, {J{\"o}nsson}, {Jullo}, {Kamble}, {Kinemuchi},
  {Kirkby}, {Kitaura}, {Klaene}, {Knapp}, {Kneib}, {Kollmeier}, {Lacerna},
  {Lane}, {Lang}, {Law}, {Lazarz}, {Lee}, {Le Goff}, {Liang}, {Li}, {Li},
  {Lian}, {Lima}, {Lin}, {Lin}, {Bertran de Lis}, {Liu}, {de Icaza Lizaola},
  {Long}, {Lucatello}, {Lundgren}, {MacDonald}, {Deconto Machado}, {MacLeod},
  {Mahadevan}, {Geimba Maia}, {Maiolino}, {Majewski}, {Malanushenko},
  {Malanushenko}, {Manchado}, {Mao}, {Maraston}, {Marques-Chaves}, {Masseron},
  {Masters}, {McBride}, {McDermid}, {McGrath}, {McGreer}, {Medina Pe{\~n}a},
  {Melendez}, {Merloni}, {Merrifield}, {Meszaros}, {Meza}, {Minchev},
  {Minniti}, {Miyaji}, {More}, {Mulchaey}, {M{\"u}ller-S{\'a}nchez}, {Muna},
  {Munoz}, {Myers}, {Nair}, {Nandra}, {Correa do Nascimento}, {Negrete},
  {Ness}, {Newman}, {Nichol}, {Nidever}, {Nitschelm}, {Ntelis}, {O'Connell},
  {Oelkers}, {Oravetz}, {Oravetz}, {Pace}, {Padilla}, {Palanque-Delabrouille},
  {Alonso Palicio}, {Pan}, {Parejko}, {Parikh}, {P{\^a}ris}, {Park}, {Patten},
  {Peirani}, {Pellejero-Ibanez}, {Penny}, {Percival}, {Perez-Fournon},
  {Petitjean}, {Pieri}, {Pinsonneault}, {Pisani}, {Poleski}, {Prada},
  {Prakash}, {Queiroz}, {Raddick}, {Raichoor}, {Barboza Rembold}, {Richstein},
  {Riffel}, {Riffel}, {Rix}, {Robin}, {Rockosi}, {Rodr{\'\i}guez-Torres},
  {Roman-Lopes}, {Rom{\'a}n-Z{\'u}{\~n}iga}, {Rosado}, {Ross}, {Rossi}, {Ruan},
  {Ruggeri}, {Rykoff}, {Salazar-Albornoz}, {Salvato}, {S{\'a}nchez}, {Aguado},
  {S{\'a}nchez-Gallego}, {Santana}, {Santiago}, {Sayres}, {Schiavon}, {da Silva
  Schimoia}, {Schlafly}, {Schlegel}, {Schneider}, {Schultheis}, {Schuster},
  {Schwope}, {Seo}, {Shao}, {Shen}, {Shetrone}, {Shull}, {Simon}, {Skinner},
  {Skrutskie}, {Slosar}, {Smith}, {Sobeck}, {Sobreira}, {Somers}, {Souto},
  {Stark}, {Stassun}, {Stauffer}, {Steinmetz}, {Storchi-Bergmann},
  {Streblyanska}, {Stringfellow}, {Su{\'a}rez}, {Sun}, {Suzuki}, {Szigeti},
  {Taghizadeh-Popp}, {Tang}, {Tao}, {Tayar}, {Tembe}, {Teske}, {Thakar},
  {Thomas}, {Thompson}, {Tinker}, {Tissera}, {Tojeiro}, {Hernandez Toledo}, {de
  la Torre}, {Tremonti}, {Troup}, {Valenzuela}, {Martinez Valpuesta},
  {Vargas-Gonz{\'a}lez}, {Vargas-Maga{\~n}a}, {Vazquez}, {Villanova}, {Vivek},
  {Vogt}, {Wake}, {Walterbos}, {Wang}, {Weaver}, {Weijmans}, {Weinberg},
  {Westfall}, {Whelan}, {Wild}, {Wilson}, {Wood-Vasey}, {Wylezalek}, {Xiao},
  {Yan}, {Yang}, {Ybarra}, {Y{\`e}che}, {Zakamska}, {Zamora}, {Zarrouk},
  {Zasowski}, {Zhang}, {Zhao}, {Zheng}, {Zheng}, {Zhou}, {Zhou}, {Zhu},
  {Zoccali}, \& {Zou}}]{Blanton2017}
{Blanton}, M.~R., {Bershady}, M.~A., {Abolfathi}, B., {et~al.} 2017, \aj, 154,
  28

\bibitem[{{Bottacini} {et~al.}(2012){Bottacini}, {Ajello}, \&
  {Greiner}}]{Bottacini2012}
{Bottacini}, E., {Ajello}, M., \& {Greiner}, J. 2012, \apjs, 201, 34

\bibitem[{{Burrows} {et~al.}(2005){Burrows}, {Hill}, {Nousek}, {Kennea},
  {Wells}, {Osborne}, {Abbey}, {Beardmore}, {Mukerjee}, {Short}, {Chincarini},
  {Campana}, {Citterio}, {Moretti}, {Pagani}, {Tagliaferri}, {Giommi},
  {Capalbi}, {Tamburelli}, {Angelini}, {Cusumano}, {Br{\"a}uninger}, {Burkert},
  \& {Hartner}}]{Burrows2005}
{Burrows}, D.~N., {Hill}, J.~E., {Nousek}, J.~A., {et~al.} 2005, \ssr, 120, 165

\bibitem[{{Churazov} {et~al.}(2007){Churazov}, {Sunyaev}, {Revnivtsev},
  {Sazonov}, {Molkov}, {Grebenev}, {Winkler}, {Parmar}, {Bazzano}, {Falanga},
  {Gros}, {Lebrun}, {Natalucci}, {Ubertini}, {Roques}, {Bouchet}, {Jourdain},
  {Kn{\"o}dlseder}, {Diehl}, {Budtz-Jorgensen}, {Brandt}, {Lund},
  {Westergaard}, {Neronov}, {T{\"u}rler}, {Chernyakova}, {Walter}, {Produit},
  {Mowlavi}, {Mas-Hesse}, {Domingo}, {Gehrels}, {Kuulkers}, {Kretschmar}, \&
  {Schmidt}}]{Churazov2007}
{Churazov}, E., {Sunyaev}, R., {Revnivtsev}, M., {et~al.} 2007, \aap, 467, 529

\bibitem[{{Cusumano} {et~al.}(2010){Cusumano}, {La Parola}, {Segreto},
  {Ferrigno}, {Maselli}, {Sbarufatti}, {Romano}, {Chincarini}, {Giommi},
  {Masetti}, {Moretti}, {Parisi}, \& {Tagliaferri}}]{Cusumano2010}
{Cusumano}, G., {La Parola}, V., {Segreto}, A., {et~al.} 2010, \aap, 524, A64

\bibitem[{{Cutri} {et~al.}(2021){Cutri}, {Wright}, {Conrow}, {Fowler},
  {Eisenhardt}, {Grillmair}, {Kirkpatrick}, {Masci}, {McCallon}, {Wheelock},
  {Fajardo-Acosta}, {Yan}, {Benford}, {Harbut}, {Jarrett}, {Lake}, {Leisawitz},
  {Ressler}, {Stanford}, {Tsai}, {Liu}, {Helou}, {Mainzer}, {Gettngs},
  {Gonzalez}, {Hoffman}, {Marsh}, {Padgett}, {Skrutskie}, {Beck}, {Papin}, \&
  {Wittman}}]{Cutri2021}
{Cutri}, R.~M., {Wright}, E.~L., {Conrow}, T., {et~al.} 2021, VizieR Online
  Data Catalog, II/328

\bibitem[{{D'Abrusco} {et~al.}(2019){D'Abrusco}, {{\'A}lvarez Crespo},
  {Massaro}, {Campana}, {Chavushyan}, {Landoni}, {La Franca}, {Masetti},
  {Milisavljevic}, {Paggi}, {Ricci}, \& {Smith}}]{D'Abrusco2019}
{D'Abrusco}, R., {{\'A}lvarez Crespo}, N., {Massaro}, F., {et~al.} 2019, \apjs,
  242, 4

\bibitem[{{D'Abrusco} {et~al.}(2012){D'Abrusco}, {Massaro}, {Ajello},
  {Grindlay}, {Smith}, \& {Tosti}}]{D'Abrusco2012}
{D'Abrusco}, R., {Massaro}, F., {Ajello}, M., {et~al.} 2012, \apj, 748, 68

\bibitem[{{D'Abrusco} {et~al.}(2014){D'Abrusco}, {Massaro}, {Paggi}, {Smith},
  {Masetti}, {Landoni}, \& {Tosti}}]{D'Abrusco2014}
{D'Abrusco}, R., {Massaro}, F., {Paggi}, A., {et~al.} 2014, \apjs, 215, 14

\bibitem[{{D'Elia} {et~al.}(2013){D'Elia}, {Perri}, {Puccetti}, {Capalbi},
  {Giommi}, {Burrows}, {Campana}, {Tagliaferri}, {Cusumano}, {Evans},
  {Gehrels}, {Kennea}, {Moretti}, {Nousek}, {Osborne}, {Romano}, \&
  {Stratta}}]{delia13}
{D'Elia}, V., {Perri}, M., {Puccetti}, S., {et~al.} 2013, \aap, 551, A142

\bibitem[{{Evans} {et~al.}(2014){Evans}, {Osborne}, {Beardmore}, {Page},
  {Willingale}, {Mountford}, {Pagani}, {Burrows}, {Kennea}, {Perri},
  {Tagliaferri}, \& {Gehrels}}]{evans14}
{Evans}, P.~A., {Osborne}, J.~P., {Beardmore}, A.~P., {et~al.} 2014, \apjs,
  210, 8

\bibitem[{{Evans} {et~al.}(2020){Evans}, {Page}, {Osborne}, {Beardmore},
  {Willingale}, {Burrows}, {Kennea}, {Perri}, {Capalbi}, {Tagliaferri}, \&
  {Cenko}}]{evans20}
{Evans}, P.~A., {Page}, K.~L., {Osborne}, J.~P., {et~al.} 2020, \apjs, 247, 54

\bibitem[{{Forman} {et~al.}(1978){Forman}, {Jones}, {Cominsky}, {Julien},
  {Murray}, {Peters}, {Tananbaum}, \& {Giacconi}}]{Forman1978}
{Forman}, W., {Jones}, C., {Cominsky}, L., {et~al.} 1978, \apjs, 38, 357

\bibitem[{{Gehrels} {et~al.}(2004){Gehrels}, {Chincarini}, {Giommi}, {Mason},
  {Nousek}, {Wells}, {White}, {Barthelmy}, {Burrows}, {Cominsky}, {Hurley},
  {Marshall}, {M{\'e}sz{\'a}ros}, {Roming}, {Angelini}, {Barbier}, {Belloni},
  {Campana}, {Caraveo}, {Chester}, {Citterio}, {Cline}, {Cropper}, {Cummings},
  {Dean}, {Feigelson}, {Fenimore}, {Frail}, {Fruchter}, {Garmire}, {Gendreau},
  {Ghisellini}, {Greiner}, {Hill}, {Hunsberger}, {Krimm}, {Kulkarni}, {Kumar},
  {Lebrun}, {Lloyd-Ronning}, {Markwardt}, {Mattson}, {Mushotzky}, {Norris},
  {Osborne}, {Paczynski}, {Palmer}, {Park}, {Parsons}, {Paul}, {Rees},
  {Reynolds}, {Rhoads}, {Sasseen}, {Schaefer}, {Short}, {Smale}, {Smith},
  {Stella}, {Tagliaferri}, {Takahashi}, {Tashiro}, {Townsley}, {Tueller},
  {Turner}, {Vietri}, {Voges}, {Ward}, {Willingale}, {Zerbi}, \&
  {Zhang}}]{Gehrels2004}
{Gehrels}, N., {Chincarini}, G., {Giommi}, P., {et~al.} 2004, \apj, 611, 1005

\bibitem[{{Giacconi} {et~al.}(1979){Giacconi}, {Branduardi}, {Briel},
  {Epstein}, {Fabricant}, {Feigelson}, {Forman}, {Gorenstein}, {Grindlay},
  {Gursky}, {Harnden}, {Henry}, {Jones}, {Kellogg}, {Koch}, {Murray},
  {Schreier}, {Seward}, {Tananbaum}, {Topka}, {Van Speybroeck}, {Holt},
  {Becker}, {Boldt}, {Serlemitsos}, {Clark}, {Canizares}, {Markert}, {Novick},
  {Helfand}, \& {Long}}]{Giacconi1979}
{Giacconi}, R., {Branduardi}, G., {Briel}, U., {et~al.} 1979, \apj, 230, 540

\bibitem[{{Giacconi} {et~al.}(1971){Giacconi}, {Kellogg}, {Gorenstein},
  {Gursky}, \& {Tananbaum}}]{Giacconi1971}
{Giacconi}, R., {Kellogg}, E., {Gorenstein}, P., {Gursky}, H., \& {Tananbaum},
  H. 1971, \apjl, 165, L27

\bibitem[{{Giommi} {et~al.}(1992{\natexlab{a}}){Giommi}, {Angelini}, {Jacobs},
  \& {Tagliaferri}}]{Giommi1992}
{Giommi}, P., {Angelini}, L., {Jacobs}, P., \& {Tagliaferri}, G.
  1992{\natexlab{a}}, in Astronomical Society of the Pacific Conference Series,
  Vol.~25, Astronomical Data Analysis Software and Systems I, ed. D.~M.
  {Worrall}, C.~{Biemesderfer}, \& J.~{Barnes}, 100

\bibitem[{{Giommi} {et~al.}(1992{\natexlab{b}}){Giommi}, {Angelini}, {Jacobs},
  \& {Tagliaferri}}]{giommi92}
{Giommi}, P., {Angelini}, L., {Jacobs}, P., \& {Tagliaferri}, G.
  1992{\natexlab{b}}, in Astronomical Society of the Pacific Conference Series,
  Vol.~25, Astronomical Data Analysis Software and Systems I, ed. D.~M.
  {Worrall}, C.~{Biemesderfer}, \& J.~{Barnes}, 100

\bibitem[{{Hasinger} {et~al.}(1999){Hasinger}, {Lehmann}, {Giacconi},
  {Schmidt}, {Tr{\"u}mper}, \& {Zamorani}}]{Hasinger1999}
{Hasinger}, G., {Lehmann}, I., {Giacconi}, R., {et~al.} 1999, in Highlights in
  X-ray Astronomy, ed. B.~{Aschenbach} \& M.~J. {Freyberg}, Vol. 272, 199

\bibitem[{{Jansen} {et~al.}(2001){Jansen}, {Lumb}, {Altieri}, {Clavel}, {Ehle},
  {Erd}, {Gabriel}, {Guainazzi}, {Gondoin}, {Much}, {Munoz}, {Santos},
  {Schartel}, {Texier}, \& {Vacanti}}]{Jansen2001}
{Jansen}, F., {Lumb}, D., {Altieri}, B., {et~al.} 2001, \aap, 365, L1

\bibitem[{{Jones} {et~al.}(2009){Jones}, {Read}, {Saunders}, {Colless},
  {Jarrett}, {Parker}, {Fairall}, {Mauch}, {Sadler}, {Watson}, {Burton},
  {Campbell}, {Cass}, {Croom}, {Dawe}, {Fiegert}, {Frankcombe}, {Hartley},
  {Huchra}, {James}, {Kirby}, {Lahav}, {Lucey}, {Mamon}, {Moore}, {Peterson},
  {Prior}, {Proust}, {Russell}, {Safouris}, {Wakamatsu}, {Westra}, \&
  {Williams}}]{Jones2009}
{Jones}, D.~H., {Read}, M.~A., {Saunders}, W., {et~al.} 2009, \mnras, 399, 683

\bibitem[{{Kewley} {et~al.}(2001){Kewley}, {Dopita}, {Sutherland}, {Heisler},
  \& {Trevena}}]{Kewley2001}
{Kewley}, L.~J., {Dopita}, M.~A., {Sutherland}, R.~S., {Heisler}, C.~A., \&
  {Trevena}, J. 2001, \apj, 556, 121

\bibitem[{{Kewley} {et~al.}(2013){Kewley}, {Maier}, {Yabe}, {Ohta}, {Akiyama},
  {Dopita}, \& {Yuan}}]{Kewley2013}
{Kewley}, L.~J., {Maier}, C., {Yabe}, K., {et~al.} 2013, \apjl, 774, L10

\bibitem[{{Kosiba} {et~al.}(2023){Kosiba}, {Pe{\~n}a-Herazo}, {Massaro},
  {Masetti}, {Paggi}, {Chavushyan}, {Bottacini}, \& {Werner}}]{Kosiba2023}
{Kosiba}, M., {Pe{\~n}a-Herazo}, H.~A., {Massaro}, F., {et~al.} 2023, \aap,
  670, A171

\bibitem[{{Koss} {et~al.}(2017){Koss}, {Trakhtenbrot}, {Ricci}, {Lamperti},
  {Oh}, {Berney}, {Schawinski}, {Balokovi{\'c}}, {Baronchelli}, {Crenshaw},
  {Fischer}, {Gehrels}, {Harrison}, {Hashimoto}, {Hogg}, {Ichikawa}, {Masetti},
  {Mushotzky}, {Sartori}, {Stern}, {Treister}, {Ueda}, {Veilleux}, \&
  {Winter}}]{Koss2017}
{Koss}, M., {Trakhtenbrot}, B., {Ricci}, C., {et~al.} 2017, \apj, 850, 74

\bibitem[{{Koss} {et~al.}(2019){Koss}, {Trakhtenbrot}, {Ricci}, {Powell},
  {Urry}, {Stern}, \& {Oh}}]{Koss2019}
{Koss}, M., {Trakhtenbrot}, B., {Ricci}, C., {et~al.} 2019, in American
  Astronomical Society Meeting Abstracts, Vol. 233, American Astronomical
  Society Meeting Abstracts \#233, 431.04

\bibitem[{{Koss} {et~al.}(2022){Koss}, {Ricci}, {Trakhtenbrot}, {Oh}, {den
  Brok}, {Mej{\'\i}a-Restrepo}, {Stern}, {Privon}, {Treister}, {Powell},
  {Mushotzky}, {Bauer}, {Ananna}, {Balokovi{\'c}}, {B{\"a}r}, {Becker},
  {Bessiere}, {Burtscher}, {Caglar}, {Congiu}, {Evans}, {Harrison}, {Heida},
  {Ichikawa}, {Kamraj}, {Lamperti}, {Pacucci}, {Ricci}, {Riffel}, {Rojas},
  {Schawinski}, {Temple}, {Urry}, {Veilleux}, \& {Williams}}]{Koss2022}
{Koss}, M.~J., {Ricci}, C., {Trakhtenbrot}, B., {et~al.} 2022, \apjs, 261, 2

\bibitem[{{Krivonos} {et~al.}(2007){Krivonos}, {Revnivtsev}, {Lutovinov},
  {Sazonov}, {Churazov}, \& {Sunyaev}}]{Krivonos2007b}
{Krivonos}, R., {Revnivtsev}, M., {Lutovinov}, A., {et~al.} 2007, \aap, 475,
  775

\bibitem[{{Krivonos} {et~al.}(2015){Krivonos}, {Tsygankov}, {Lutovinov},
  {Revnivtsev}, {Churazov}, \& {Sunyaev}}]{Krivonos2015}
{Krivonos}, R., {Tsygankov}, S., {Lutovinov}, A., {et~al.} 2015, \mnras, 448,
  3766

\bibitem[{{Krivonos} {et~al.}(2021){Krivonos}, {Bird}, {Churazov}, {Tomsick},
  {Bazzano}, {Beckmann}, {B{\'e}langer}, {Bodaghee}, {Chaty}, {Kuulkers},
  {Lutovinov}, {Malizia}, {Masetti}, {Mereminskiy}, {Sunyaev}, {Tsygankov},
  {Ubertini}, \& {Winkler}}]{Krivonos2021}
{Krivonos}, R.~A., {Bird}, A.~J., {Churazov}, E.~M., {et~al.} 2021, \nar, 92,
  101612

\bibitem[{{Krivonos} {et~al.}(2022){Krivonos}, {Sazonov}, {Kuznetsova},
  {Lutovinov}, {Mereminskiy}, \& {Tsygankov}}]{Krivonos2022}
{Krivonos}, R.~A., {Sazonov}, S.~Y., {Kuznetsova}, E.~A., {et~al.} 2022,
  \mnras, 510, 4796

\bibitem[{{Krivonos} {et~al.}(2017){Krivonos}, {Tsygankov}, {Mereminskiy},
  {Lutovinov}, {Sazonov}, \& {Sunyaev}}]{Krivonos2017}
{Krivonos}, R.~A., {Tsygankov}, S.~S., {Mereminskiy}, I.~A., {et~al.} 2017,
  \mnras, 470, 512

\bibitem[{{Landi} {et~al.}(2017){Landi}, {Bassani}, {Bazzano}, {Bird},
  {Fiocchi}, {Malizia}, {Panessa}, {Sguera}, \& {Ubertini}}]{Landi2017}
{Landi}, R., {Bassani}, L., {Bazzano}, A., {et~al.} 2017, \mnras, 470, 1107

\bibitem[{{Levine} {et~al.}(1984){Levine}, {Lang}, {Lewin}, {Primini},
  {Dobson}, {Doty}, {Hoffman}, {Howe}, {Scheepmaker}, {Wheaton}, {Matteson},
  {Baity}, {Gruber}, {Knight}, {Nolan}, {Pelling}, {Rothschild}, \&
  {Peterson}}]{Levine1984}
{Levine}, A.~M., {Lang}, F.~L., {Lewin}, W.~H.~G., {et~al.} 1984, \apjs, 54,
  581

\bibitem[{{Lien} {et~al.}(2023){Lien}, {Krimm}, {Markwardt}, {Collins},
  {Barthelmy}, {Oh}, {Koss}, {Parsotan}, \& {Cenko}}]{Lien2023}
{Lien}, A., {Krimm}, H., {Markwardt}, C., {et~al.} 2023, in American
  Astronomical Society Meeting Abstracts, Vol.~55, American Astronomical
  Society Meeting Abstracts, 254.07

\bibitem[{{Malizia} {et~al.}(2010){Malizia}, {Bassani}, {Sguera}, {Stephen},
  {Bazzano}, {Fiocchi}, \& {Bird}}]{Malizia2010}
{Malizia}, A., {Bassani}, L., {Sguera}, V., {et~al.} 2010, \mnras, 408, 975

\bibitem[{{Marchesini} {et~al.}(2019{\natexlab{a}}){Marchesini}, {Masetti},
  {Palazzi}, {Chavushyan}, {Jim{\'e}nez-Bail{\'o}n}, {Pati{\~n}o-{\'A}lvarez},
  {Reynaldi}, {Rojas}, {Saviane}, {Andruchow}, {Bassani}, {Bazzano}, {Bird},
  {Malizia}, {Minniti}, {Monaco}, {Stephen}, \& {Ubertini}}]{Marchesini2019}
{Marchesini}, E.~J., {Masetti}, N., {Palazzi}, E., {et~al.} 2019{\natexlab{a}},
  \apss, 364, 153

\bibitem[{{Marchesini} {et~al.}(2020){Marchesini}, {Paggi}, {Massaro},
  {Masetti}, {D'Abrusco}, \& {Andruchow}}]{marchesini20}
{Marchesini}, E.~J., {Paggi}, A., {Massaro}, F., {et~al.} 2020, \aap, 638, A128

\bibitem[{{Marchesini} {et~al.}(2019{\natexlab{b}}){Marchesini}, {Paggi},
  {Massaro}, {Masetti}, {D'Abrusco}, {Andruchow}, \& {de
  Menezes}}]{marchesini19}
{Marchesini}, E.~J., {Paggi}, A., {Massaro}, F., {et~al.} 2019{\natexlab{b}},
  \aap, 631, A150

\bibitem[{{Markwardt} {et~al.}(2005){Markwardt}, {Tueller}, {Skinner},
  {Gehrels}, {Barthelmy}, \& {Mushotzky}}]{Markwardt2005}
{Markwardt}, C.~B., {Tueller}, J., {Skinner}, G.~K., {et~al.} 2005, \apjl, 633,
  L77

\bibitem[{{Marscher} \& {Gear}(1985)}]{Marscher1985}
{Marscher}, A.~P. \& {Gear}, W.~K. 1985, \apj, 298, 114

\bibitem[{{Marscher} \& {Travis}(1996)}]{Marscher1996}
{Marscher}, A.~P. \& {Travis}, J.~P. 1996, \aaps, 120, 537

\bibitem[{{Masetti} {et~al.}(2006{\natexlab{a}}){Masetti}, {Mason}, {Bassani},
  {Bird}, {Maiorano}, {Malizia}, {Palazzi}, {Stephen}, {Bazzano}, {Dean},
  {Ubertini}, \& {Walter}}]{Masetti2006a}
{Masetti}, N., {Mason}, E., {Bassani}, L., {et~al.} 2006{\natexlab{a}}, \aap,
  448, 547

\bibitem[{{Masetti} {et~al.}(2008){Masetti}, {Mason}, {Morelli}, {Cellone},
  {McBride}, {Palazzi}, {Bassani}, {Bazzano}, {Bird}, {Charles}, {Dean},
  {Galaz}, {Gehrels}, {Landi}, {Malizia}, {Minniti}, {Panessa}, {Romero},
  {Stephen}, {Ubertini}, \& {Walter}}]{Masetti2008}
{Masetti}, N., {Mason}, E., {Morelli}, L., {et~al.} 2008, \aap, 482, 113

\bibitem[{{Masetti} {et~al.}(2006{\natexlab{b}}){Masetti}, {Morelli},
  {Palazzi}, {Galaz}, {Bassani}, {Bazzano}, {Bird}, {Dean}, {Israel}, {Landi},
  {Malizia}, {Minniti}, {Schiavone}, {Stephen}, {Ubertini}, \&
  {Walter}}]{Masetti2006b}
{Masetti}, N., {Morelli}, L., {Palazzi}, E., {et~al.} 2006{\natexlab{b}}, \aap,
  459, 21

\bibitem[{{Masetti} {et~al.}(2012){Masetti}, {Parisi},
  {Jim{\'e}nez-Bail{\'o}n}, {Palazzi}, {Chavushyan}, {Bassani}, {Bazzano},
  {Bird}, {Dean}, {Galaz}, {Landi}, {Malizia}, {Minniti}, {Morelli},
  {Schiavone}, {Stephen}, \& {Ubertini}}]{Masetti2012}
{Masetti}, N., {Parisi}, P., {Jim{\'e}nez-Bail{\'o}n}, E., {et~al.} 2012, \aap,
  538, A123

\bibitem[{{Masetti} {et~al.}(2013){Masetti}, {Parisi}, {Palazzi},
  {Jim{\'e}nez-Bail{\'o}n}, {Chavushyan}, {McBride}, {Rojas}, {Steward},
  {Bassani}, {Bazzano}, {Bird}, {Charles}, {Galaz}, {Landi}, {Malizia},
  {Mason}, {Minniti}, {Morelli}, {Schiavone}, {Stephen}, \&
  {Ubertini}}]{Masetti2013}
{Masetti}, N., {Parisi}, P., {Palazzi}, E., {et~al.} 2013, \aap, 556, A120

\bibitem[{{Massaro} {et~al.}(2006){Massaro}, {Tramacere}, {Perri}, {Giommi}, \&
  {Tosti}}]{Massaro2006}
{Massaro}, E., {Tramacere}, A., {Perri}, M., {Giommi}, P., \& {Tosti}, G. 2006,
  \aap, 448, 861

\bibitem[{{Massaro} \& {D'Abrusco}(2016)}]{Massaro2016}
{Massaro}, F. \& {D'Abrusco}, R. 2016, \apj, 827, 67

\bibitem[{{Massaro} {et~al.}(2008{\natexlab{a}}){Massaro}, {Giommi}, {Tosti},
  {Cassetti}, {Nesci}, {Perri}, {Burrows}, \& {Gerehls}}]{massaro08b}
{Massaro}, F., {Giommi}, P., {Tosti}, G., {et~al.} 2008{\natexlab{a}}, \aap,
  489, 1047

\bibitem[{{Massaro} {et~al.}(2008{\natexlab{b}}){Massaro}, {Tramacere},
  {Cavaliere}, {Perri}, \& {Giommi}}]{massaro08a}
{Massaro}, F., {Tramacere}, A., {Cavaliere}, A., {Perri}, M., \& {Giommi}, P.
  2008{\natexlab{b}}, \aap, 478, 395

\bibitem[{{Massaro} {et~al.}(2023{\natexlab{a}}){Massaro}, {White},
  {Garc{\'\i}a-P{\'e}rez}, {Jimenez-Gallardo}, {Capetti}, {Cheung}, {Forman},
  {Mazzucchelli}, {Paggi}, {Nesvadba}, {Madrid}, {Andruchow}, {Cellone},
  {Pe{\~n}a-Herazo}, {Grossov{\'a}}, {Balmaverde}, {Sani}, {Chavushyan},
  {Kraft}, {Reynaldi}, \& {Leto}}]{massaro23}
{Massaro}, F., {White}, S.~V., {Garc{\'\i}a-P{\'e}rez}, A., {et~al.}
  2023{\natexlab{a}}, \apjs, 265, 32

\bibitem[{{Massaro} {et~al.}(2023{\natexlab{b}}){Massaro}, {White}, {Paggi},
  {Jimenez-Gallardo}, {Madrid}, {Mazzucchelli}, {Forman}, {Capetti}, {Leto},
  {Garc{\'\i}a-P{\'e}rez}, {Cheung}, {Chavushyan}, {Nesvadba}, {Andruchow},
  {Pe{\~n}a-Herazo}, {Sani}, {Grossov{\'a}}, {Reynaldi}, {Kraft}, {Balmaverde},
  \& {Cellone}}]{Massaro2023}
{Massaro}, F., {White}, S.~V., {Paggi}, A., {et~al.} 2023{\natexlab{b}}, \apjs,
  268, 32

\bibitem[{{Mereminskiy} {et~al.}(2016){Mereminskiy}, {Krivonos}, {Lutovinov},
  {Sazonov}, {Revnivtsev}, \& {Sunyaev}}]{Mereminskiy2016}
{Mereminskiy}, I.~A., {Krivonos}, R.~A., {Lutovinov}, A.~A., {et~al.} 2016,
  \mnras, 459, 140

\bibitem[{{Nasa High Energy Astrophysics Science Archive Research Center
  (Heasarc)}(2014)}]{HEAsoft2014}
{Nasa High Energy Astrophysics Science Archive Research Center (Heasarc)}.
  2014, {HEAsoft: Unified Release of FTOOLS and XANADU}, Astrophysics Source
  Code Library, record ascl:1408.004

\bibitem[{{Oh} {et~al.}(2018){Oh}, {Koss}, {Markwardt}, {Schawinski},
  {Baumgartner}, {Barthelmy}, {Cenko}, {Gehrels}, {Mushotzky}, {Petulante},
  {Ricci}, {Lien}, \& {Trakhtenbrot}}]{Oh2018}
{Oh}, K., {Koss}, M., {Markwardt}, C.~B., {et~al.} 2018, \apjs, 235, 4

\bibitem[{{Paggi} {et~al.}(2013){Paggi}, {Massaro}, {D'Abrusco}, {Smith},
  {Masetti}, {Giroletti}, {Tosti}, \& {Funk}}]{paggi13}
{Paggi}, A., {Massaro}, F., {D'Abrusco}, R., {et~al.} 2013, \apjs, 209, 9

\bibitem[{{Parisi} {et~al.}(2014){Parisi}, {Masetti}, {Rojas},
  {Jim{\'e}nez-Bail{\'o}n}, {Chavushyan}, {Palazzi}, {Bassani}, {Bazzano},
  {Bird}, {Galaz}, {Minniti}, {Morelli}, \& {Ubertini}}]{Parisi2014}
{Parisi}, P., {Masetti}, N., {Rojas}, A.~F., {et~al.} 2014, \aap, 561, A67

\bibitem[{{Pe{\~n}a-Herazo} {et~al.}(2020){Pe{\~n}a-Herazo},
  {Amaya-Almaz{\'a}n}, {Massaro}, {de Menezes}, {Marchesini}, {Chavushyan},
  {Paggi}, {Landoni}, {Masetti}, {Ricci}, {D'Abrusco}, {Cheung}, {La Franca},
  {Smith}, {Milisavljevic}, {Jim{\'e}nez-Bail{\'o}n}, {Pati{\~n}o-{\'A}lvarez},
  \& {Tosti}}]{Herazo2020}
{Pe{\~n}a-Herazo}, H.~A., {Amaya-Almaz{\'a}n}, R.~A., {Massaro}, F., {et~al.}
  2020, \aap, 643, A103

\bibitem[{{Pe{\~n}a-Herazo} {et~al.}(2022){Pe{\~n}a-Herazo}, {Massaro},
  {Chavushyan}, {Masetti}, {Paggi}, \& {Capetti}}]{Herazo2022}
{Pe{\~n}a-Herazo}, H.~A., {Massaro}, F., {Chavushyan}, V., {et~al.} 2022, \aap,
  659, A32

\bibitem[{{Rojas} {et~al.}(2017){Rojas}, {Masetti}, {Minniti},
  {Jim{\'e}nez-Bail{\'o}n}, {Chavushyan}, {Hau}, {McBride}, {Bassani},
  {Bazzano}, {Bird}, {Galaz}, {Gavignaud}, {Landi}, {Malizia}, {Morelli},
  {Palazzi}, {Pati{\~n}o-{\'A}lvarez}, {Stephen}, \& {Ubertini}}]{Rojas2017}
{Rojas}, A.~F., {Masetti}, N., {Minniti}, D., {et~al.} 2017, \aap, 602, A124

\bibitem[{{Rothschild} {et~al.}(1979){Rothschild}, {Boldt}, {Holt},
  {Serlemitsos}, {Garmire}, {Agrawal}, {Riegler}, {Bowyer}, \&
  {Lampton}}]{Rothschild1979}
{Rothschild}, R., {Boldt}, E., {Holt}, S., {et~al.} 1979, Space Science
  Instrumentation, 4, 269

\bibitem[{{Sazonov} {et~al.}(2007){Sazonov}, {Revnivtsev}, {Krivonos},
  {Churazov}, \& {Sunyaev}}]{Sazonov2007}
{Sazonov}, S., {Revnivtsev}, M., {Krivonos}, R., {Churazov}, E., \& {Sunyaev},
  R. 2007, \aap, 462, 57

\bibitem[{{Smith} {et~al.}(2020){Smith}, {Mushotzky}, {Koss}, {Trakhtenbrot},
  {Ricci}, {Wong}, {Bauer}, {Ricci}, {Vogel}, {Stern}, {Powell}, {Urry},
  {Harrison}, {Mejia-Restrepo}, {Oh}, {Baek}, \& {Chung}}]{Smith2020}
{Smith}, K.~L., {Mushotzky}, R.~F., {Koss}, M., {et~al.} 2020, \mnras, 492,
  4216

\bibitem[{{Strauss} {et~al.}(1992){Strauss}, {Huchra}, {Davis}, {Yahil},
  {Fisher}, \& {Tonry}}]{Strauss1992}
{Strauss}, M.~A., {Huchra}, J.~P., {Davis}, M., {et~al.} 1992, \apjs, 83, 29

\bibitem[{Tomsick {et~al.}(2020)Tomsick, Bodaghee, Chaty, Clavel, Fornasini,
  Hare, Krivonos, Rahoui, \& Rodriguez}]{Tomsick2020}
Tomsick, J.~A., Bodaghee, A., Chaty, S., {et~al.} 2020, The Astrophysical
  Journal, 889, 53

\bibitem[{{Tueller} {et~al.}(2008){Tueller}, {Mushotzky}, {Barthelmy},
  {Cannizzo}, {Gehrels}, {Markwardt}, {Skinner}, \& {Winter}}]{Tueller2008}
{Tueller}, J., {Mushotzky}, R.~F., {Barthelmy}, S., {et~al.} 2008, \apj, 681,
  113

\bibitem[{{Ubertini} {et~al.}(2003){Ubertini}, {Lebrun}, {Di Cocco}, {Bazzano},
  {Bird}, {Broenstad}, {Goldwurm}, {La Rosa}, {Labanti}, {Laurent}, {Mirabel},
  {Quadrini}, {Ramsey}, {Reglero}, {Sabau}, {Sacco}, {Staubert}, {Vigroux},
  {Weisskopf}, \& {Zdziarski}}]{Ubertini2003}
{Ubertini}, P., {Lebrun}, F., {Di Cocco}, G., {et~al.} 2003, \aap, 411, L131

\bibitem[{{Weisskopf} {et~al.}(2000){Weisskopf}, {Tananbaum}, {Van Speybroeck},
  \& {O'Dell}}]{Weisskopf2000}
{Weisskopf}, M.~C., {Tananbaum}, H.~D., {Van Speybroeck}, L.~P., \& {O'Dell},
  S.~L. 2000, in Society of Photo-Optical Instrumentation Engineers (SPIE)
  Conference Series, Vol. 4012, X-Ray Optics, Instruments, and Missions III,
  ed. J.~E. {Truemper} \& B.~{Aschenbach}, 2--16

\bibitem[{{Winkler} {et~al.}(2003){Winkler}, {Courvoisier}, {Di Cocco},
  {Gehrels}, {Gim{\'e}nez}, {Grebenev}, {Hermsen}, {Mas-Hesse}, {Lebrun},
  {Lund}, {Palumbo}, {Paul}, {Roques}, {Schnopper}, {Sch{\"o}nfelder},
  {Sunyaev}, {Teegarden}, {Ubertini}, {Vedrenne}, \& {Dean}}]{Winkler2003}
{Winkler}, C., {Courvoisier}, T.~J.~L., {Di Cocco}, G., {et~al.} 2003, \aap,
  411, L1

\bibitem[{{Wright} {et~al.}(2010){Wright}, {Eisenhardt}, {Mainzer}, {Ressler},
  {Cutri}, {Jarrett}, {Kirkpatrick}, {Padgett}, {McMillan}, {Skrutskie},
  {Stanford}, {Cohen}, {Walker}, {Mather}, {Leisawitz}, {Gautier}, {McLean},
  {Benford}, {Lonsdale}, {Blain}, {Mendez}, {Irace}, {Duval}, {Liu}, {Royer},
  {Heinrichsen}, {Howard}, {Shannon}, {Kendall}, {Walsh}, {Larsen}, {Cardon},
  {Schick}, {Schwalm}, {Abid}, {Fabinsky}, {Naes}, \& {Tsai}}]{Wright2010}
{Wright}, E.~L., {Eisenhardt}, P. R.~M., {Mainzer}, A.~K., {et~al.} 2010, \aj,
  140, 1868

\end{thebibliography}

\appendix


\section{1. Catalog of soft Swift-XRT candidate counterparts of the UHXs}
\label{app:Appendix_1}

Here we report the first 10 lines of the two main tables, showing parameters derived from our analysis, for soft X-ray sources detected by \textit{Swift}-XRT that could be candidate counterparts of the sample of unidentified 3PBC hard X-ray sources, 

Here, we report our final catalog table, showing parameters derived from our analysis for 93 soft X-ray sources detected by \textit{Swift}-XRT that could be candidate counterparts of the sample of unidentified 3PBC hard X-ray sources we analyzed in our study. The catalog table is split into 4 tables due to its size. Tab. \ref{table:main_table_first_part} lists the first 7 columns of the first 50 rows, the Tab. \ref{table:main_table_first_part_last_half} lists the first 7 columns of the second half, the last 43 rows. Tab.\,\ref{table:main_table_second_part} lists the last 8 columns of the first 50 rows, and the Tab.\,\ref{table:main_table_second_part_last_half} lists the remaining 43 rows of the last 8 columns. Together, the catalog table contains 15 columns and 93 rows.

Complementary to the Fig.\,\ref{fig:spectra} and Sec.\,\ref{sec:sdss}, we provide here a table of our spectral classification of the 10 sources for which we found available spectra in the SDSS archive (Tab.\,\ref{table:spectral_classification}).



\begin{table*}[h]

\caption{First fifty lines of our final table of all 93 detected soft X-ray \textit{Swift}-XRT sources within the BAT positional uncertainty in the total of 73 3PBC sources marked as unidentified in our previous analyses (paper I). Here are the first seven columns listing the 3PBC name with '3PBC' omitted to keep the table shorter, the name of an XRT candidate counterpart, which always starts with 'SWXRT' but was omitted here to keep the table smaller, the X-ray detection flag, the right ascension of the centroid, the declination of the centroid, the $ \sigma_{XRT}$ stands for XRT centroid positional uncertainty and $\vartheta_{XRT-BAT}$ is the angular separation between the XRT centroid and BAT centroid. The first row is the header, and the second row shows the units, if applicable.}     
\label{table:main_table_first_part}      

\begin{tabular}{cccrrcc}
 \hline 
 \hline
  3PBC name & XRT detection   & XDF &  RA centroid & DEC centroid &  $\sigma_{XRT}$ & $\vartheta_{XRT-BAT}$  \\
       &                    &     &  hh:mm:ss.ss & dd:mm:ss.ss  &        arcsec &  arcsec             \\

\hline

J0016.7-2611 & J001637.005-261425.118 & x & 00:16:37.14 & -26:14:26.22 & 4.8 & 193.5 \\ 
J0022.2+2539 & J002203.037+254003.324 & m & 00:22:03.09 & 25:40:02.51 & 3.6 & 134.1 \\ 
J0022.2+2539 & J002208.994+254004.792 & m & 00:22:09.44 & 25:39:57.24 & 4.3 & 59.9 \\ 
J0022.2+2539 & J002211.671+253830.384 & m & 00:22:11.73 & 25:38:31.64 & 5.0 & 39.8 \\ 
J0024.1-6823 & J002406.457-682052.549 & x & 00:24:06.47 & -68:20:52.42 & 3.8 & 141.1 \\ 
J0040.5+2542 & J004024.511+254302.390 & x & 00:40:24.5 & 25:43:03.91 & 5.6 & 133.0 \\ 
J0122.0-6105 & J012204.938-610706.365 & x & 01:22:05.08 & -61:07:07.16 & 4.0 & 103.4 \\ 
J0132.5-7426 & J013251.760-742547.319 & x & 01:32:52.0 & -74:25:47.2 & 3.7 & 102.6 \\ 
J0154.1-5034 & J015422.390-503235.190 & x & 01:54:22.34 & -50:32:41.76 & 5.4 & 166.4 \\ 
J0158.9+2644 & J015848.911+264247.120 & x & 01:58:48.68 & 26:42:47.38 & 5.8 & 115.1 \\ 
J0208.0-2904 & J020803.559-290409.558 & x & 02:08:03.71 & -29:04:07.11 & 5.4 & 30.3 \\ 
J0218.5-5005 & J021820.926-500642.137 & m & 02:18:21.06 & -50:06:43.34 & 3.8 & 135.7 \\ 
J0218.5-5005 & J021822.887-500557.535 & m & 02:18:23.02 & -50:05:58.82 & 3.8 & 95.4 \\ 
J0218.5-5005 & J021832.929-500650.270 & m & 02:18:32.95 & -50:06:51.49 & 3.9 & 95.2 \\ 
J0224.8+4525 & J022454.536+452747.735 & x & 02:24:54.54 & 45:27:47.14 & 3.7 & 132.0 \\ 
J0251.6+4335 & J025150.892+433654.523 & x & 02:51:50.83 & 43:36:53.45 & 4.9 & 146.6 \\ 
J0258.8+3545 & J025846.635+354358.068 & x & 02:58:46.87 & 35:43:59.08 & 5.6 & 115.1 \\ 
J0334.2-5201 & J033423.045-520421.808 & x & 03:34:23.01 & -52:04:20.45 & 5.5 & 173.0 \\ 
J0408.0-7912 & J040847.166-791404.530 & x & 04:08:47.44 & -79:14:05.3 & 3.8 & 149.5 \\ 
J0412.2+3051 & J041222.483+305236.417 & x & 04:12:22.48 & 30:52:35.48 & 4.3 & 152.9 \\ 
J0421.1+2603 & J042113.182+260356.044 & x & 04:21:13.26 & 26:03:53.29 & 5.4 & 87.4 \\ 
J0433.2-5107 & J043335.935-510604.141 & x & 04:33:35.84 & -51:06:05.82 & 5.8 & 195.8 \\ 
J0536.1-3205 & J053601.472-320423.246 & m & 05:36:01.33 & -32:04:23.66 & 4.6 & 104.9 \\ 
J0536.1-3205 & J053618.640-320530.341 & m & 05:36:18.7 & -32:05:30.4 & 4.7 & 127.1 \\ 
J0606.8+5651 & J060653.591+565203.444 & x & 06:06:53.36 & 56:52:03.08 & 4.8 & 60.3 \\ 
J0709.5-3538 & J070931.970-353745.100 & m & 07:09:31.99 & -35:37:45.25 & 3.6 & 49.6 \\ 
J0709.5-3538 & J070937.139-353821.510 & m & 07:09:35.92 & -35:38:18.84 & 3.6 & 51.3 \\ 
J0712.1+1541 & J071215.621+153930.156 & x & 07:12:15.71 & 15:39:30.07 & 3.6 & 157.7 \\ 
J0741.6-6724 & J074122.480-672350.269 & x & 07:41:22.24 & -67:23:51.38 & 5.4 & 103.7 \\ 
J0751.7+6449 & J075145.091+644901.618 & x & 07:51:45.14 & 64:49:01.89 & 4.0 & 14.1 \\ 
J0800.7-4308 & J080040.137-431105.341 & m & 08:00:40.16 & -43:11:07.73 & 3.6 & 169.4 \\ 
J0800.7-4308 & J080045.827-430937.585 & m & 08:00:45.86 & -43:09:39.1 & 4.3 & 68.0 \\ 
J0819.2-2509 & J081914.849-251116.953 & m & 08:19:14.91 & -25:11:16.67 & 3.8 & 84.3 \\ 
J0819.2-2509 & J081916.504-250706.396 & m & 08:19:16.51 & -25:07:06.23 & 4.4 & 172.1 \\ 
J0838.7+2613 & J083849.833+261106.829 & x & 08:38:50.05 & 26:11:08.05 & 5.4 & 120.0 \\ 
J0857.2+6703 & J085656.215+670256.800 & m & 08:56:56.35 & 67:02:56.11 & 3.7 & 111.6 \\ 
J0857.2+6703 & J085655.487+670154.512 & m & 08:56:57.42 & 67:01:53.1 & 4.3 & 140.4 \\ 
J0900.4-3333 & J090036.271-333420.119 & x & 09:00:36.35 & -33:34:15.99 & 5.4 & 153.0 \\ 
J0905.4-1502 & J090522.736-150340.357 & m & 09:05:22.53 & -15:03:40.63 & 6.6 & 116.0 \\ 
J0905.4-1502 & J090533.705-145953.170 & m & 09:05:33.88 & -14:59:55.68 & 6.2 & 165.7 \\ 
J0910.9+1033 & J091038.484+103448.245 & x & 09:10:38.48 & 10:34:51.52 & 5.1 & 274.3 \\ 
J1039.4-4903 & J103915.366-490302.019 & x & 10:39:15.35 & -49:03:02.23 & 3.7 & 101.9 \\ 
J1041.2+0451 & J104115.614+045314.284 & x & 10:41:15.74 & 04:53:15.52 & 5.4 & 88.2 \\ 
J1057.1-0323 & J105656.786-032141.218 & x & 10:56:56.76 & -03:21:43.11 & 5.2 & 188.2 \\ 
J1126.8+1851 & J112654.918+184956.112 & x & 11:26:54.87 & 18:49:54.97 & 4.3 & 105.4 \\ 
J1140.8+3611 & J114054.585+360956.220 & x & 11:40:54.62 & 36:09:56.37 & 4.6 & 100.0 \\ 
J1154.2-5018 & J115416.180-501804.204 & x & 11:54:16.23 & -50:18:04.95 & 3.9 & 50.2 \\ 
J1158.4+5704 & J115843.318+570538.699 & x & 11:58:43.38 & 57:05:39.5 & 4.8 & 151.3 \\ 
J1228.0-0926 & J122810.146-092703.684 & x & 12:28:10.2 & -09:27:04.36 & 3.8 & 99.8 \\ 
J1236.2+2216 & J123622.249+221833.024 & x & 12:36:22.36 & 22:18:31.42 & 5.4 & 153.9 \\

\hline
\end{tabular}
\end{table*}

\begin{table*}[h]

\caption{The second half (lines 51-93, including) of our final table of all 93 detected soft X-ray \textit{Swift}-XRT sources within the BAT positional uncertainty in the total of 73 3PBC sources marked as unidentified in our previous analyses (paper I). Here are the first seven columns listing the 3PBC name with '3PBC' omitted to keep the table shorter, the name of an XRT candidate counterpart, which always starts with 'SWXRT' but was omitted here to keep the table smaller, the X-ray detection flag, the right ascension of the centroid, the declination of the centroid, the $ \sigma_{XRT}$ stands for XRT centroid positional uncertainty and $\vartheta_{XRT-BAT}$ is the angular separation between the XRT centroid and BAT centroid. The first row is the header, and the second row shows the units, if applicable.}     
\label{table:main_table_first_part_last_half}      

\begin{tabular}{cccrrcc}
 \hline 
 \hline
  3PBC name & XRT detection   & XDF &  RA centroid & DEC centroid &  $\sigma_{XRT}$ & $\vartheta_{XRT-BAT}$  \\
       &                    &     &  hh:mm:ss.ss & dd:mm:ss.ss  &        arcsec &  arcsec             \\

\hline

J1329.7-1052 & J132949.669-105255.343 & x & 13:29:49.47 & -10:52:54.64 & 5.2 & 53.6 \\ 
J1430.3+2303 & J143008.784+230620.768 & m & 14:30:08.81 & 23:06:20.16 & 3.5 & 219.2 \\ 
J1430.3+2303 & J143010.838+230134.277 & m & 14:30:10.86 & 23:01:33.84 & 3.7 & 184.8 \\ 
J1430.3+2303 & J143016.094+230343.862 & m & 14:30:16.1 & 23:03:42.97 & 3.5 & 56.1 \\ 
J1430.3+2303 & J143024.841+230009.905 & m & 14:30:24.76 & 23:00:09.77 & 3.7 & 226.1 \\ 
J1430.3+2303 & J143035.788+230407.712 & m & 14:30:35.82 & 23:04:06.32 & 3.6 & 217.0 \\ 
J1451.1-8035 & J145054.047-803518.532 & x & 14:50:54.84 & -80:35:21.85 & 4.7 & 40.4 \\ 
J1504.1-6019 & J150416.037-602121.037 & x & 15:04:16.02 & -60:21:21.44 & 4.6 & 123.1 \\ 
J1506.4-2923 & J150632.003-292237.960 & x & 15:06:32.14 & -29:22:35.8 & 6.0 & 112.0 \\ 
J1529.1-6523 & J152930.220-652229.724 & x & 15:29:30.55 & -65:22:29.86 & 5.0 & 162.5 \\ 
J1541.6+1113 & J154132.811+111340.924 & x & 15:41:32.8 & 11:13:40.28 & 4.4 & 114.2 \\ 
J1547.3+3146 & J154706.288+314557.923 & x & 15:47:06.24 & 31:45:57.8 & 5.6 & 171.6 \\ 
J1619.9+6510 & J161956.364+651301.448 & x & 16:19:56.37 & 65:12:59.97 & 5.2 & 168.1 \\ 
J1655.7-4958 & J165551.841-495732.542 & x & 16:55:51.93 & -49:57:32.35 & 4.1 & 90.0 \\ 
J1701.2-4212 & J170117.822-421335.370 & x & 17:01:17.96 & -42:13:35.62 & 4.1 & 101.4 \\ 
J1732.0-3439 & J173215.277-343943.796 & m & 17:32:15.44 & -34:39:46.06 & 4.1 & 148.1 \\ 
J1732.0-3439 & J173217.430-343911.830 & m & 17:32:17.34 & -34:39:13.7 & 4.1 & 169.6 \\ 
J1735.9-1528 & J173600.799-152945.813 & x & 17:36:00.8 & -15:29:46.09 & 3.6 & 104.9 \\ 
J1742.4+1502 & J174230.735+150028.760 & x & 17:42:30.83 & 15:00:27.12 & 5.1 & 147.7 \\ 
J1757.9+0427 & J175757.964+042729.167 & x & 17:57:57.89 & 04:27:31.13 & 6.0 & 51.1 \\ 
J1807.3-5935 & J180715.465-593630.848 & x & 18:07:15.54 & -59:36:32.54 & 4.3 & 83.7 \\ 
J1846.1-0226 & J184610.858-022451.834 & m & 18:46:10.91 & -02:24:53.32 & 3.8 & 102.8 \\ 
J1846.1-0226 & J184611.124-022822.595 & m & 18:46:11.35 & -02:28:24.34 & 3.8 & 110.0 \\ 
J1846.1-0226 & J184617.093-022753.462 & m & 18:46:17.09 & -02:27:53.39 & 3.8 & 109.5 \\ 
J1851.8+1846 & J185152.906+184811.717 & x & 18:51:52.96 & 18:48:11.34 & 4.2 & 120.2 \\ 
J1854.4-3436 & J185421.922-343641.701 & x & 18:54:21.88 & -34:36:38.85 & 5.8 & 54.6 \\ 
J1856.5+2836 & J185627.117+283809.639 & m & 18:56:27.06 & 28:38:09.47 & 6.2 & 117.1 \\ 
J1856.5+2836 & J185632.226+283626.262 & m & 18:56:32.15 & 28:36:27.37 & 4.4 & 25.2 \\ 
J1856.5+2836 & J185634.697+283529.554 & m & 18:56:34.74 & 28:35:29.05 & 4.1 & 79.8 \\ 
J1911.4+1412 & J191124.874+141143.793 & x & 19:11:24.9 & 14:11:43.93 & 3.9 & 95.2 \\ 
J1929.5-5606 & J192933.020-560342.697 & x & 19:29:33.02 & -56:03:43.43 & 4.3 & 161.8 \\ 
J1952.3+3803 & J195225.218+380027.051 & x & 19:52:25.27 & 38:00:26.57 & 4.0 & 184.5 \\ 
J1958.2+1941 & J195815.495+194131.811 & x & 19:58:15.57 & 19:41:31.17 & 4.4 & 41.9 \\ 
J2028.0+2221 & J202807.882+222324.863 & x & 20:28:07.87 & 22:23:24.23 & 4.2 & 124.9 \\ 
J2030.1+7608 & J202920.605+760812.618 & x & 20:29:20.78 & 76:08:12.69 & 3.9 & 171.3 \\ 
J2030.8+3833 & J203055.241+383346.714 & x & 20:30:55.21 & 38:33:45.56 & 4.2 & 28.3 \\ 
J2136.1+2002 & J213615.094+200208.011 & x & 21:36:15.16 & 20:02:07.27 & 3.8 & 51.5 \\ 
J2155.3+6204 & J215515.232+620650.707 & x & 21:55:15.39 & 62:06:51.22 & 4.2 & 159.6 \\ 
J2155.1+7017 & J215532.508+701845.088 & x & 21:55:32.58 & 70:18:44.61 & 3.7 & 149.5 \\ 
J2201.4+7546 & J220149.675+754429.190 & x & 22:01:49.87 & 75:44:29.72 & 3.9 & 138.5 \\ 
J2238.8+4050 & J223856.644+405141.294 & x & 22:38:56.86 & 40:51:42.62 & 4.7 & 70.1 \\ 
J2313.3-3402 & J231313.188-340056.922 & m & 23:13:13.24 & -34:00:57.52 & 4.1 & 133.1 \\ 
J2313.3-3402 & J231336.783-340305.733 & m & 23:13:36.79 & -34:03:03.43 & 4.8 & 198.0 \\ 

\hline
\end{tabular}
\end{table*}

\begin{table*}[h]

\caption{First fifty lines of our final table of all 93 detected soft X-ray \textit{Swift}-XRT sources within the BAT positional uncertainty in the total of 73 3PBC sources marked as unidentified in our previous analyses (paper I). Here are the last eight columns list the following parameters: the count rate, the delta of the count rate, Signal to Noise (S/N) in full band ([0.5, 10] keV), hard band ([2, 10] keV) and soft band ([0.5, 2] keV), the number of observations, N$_{obs}$, the approximated total integrated exposure time, T$_{exp}$ and the name of the source if found in the WISE catalog. The first row is the header, and the second row shows the units, if applicable. The S/N threshold for the full band is 3, while we also report hard and soft bands but with a threshold of 1. In the case where no S/N was reported in either the hard or soft band, the S/N was below the threshold of 1 in that particular band.}
\label{table:main_table_second_part}     

\begin{tabular}{rrccccrc}
 \hline 
 \hline
 count rate & $\delta_{count\,rate}$ & S/N full & S/N hard & S/N soft & N$_{obs}$ & T$_{exp}$ & WISE Name \\
 $photons/s$ & $photons/s$ & [0.5, 10] keV & [2, 10] keV & [0.5, 2] keV & & s & \\
\hline


3.33e-03 & 7.10e-04 & 4.7 &  & 3.9 & 4 & 8718 & J001637.05-261426.7 \\ 
1.58e-01 & 4.30e-03 & 37.1 & 22.4 & 31.4 & 6 & 12211 & J002203.09+254003.2 \\ 
1.35e-03 & 4.30e-04 & 3.1 & 2.4 &  & 6 & 12211 &  \\ 
2.63e-03 & 5.60e-04 & 4.7 & 2.4 & 3.8 & 6 & 12211 &  \\ 
4.77e-02 & 3.50e-03 & 13.8 & 7.6 & 11.9 & 5 & 5326 & J002406.72-682054.5 \\ 
2.89e-03 & 9.00e-04 & 3.2 &  & 3.0 & 5 & 4596 & J004024.38+254303.4 \\ 
1.90e-01 & 1.60e-02 & 11.9 & 6.6 & 10.3 & 1 & 975 & J012205.12-610705.0 \\ 
2.69e-02 & 1.50e-03 & 18.3 & 14.2 & 11.5 & 5 & 15292 & J013251.50-742545.4 \\ 
3.68e-03 & 9.10e-04 & 4.0 &  & 3.6 & 4 & 6035 & J015422.40-503240.0 \\ 
1.78e-03 & 5.70e-04 & 3.2 &  & 2.9 & 5 & 8247 & J015848.81+264250.6 \\ 
4.08e-03 & 9.10e-04 & 4.5 & 3.9 &  & 3 & 6619 & J020803.46-290408.3 \\ 
2.39e-04 & 5.50e-05 & 4.4 & 3.6 & 3.0 & 14 & 166916 &  \\ 
1.98e-04 & 5.20e-05 & 3.8 & 3.2 &  & 14 & 166916 & J021822.70-500557.5 \\ 
4.35e-04 & 7.40e-05 & 5.9 & 3.9 & 4.6 & 14 & 166916 &  \\ 
1.13e-01 & 6.00e-03 & 19.0 & 15.1 & 14.3 & 4 & 4440 & J022454.49+452747.1 \\ 
3.69e-02 & 5.50e-03 & 6.7 & 5.1 & 4.8 & 1 & 1507 & J025150.95+433653.5 \\ 
5.27e-03 & 1.20e-03 & 4.3 & 4.0 &  & 5 & 4651 & J025847.30+354357.3 \\ 
2.92e-03 & 8.80e-04 & 3.3 &  & 3.4 & 3 & 5549 & J033423.01-520416.0 \\ 
2.48e-02 & 1.90e-03 & 13.2 & 12.8 & 5.5 & 9 & 10395 & J040847.92-791405.6 \\ 
4.95e-02 & 5.60e-03 & 8.9 & 6.1 & 6.8 & 2 & 2086 & J041222.58+305234.4 \\ 
1.92e-03 & 5.90e-04 & 3.3 &  & 3.0 & 4 & 8089 & J042113.19+260352.6 \\ 
3.80e-03 & 1.10e-03 & 3.4 &  & 2.9 & 3 & 4345 & J043336.20-510607.9 \\ 
1.03e-03 & 2.60e-04 & 4.0 & 2.6 & 3.3 & 2 & 24394 &  \\ 
1.27e-03 & 3.10e-04 & 4.1 &  & 3.5 & 2 & 24394 & J053618.88-320533.0 \\ 
9.32e-03 & 1.60e-03 & 5.9 & 5.8 &  & 4 & 4847 & J060653.35+565204.1 \\ 
1.76e-01 & 6.20e-03 & 28.6 & 18.7 & 24.2 & 4 & 6338 & J070932.05-353746.5 \\ 
3.02e-03 & 8.40e-04 & 3.6 &  &  & 4 & 6338 &  \\ 
1.56e-01 & 5.90e-03 & 26.5 & 17.2 & 21.5 & 7 & 6504 & J071215.60+153930.1 \\ 
3.50e-03 & 8.60e-04 & 4.1 &  & 3.4 & 6 & 7454 & J074121.92-672352.0 \\ 
4.56e-02 & 4.00e-03 & 11.4 & 7.7 & 8.6 & 5 & 3861 & J075145.40+644903.0 \\ 
6.51e-02 & 2.40e-03 & 27.6 & 22.5 & 16.1 & 4 & 13050 & J080039.96-431107.2 \\ 
3.63e-03 & 6.20e-04 & 5.8 & 3.7 & 4.6 & 4 & 13050 & J080045.83-430939.3 \\ 
4.42e-02 & 3.20e-03 & 14.0 & 5.0 & 13.1 & 4 & 6692 & J081914.73-251116.6 \\ 
1.13e-02 & 1.50e-03 & 7.4 & 5.0 & 5.8 & 4 & 6692 & J081916.20-250706.4 \\ 
4.52e-03 & 1.10e-03 & 4.0 &  & 3.8 & 3 & 4949 & J083850.15+261105.5 \\ 
8.52e-02 & 4.80e-03 & 17.6 & 10.5 & 15.4 & 5 & 4992 & J085656.49+670257.3 \\ 
2.67e-03 & 8.70e-04 & 3.1 &  &  & 5 & 4992 &  \\ 
4.41e-03 & 1.20e-03 & 3.6 & 2.9 &  & 1 & 4272 & J090036.26-333415.1 \\ 
1.66e-02 & 4.40e-03 & 3.8 &  & 4.0 & 2 & 1158 & J090522.48-150344.6 \\ 
1.77e-02 & 4.50e-03 & 3.9 & 3.5 &  & 2 & 1158 & J090533.64-145956.2 \\ 
1.98e-03 & 5.50e-04 & 3.6 &  & 3.6 & 13 & 9831 & J091038.47+103450.4 \\ 
8.41e-02 & 4.50e-03 & 18.7 & 13.0 & 14.5 & 1 & 5483 & J103915.25-490303.3 \\ 
1.04e-02 & 2.00e-03 & 5.2 & 5.3 &  & 3 & 3570 & J104115.62+045313.8 \\ 
2.38e-03 & 7.30e-04 & 3.3 &  & 3.3 & 14 & 6434 & J105656.82-032143.0 \\ 
2.07e-02 & 2.30e-03 & 8.8 & 8.8 &  & 8 & 5127 & J112655.08+184957.2 \\ 
1.38e-02 & 2.00e-03 & 6.9 & 7.1 &  & 8 & 4533 & J114054.57+360957.2 \\ 
3.03e-02 & 2.30e-03 & 13.4 & 9.4 & 10.4 & 10 & 9089 &  \\ 
4.01e-03 & 7.30e-04 & 5.5 & 2.9 & 4.5 & 10 & 10062 & J115843.34+570536.5 \\ 
2.69e-02 & 1.60e-03 & 16.3 & 15.4 & 5.6 & 10 & 11616 & J122810.13-092703.4 \\ 
2.46e-03 & 6.90e-04 & 3.6 &  & 3.6 & 8 & 8292 & J123622.59+221834.6 \\

\hline
\end{tabular}
\end{table*}

\begin{table*}[h]

\caption{The second half (lines 51-93, including) of our final table of all 93 detected soft X-ray \textit{Swift}-XRT sources within the BAT positional uncertainty in the total of 73 3PBC sources marked as unidentified in our previous analyses (paper I). Here are the last eight columns list the following parameters: the count rate, the delta of the count rate, Signal to Noise (S/N) in full band ([0.5, 10] keV), hard band ([2, 10] keV) and soft band ([0.5, 2] keV), the number of observations, N$_{obs}$, the approximated total integrated exposure time, T$_{exp}$ and the name of the source if found in the WISE catalog. The first row is the header, and the second row shows the units, if applicable. The S/N threshold for the full band is 3, while we also report hard and soft bands but with a threshold of 1. In the case where no S/N was reported in either the hard or soft band, the S/N was below the threshold of 1 in that particular band.}
\label{table:main_table_second_part_last_half}     

\begin{tabular}{rrccccrc}
 \hline 
 \hline
 count rate & $\delta_{count\,rate}$ & S/N full & S/N hard & S/N soft & N$_{obs}$ & T$_{exp}$ & WISE Name \\
 $photons/s$ & $photons/s$ & [0.5, 10] keV & [2, 10] keV & [0.5, 2] keV & & s & \\
\hline

3.29e-03 & 7.70e-04 & 4.3 & 3.9 &  & 20 & 8500 & J132949.54-105255.3 \\ 
3.10e-02 & 5.70e-04 & 54.4 & 38.7 & 42.3 & 118 & 150511 & J143008.64+230621.6 \\ 
4.50e-04 & 8.20e-05 & 5.5 &  & 5.4 & 118 & 150511 & J143010.96+230134.8 \\ 
6.15e-02 & 8.10e-04 & 76.3 & 47.4 & 61.0 & 118 & 150511 & J143016.03+230344.2 \\ 
3.66e-04 & 7.40e-05 & 5.0 & 3.1 & 4.0 & 118 & 150511 & J143024.88+230012.7 \\ 
1.18e-03 & 1.10e-04 & 10.4 & 5.8 & 8.4 & 118 & 150511 & J143035.69+230406.5 \\ 
1.67e-03 & 4.50e-04 & 3.7 &  & 3.2 & 16 & 11591 & J145054.14-803520.7 \\ 
2.79e-02 & 3.60e-03 & 7.8 & 6.6 & 4.1 & 4 & 2683 &  \\ 
4.15e-03 & 1.20e-03 & 3.5 &  & 3.0 & 7 & 3999 & J150631.92-292235.4 \\ 
3.55e-03 & 1.10e-03 & 3.3 & 2.8 &  & 4 & 4433 & J152931.06-652229.9 \\ 
1.56e-03 & 3.30e-04 & 4.7 &  & 4.5 & 16 & 22633 & J154132.85+111339.7 \\ 
4.11e-03 & 1.00e-03 & 3.9 & 3.0 & 2.6 & 4 & 4746 & J154706.47+314558.5 \\ 
2.95e-03 & 6.80e-04 & 4.3 & 2.8 & 3.3 & 7 & 10074 & J161955.84+651301.0 \\ 
3.42e-02 & 3.20e-03 & 10.7 & 9.4 & 6.0 & 4 & 4232 & J165551.96-495732.2 \\ 
9.44e-03 & 1.10e-03 & 8.4 & 8.3 &  & 8 & 9716 &  \\ 
4.43e-03 & 1.30e-03 & 3.3 &  & 2.8 & 4 & 4187 &  \\ 
8.67e-03 & 1.80e-03 & 4.7 & 4.1 & 3.4 & 4 & 4187 &  \\ 
1.72e-01 & 6.80e-03 & 25.3 & 17.5 & 20.7 & 3 & 5516 & J173600.69-152946.6 \\ 
3.25e-03 & 1.00e-03 & 3.2 &  & 2.5 & 4 & 4400 &  \\ 
3.03e-03 & 9.10e-04 & 3.3 & 2.2 & 2.5 & 2 & 5586 & J175757.68+042735.3 \\ 
3.59e-03 & 1.00e-03 & 3.6 &  & 3.8 & 4 & 5105 & J180715.81-593632.5 \\ 
4.82e-04 & 1.20e-04 & 4.2 &  & 3.7 & 81 & 83400 &  \\ 
5.52e-04 & 1.10e-04 & 4.9 & 4.2 & 3.0 & 81 & 83400 &  \\ 
5.01e-04 & 1.10e-04 & 4.5 &  & 4.0 & 81 & 83400 &  \\ 
2.41e-02 & 2.50e-03 & 9.7 & 9.0 & 4.5 & 4 & 5436 & J185152.92+184812.2 \\ 
3.80e-03 & 1.20e-03 & 3.2 & 3.3 &  & 4 & 3773 & J185422.21-343640.3 \\ 
4.11e-03 & 1.10e-03 & 3.6 &  & 3.8 & 1 & 5050 & J185626.89+283809.3 \\ 
3.44e-03 & 9.90e-04 & 3.5 & 2.8 & 2.4 & 1 & 5050 & J185632.13+283628.8 \\ 
2.65e-02 & 2.70e-03 & 9.9 & 9.9 &  & 1 & 5050 & J185634.58+283531.3 \\ 
5.32e-02 & 4.00e-03 & 13.4 & 10.8 & 9.1 & 6 & 4428 &  \\ 
2.61e-02 & 2.80e-03 & 9.3 & 6.1 & 7.0 & 5 & 4282 & J192933.07-560343.1 \\ 
1.38e-01 & 1.20e-02 & 11.6 & 7.6 & 9.0 & 1 & 1299 & J195225.09+380026.8 \\ 
2.16e-02 & 2.60e-03 & 8.3 & 8.6 &  & 4 & 4408 & J195815.52+194131.0 \\ 
3.04e-02 & 3.20e-03 & 9.4 & 6.8 & 7.3 & 2 & 3866 & J202807.84+222324.7 \\ 
4.83e-02 & 3.90e-03 & 12.5 & 9.0 & 9.6 & 5 & 4644 & J202920.02+760810.4 \\ 
2.10e-02 & 2.30e-03 & 9.3 & 9.0 & 2.9 & 6 & 5499 &  \\ 
8.60e-02 & 5.00e-03 & 17.2 & 10.7 & 13.4 & 4 & 4165 & J213615.30+200207.2 \\ 
2.72e-02 & 2.60e-03 & 10.4 & 6.6 & 8.0 & 2 & 4583 &  \\ 
9.21e-02 & 5.20e-03 & 17.8 & 12.9 & 12.9 & 3 & 4844 & J215532.33+701844.5 \\ 
4.78e-02 & 3.60e-03 & 13.2 & 10.2 & 8.6 & 3 & 4395 & J220149.81+754428.7 \\ 
1.50e-02 & 2.10e-03 & 7.0 & 4.8 & 5.1 & 5 & 4323 & J223857.02+405141.9 \\ 
6.73e-03 & 8.00e-04 & 8.4 & 4.4 & 7.5 & 10 & 23147 & J231313.21-340056.2 \\ 
1.88e-03 & 5.20e-04 & 3.6 & 2.3 & 2.8 & 10 & 23147 & J231337.01-340302.1 \\ 

\hline
\end{tabular}
\end{table*}

 \begin{table}[h]
\caption{Multiwavelength classification of the ten counterparts with available spectra in the SDSS. There are three sources with the same 3PBC name, each corresponding to a different \textit{Swift}-XRT soft X-ray detection.}     
\label{table:spectral_classification}      

\begin{tabular}{cccc}
 \hline 
 \hline
  3PBC name    & SDSS name   & $z$ &  Class \\
\hline
3PBC J0040.5+2542 & J004024.39+254303.4 & 1.0015 & QSO \\ 
3PBC J0838.7+2613 & J083850.15+261105.4 & 1.6139 & QSO \\ 
3PBC J1041.2+0451 & J104115.61+045313.8 & 0.0683 & Sy 2 \\ 
3PBC J1126.8+1851 & J112655.08+184957.4 & 0.0188 & Sy 2 \\ 
3PBC J1140.8+3611 & J114054.58+360957.0 & 0.0701 & Sy 2 \\ 
3PBC J1158.4+5704 & J115843.32+570536.6 & 1.1202 & QSO \\ 
3PBC J1236.2+2216 & J123622.60+221834.8 & 1.8559 & QSO \\ 
3PBC J1430.3+2303 & J143016.05+230344.4 & 0.0811 & Sy 2 \\ 
3PBC J1430.3+2303 & J143008.64+230621.5 & 0.0791 & Sy 1 \\ 
3PBC J1430.3+2303 & J143010.96+230134.7 & 0.0173 & Sy 2\\  
\hline
\end{tabular}
\end{table}


\section{2. Images of all x flagged 3PBX sources}
\label{app:Appendix_x_flag_3PBC_images}

This section shows images of all the 3PBC sources with exactly one soft \textit{Swift}-XRT detection inside the BAT positional uncertainty region (red dashed circle). The green cross, if present, indicated that the source has a WISE counterpart. The black circle only highlights the position of the \textit{Swift}-XRT detection, not its positional uncertainty region. One exception is 3PBCJ1730.0-3436, which has a cyan circle instead of black because the black would not be visible on top of the source's emission.

\begin{figure*}[!th]
\begin{center}
    \includegraphics[height=4.2cm,width=6cm,angle=0]{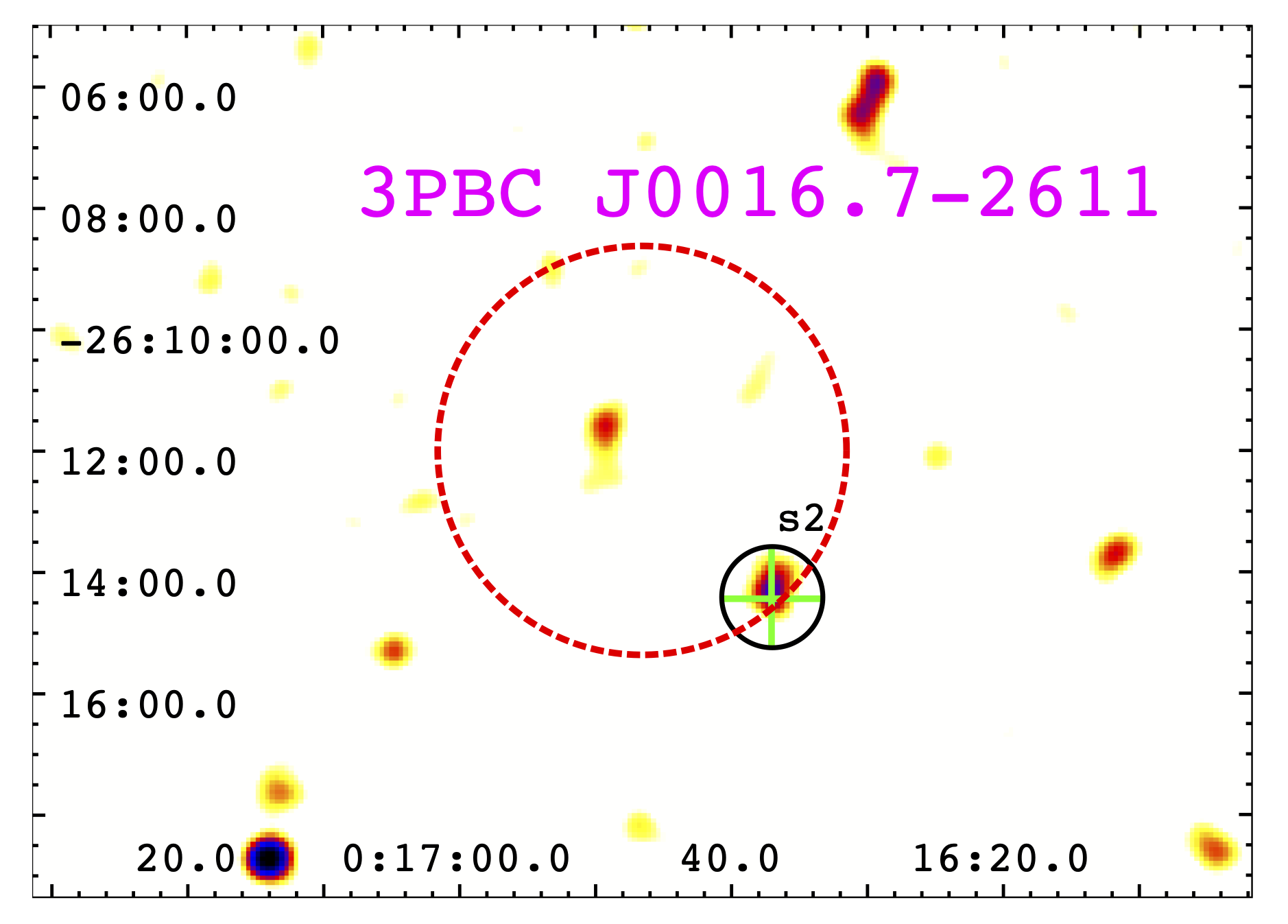}
    \includegraphics[height=4.2cm,width=6cm,angle=0]{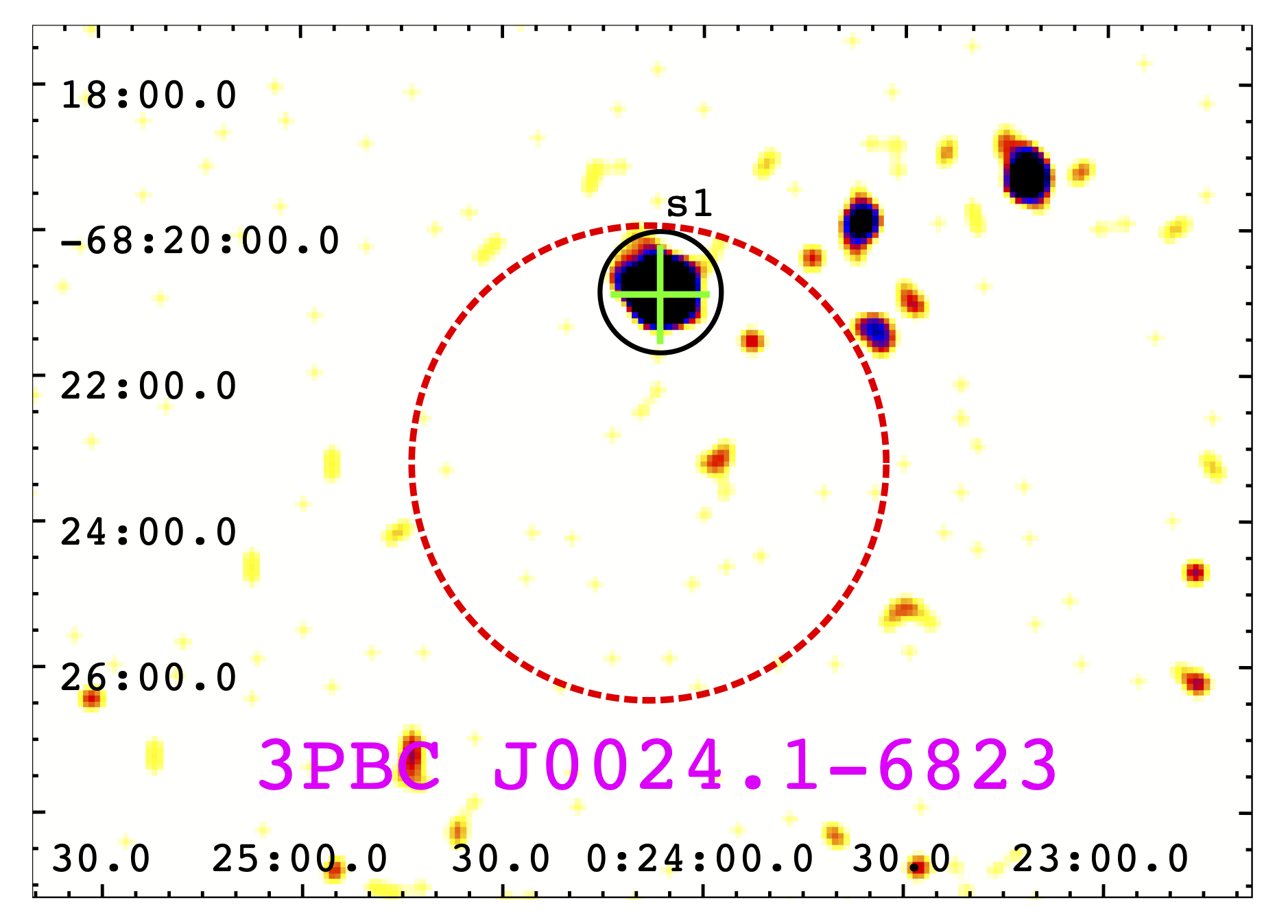}
    \includegraphics[height=4.2cm,width=6cm,angle=0]{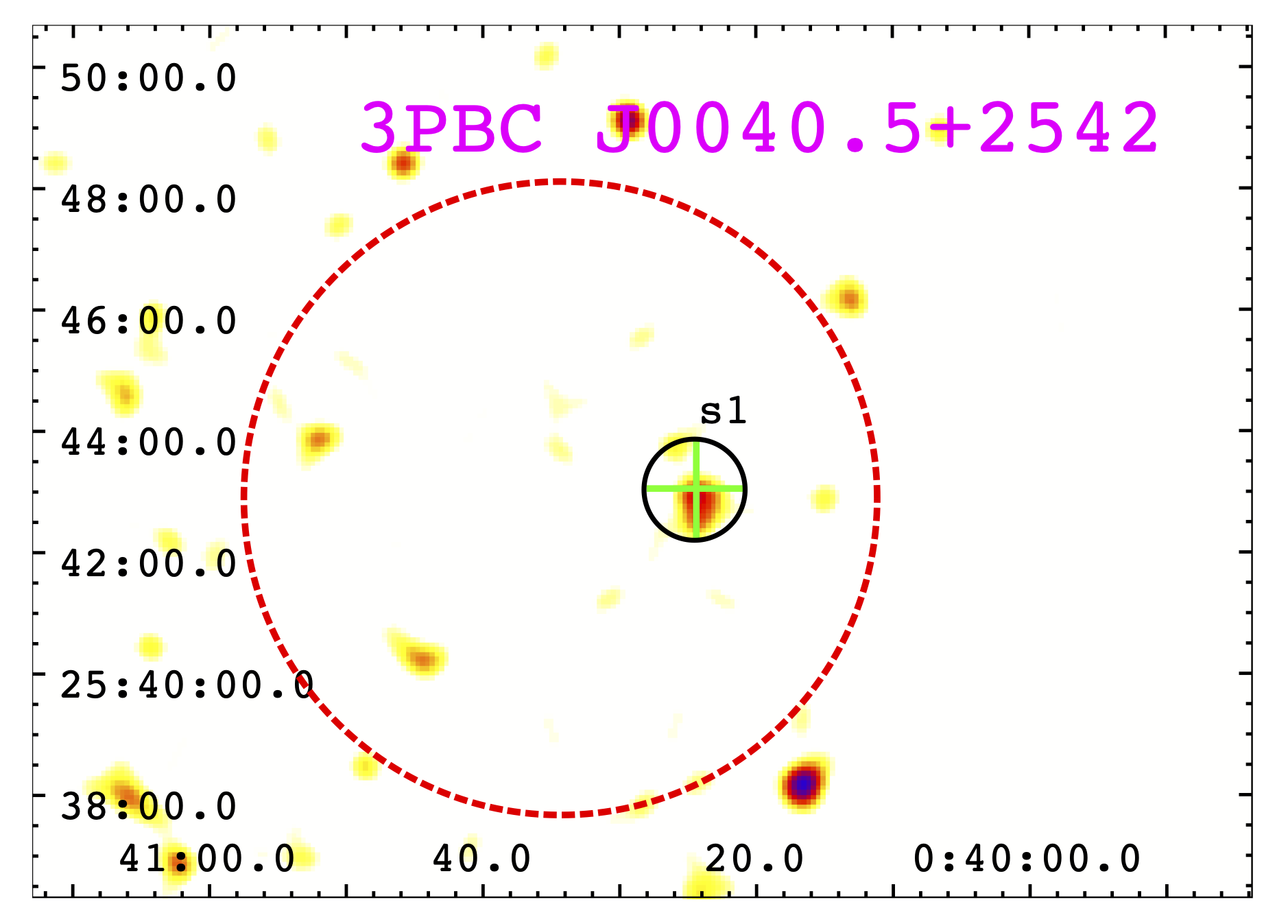}
    
    \includegraphics[height=4.2cm,width=6cm,angle=0]{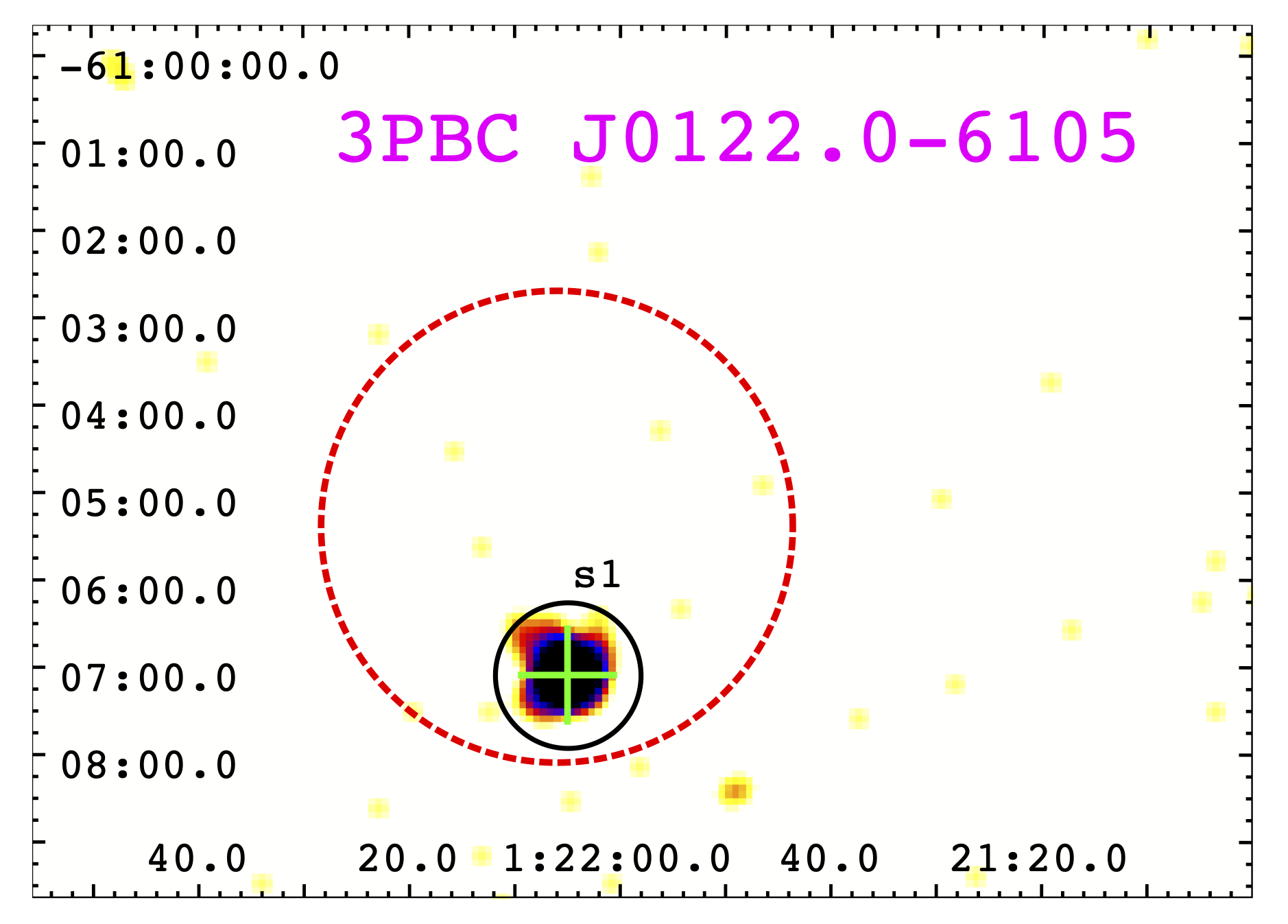}
    \includegraphics[height=4.2cm,width=6cm,angle=0]{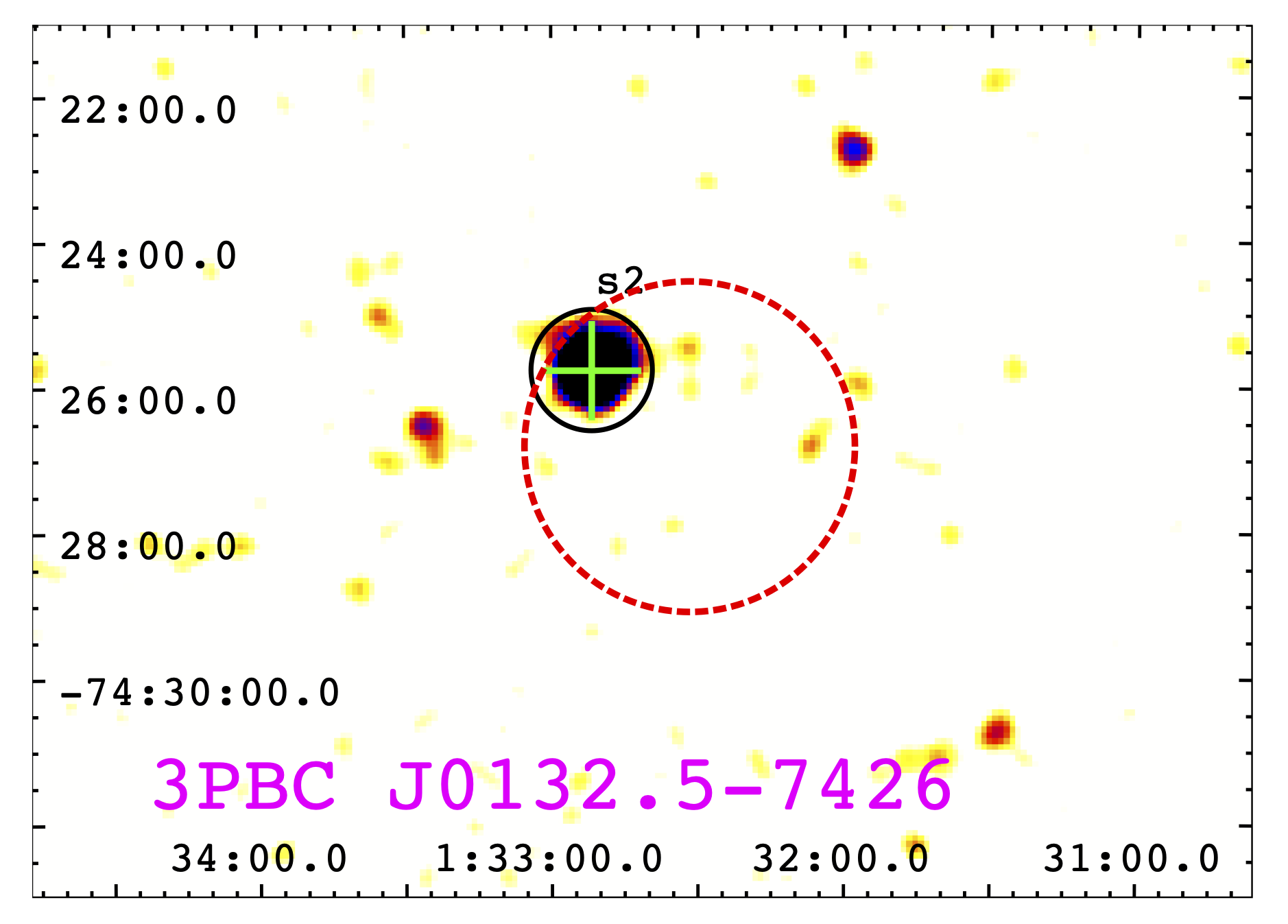}
    \includegraphics[height=4.2cm,width=6cm,angle=0]{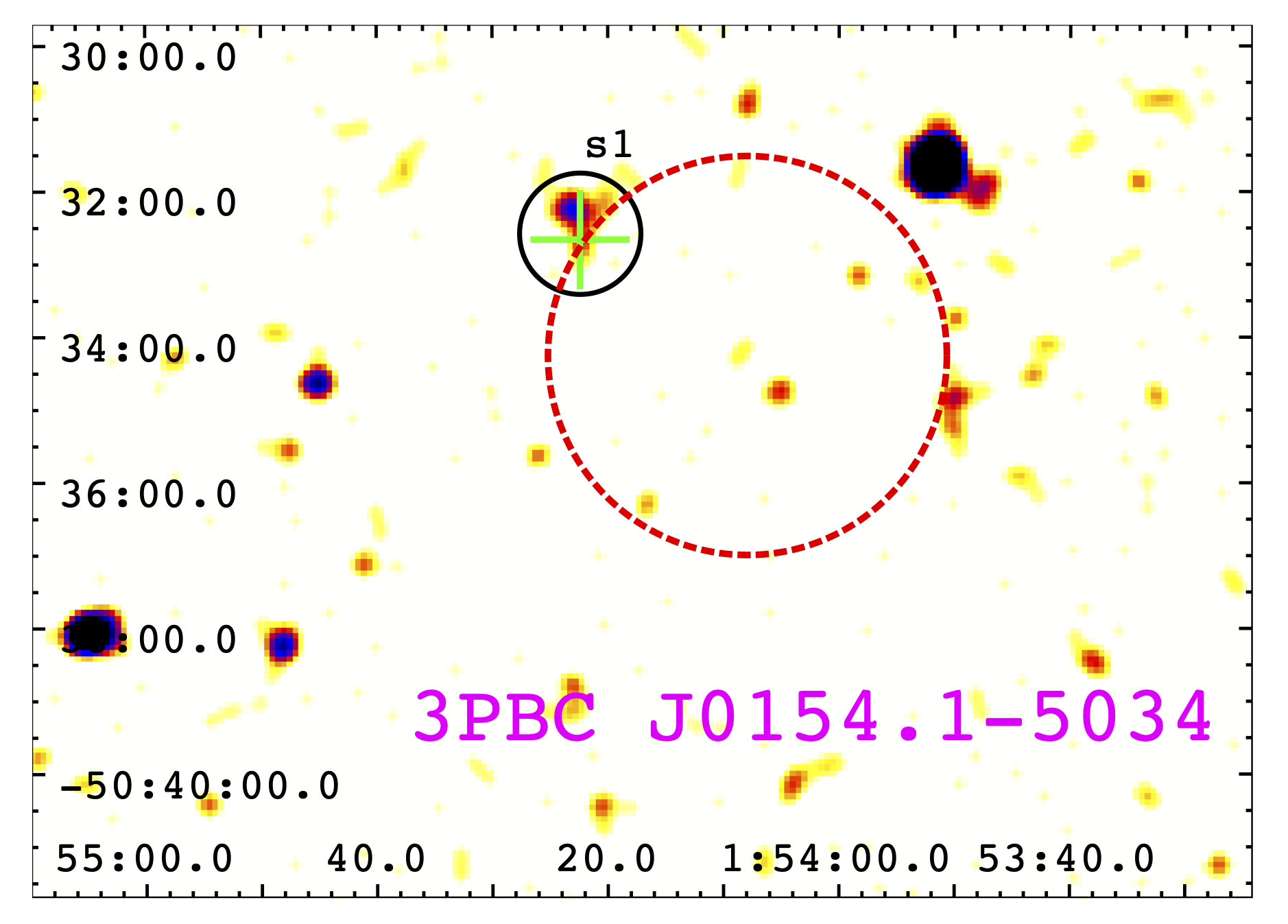}

    \includegraphics[height=4.2cm,width=6cm,angle=0]{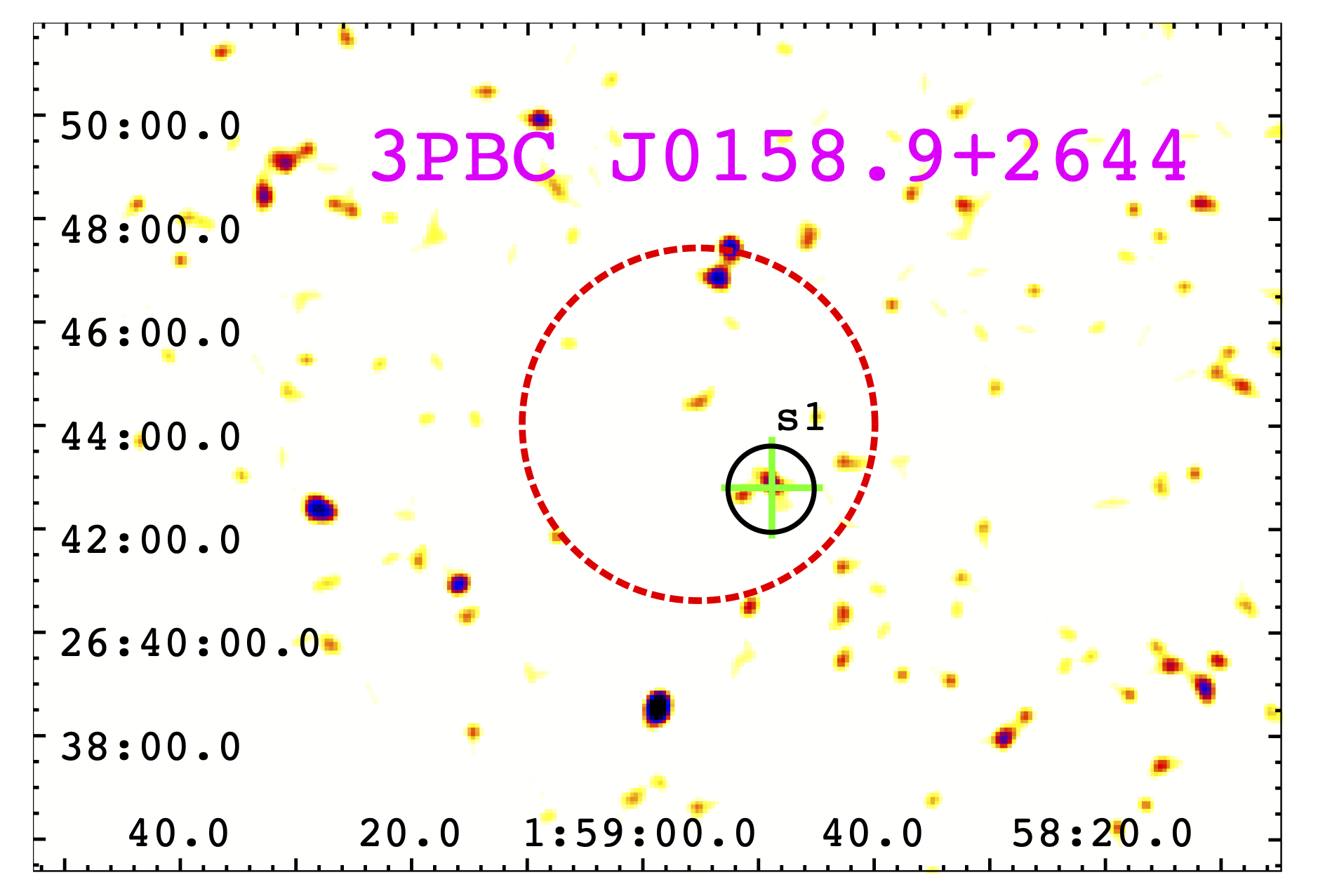}
    \includegraphics[height=4.2cm,width=6cm,angle=0]{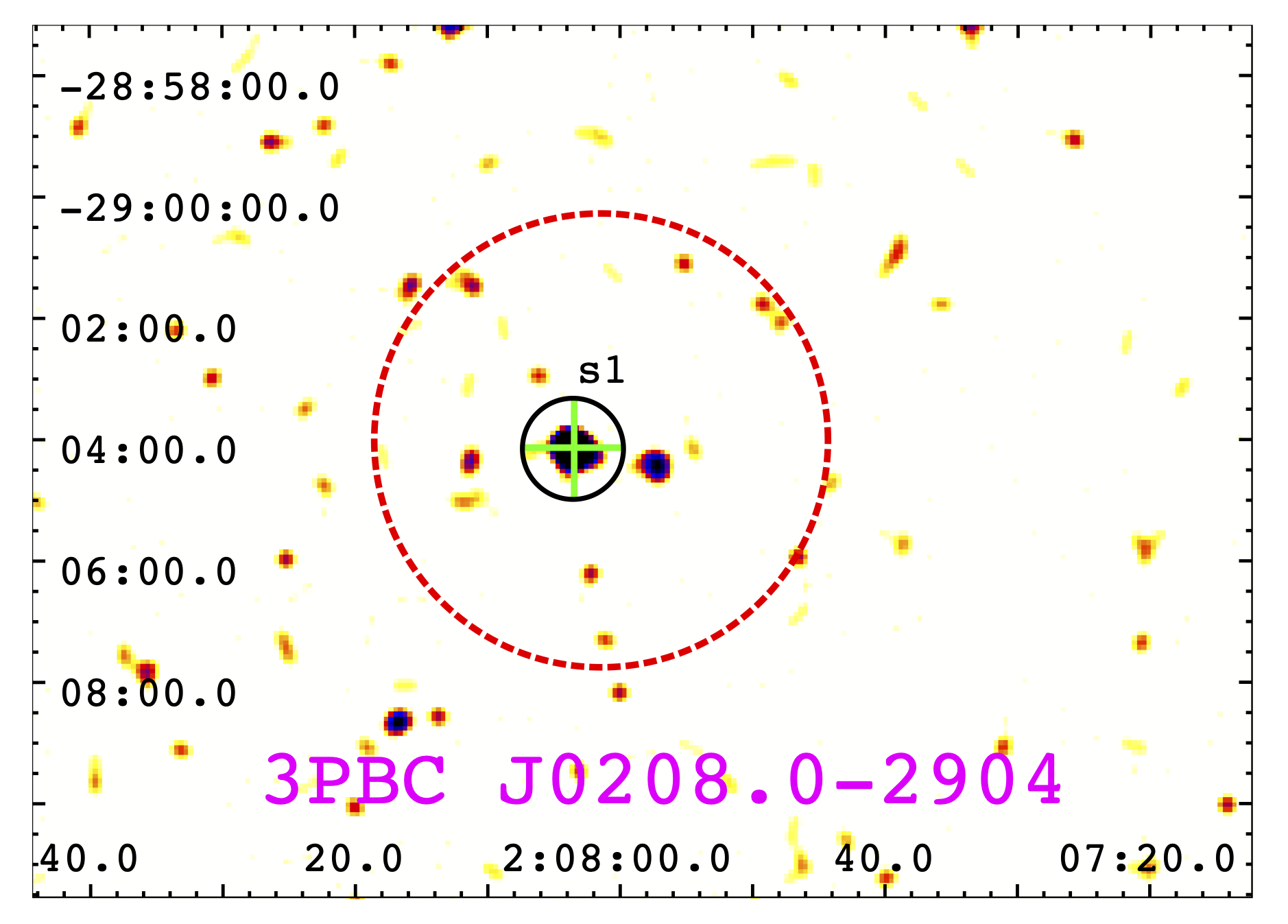}
    \includegraphics[height=4.2cm,width=6cm,angle=0]{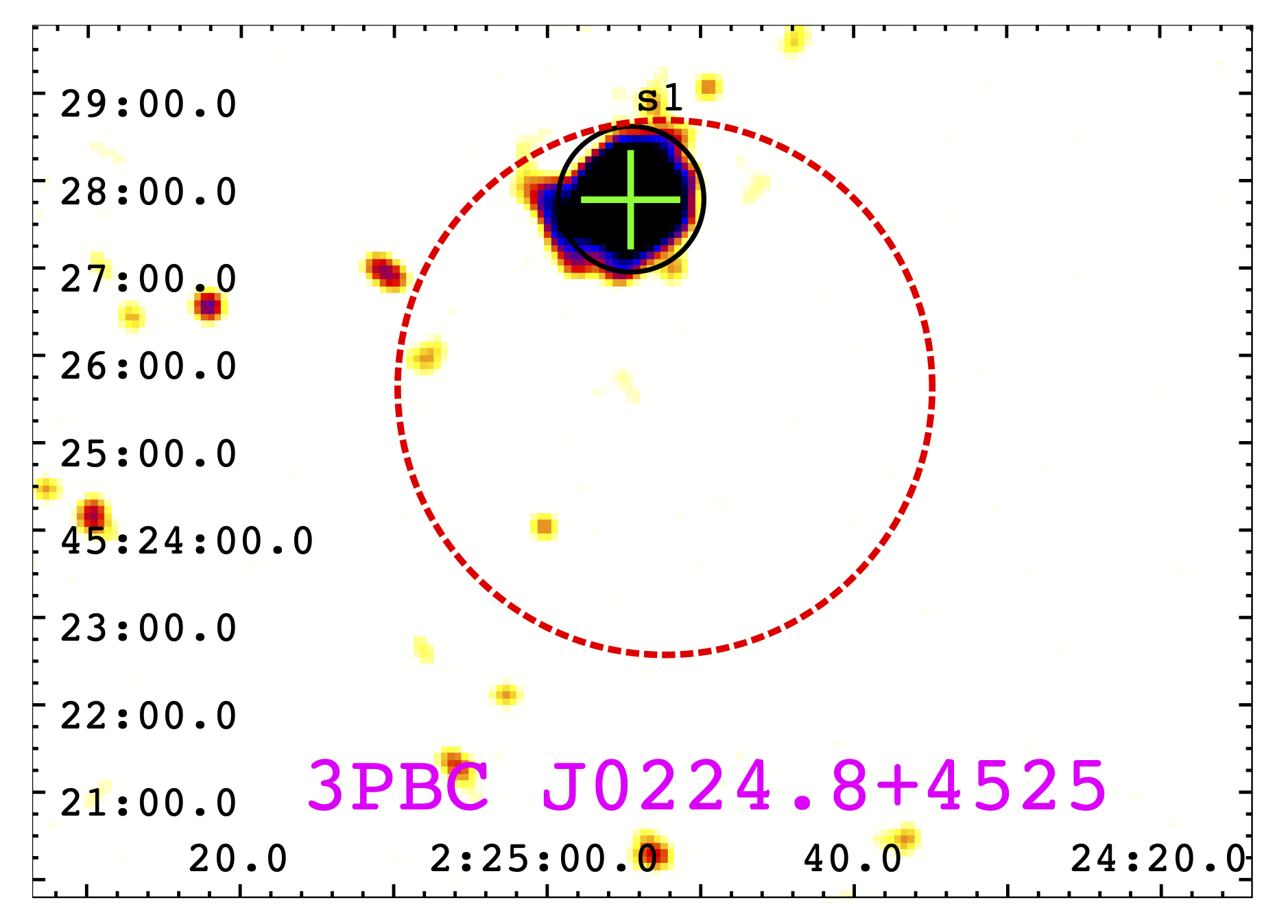}
    
    \includegraphics[height=4.2cm,width=6cm,angle=0]{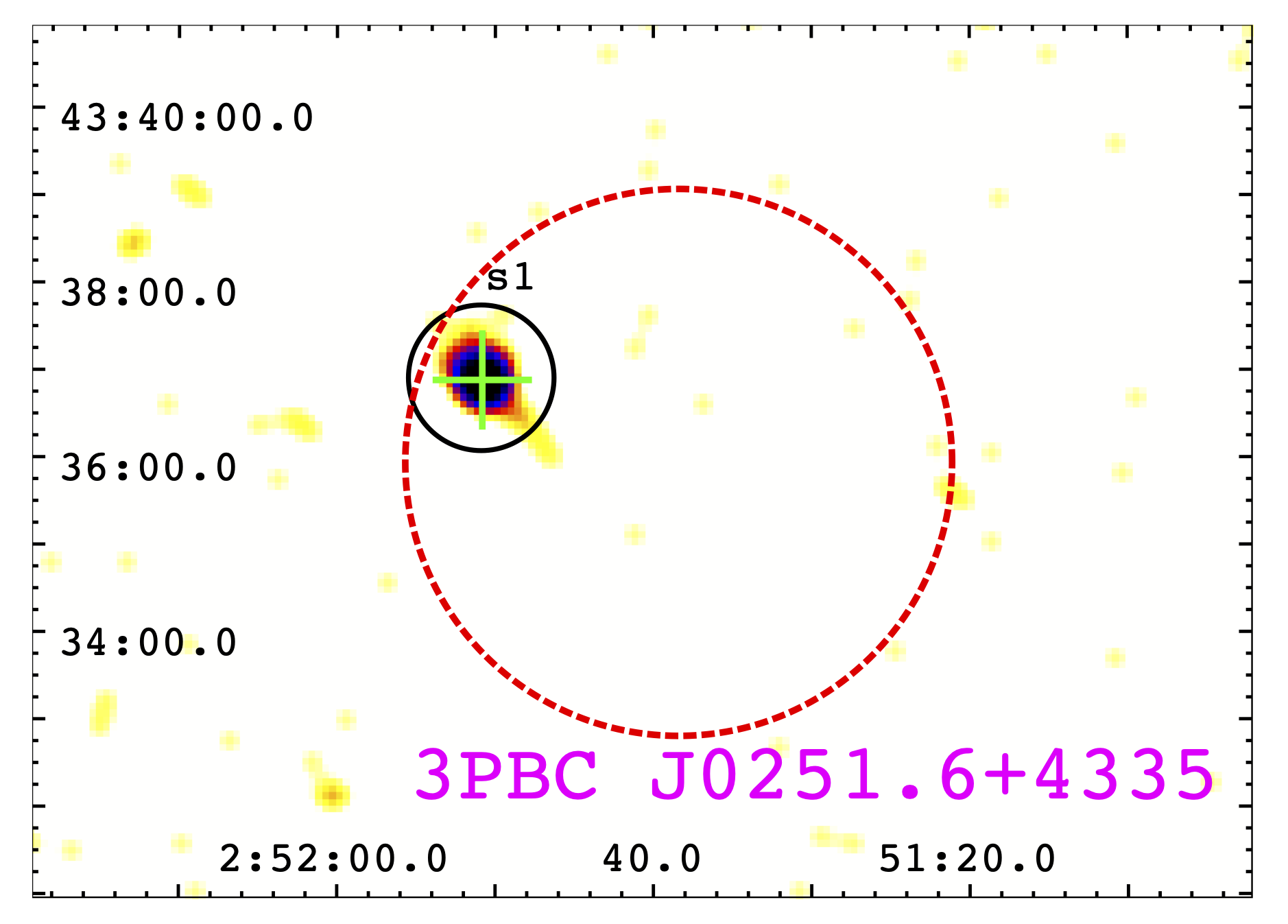}
    \includegraphics[height=4.2cm,width=6cm,angle=0]{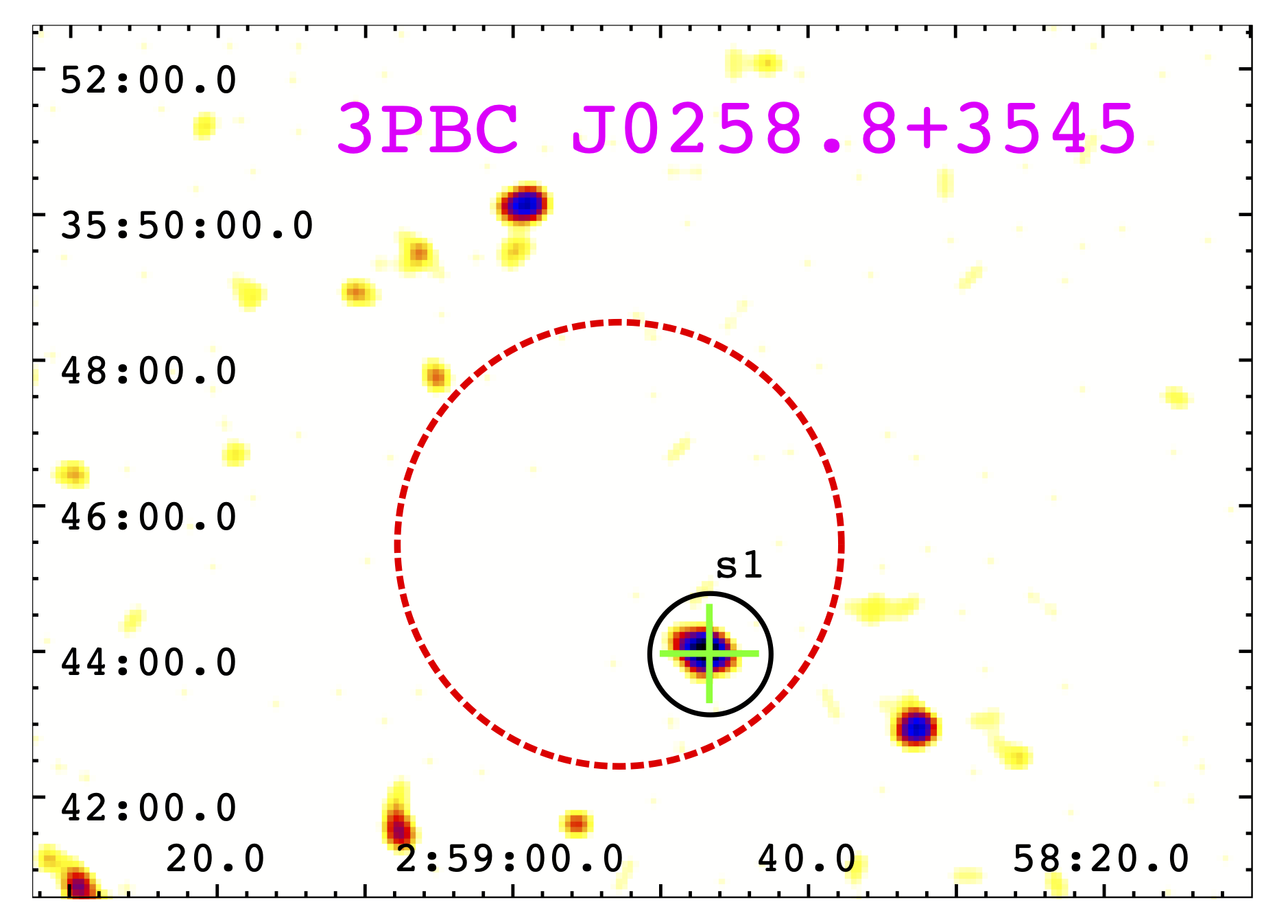}
    \includegraphics[height=4.2cm,width=6cm,angle=0]{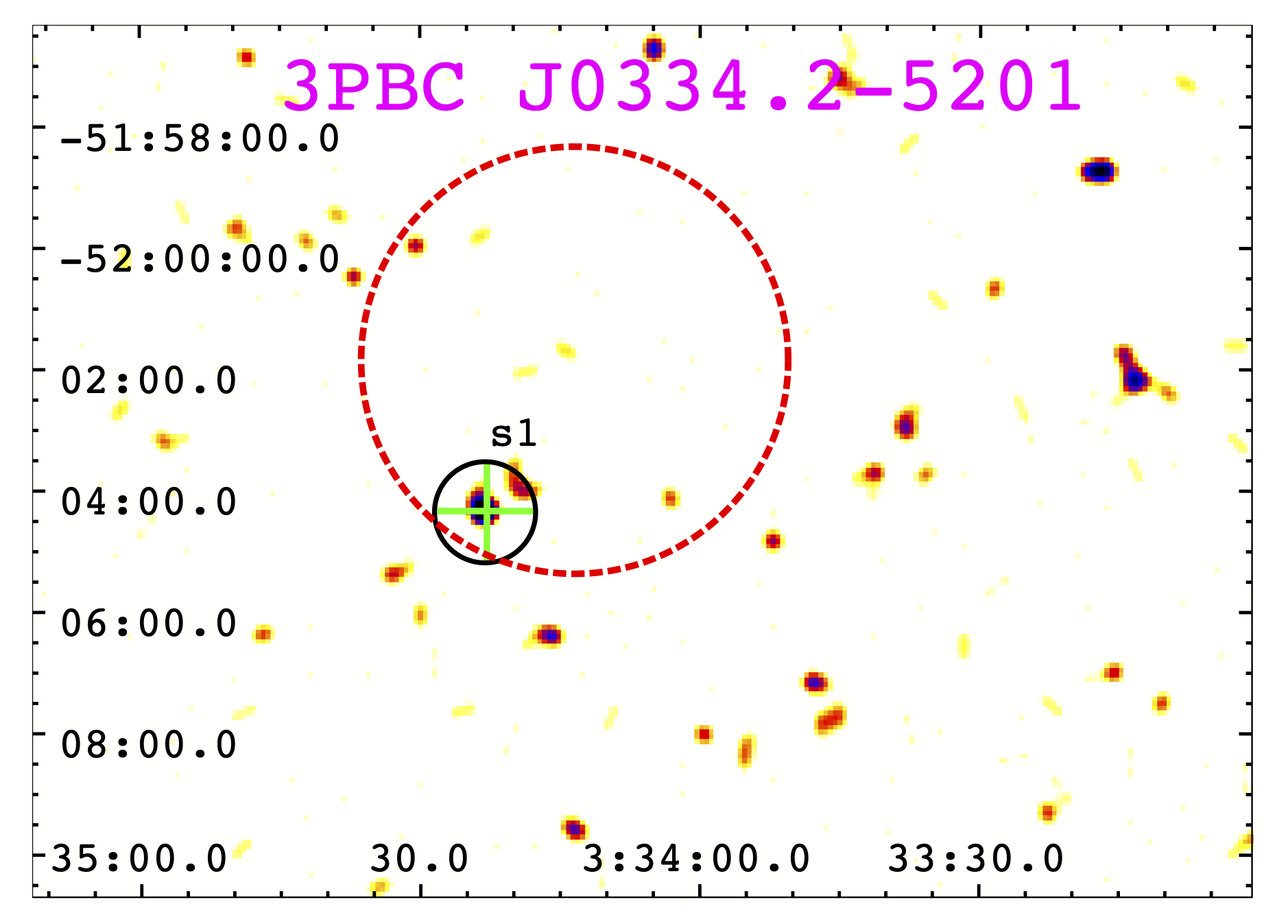}
    
    \includegraphics[height=4.2cm,width=6cm,angle=0]{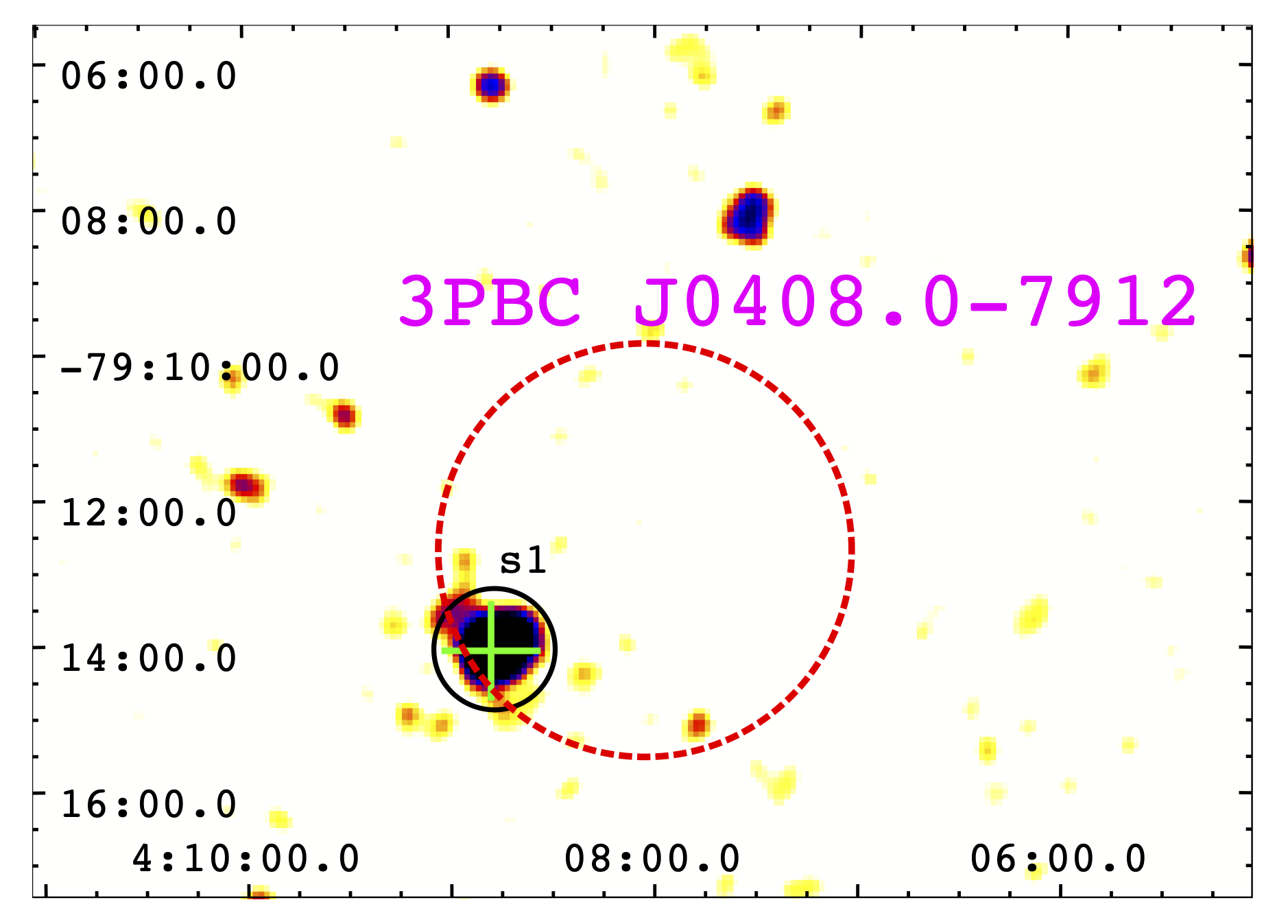}
    \includegraphics[height=4.2cm,width=6cm,angle=0]{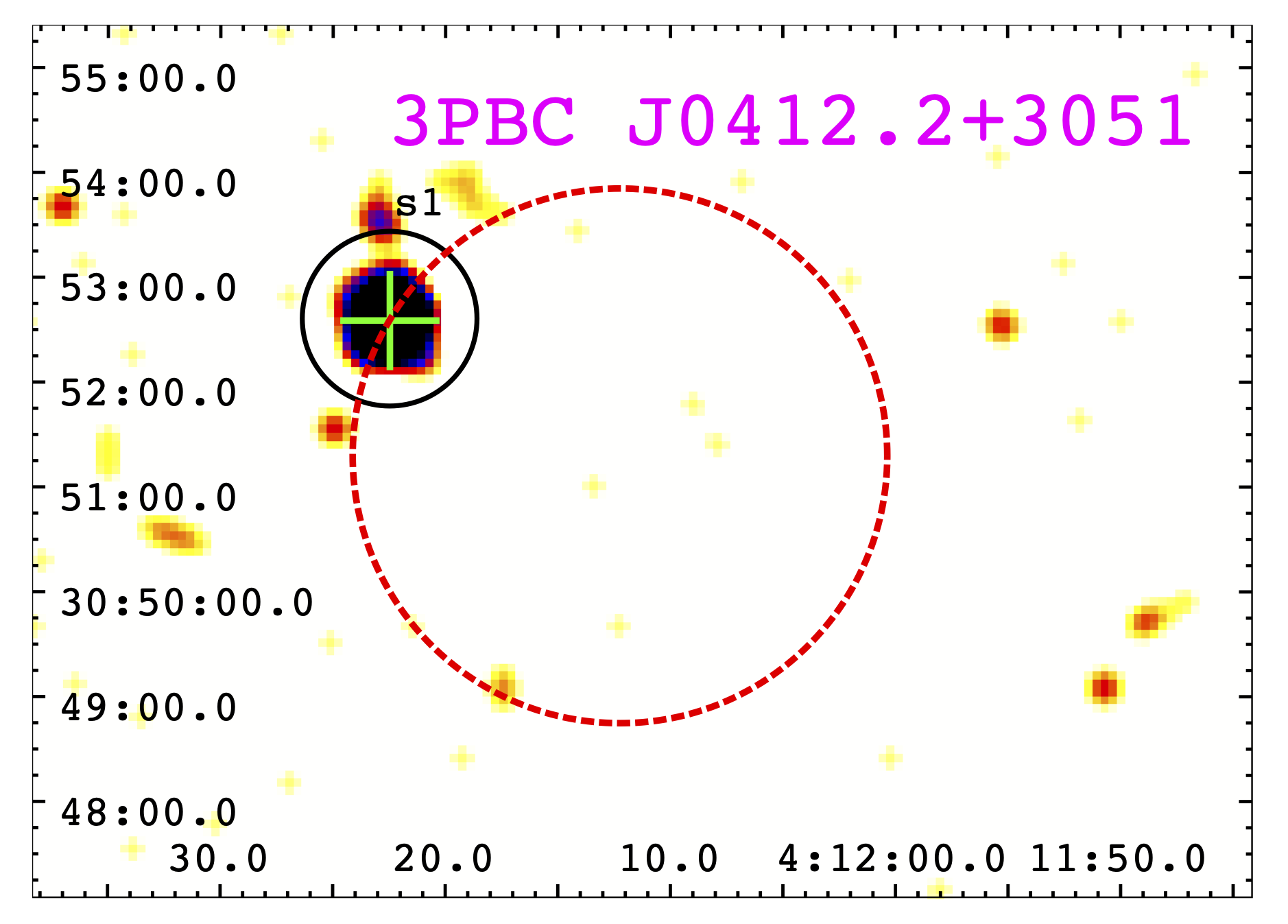}
    \includegraphics[height=4.2cm,width=6cm,angle=0]{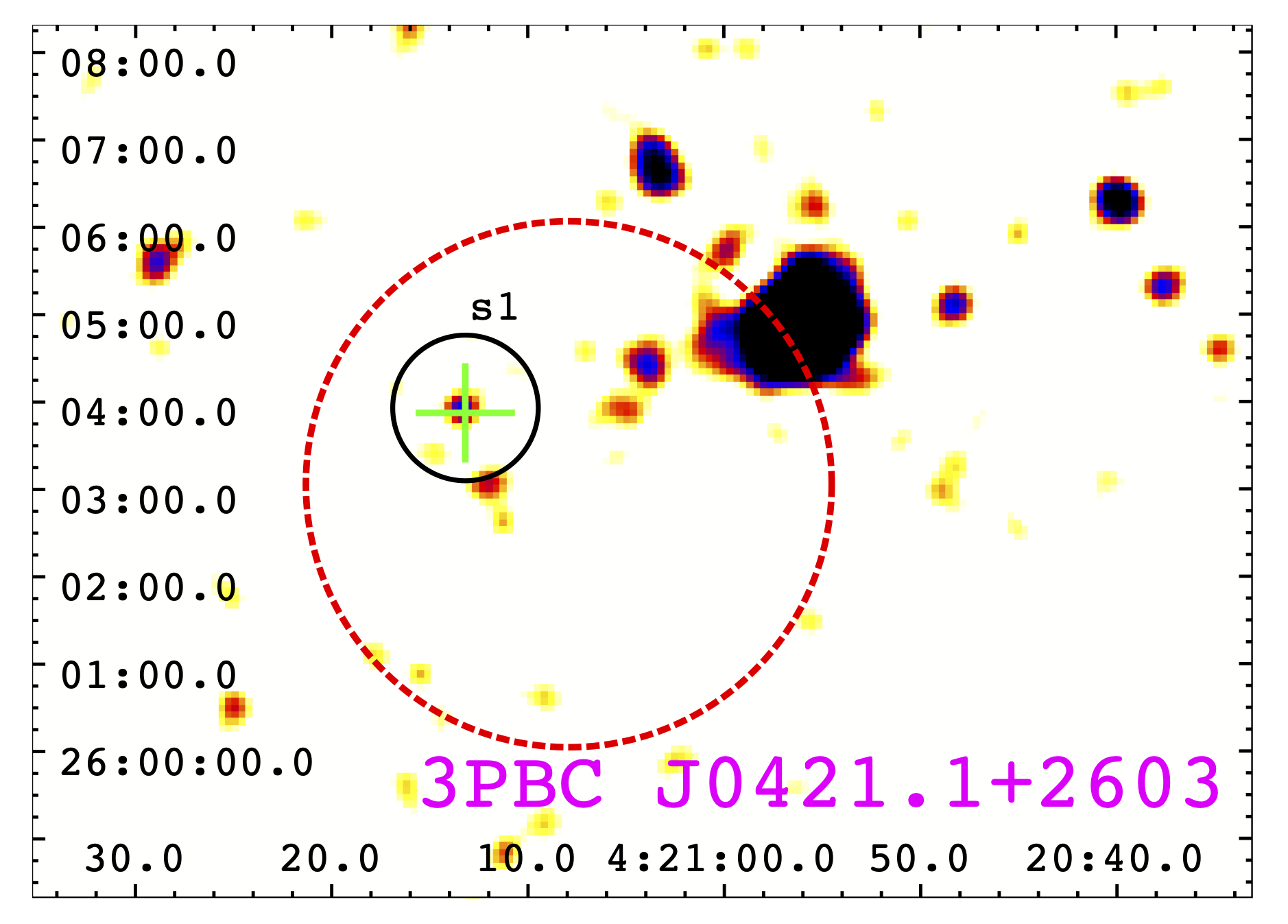}

    \caption{Images of 3PBC sources with exactly one soft \textit{Swift}-XRT source (XDF flag \textit{x}) detected inside of the BAT positional uncertainty region (red dashed circle). The soft X-ray detections are indicated with a black circle. The black circle indicates the position. It does not show the positional uncertainty of the source. If the soft X-ray detection is also marked with a green cross, it indicates that it has a WISE counterpart. }
    \label{fig:u_flagged_sources_no1}
\end{center}
\end{figure*}

\begin{figure*}[!th]
\begin{center}
    \includegraphics[height=4.2cm,width=6cm,angle=0]{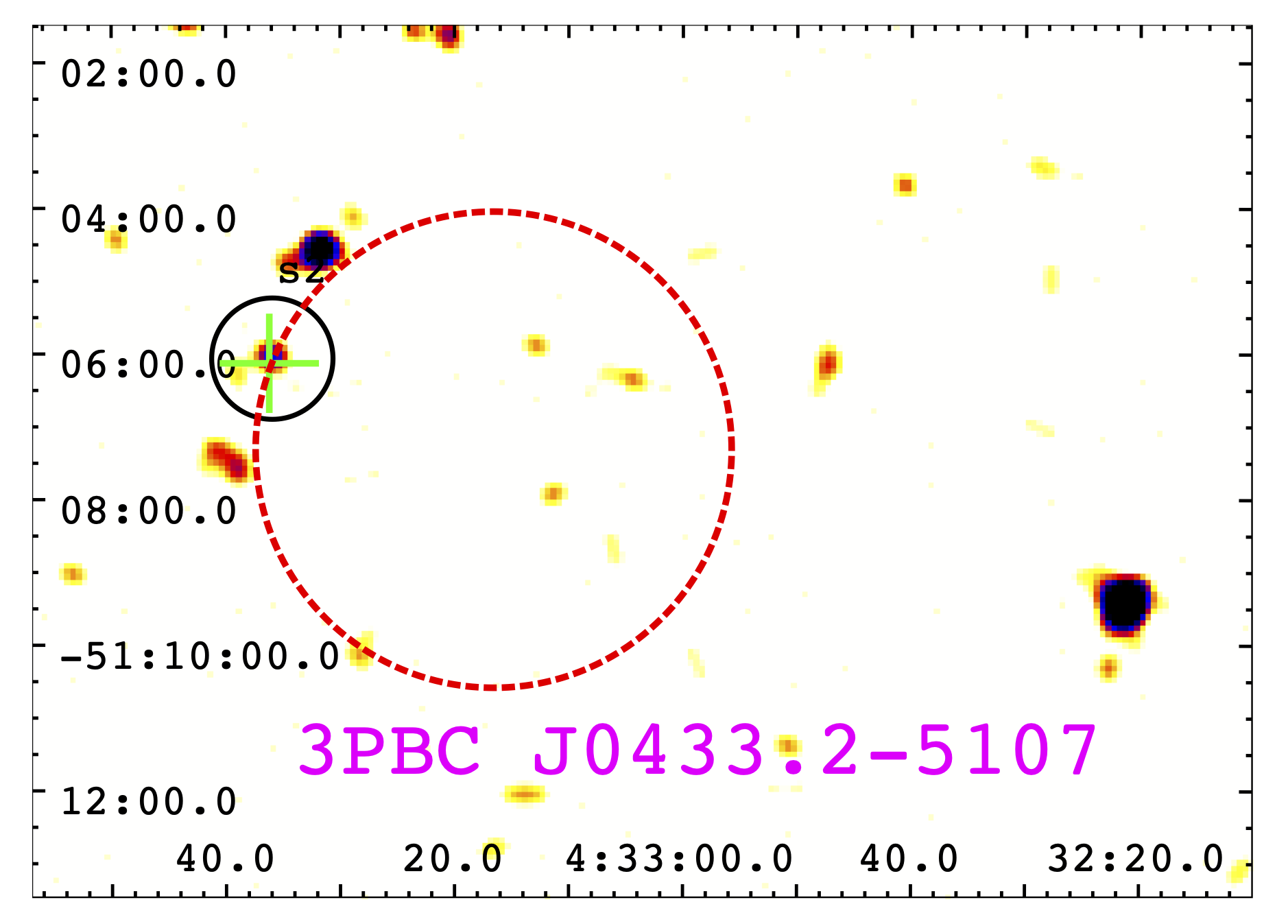}
    \includegraphics[height=4.2cm,width=6cm,angle=0]{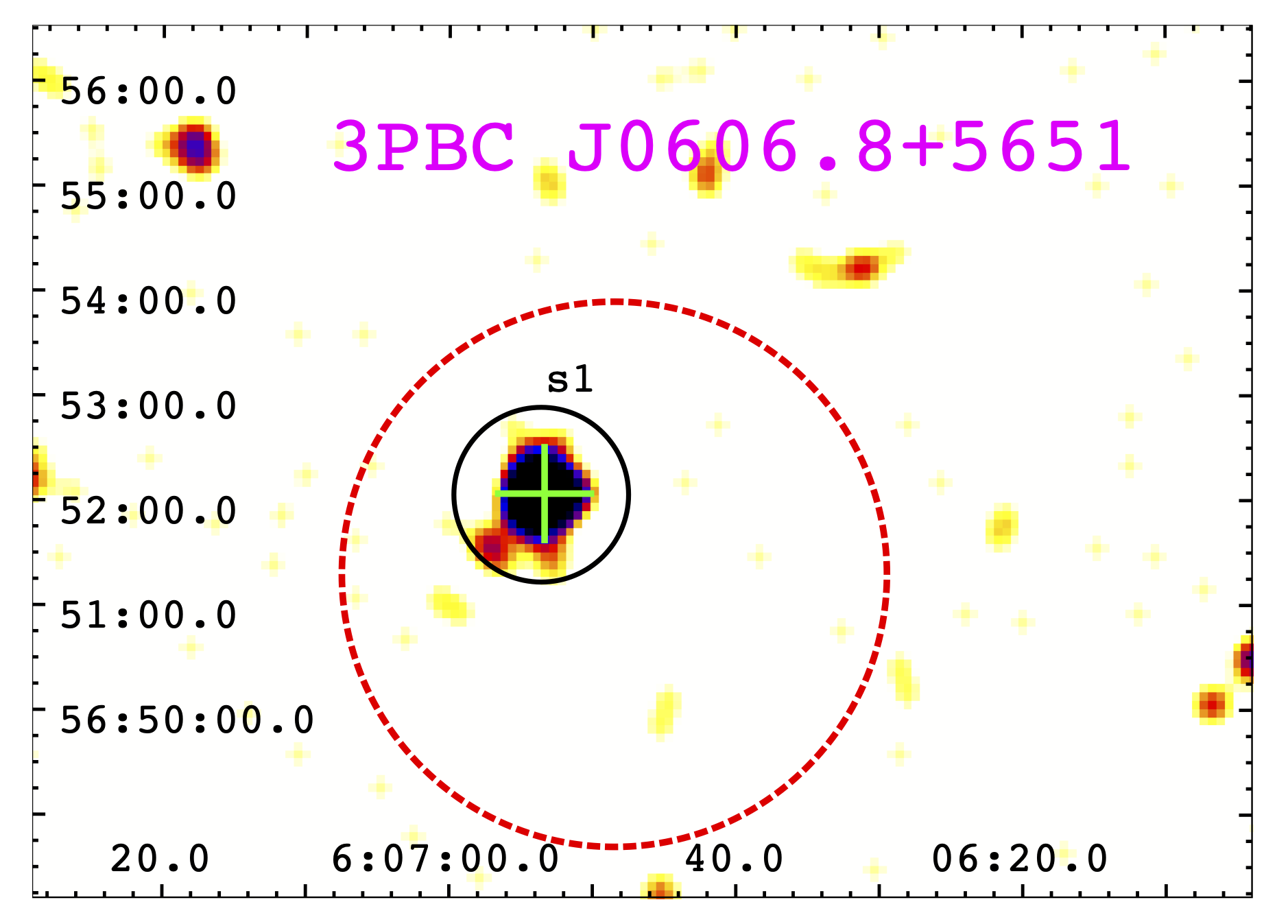}
    \includegraphics[height=4.2cm,width=6cm,angle=0]{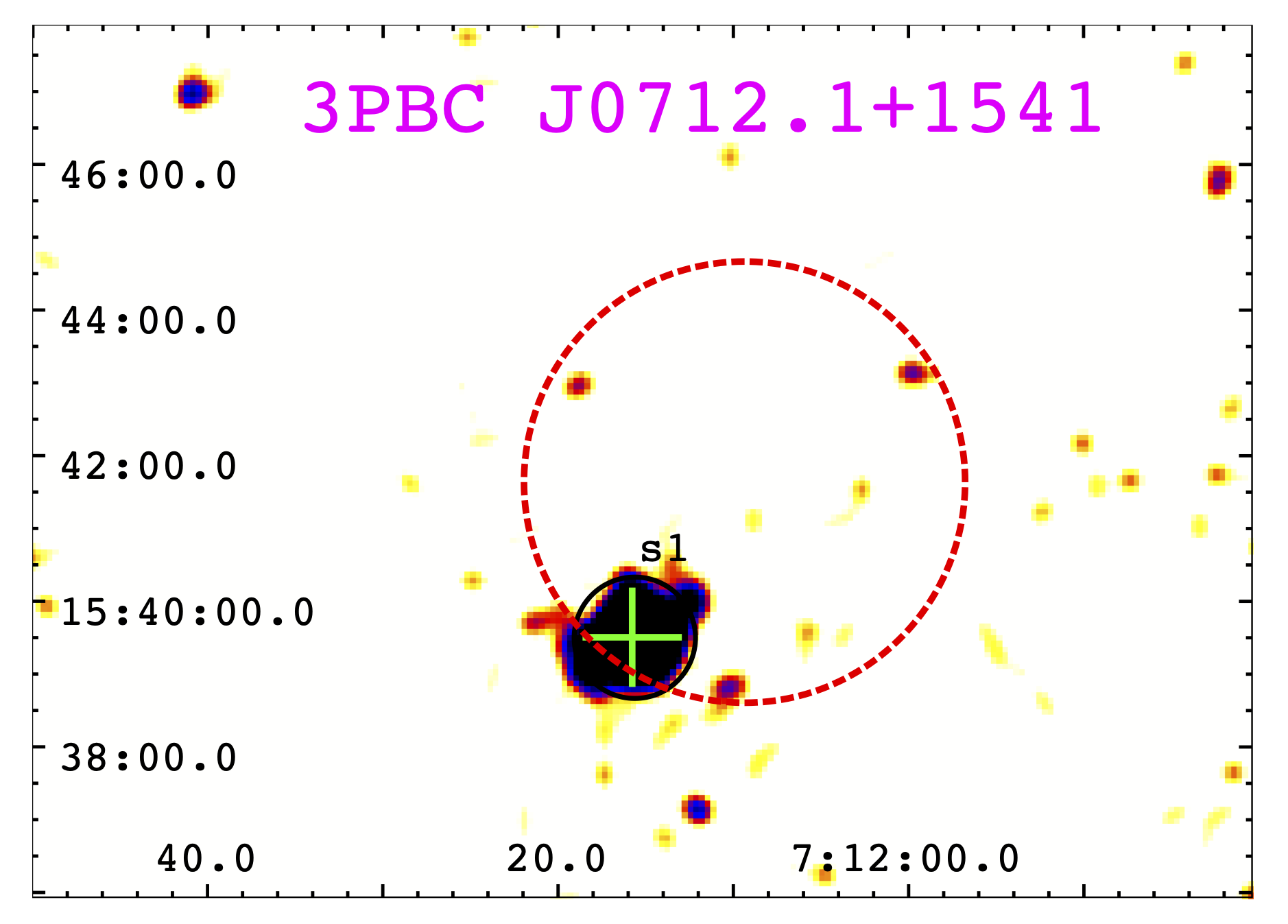}
    \includegraphics[height=4.2cm,width=6cm,angle=0]{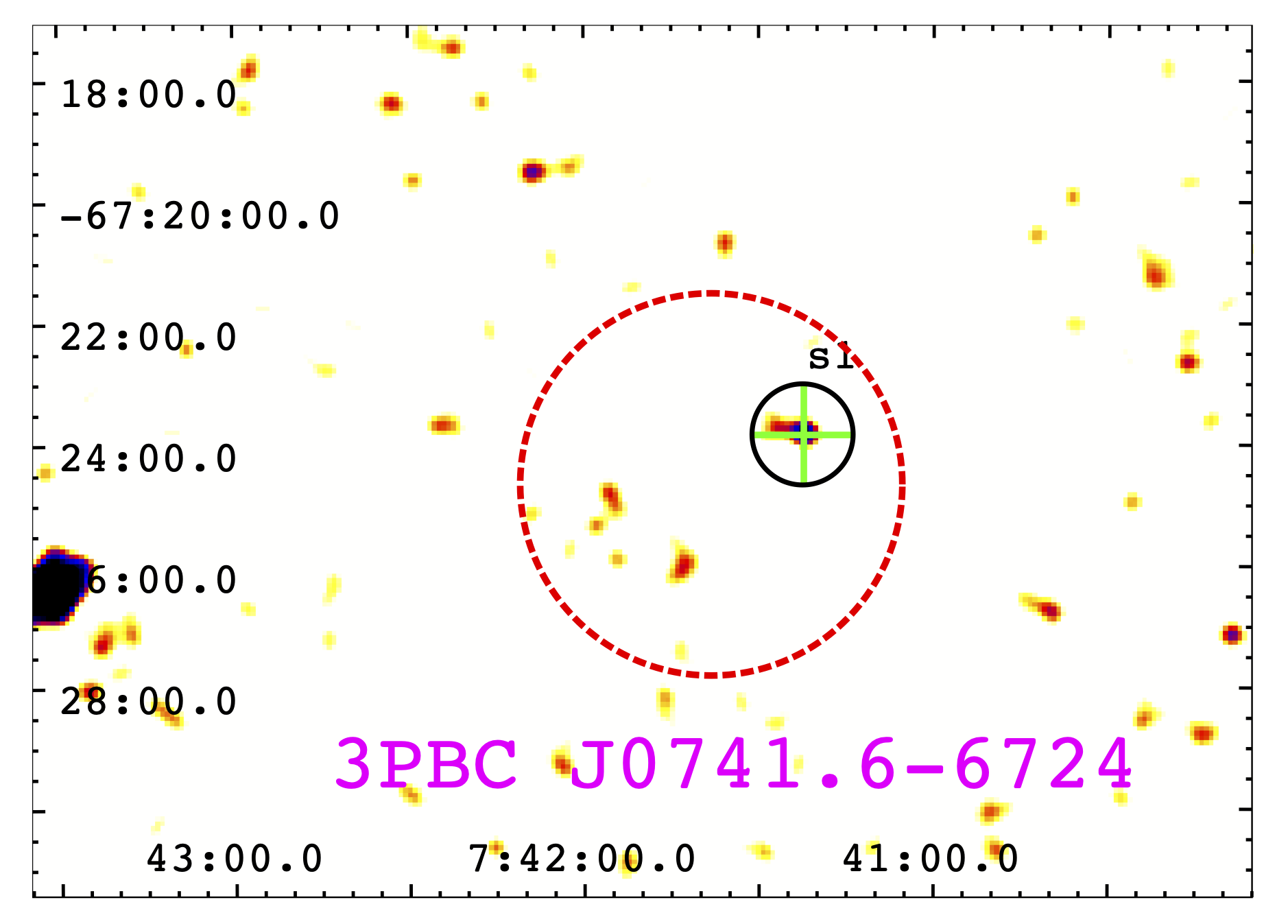}
    \includegraphics[height=4.2cm,width=6cm,angle=0]{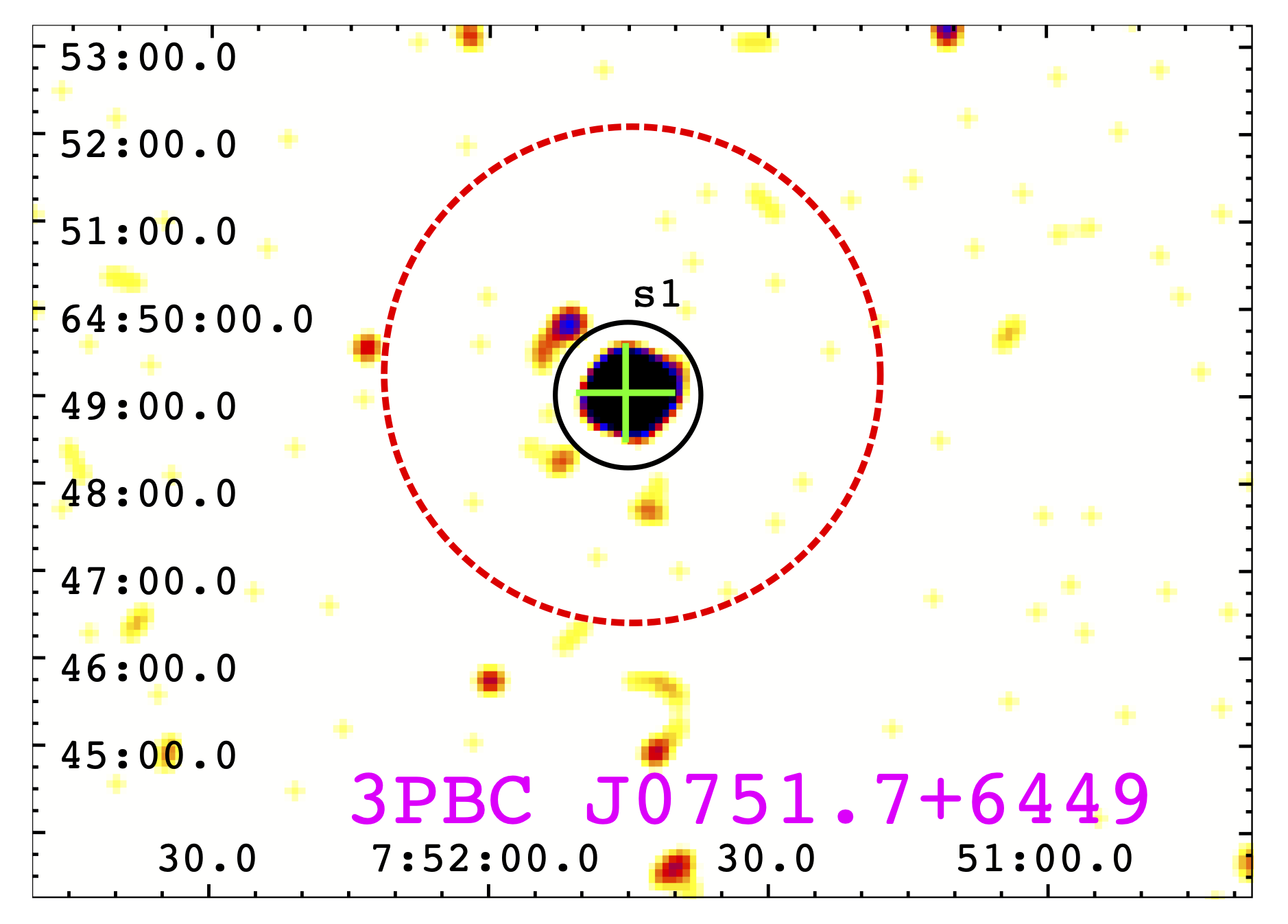}
    \includegraphics[height=4.2cm,width=6cm,angle=0]{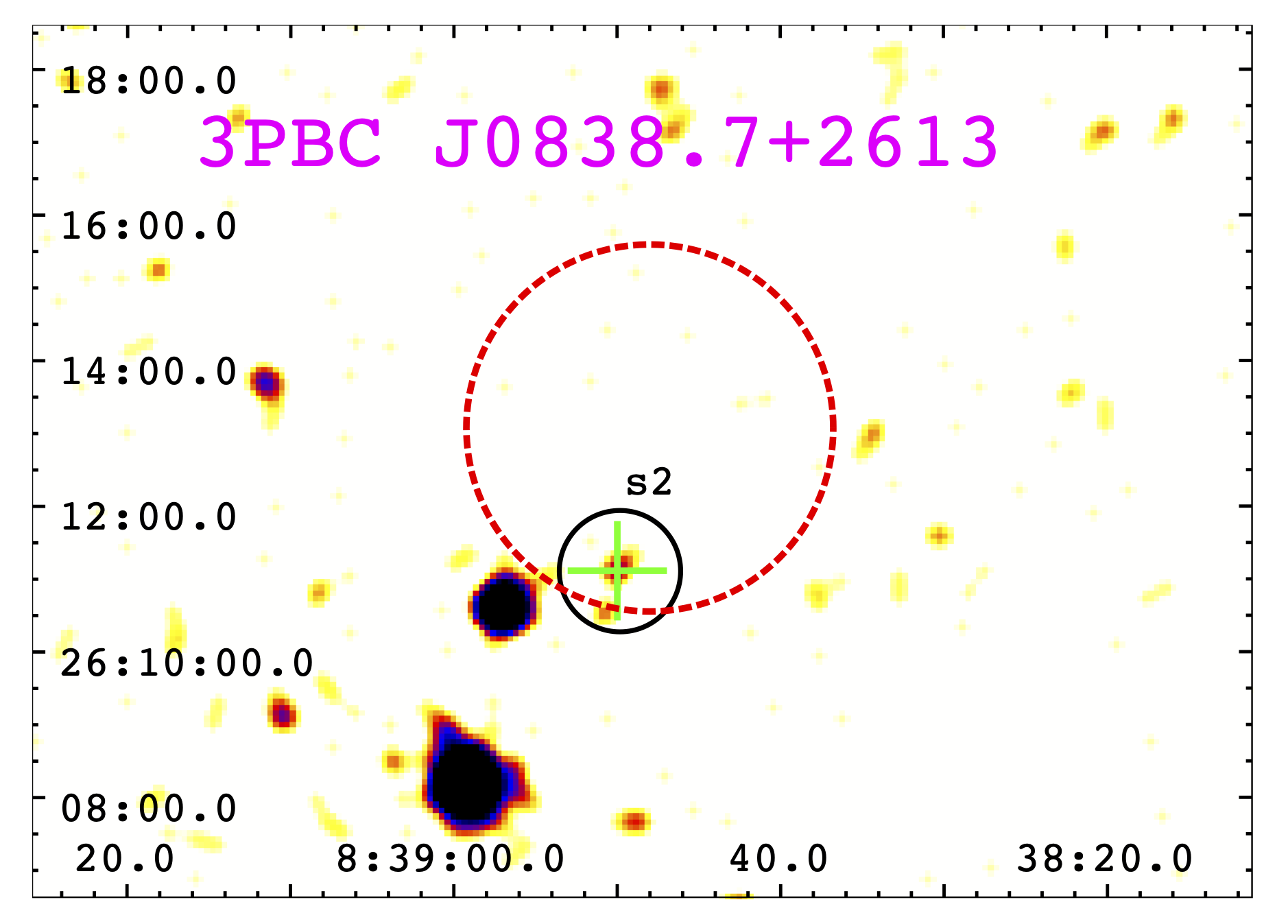}
    \includegraphics[height=4.2cm,width=6cm,angle=0]{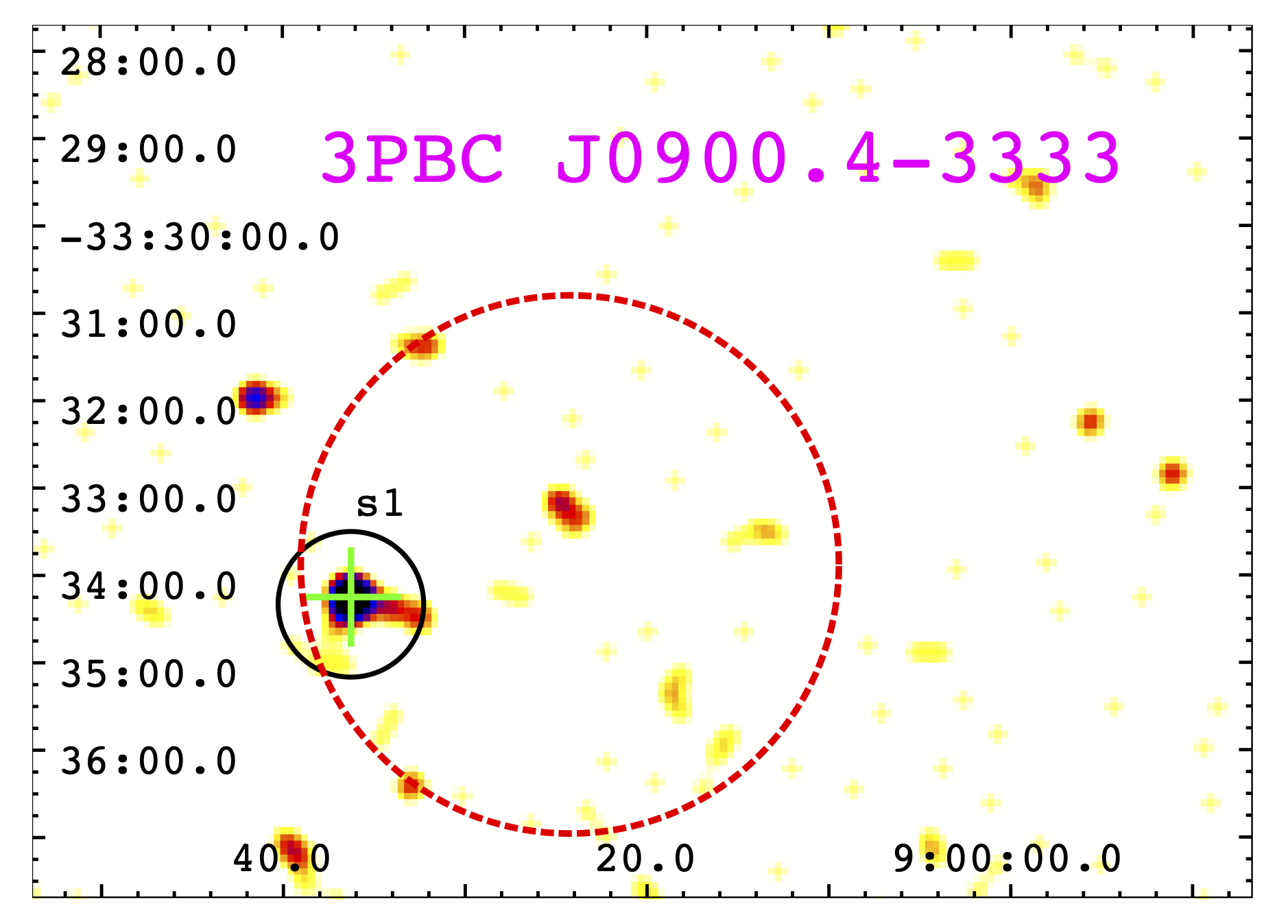}
    \includegraphics[height=4.2cm,width=6cm,angle=0]{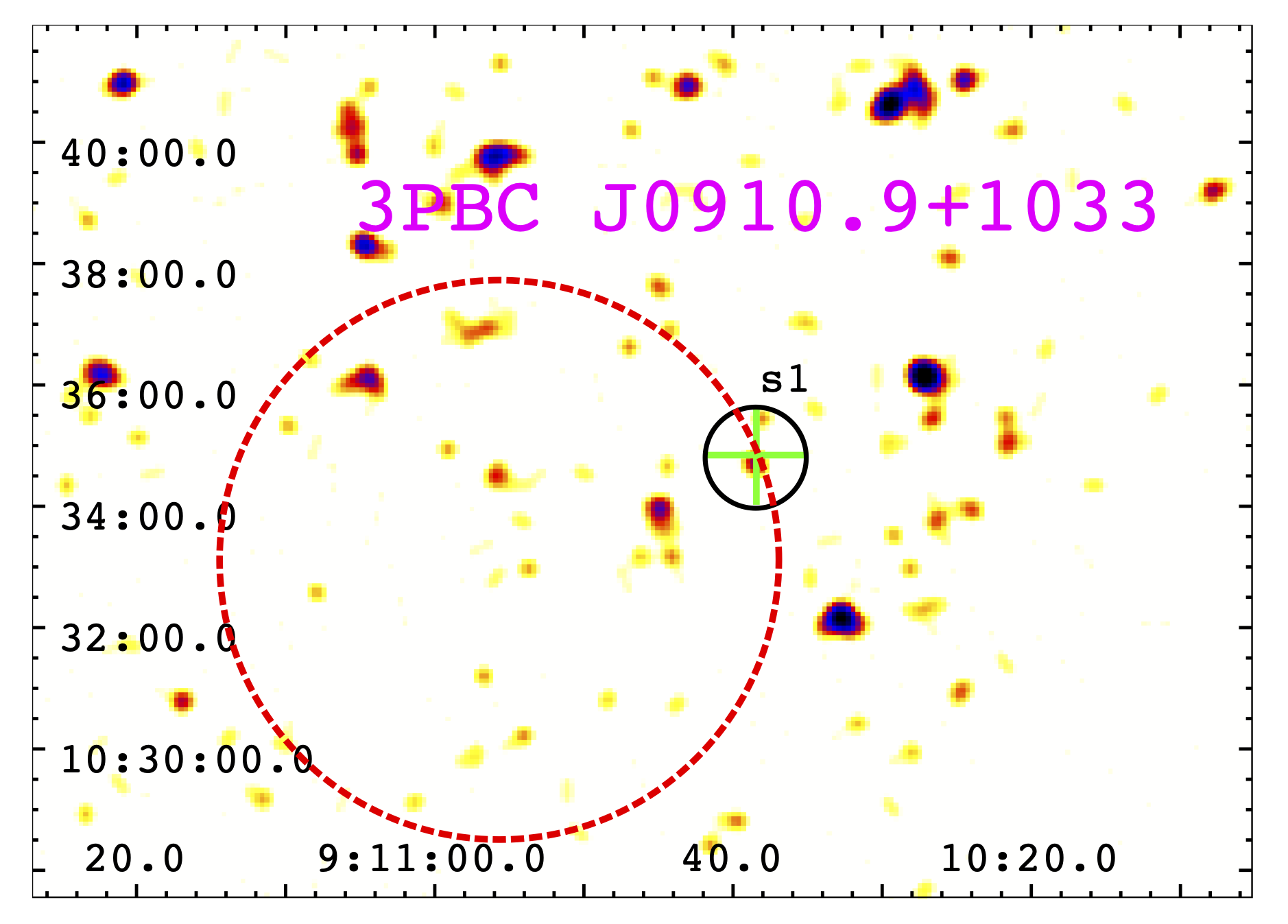}
    \includegraphics[height=4.2cm,width=6cm,angle=0]{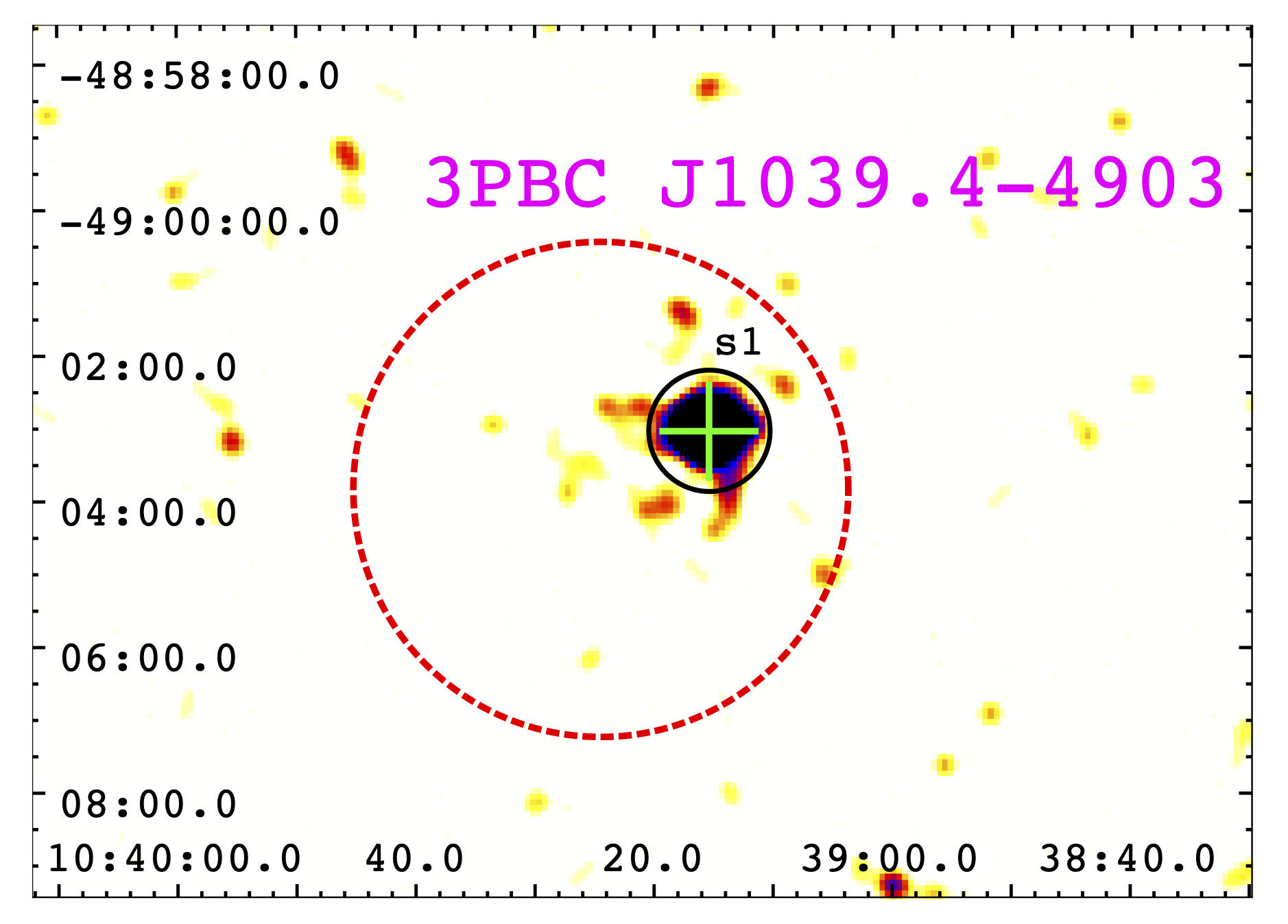}
    \includegraphics[height=4.2cm,width=6cm,angle=0]{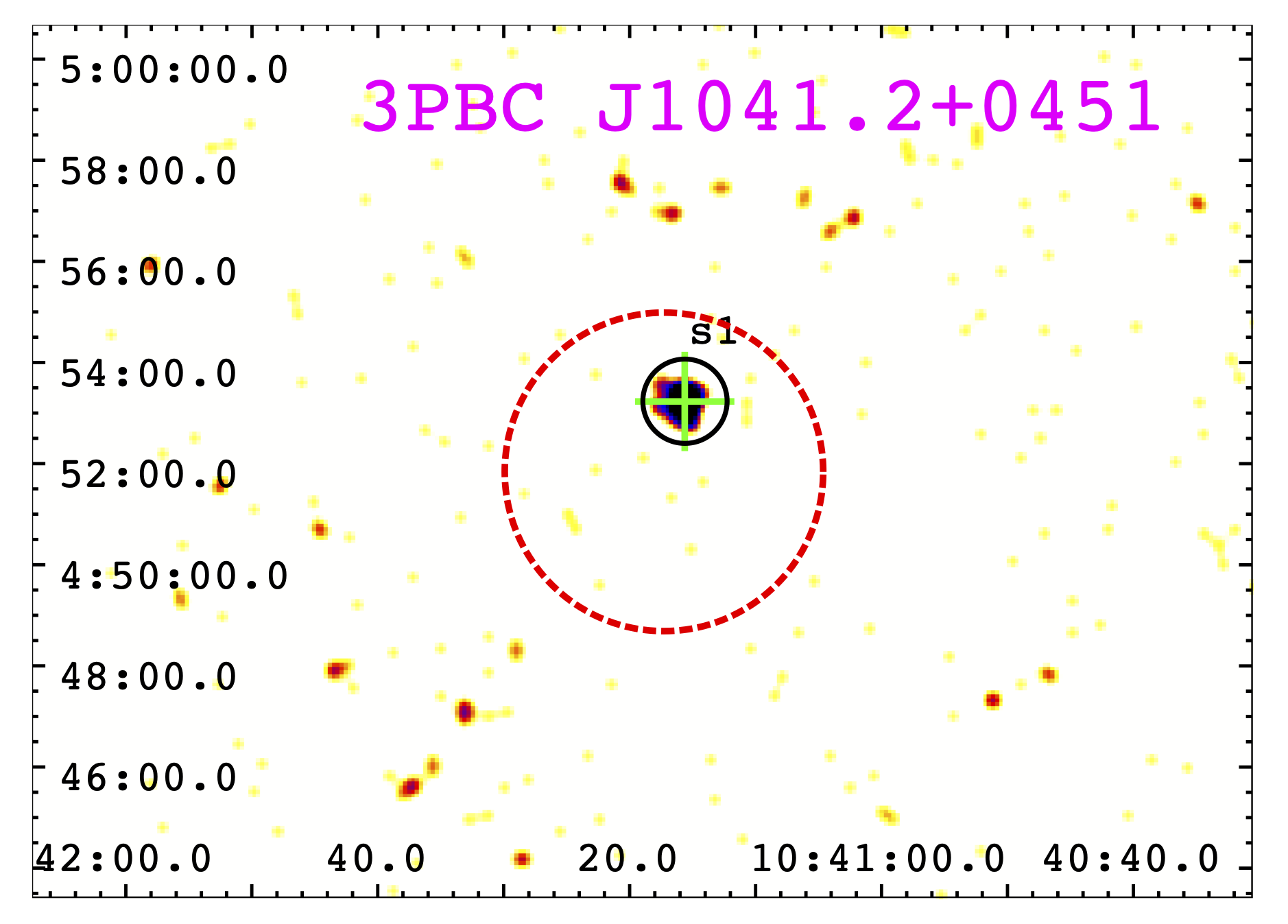}
    \includegraphics[height=4.2cm,width=6cm,angle=0]{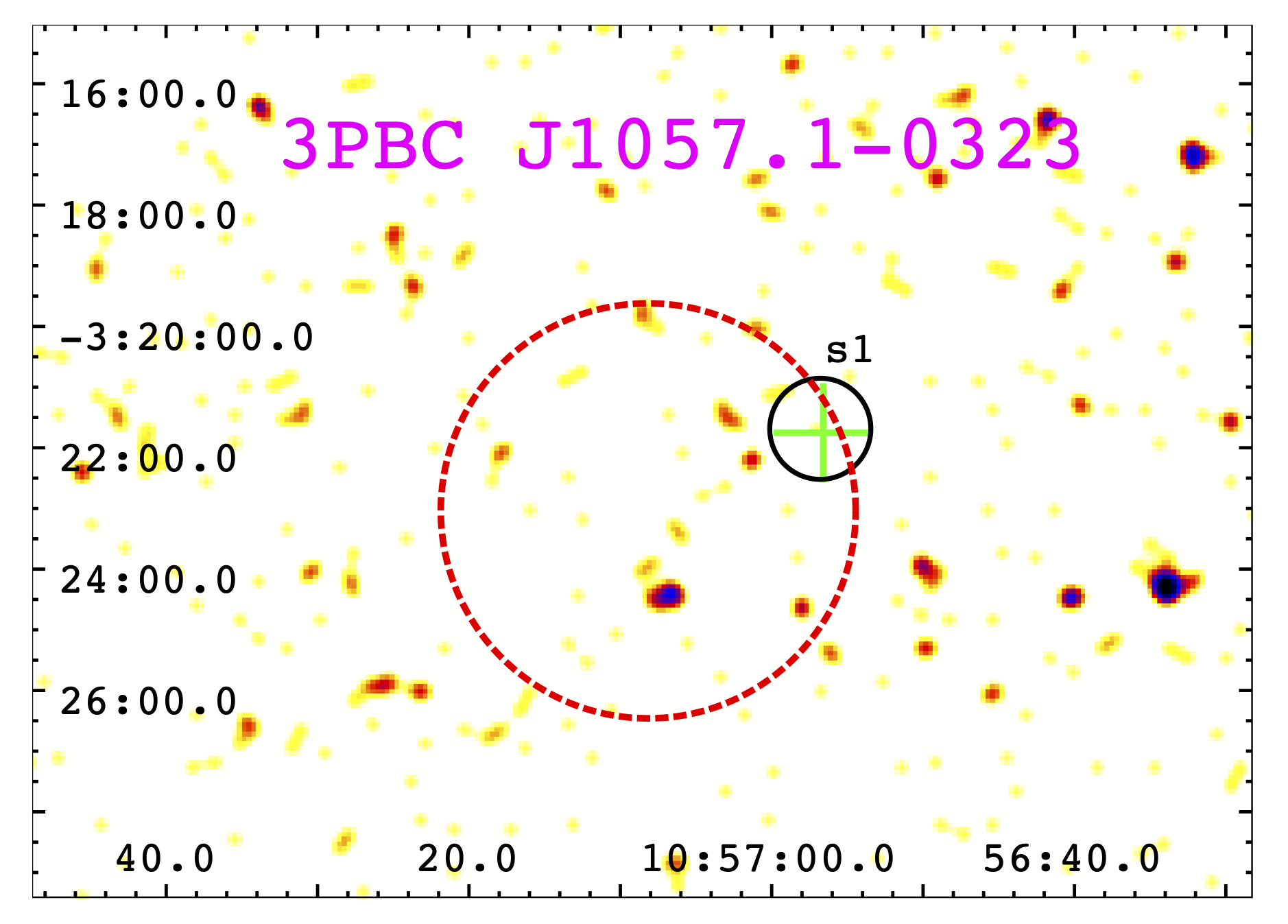}
    \includegraphics[height=4.2cm,width=6cm,angle=0]{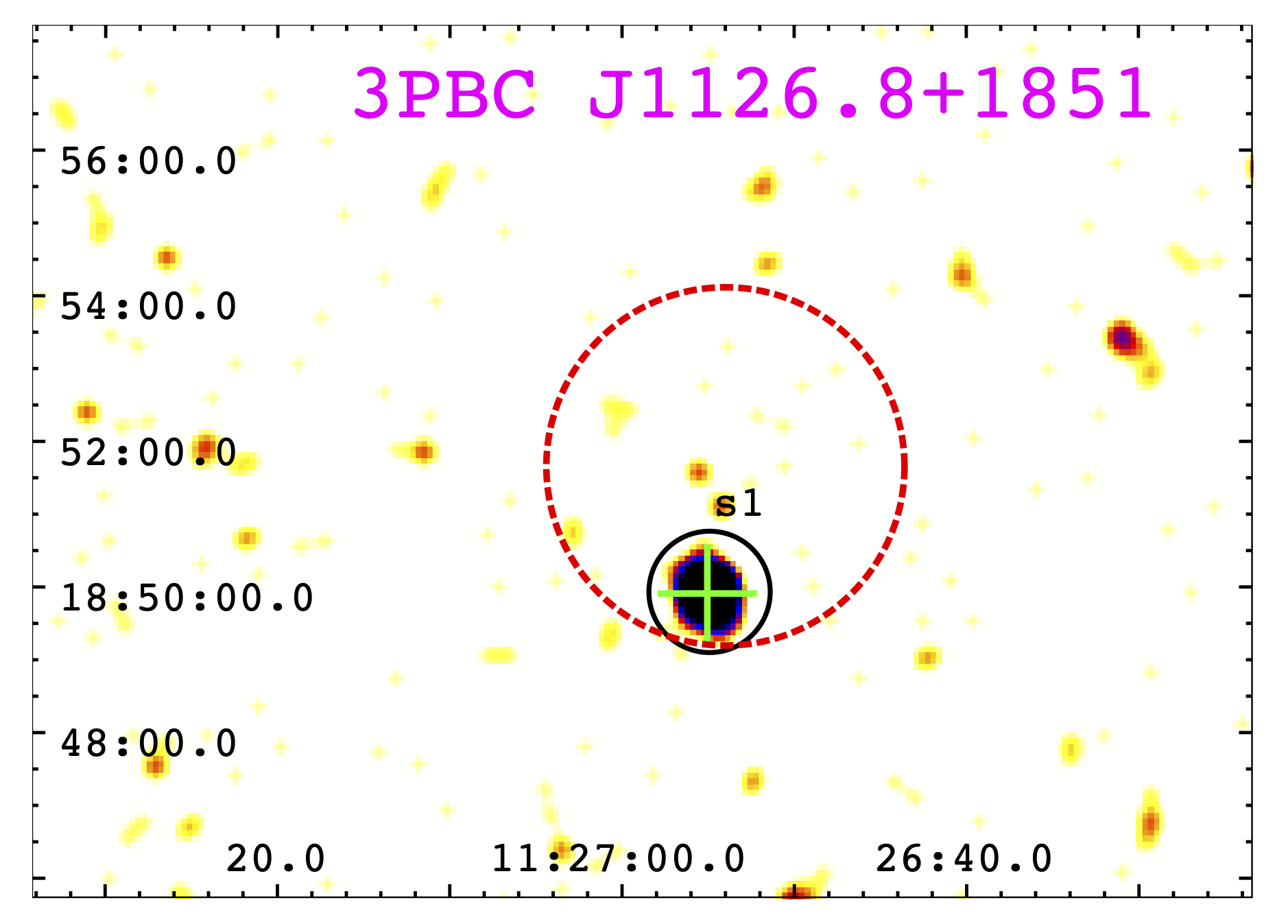}
    \includegraphics[height=4.2cm,width=6cm,angle=0]{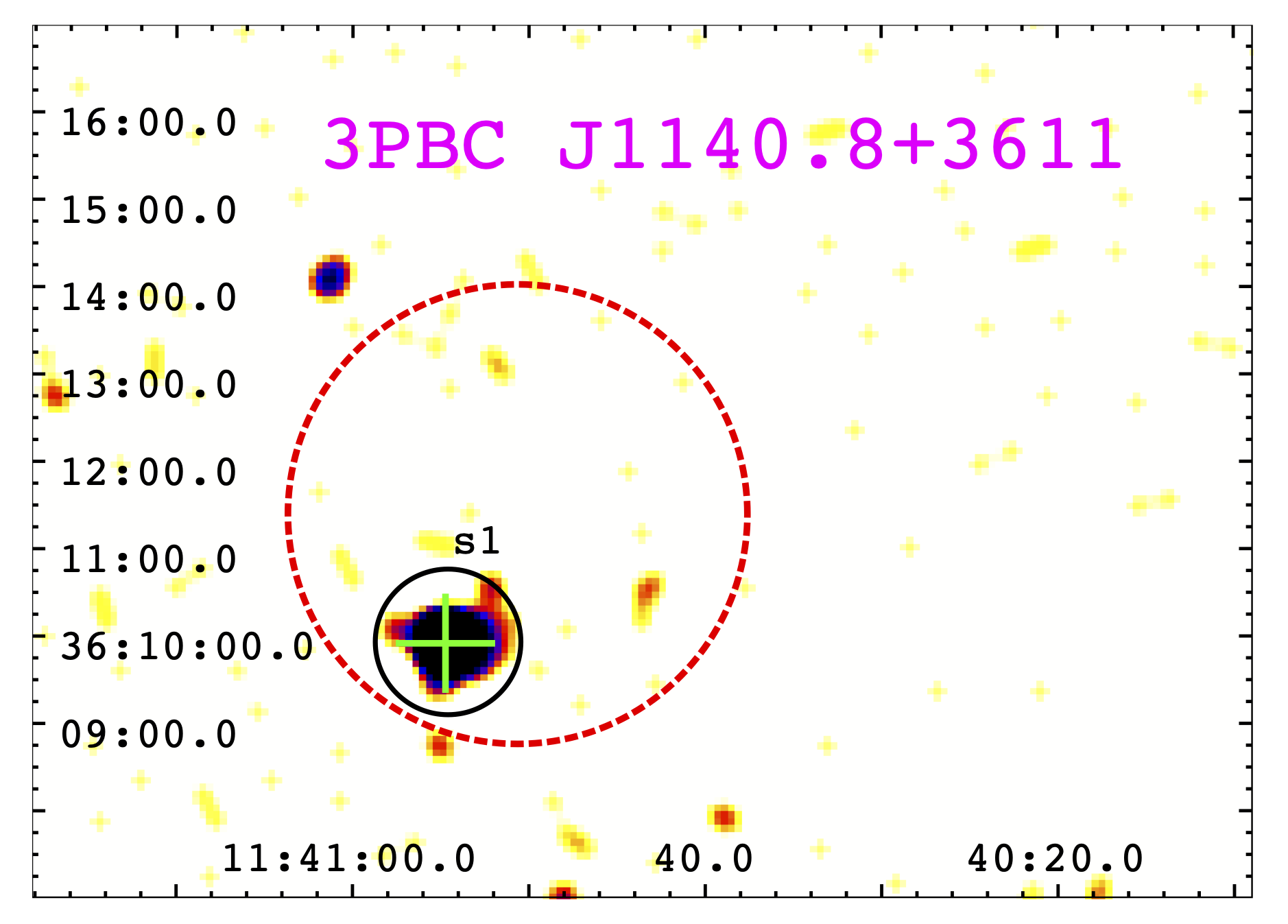}
    \includegraphics[height=4.2cm,width=6cm,angle=0]{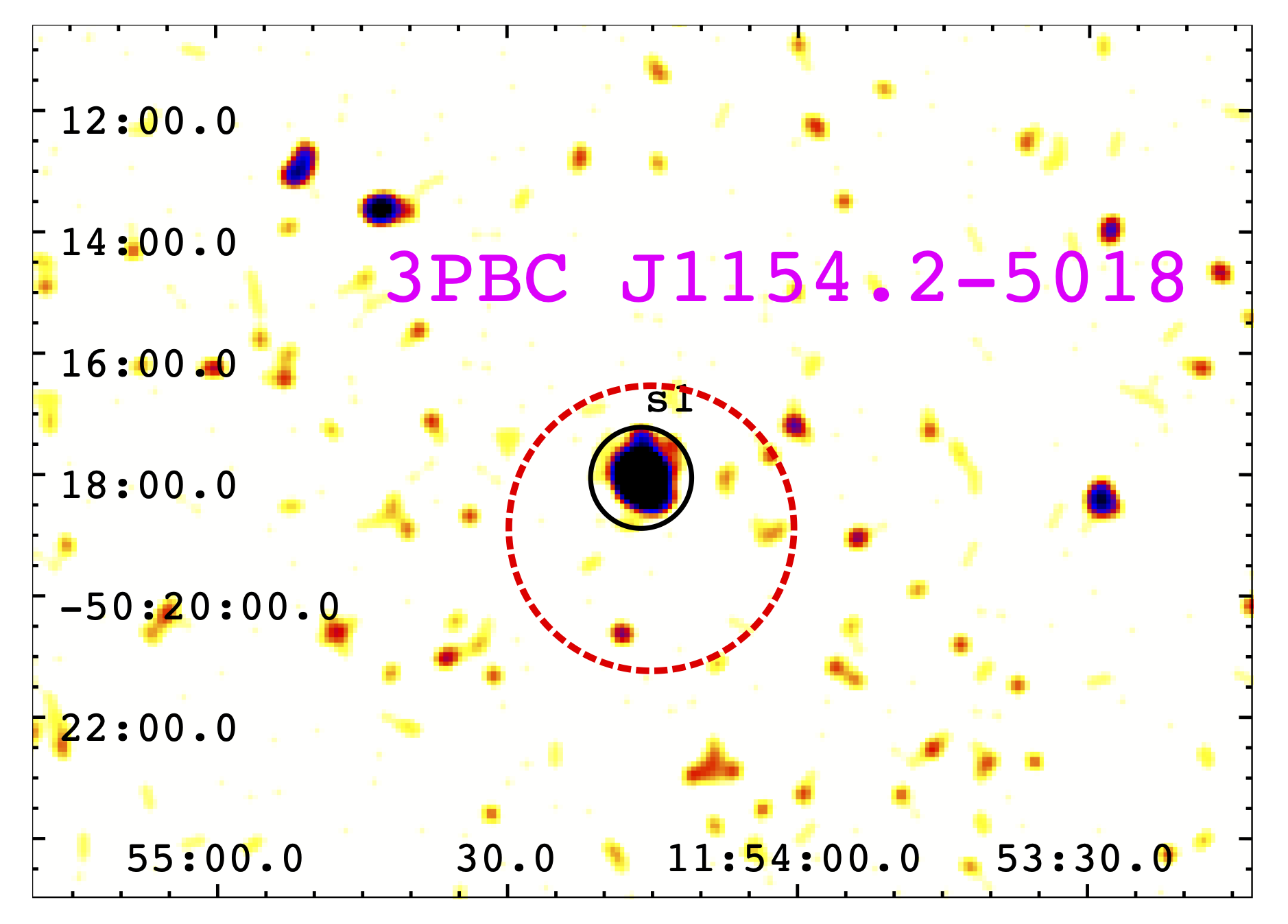}
    \includegraphics[height=4.2cm,width=6cm,angle=0]{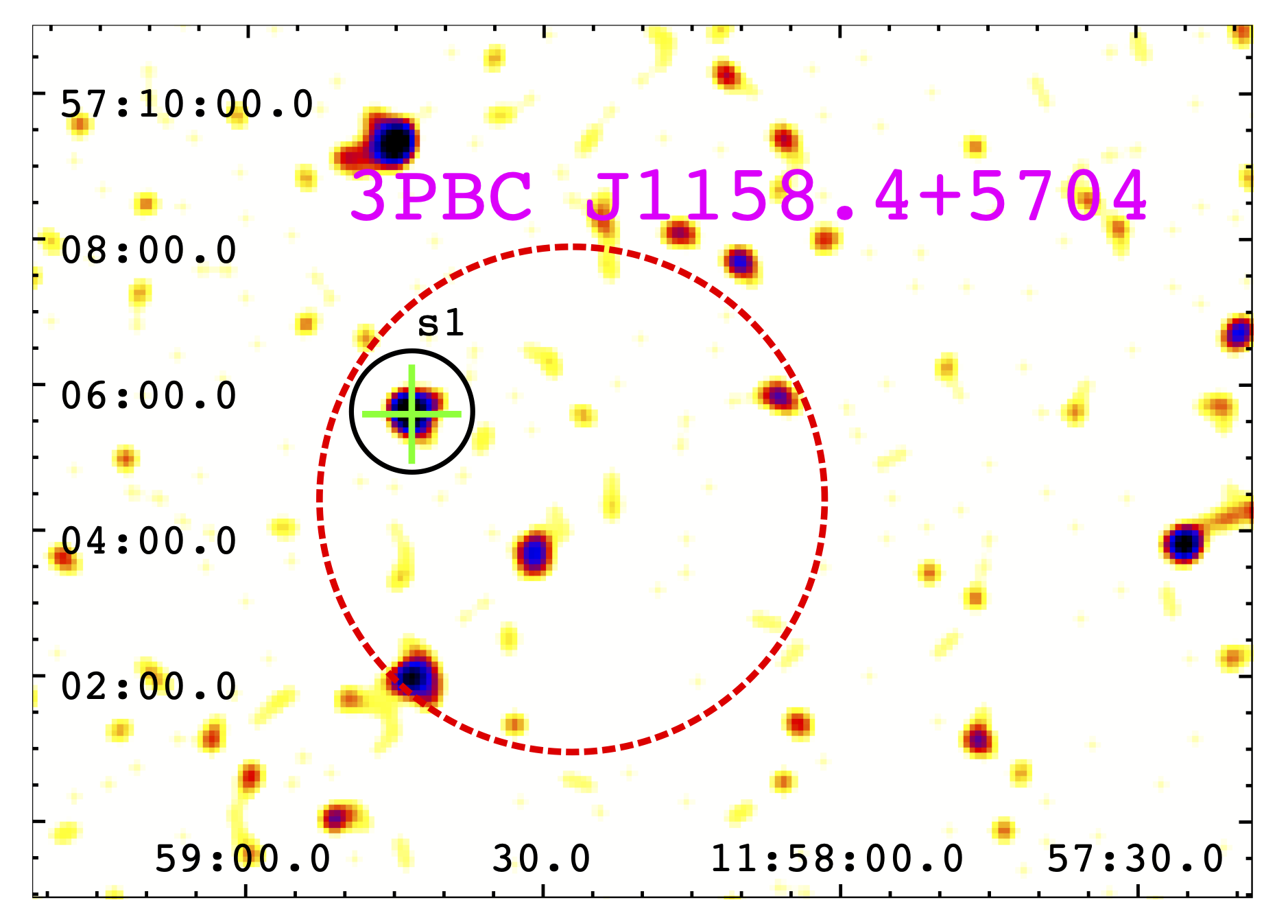}

    \caption{Images of 3PBC sources with exactly one soft \textit{Swift}-XRT source (XDF flag \textit{x}) detected inside of the BAT positional uncertainty region (red dashed circle). The soft X-ray detections are indicated with a black circle. The black circle indicates the position. It does not show the positional uncertainty of the source. If the soft X-ray detection is also marked with a green cross, it indicates that it has a WISE counterpart. }
    \label{fig:u_flagged_sources_no2}
\end{center}
\end{figure*}

\begin{figure*}[!th]
\begin{center}
    \includegraphics[height=4.2cm,width=6cm,angle=0]{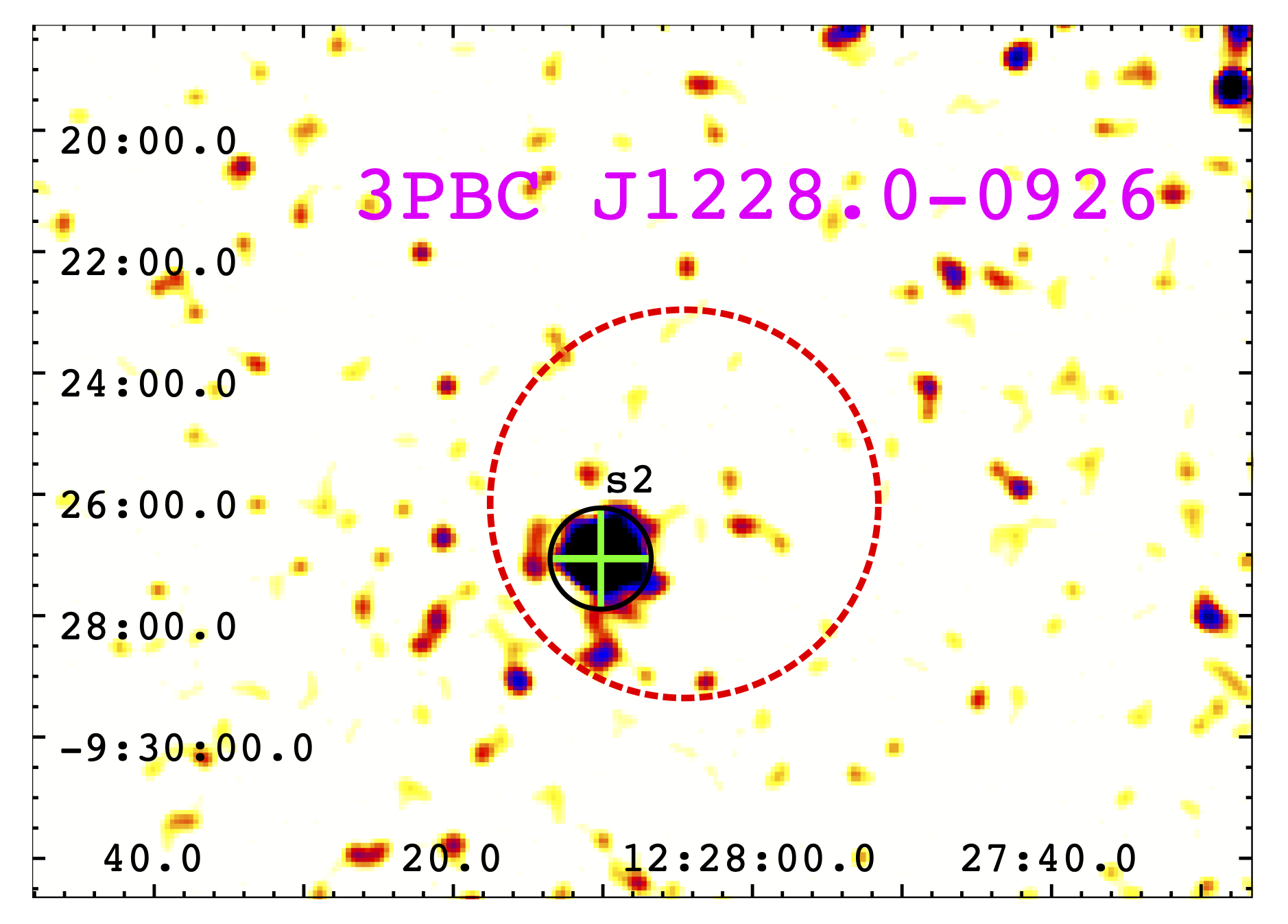}
    \includegraphics[height=4.2cm,width=6cm,angle=0]{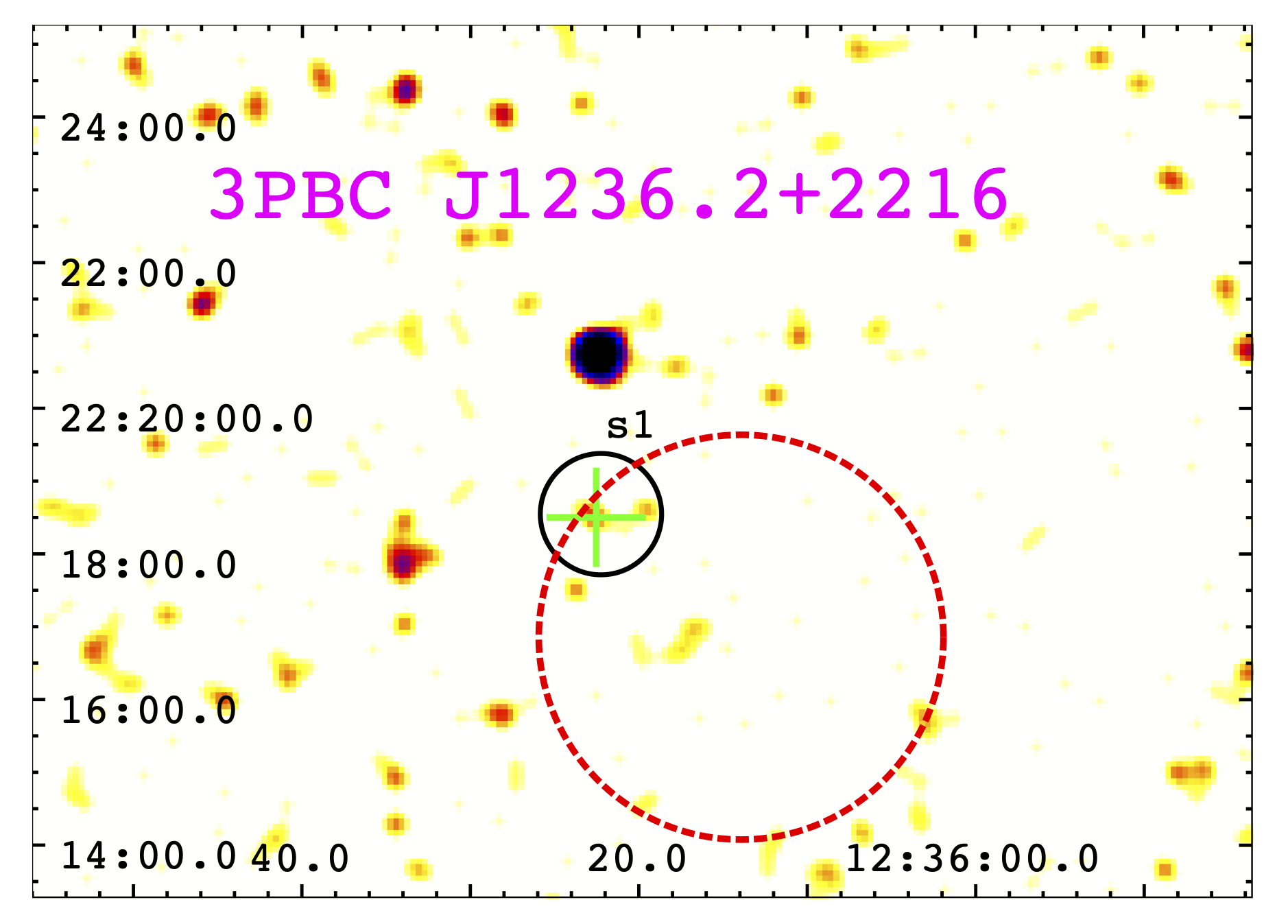}
    \includegraphics[height=4.2cm,width=6cm,angle=0]{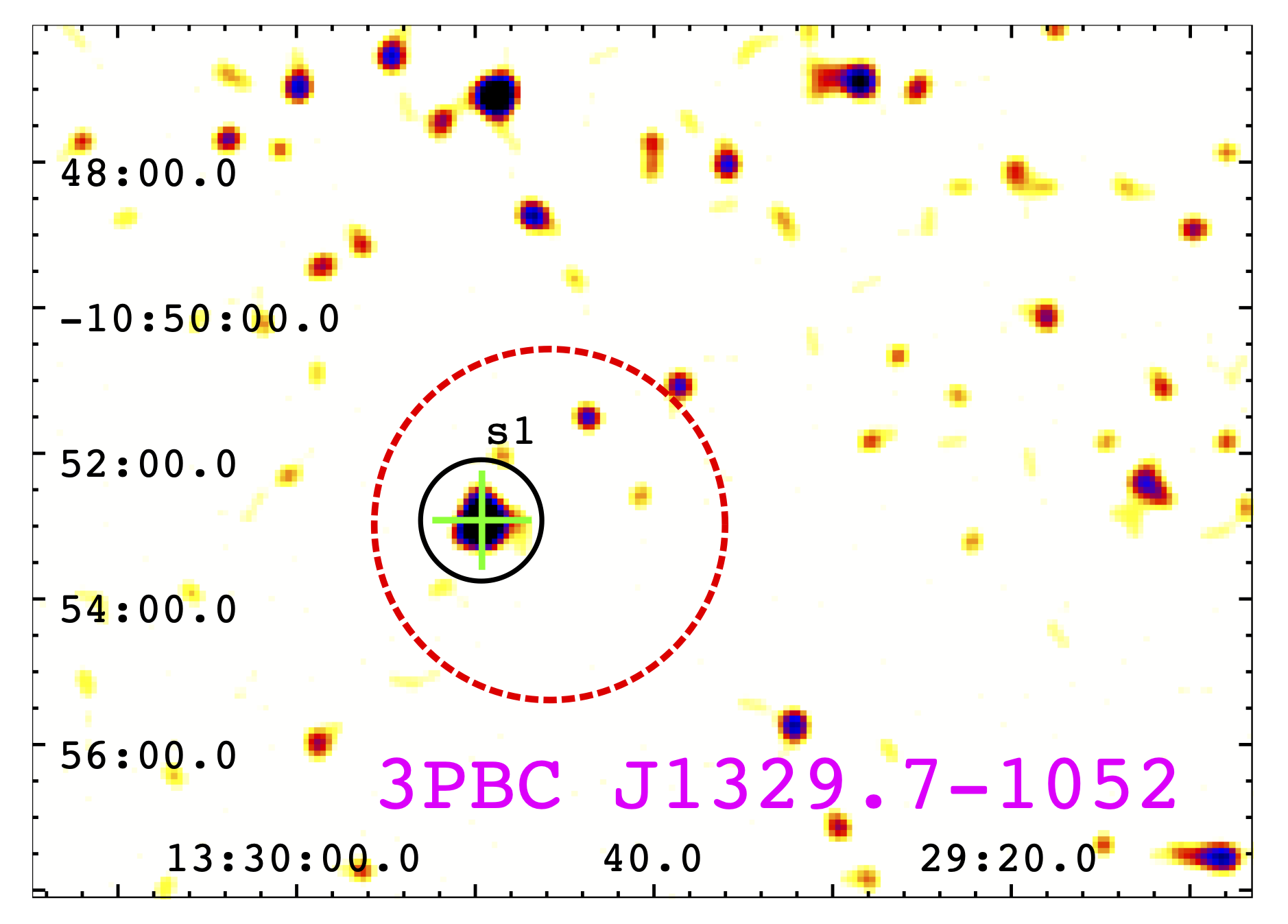}
    \includegraphics[height=4.2cm,width=6cm,angle=0]{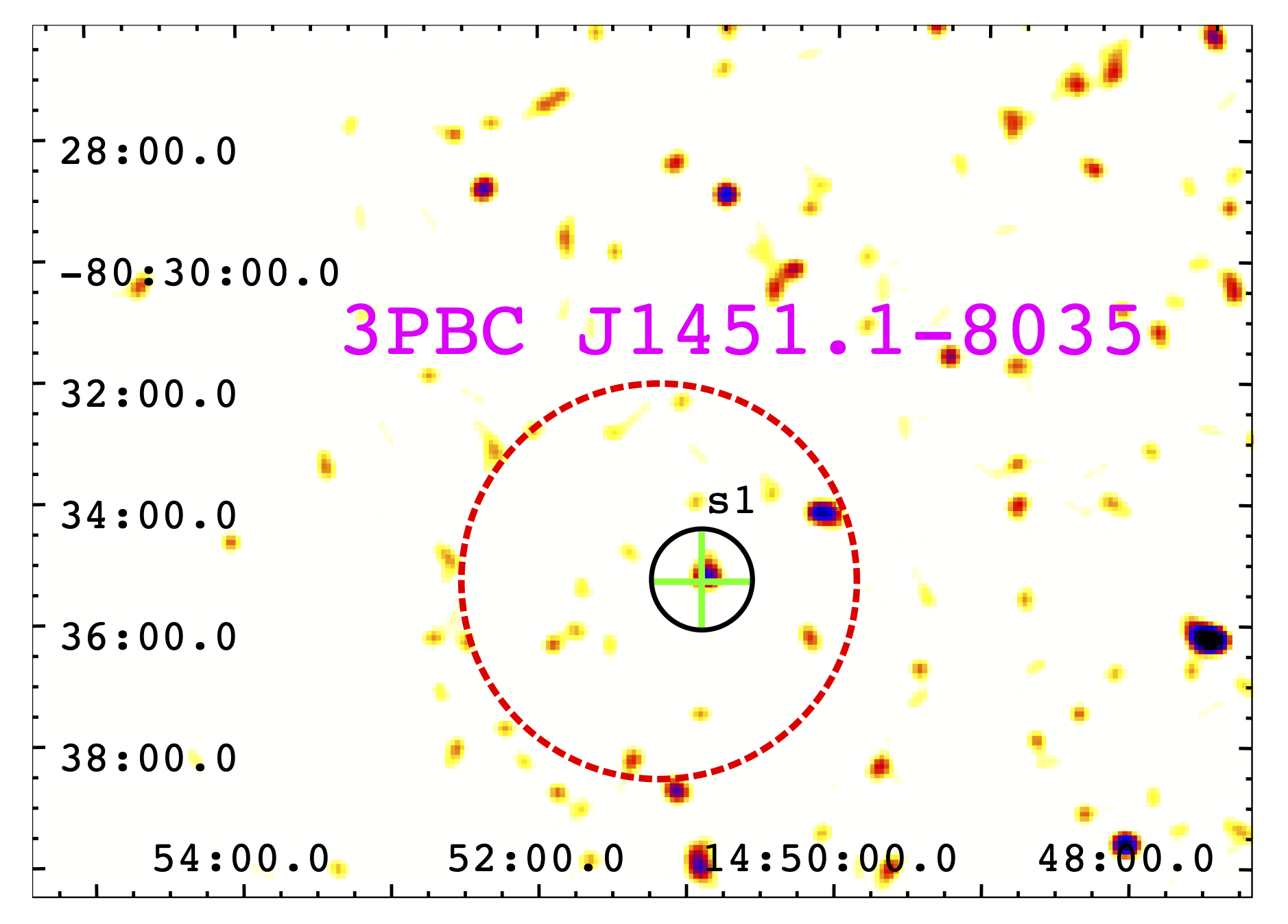}
    \includegraphics[height=4.2cm,width=6cm,angle=0]{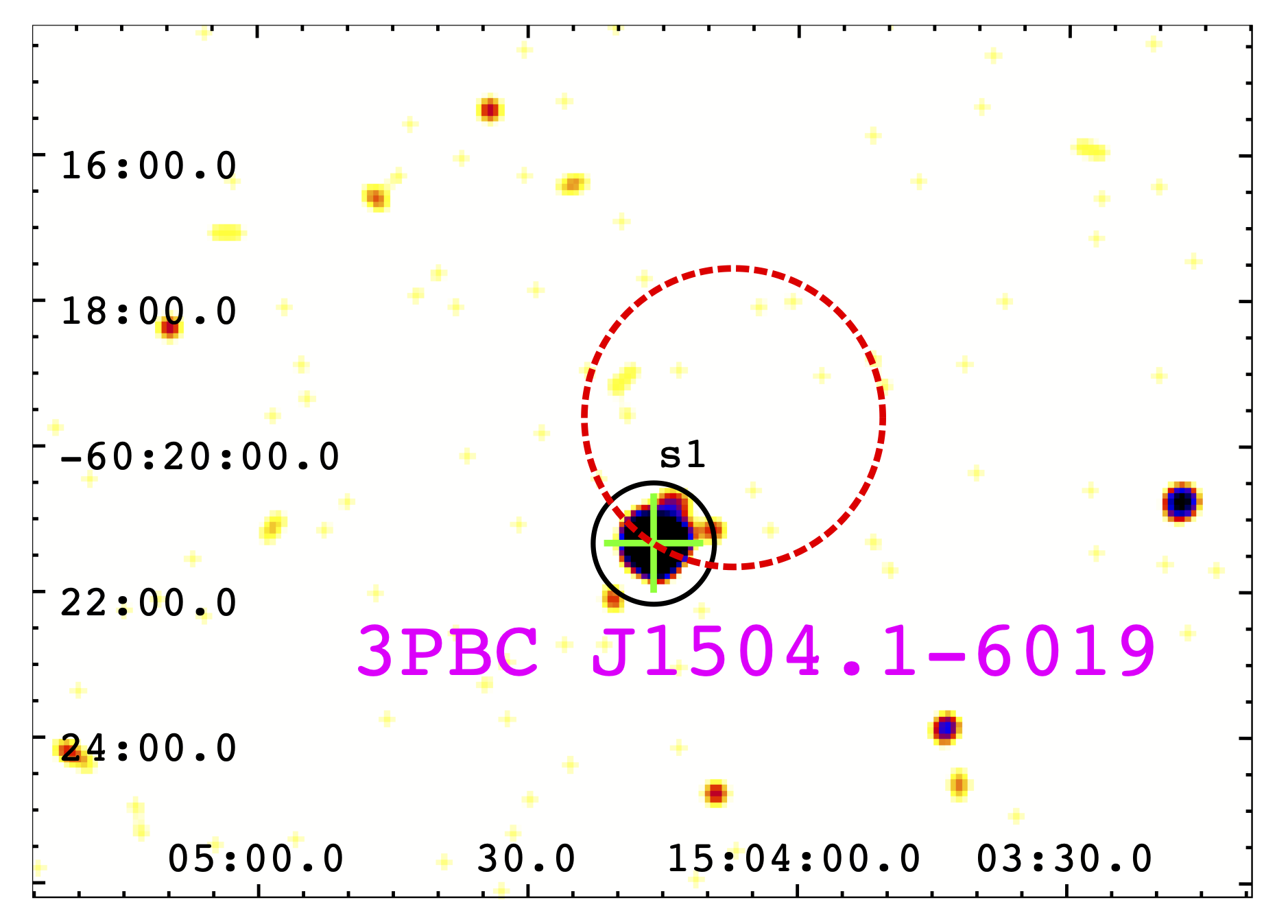}
    \includegraphics[height=4.2cm,width=6cm,angle=0]{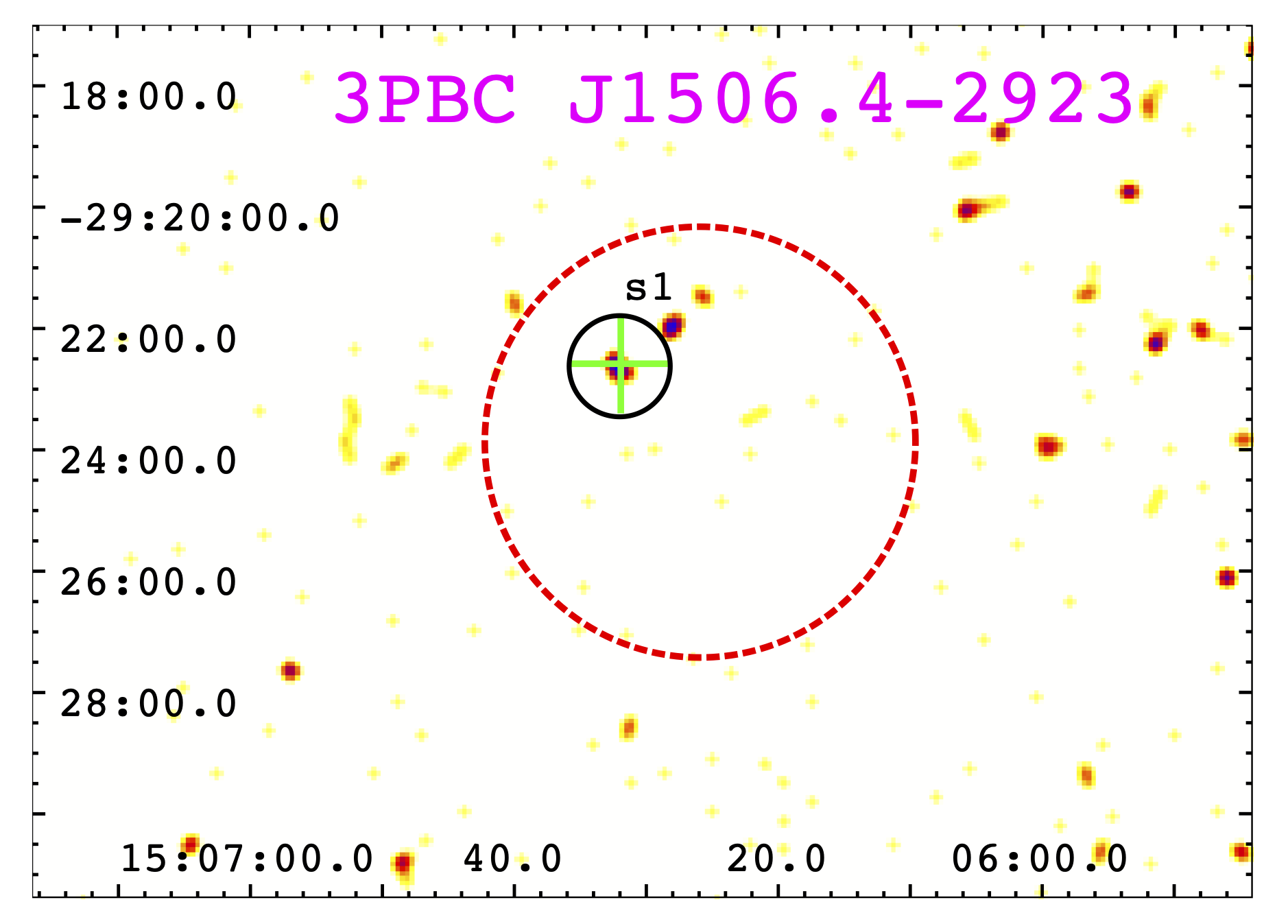}
    \includegraphics[height=4.2cm,width=6cm,angle=0]{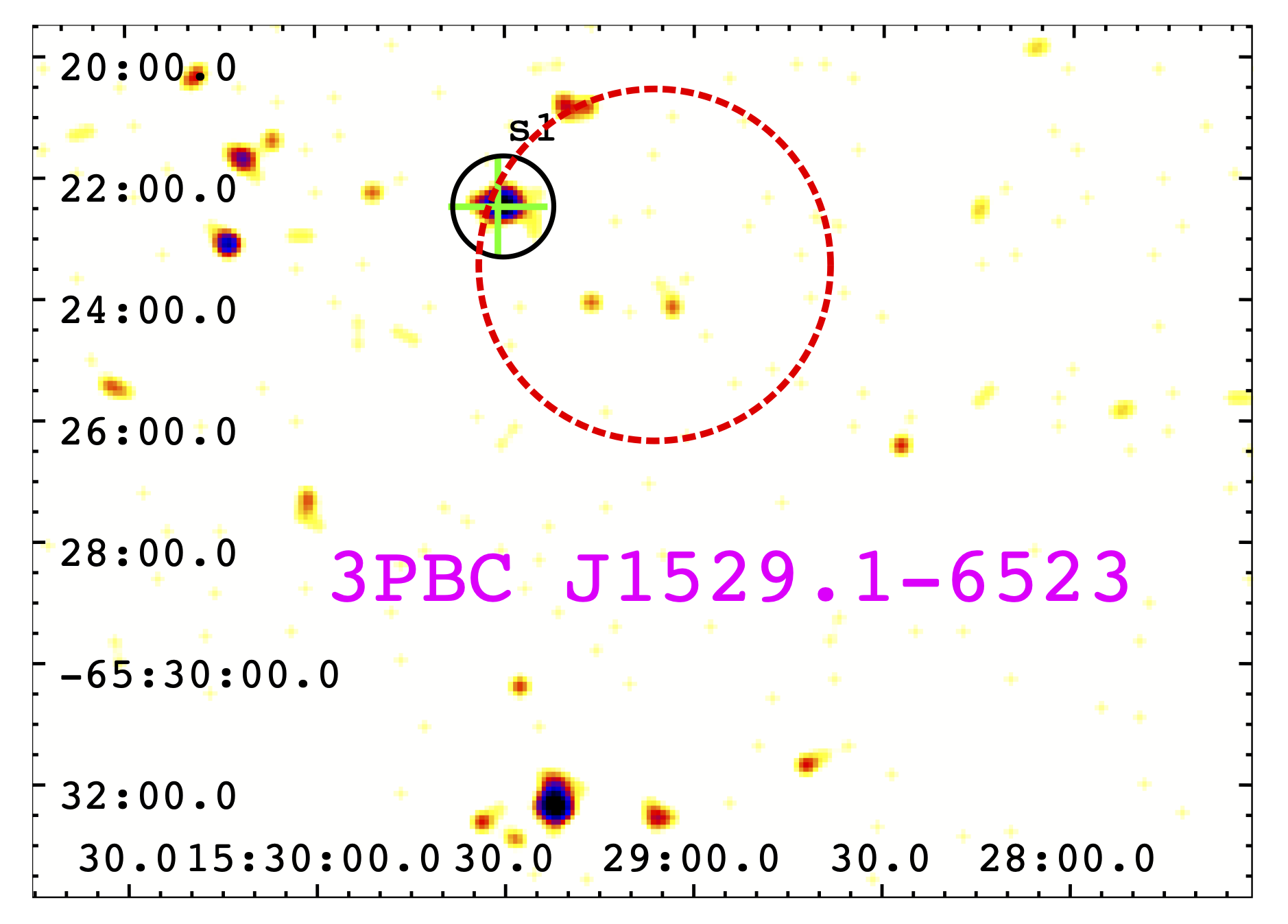}
    \includegraphics[height=4.2cm,width=6cm,angle=0]{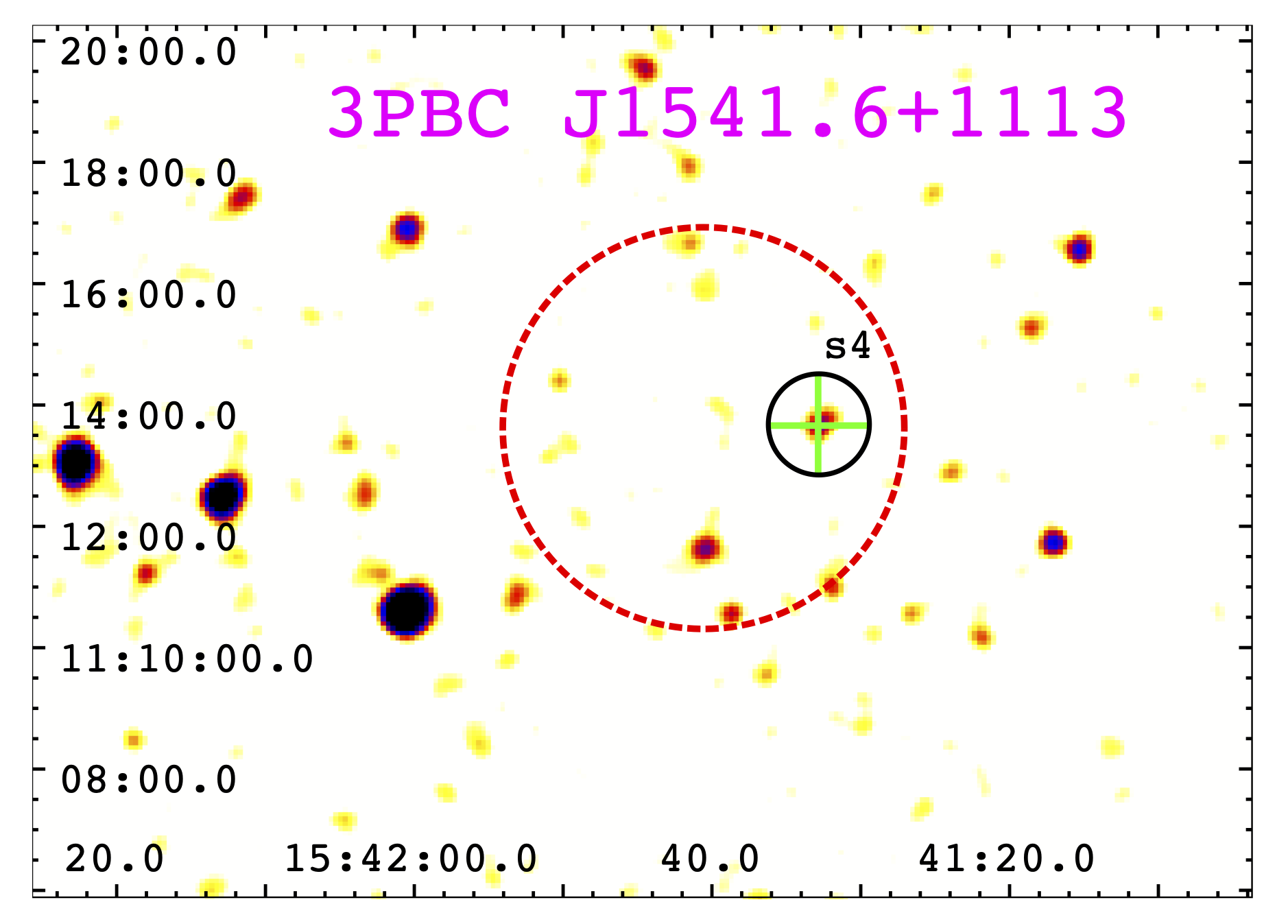}
    \includegraphics[height=4.2cm,width=6cm,angle=0]{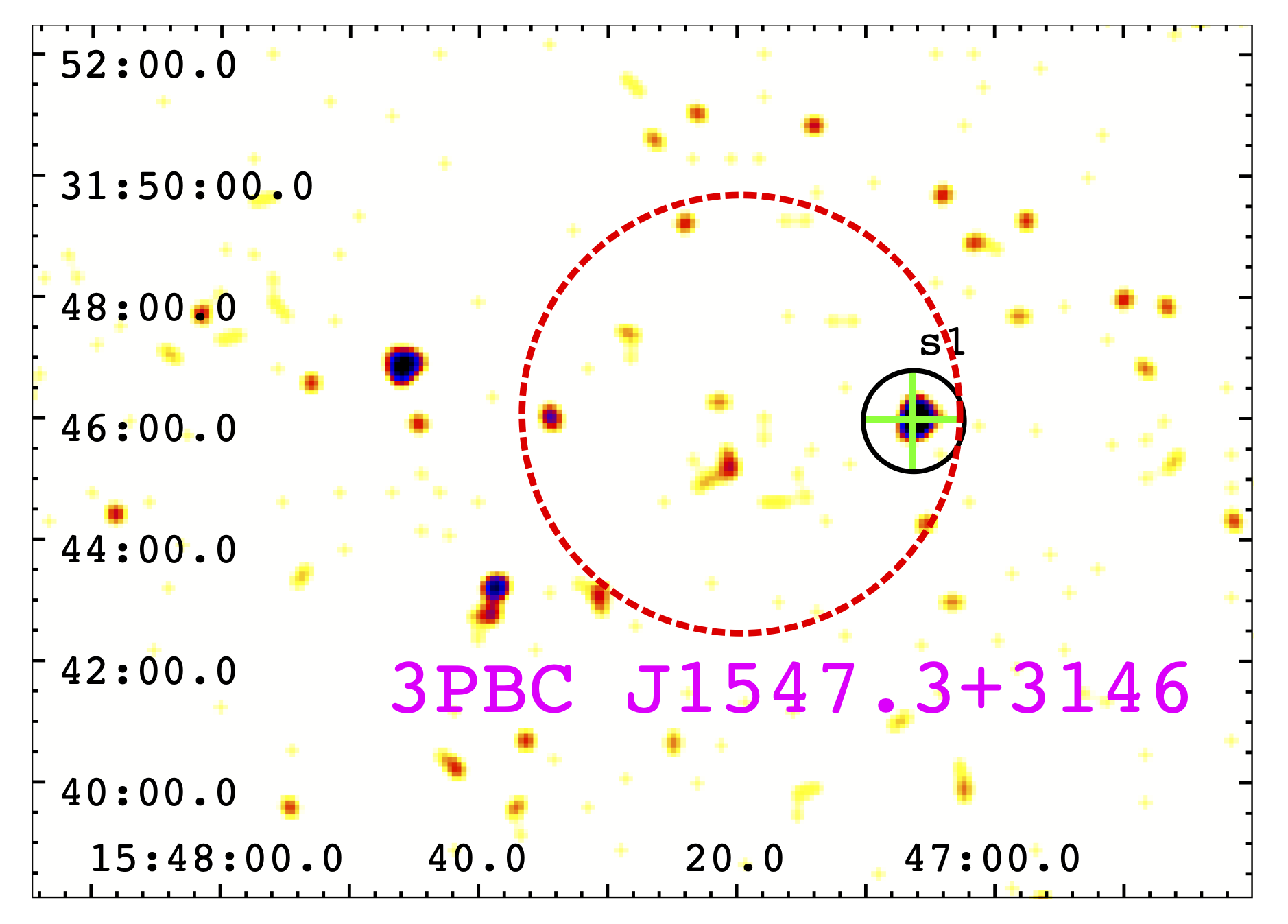}
    \includegraphics[height=4.2cm,width=6cm,angle=0]{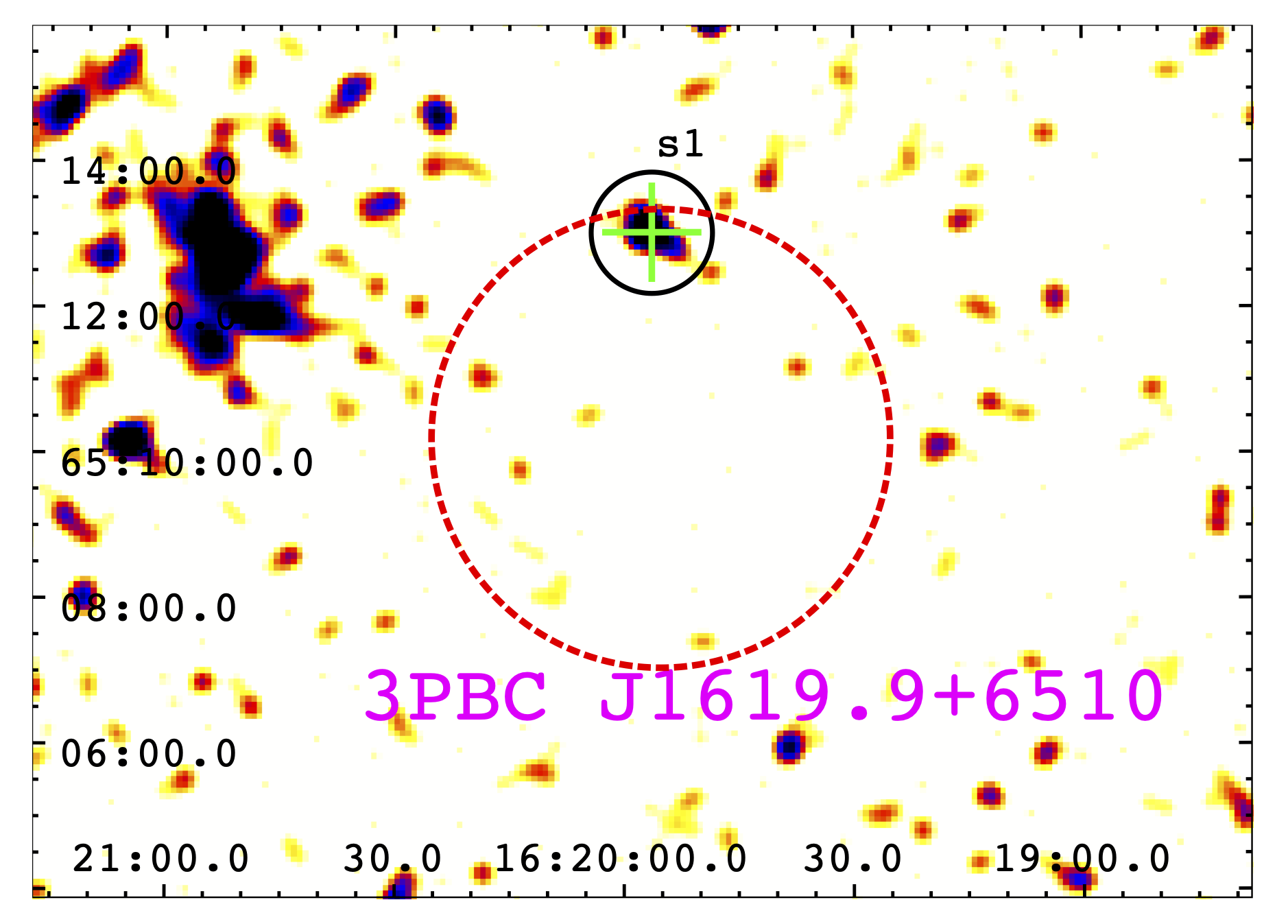}
    \includegraphics[height=4.2cm,width=6cm,angle=0]{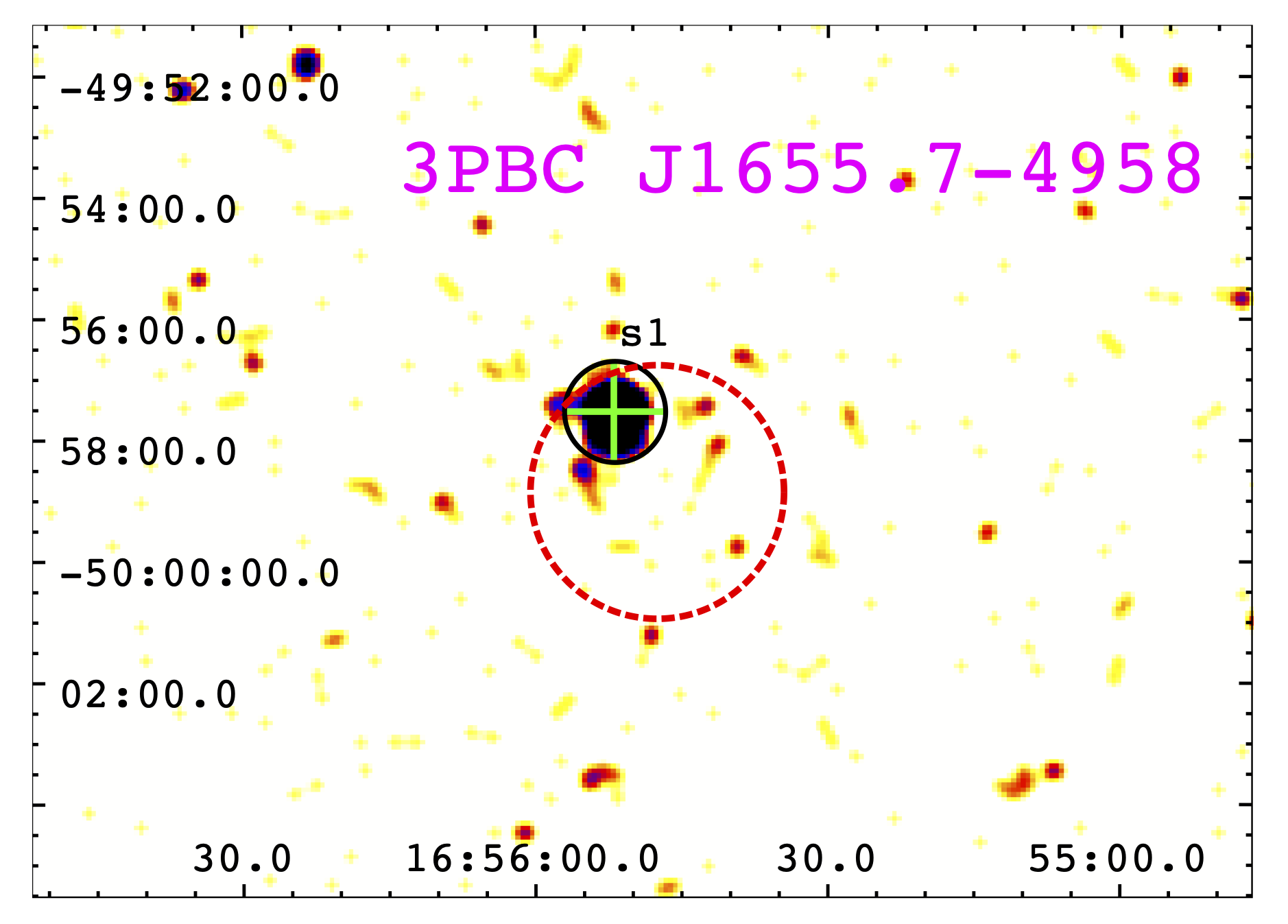}
    \includegraphics[height=4.2cm,width=6cm,angle=0]{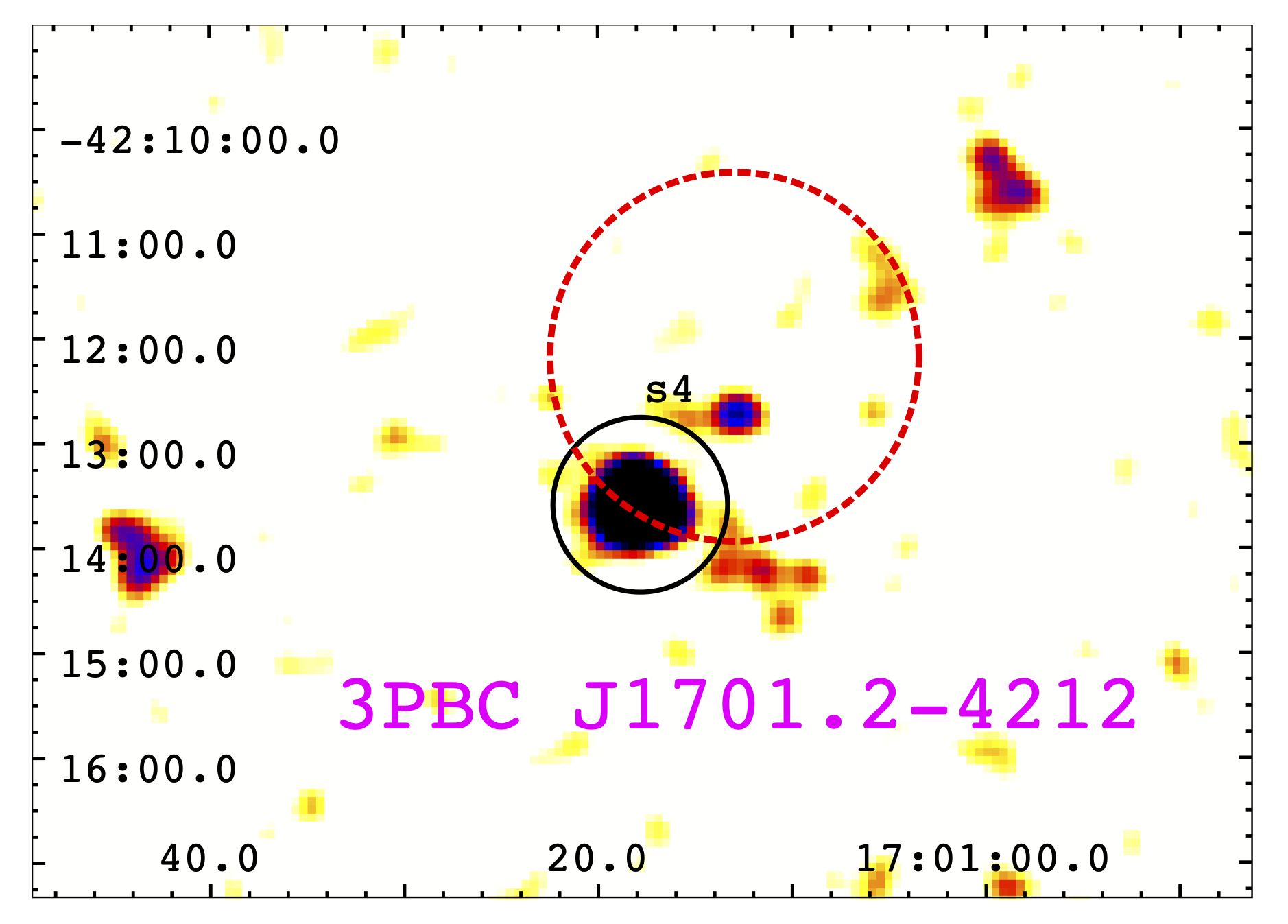}
    \includegraphics[height=4.2cm,width=6cm,angle=0]{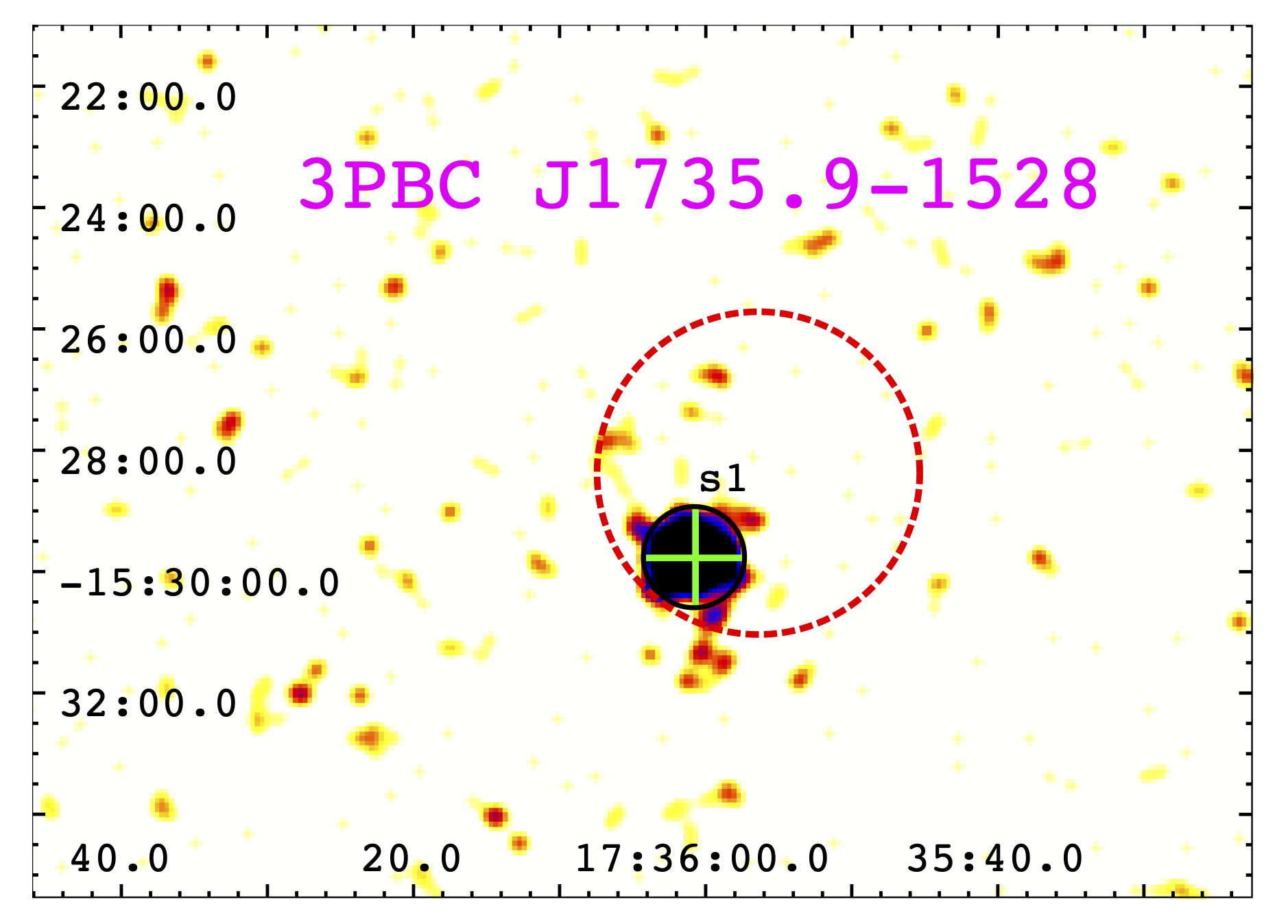}
    \includegraphics[height=4.2cm,width=6cm,angle=0]{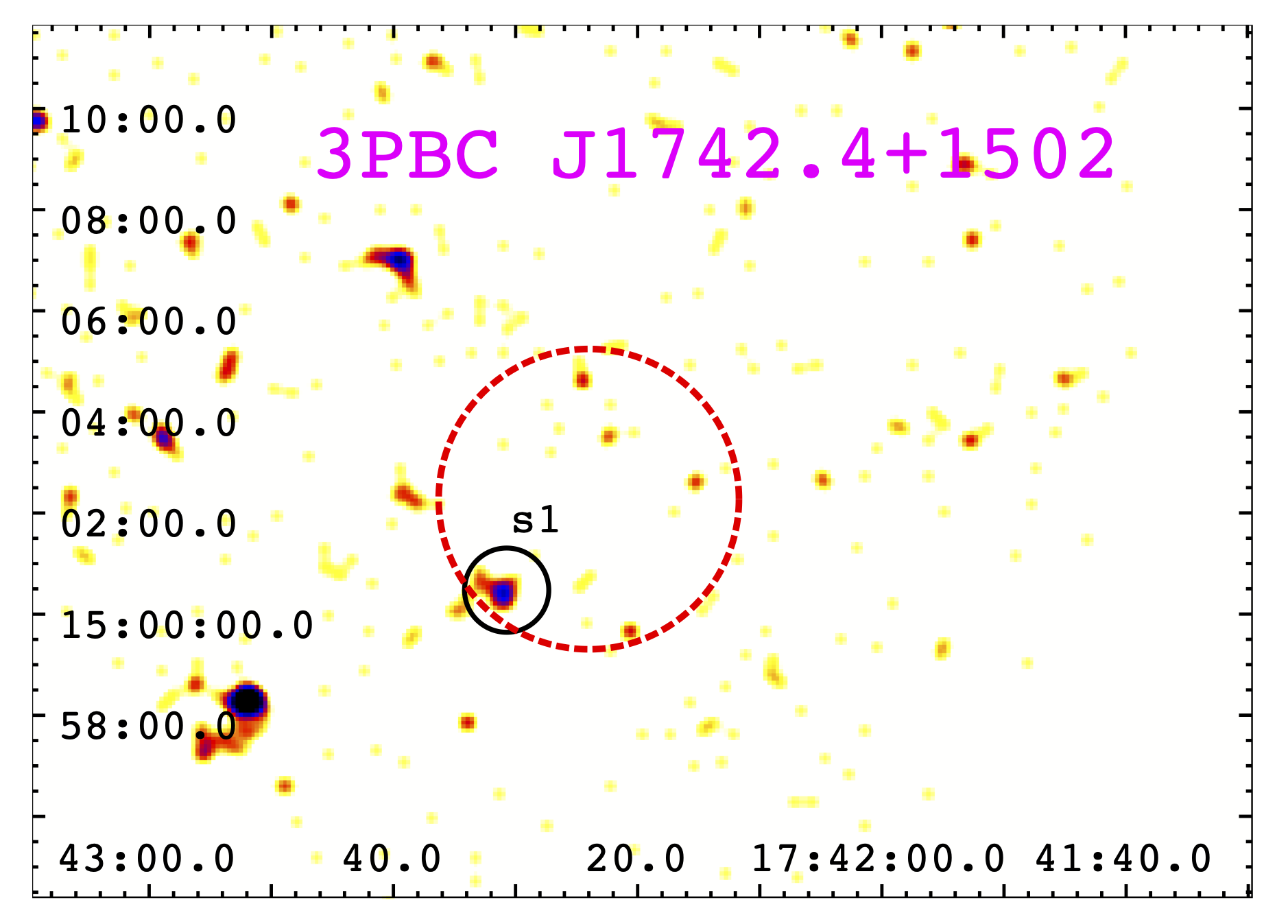}
    \includegraphics[height=4.2cm,width=6cm,angle=0]{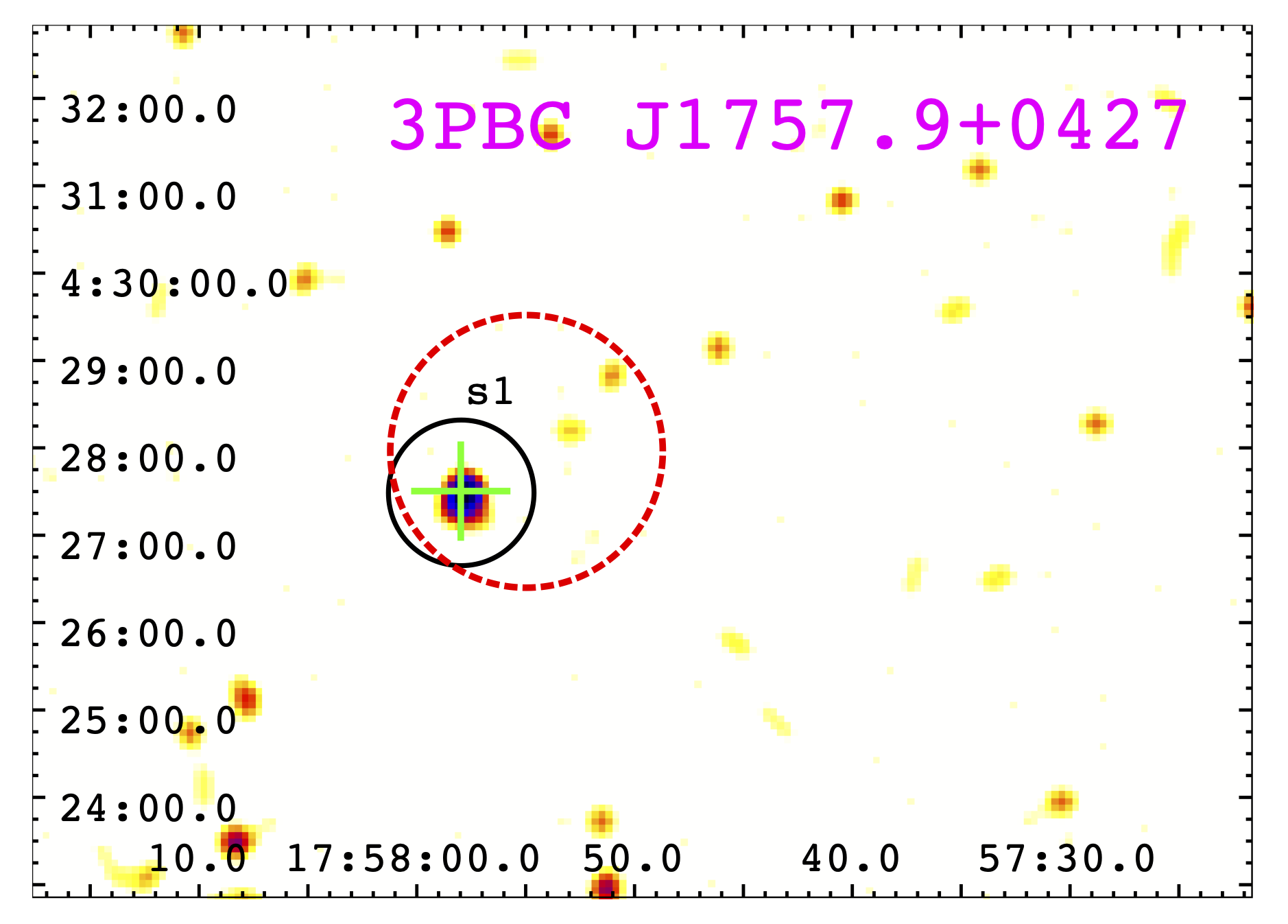}

    \caption{Images of 3PBC sources with exactly one soft \textit{Swift}-XRT source (XDF flag \textit{x}) detected inside of the BAT positional uncertainty region (red dashed circle). The soft X-ray detections are indicated with a black circle. The black circle indicates the position. It does not show the positional uncertainty of the source. If the soft X-ray detection is also marked with a green cross, it indicates that it has a WISE counterpart. }
    \label{fig:u_flagged_sources_no3}
\end{center}
\end{figure*}

\begin{figure*}[!th]
\begin{center}
    
    \includegraphics[height=4.2cm,width=6cm,angle=0]{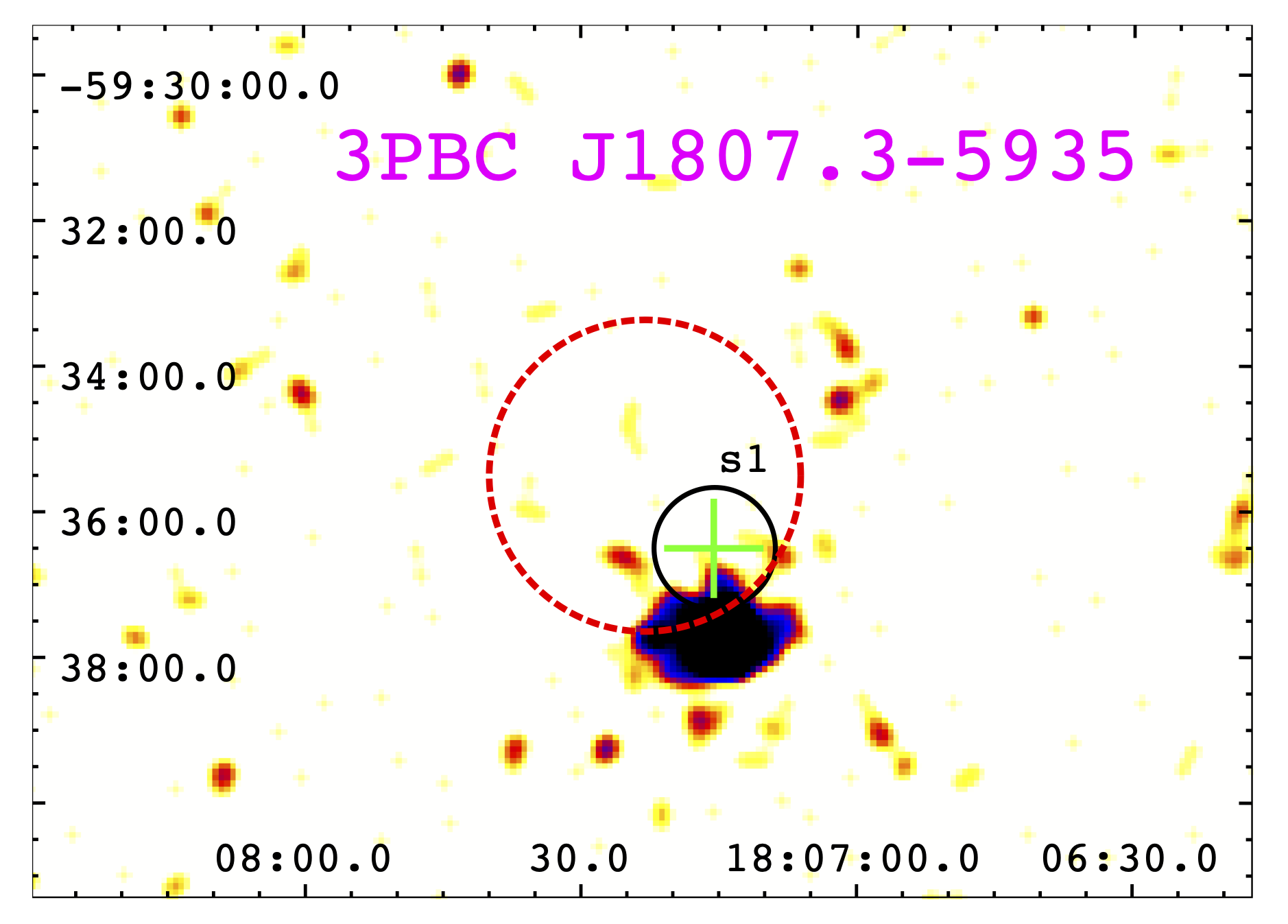}
    \includegraphics[height=4.2cm,width=6cm,angle=0]{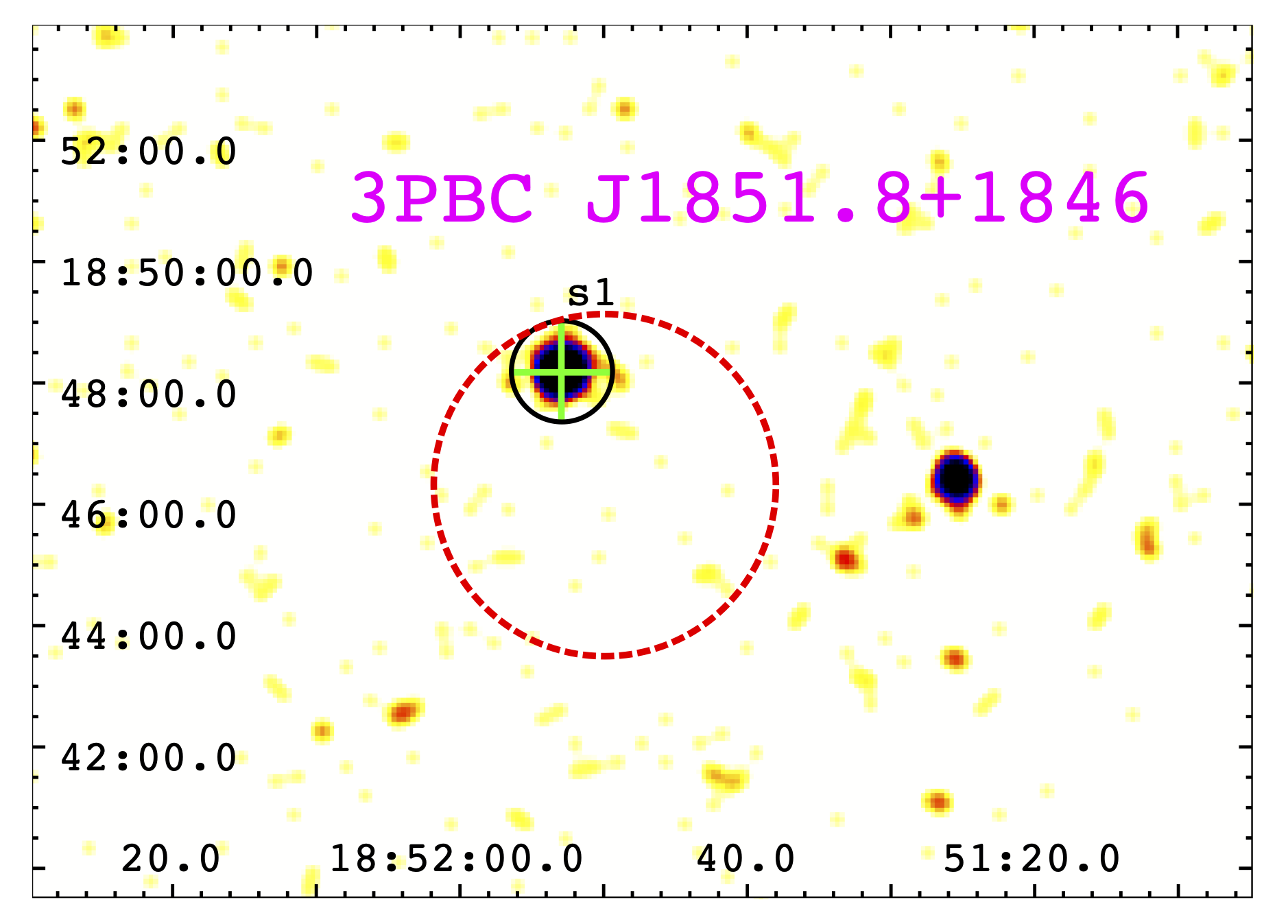}
    \includegraphics[height=4.2cm,width=6cm,angle=0]{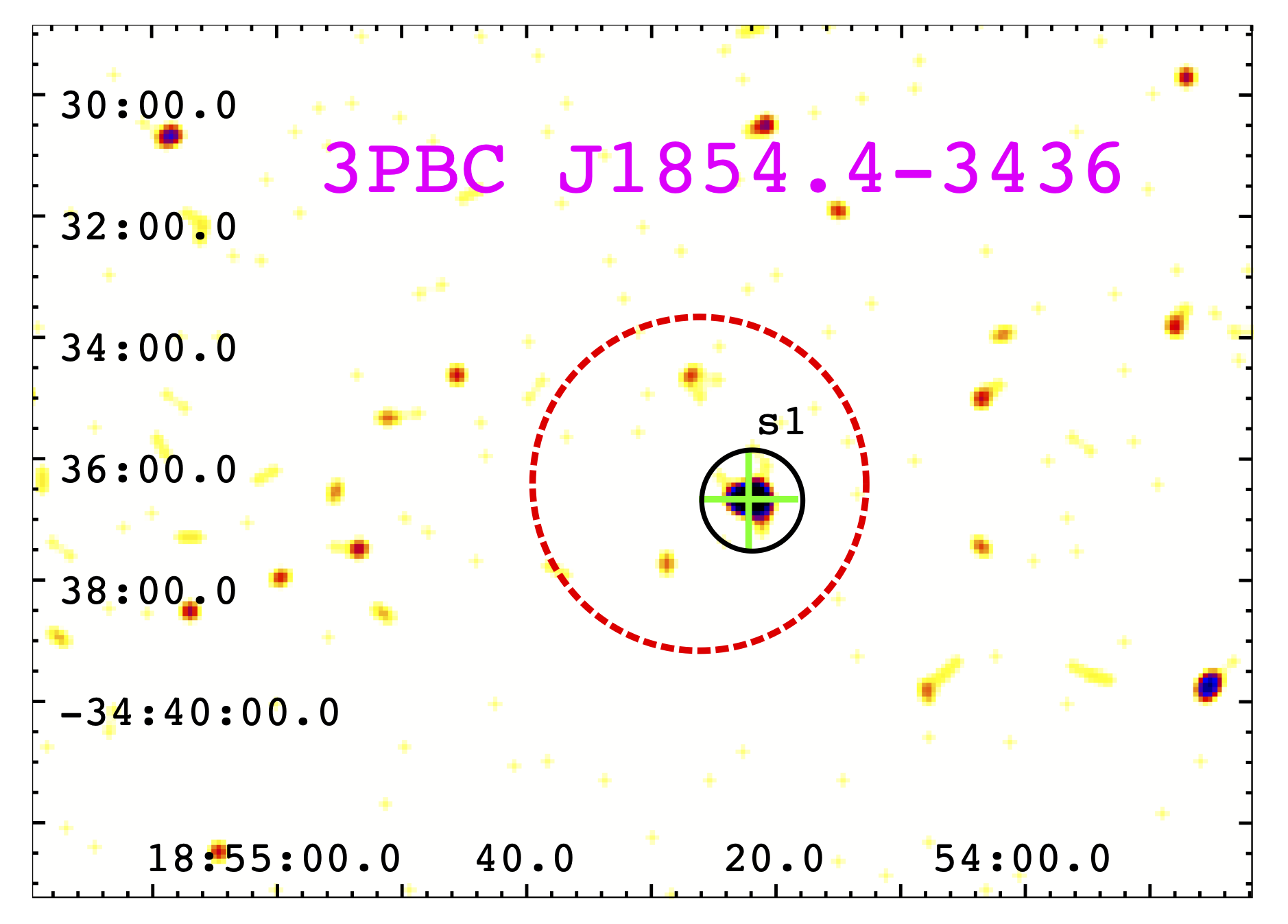}
    \includegraphics[height=4.2cm,width=6cm,angle=0]{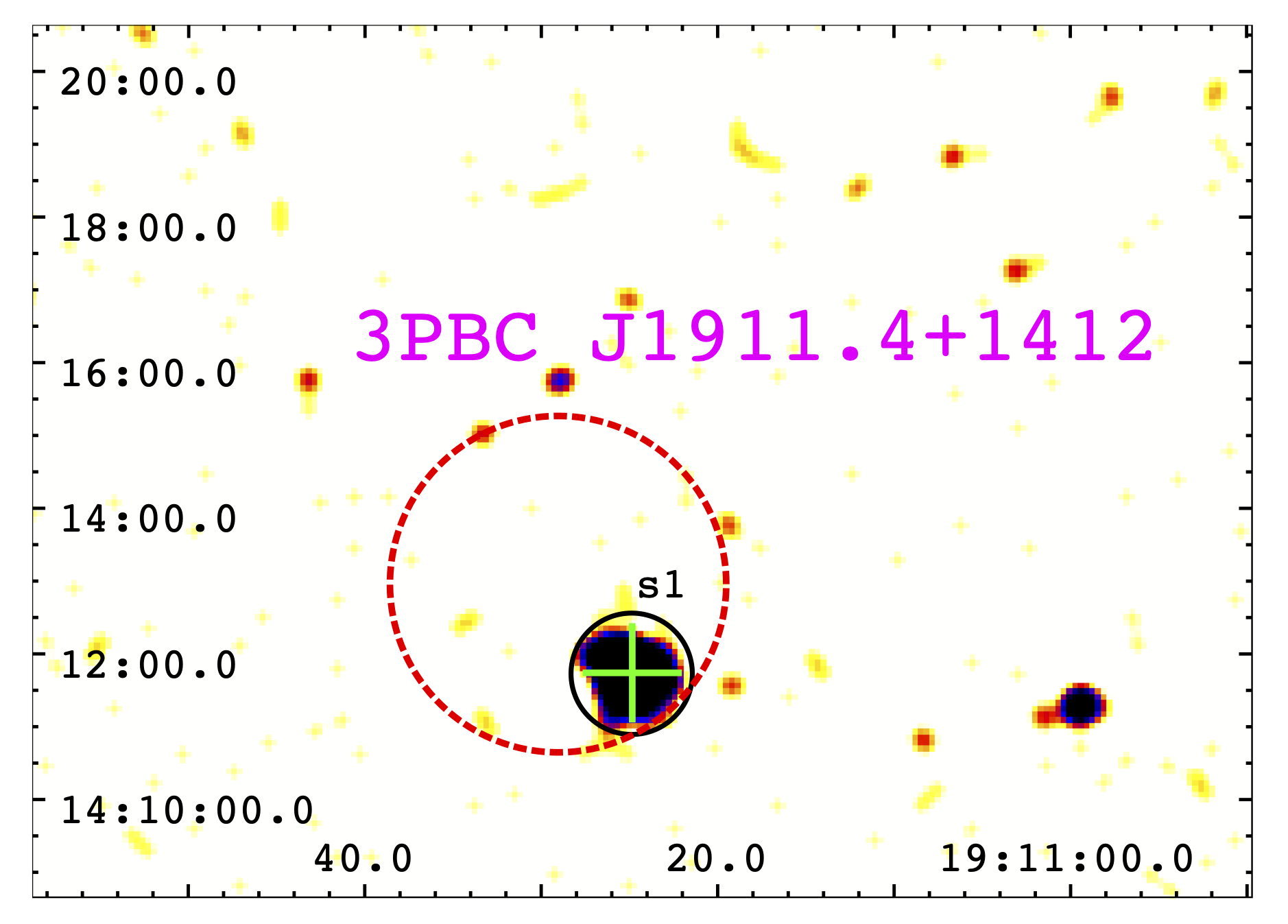}
    \includegraphics[height=4.2cm,width=6cm,angle=0]{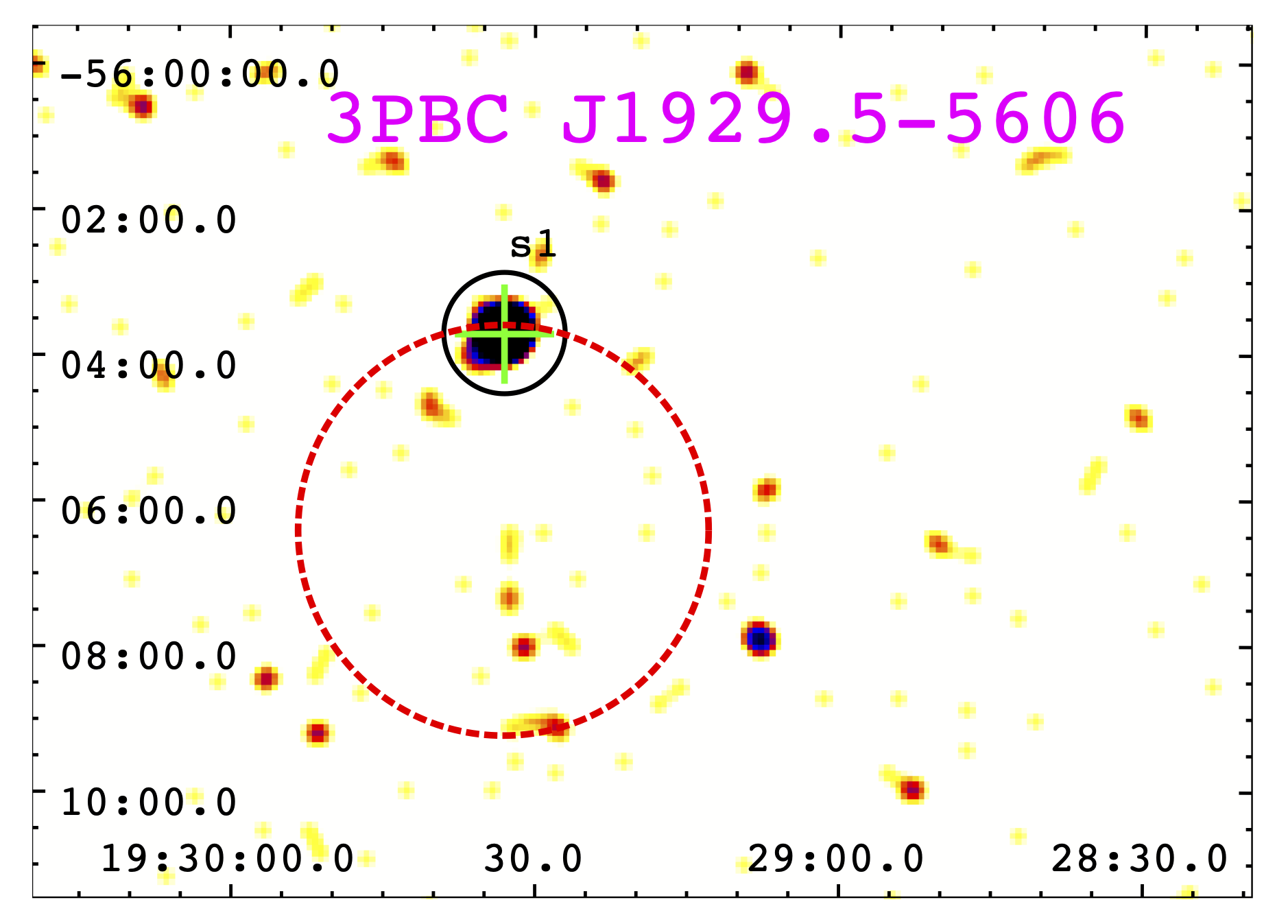}
    \includegraphics[height=4.2cm,width=6cm,angle=0]{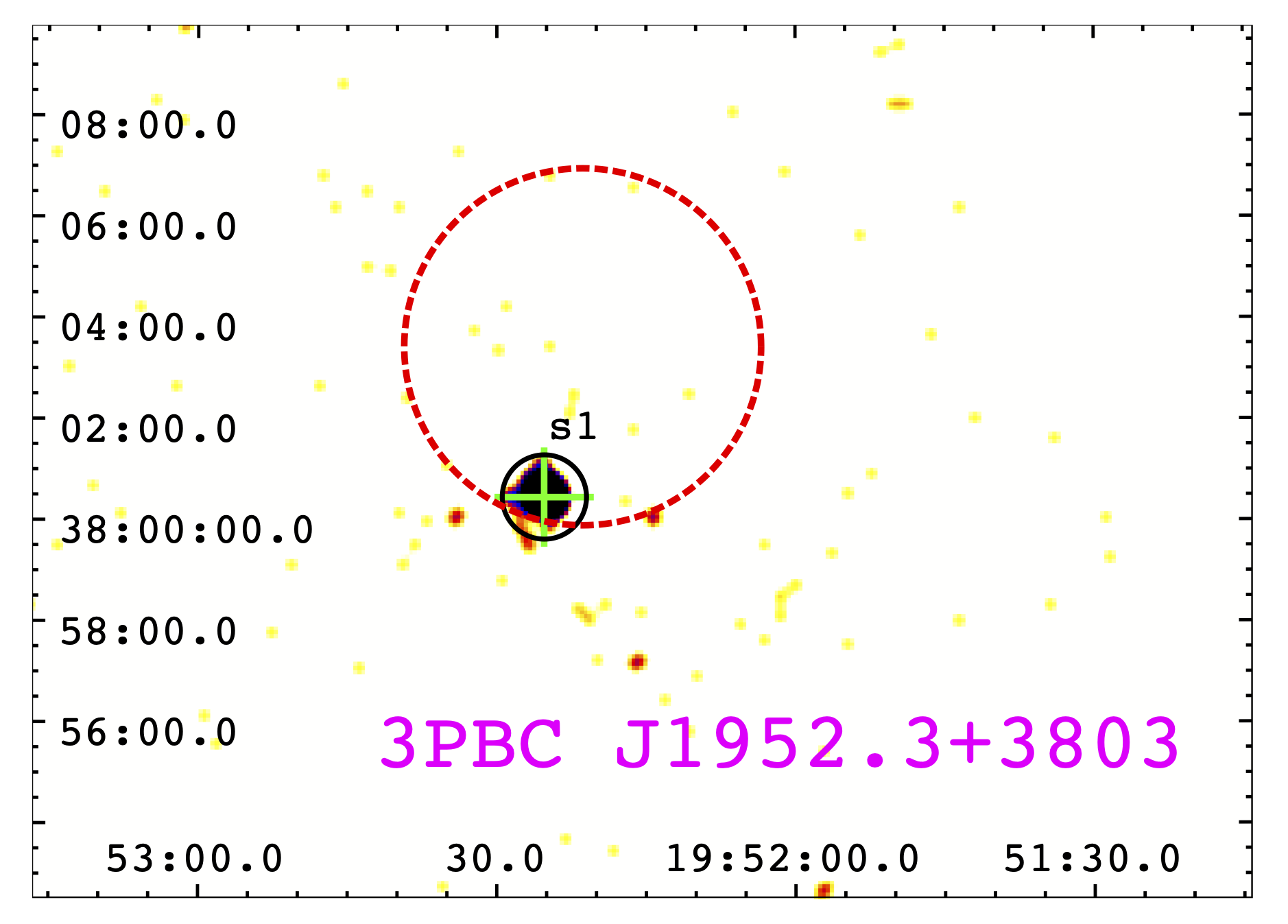}
    \includegraphics[height=4.2cm,width=6cm,angle=0]{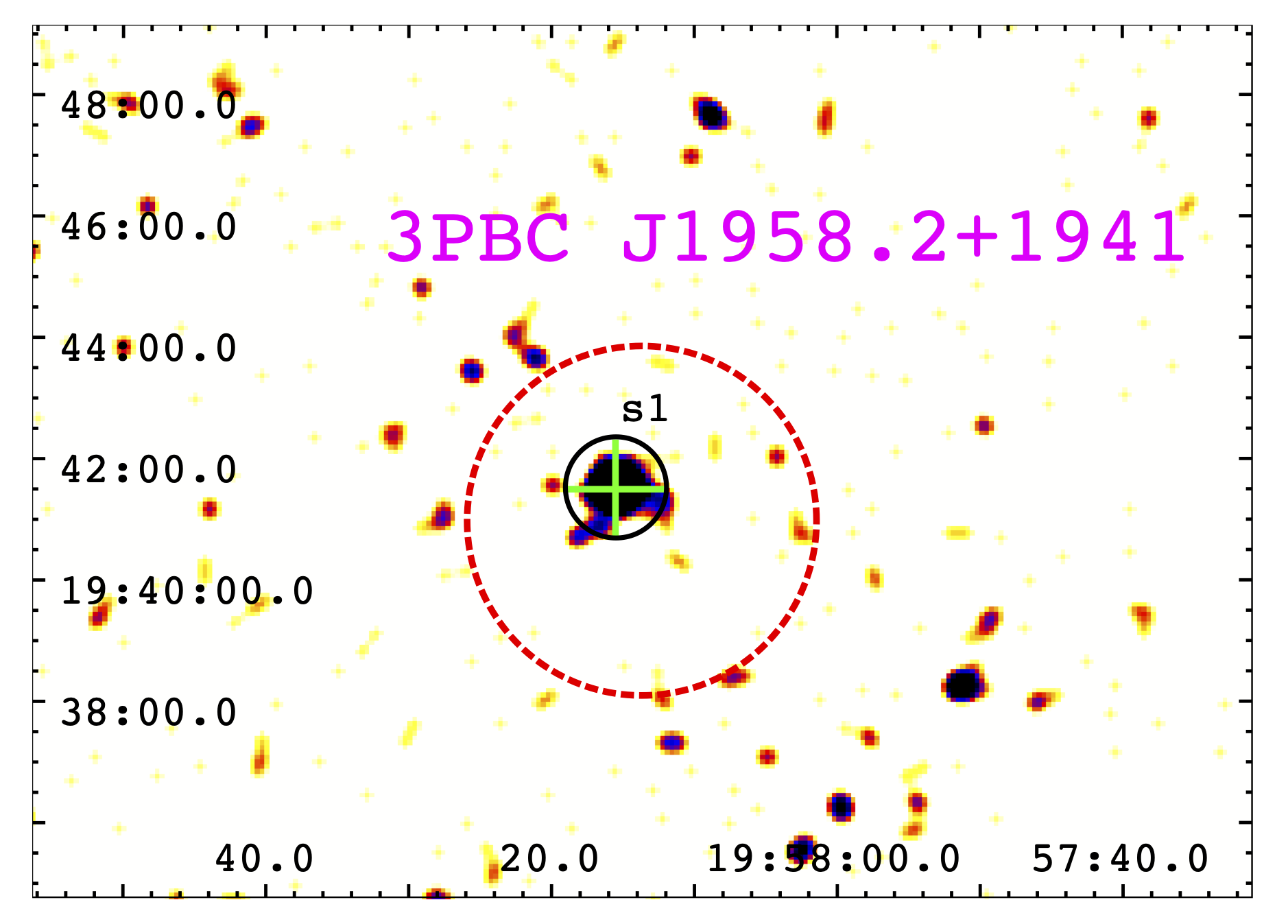}
    \includegraphics[height=4.2cm,width=6cm,angle=0]{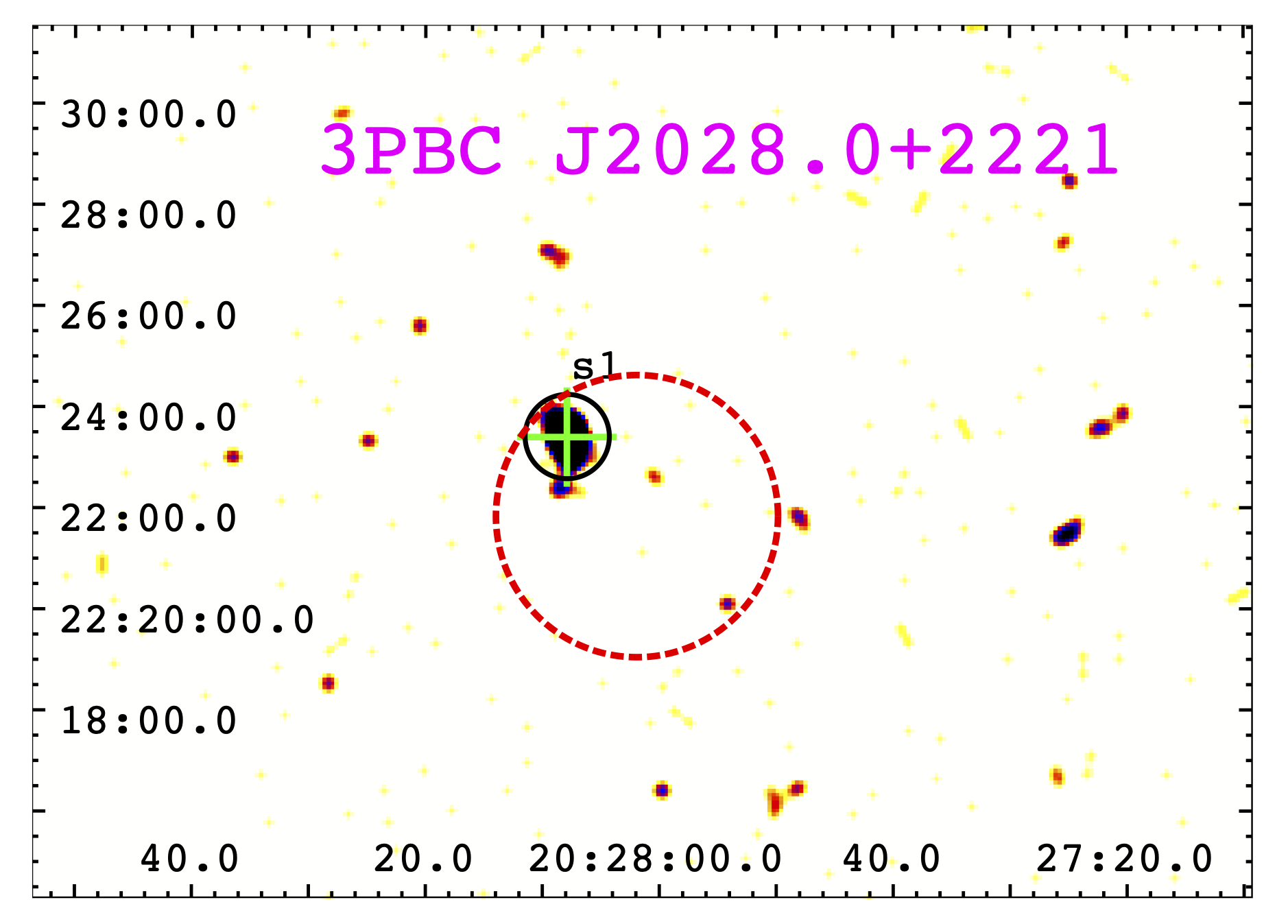}
    \includegraphics[height=4.2cm,width=6cm,angle=0]{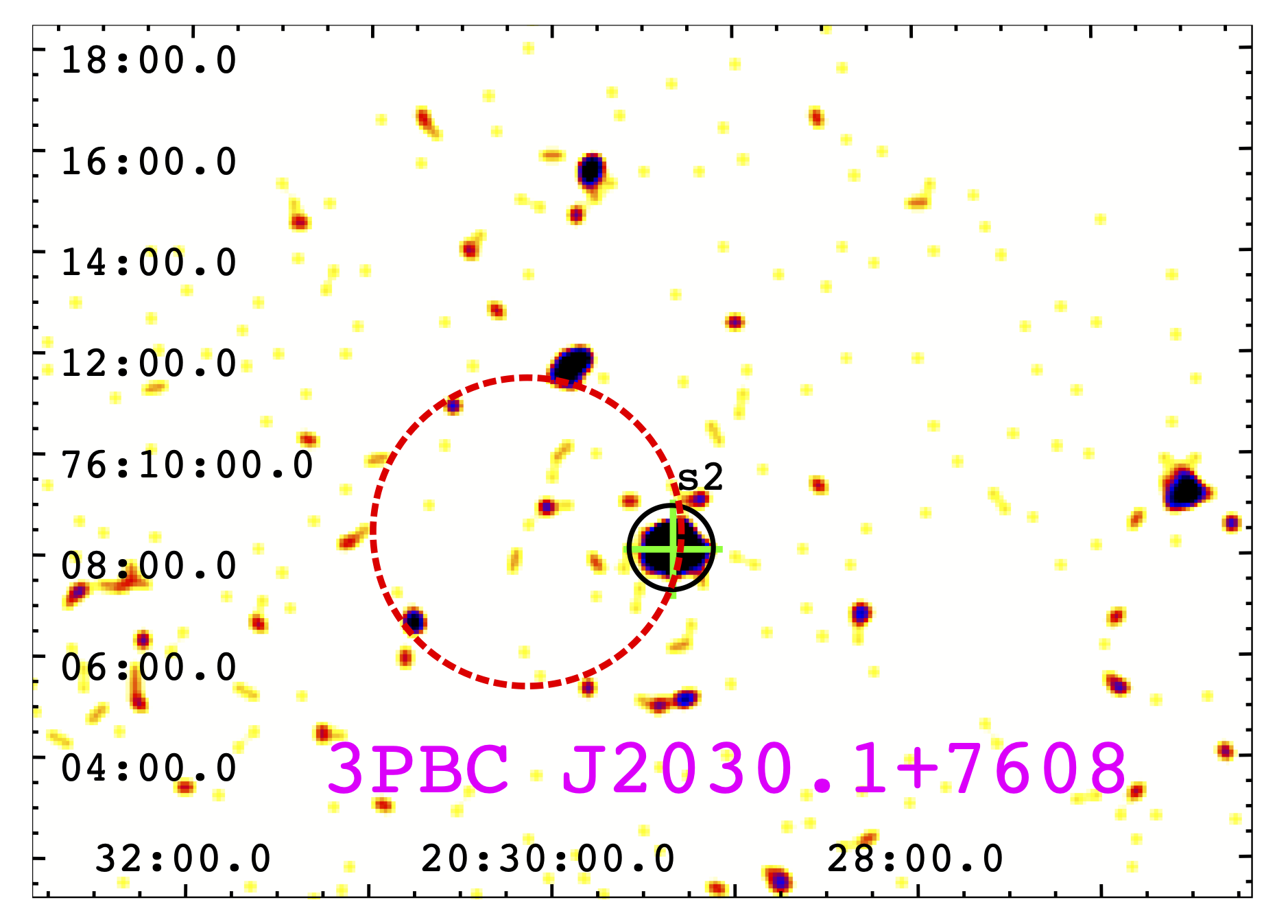}
    \includegraphics[height=4.2cm,width=6cm,angle=0]{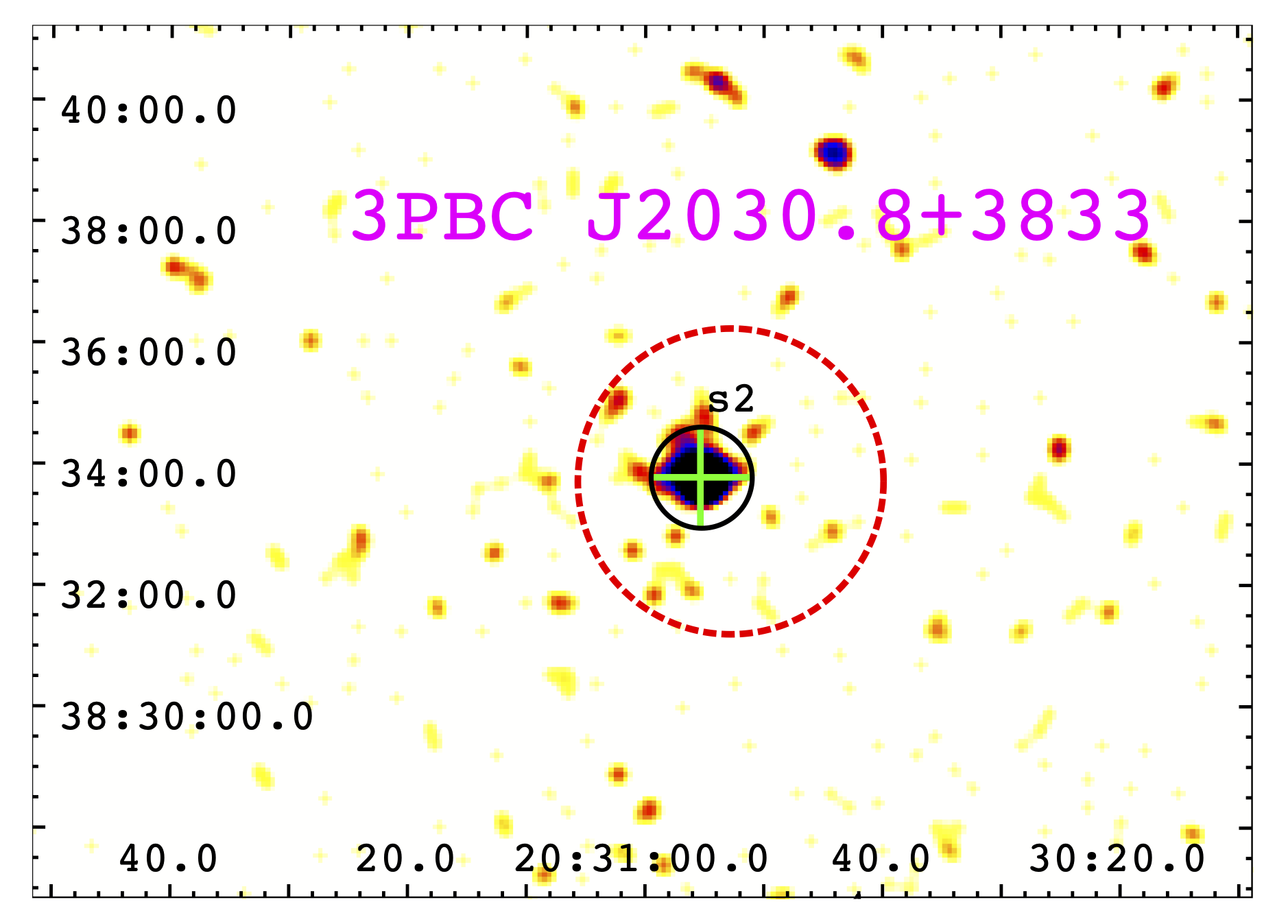}
    \includegraphics[height=4.2cm,width=6cm,angle=0]{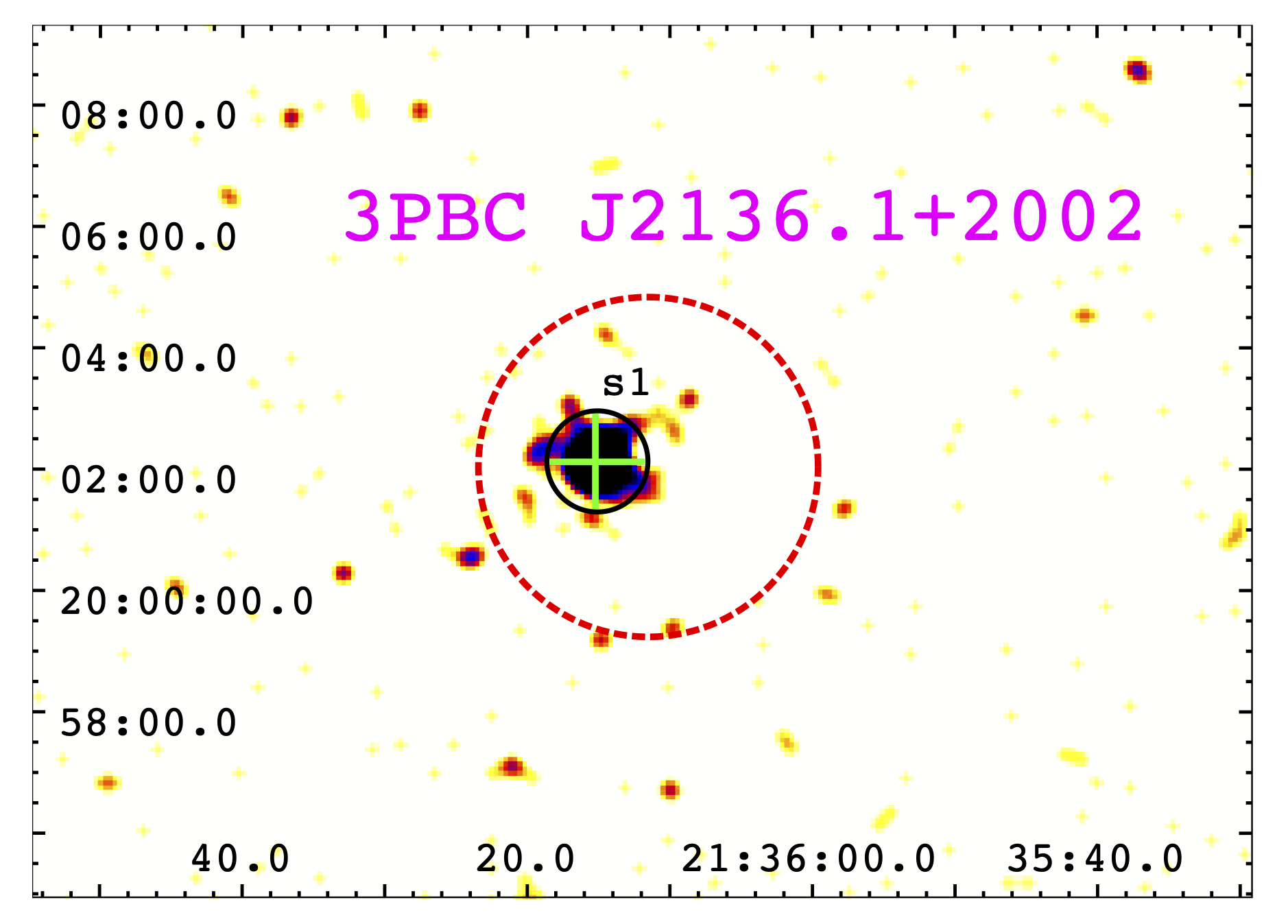}
    \includegraphics[height=4.2cm,width=6cm,angle=0]{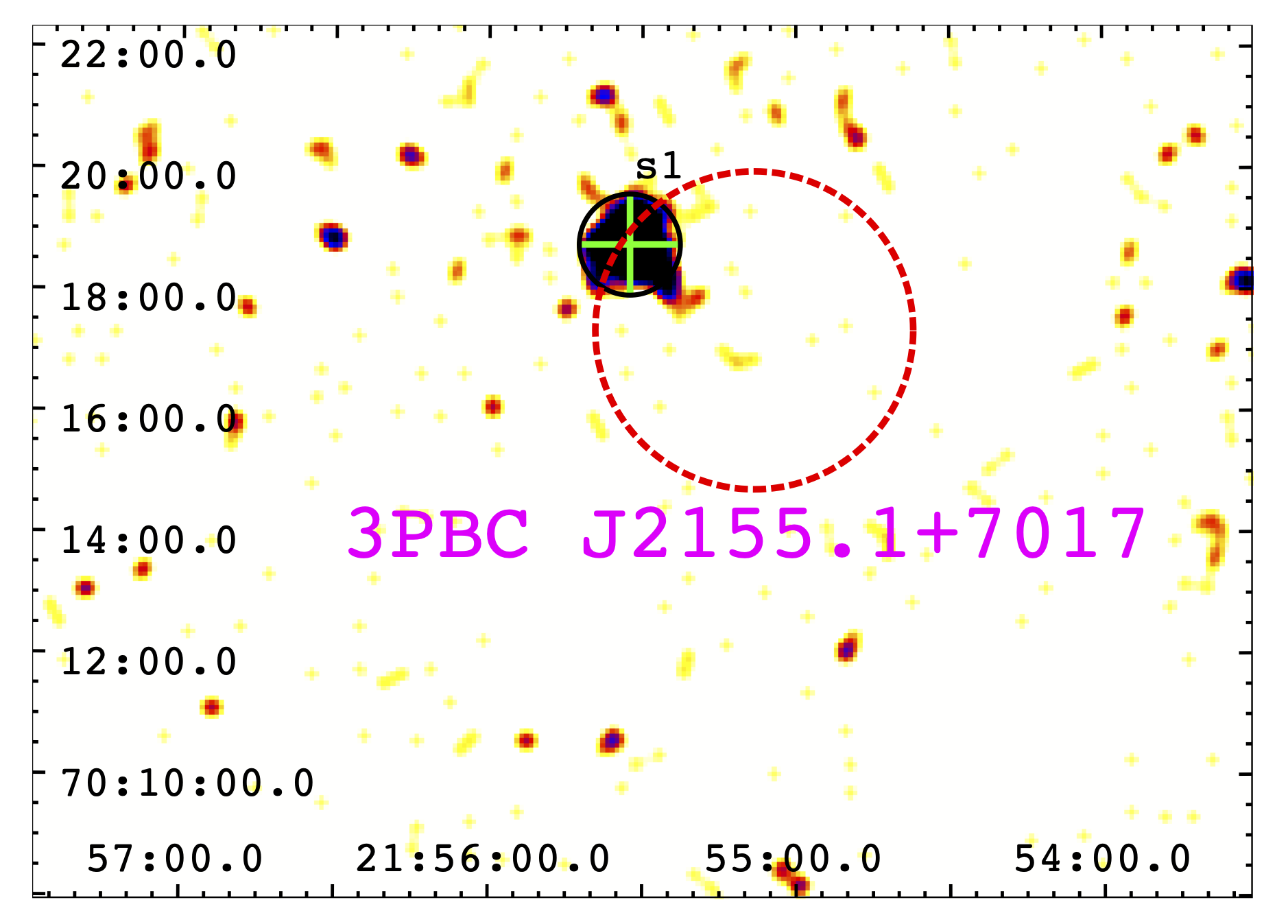}
    \includegraphics[height=4.2cm,width=6cm,angle=0]{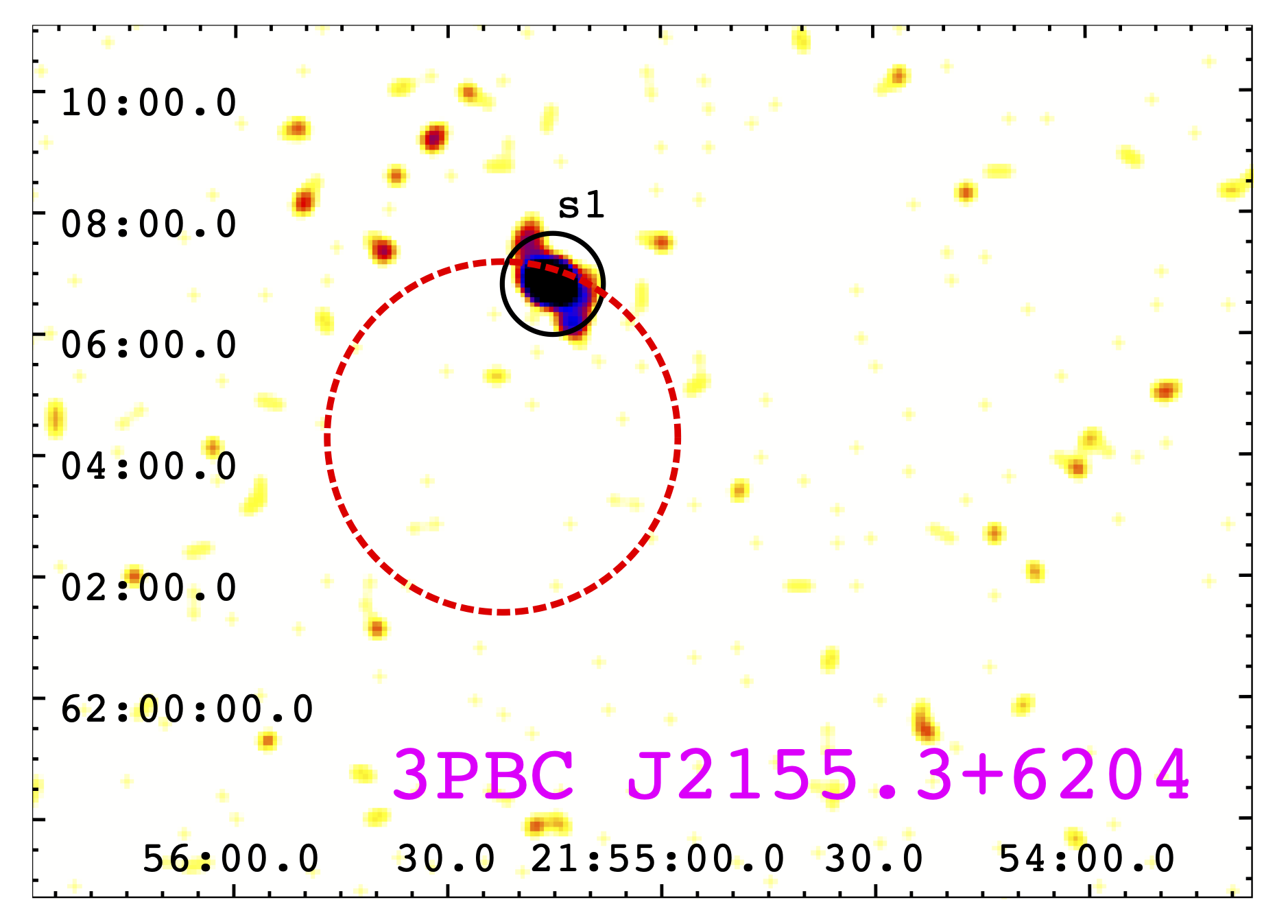}
    \includegraphics[height=4.2cm,width=6cm,angle=0]{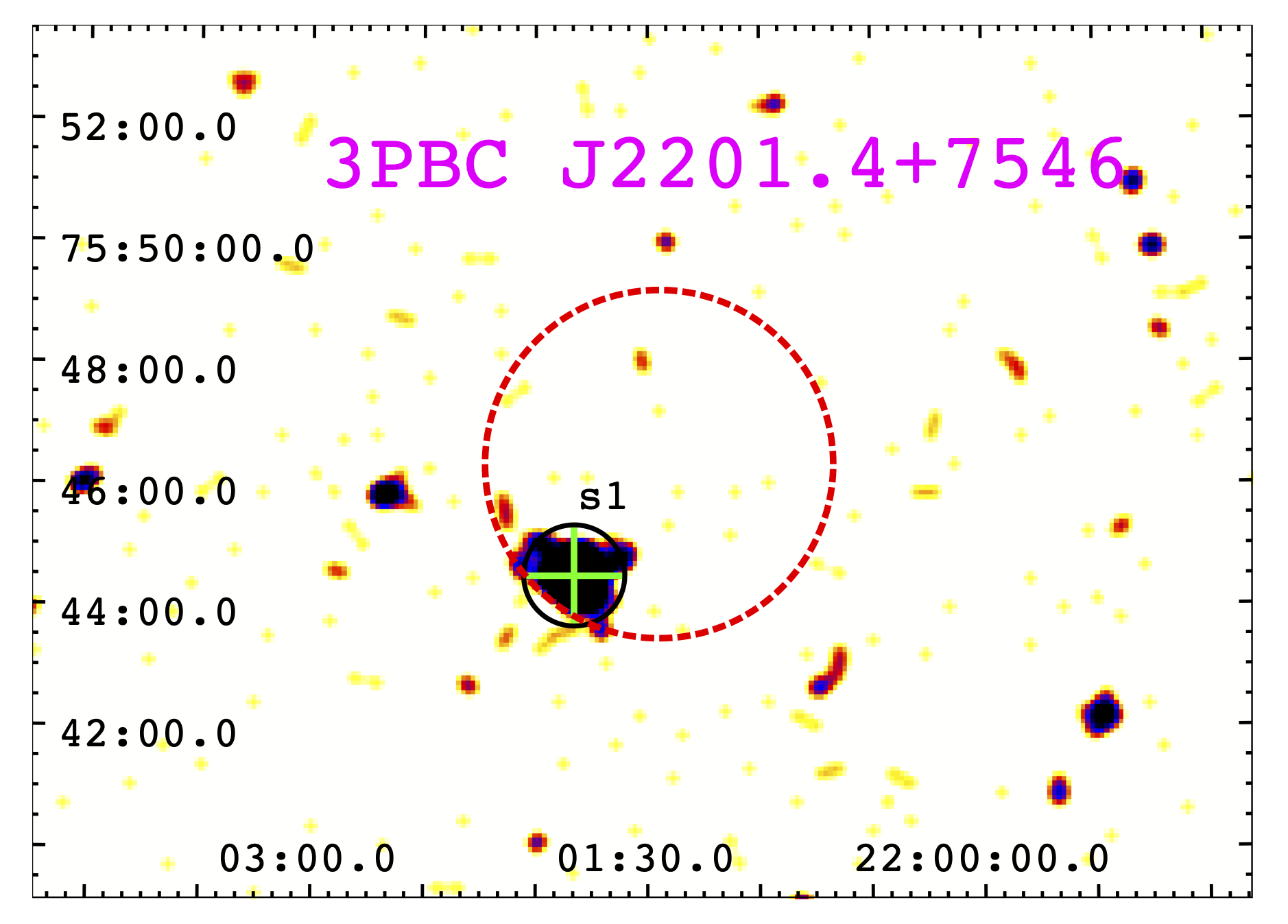}
    \includegraphics[height=4.2cm,width=6cm,angle=0]{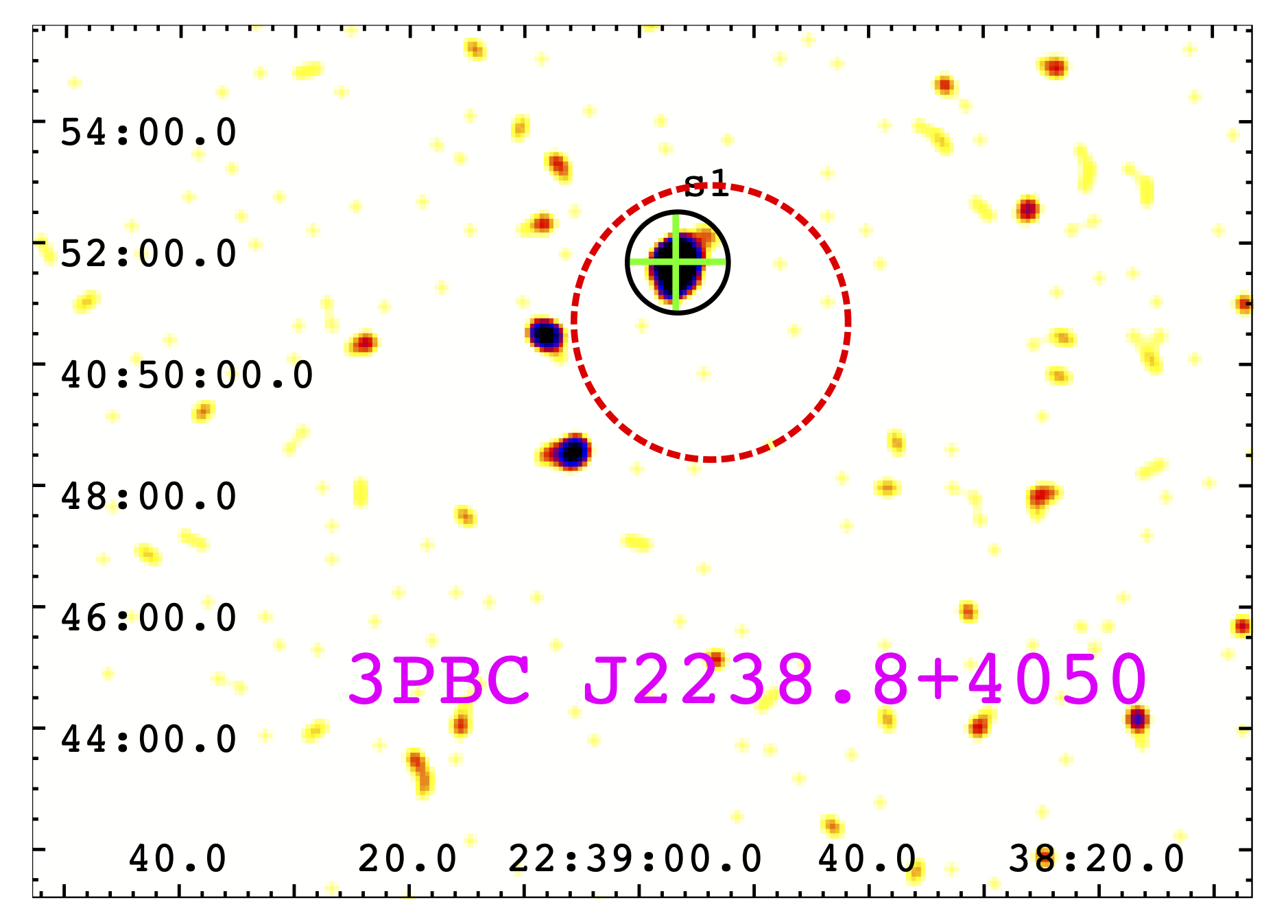}
    
    \caption{Images of 3PBC sources with exactly one soft \textit{Swift}-XRT source (XDF flag \textit{x}) detected inside of the BAT positional uncertainty region (red dashed circle). The soft X-ray detections are indicated with a black circle. The black circle indicates the position. It does not show the positional uncertainty of the source. If the soft X-ray detection is also marked with a green cross, it indicates that it has a WISE counterpart. }
    \label{fig:u_flagged_sources_no4}
\end{center}
\end{figure*}

\end{document}